\documentclass[a4paper]{svjour3}

\usepackage[numbers]{natbib}
\usepackage{times}

\usepackage[includeheadfoot, head=13pt, foot=2pc, top=58pt, bottom=44pt, inner=60pt, outer=60pt, marginparwidth=2pc, heightrounded]{geometry}

\def\makeheadbox{}

\usepackage{mathpazo}
\usepackage[T1]{fontenc}
\usepackage[utf8]{inputenc}
\usepackage[scaled=0.81]{beramono}  %

\usepackage{fontawesome}

\newcommand{\confOK}[1][green]{\text{\color{#1}\faCheck}\xspace}
\newcommand{\confNO}[1][red]{\text{\color{#1}\faClose}\xspace}

\newcommand{\nay}{\confNO}

\usepackage[multiple]{footmisc} %
\usepackage{tabularx}
\usepackage[yyyymmdd]{datetime}

\usepackage{xparse,xstring,xspace}
\usepackage{dcolumn}
\usepackage{caption}

\usepackage[inline]{enumitem}
\setlist[enumerate]{label=\emph{\roman*})}
\setlist[description]{font=\normalfont\bfseries}

\usepackage{ragged2e}

\usepackage{tikz}
\usetikzlibrary{shapes,arrows,positioning,calc,fit,intersections,through}

\definecolor{acol}{RGB}{27,158,119}
\definecolor{bcol}{RGB}{217,95,2}
\definecolor{ccol}{RGB}{117,112,179}
\definecolor{dcol}{RGB}{231,41,138}
\definecolor{ecol}{RGB}{102,166,30}
\definecolor{fcol}{RGB}{230,171,2}
\definecolor{gcol}{RGB}{166,118,29}

\definecolor{cole}{HTML}{e41a1c}
\definecolor{colt}{HTML}{377eb8}

\usepackage{pgfkeys,pgfplots,numprint,relsize}
\xspaceaddexceptions{\%}  %
\def\NAN{??}              %
\def\keyfamily{/flpy/}    %

\usepackage{fmtcount}     %

\newcommand{\examdec}{4}
\newcommand{\examdecjava}{4}
\newcommand{\cddec}{1}
\pgfkeyssetvalue{/flpy/num projects}{4}
\pgfkeyssetvalue{/flpy/statement/einspect/sbfl/keras/5}{12.5}
\pgfkeyssetvalue{/flpy/statement/einspect/sbfl/keras/7}{1.436545}
\pgfkeyssetvalue{/flpy/statement/einspect/pvalue}{1.66433659543306e-02}
\pgfkeyssetvalue{/flpy/project:4}{Keras}

\pgfkeyssetvalue{/flpy/func/avfl/alfa/all/time}{-1}
\pgfkeyssetvalue{/flpy/func/avfl/alfa/all/einspect}{66}
\pgfkeyssetvalue{/flpy/func/avfl/alfa/all/@1}{53}
\pgfkeyssetvalue{/flpy/func/avfl/alfa/all/@3}{71}
\pgfkeyssetvalue{/flpy/func/avfl/alfa/all/@5}{76}
\pgfkeyssetvalue{/flpy/func/avfl/alfa/all/@10}{84}
\pgfkeyssetvalue{/flpy/func/avfl/alfa/all/exam}{0.0720454802458773}
\pgfkeyssetvalue{/flpy/func/avfl/alfa/all/javaexam}{0.013006330013549632}
\pgfkeyssetvalue{/flpy/func/avfl/alfa/all/outputlength}{296.34814814814814}
\pgfkeyssetvalue{/flpy/func/avfl/sbst/all/time}{-1}
\pgfkeyssetvalue{/flpy/func/avfl/sbst/all/einspect}{66}
\pgfkeyssetvalue{/flpy/func/avfl/sbst/all/@1}{44}
\pgfkeyssetvalue{/flpy/func/avfl/sbst/all/@3}{64}
\pgfkeyssetvalue{/flpy/func/avfl/sbst/all/@5}{73}
\pgfkeyssetvalue{/flpy/func/avfl/sbst/all/@10}{79}
\pgfkeyssetvalue{/flpy/func/avfl/sbst/all/exam}{0.07414000002690645}
\pgfkeyssetvalue{/flpy/func/avfl/sbst/all/javaexam}{0.015268411377061086}
\pgfkeyssetvalue{/flpy/func/avfl/sbst/all/outputlength}{296.34814814814814}
\pgfkeyssetvalue{/flpy/func/avfl/favg/all/time}{-1}
\pgfkeyssetvalue{/flpy/func/avfl/favg/all/einspect}{66}
\pgfkeyssetvalue{/flpy/func/avfl/favg/all/@1}{48}
\pgfkeyssetvalue{/flpy/func/avfl/favg/all/@3}{67}
\pgfkeyssetvalue{/flpy/func/avfl/favg/all/@5}{74}
\pgfkeyssetvalue{/flpy/func/avfl/favg/all/@10}{82}
\pgfkeyssetvalue{/flpy/func/avfl/favg/all/exam}{0.07309274013639187}
\pgfkeyssetvalue{/flpy/func/avfl/favg/all/javaexam}{0.014137370695305358}
\pgfkeyssetvalue{/flpy/func/avfl/favg/all/outputlength}{296.34814814814814}
\pgfkeyssetvalue{/flpy/func/sbfl/tarantula/all/time}{589}
\pgfkeyssetvalue{/flpy/func/sbfl/tarantula/all/einspect}{67}
\pgfkeyssetvalue{/flpy/func/sbfl/tarantula/all/@1}{36}
\pgfkeyssetvalue{/flpy/func/sbfl/tarantula/all/@3}{61}
\pgfkeyssetvalue{/flpy/func/sbfl/tarantula/all/@5}{71}
\pgfkeyssetvalue{/flpy/func/sbfl/tarantula/all/@10}{78}
\pgfkeyssetvalue{/flpy/func/sbfl/tarantula/all/exam}{0.07434851496984929}
\pgfkeyssetvalue{/flpy/func/sbfl/tarantula/all/javaexam}{0.015648646275129314}
\pgfkeyssetvalue{/flpy/func/sbfl/tarantula/all/outputlength}{296.31111111111113}
\pgfkeyssetvalue{/flpy/func/sbfl/ochiai/all/time}{589}
\pgfkeyssetvalue{/flpy/func/sbfl/ochiai/all/einspect}{67}
\pgfkeyssetvalue{/flpy/func/sbfl/ochiai/all/@1}{38}
\pgfkeyssetvalue{/flpy/func/sbfl/ochiai/all/@3}{61}
\pgfkeyssetvalue{/flpy/func/sbfl/ochiai/all/@5}{72}
\pgfkeyssetvalue{/flpy/func/sbfl/ochiai/all/@10}{79}
\pgfkeyssetvalue{/flpy/func/sbfl/ochiai/all/exam}{0.07429098916267678}
\pgfkeyssetvalue{/flpy/func/sbfl/ochiai/all/javaexam}{0.015586518403382993}
\pgfkeyssetvalue{/flpy/func/sbfl/ochiai/all/outputlength}{296.31111111111113}
\pgfkeyssetvalue{/flpy/func/sbfl/dstar/all/time}{589}
\pgfkeyssetvalue{/flpy/func/sbfl/dstar/all/einspect}{67}
\pgfkeyssetvalue{/flpy/func/sbfl/dstar/all/@1}{37}
\pgfkeyssetvalue{/flpy/func/sbfl/dstar/all/@3}{61}
\pgfkeyssetvalue{/flpy/func/sbfl/dstar/all/@5}{72}
\pgfkeyssetvalue{/flpy/func/sbfl/dstar/all/@10}{79}
\pgfkeyssetvalue{/flpy/func/sbfl/dstar/all/exam}{0.0742769463179623}
\pgfkeyssetvalue{/flpy/func/sbfl/dstar/all/javaexam}{0.01557135213109137}
\pgfkeyssetvalue{/flpy/func/sbfl/dstar/all/outputlength}{296.31111111111113}
\pgfkeyssetvalue{/flpy/func/mbfl/metallaxis/all/time}{15774}
\pgfkeyssetvalue{/flpy/func/mbfl/metallaxis/all/einspect}{93}
\pgfkeyssetvalue{/flpy/func/mbfl/metallaxis/all/@1}{34}
\pgfkeyssetvalue{/flpy/func/mbfl/metallaxis/all/@3}{56}
\pgfkeyssetvalue{/flpy/func/mbfl/metallaxis/all/@5}{64}
\pgfkeyssetvalue{/flpy/func/mbfl/metallaxis/all/@10}{70}
\pgfkeyssetvalue{/flpy/func/mbfl/metallaxis/all/exam}{0.11440468766666244}
\pgfkeyssetvalue{/flpy/func/mbfl/metallaxis/all/javaexam}{0.013497065802816822}
\pgfkeyssetvalue{/flpy/func/mbfl/metallaxis/all/outputlength}{30.666666666666668}
\pgfkeyssetvalue{/flpy/func/mbfl/muse/all/time}{15774}
\pgfkeyssetvalue{/flpy/func/mbfl/muse/all/einspect}{97}
\pgfkeyssetvalue{/flpy/func/mbfl/muse/all/@1}{27}
\pgfkeyssetvalue{/flpy/func/mbfl/muse/all/@3}{46}
\pgfkeyssetvalue{/flpy/func/mbfl/muse/all/@5}{57}
\pgfkeyssetvalue{/flpy/func/mbfl/muse/all/@10}{64}
\pgfkeyssetvalue{/flpy/func/mbfl/muse/all/exam}{0.11691600104235392}
\pgfkeyssetvalue{/flpy/func/mbfl/muse/all/javaexam}{0.016579132218438226}
\pgfkeyssetvalue{/flpy/func/mbfl/muse/all/outputlength}{30.666666666666668}
\pgfkeyssetvalue{/flpy/func/ps/favg/all/time}{9751}
\pgfkeyssetvalue{/flpy/func/ps/favg/all/einspect}{618}
\pgfkeyssetvalue{/flpy/func/ps/favg/all/@1}{8}
\pgfkeyssetvalue{/flpy/func/ps/favg/all/@3}{13}
\pgfkeyssetvalue{/flpy/func/ps/favg/all/@5}{13}
\pgfkeyssetvalue{/flpy/func/ps/favg/all/@10}{15}
\pgfkeyssetvalue{/flpy/func/ps/favg/all/exam}{0.4047228068308939}
\pgfkeyssetvalue{/flpy/func/ps/favg/all/javaexam}{0.002537295786867146}
\pgfkeyssetvalue{/flpy/func/ps/favg/all/outputlength}{0.562962962962963}
\pgfkeyssetvalue{/flpy/func/st/favg/all/time}{2}
\pgfkeyssetvalue{/flpy/func/st/favg/all/einspect}{451}
\pgfkeyssetvalue{/flpy/func/st/favg/all/@1}{21}
\pgfkeyssetvalue{/flpy/func/st/favg/all/@3}{27}
\pgfkeyssetvalue{/flpy/func/st/favg/all/@5}{27}
\pgfkeyssetvalue{/flpy/func/st/favg/all/@10}{29}
\pgfkeyssetvalue{/flpy/func/st/favg/all/exam}{0.3375483217008306}
\pgfkeyssetvalue{/flpy/func/st/favg/all/javaexam}{0.00448981826856575}
\pgfkeyssetvalue{/flpy/func/st/favg/all/outputlength}{0.9925925925925926}
\pgfkeyssetvalue{/flpy/func/sbfl/favg/all/time}{589}
\pgfkeyssetvalue{/flpy/func/sbfl/favg/all/einspect}{67}
\pgfkeyssetvalue{/flpy/func/sbfl/favg/all/@1}{37}
\pgfkeyssetvalue{/flpy/func/sbfl/favg/all/@3}{61}
\pgfkeyssetvalue{/flpy/func/sbfl/favg/all/@5}{72}
\pgfkeyssetvalue{/flpy/func/sbfl/favg/all/@10}{79}
\pgfkeyssetvalue{/flpy/func/sbfl/favg/all/exam}{0.07430548348349612}
\pgfkeyssetvalue{/flpy/func/sbfl/favg/all/javaexam}{0.015602172269867891}
\pgfkeyssetvalue{/flpy/func/sbfl/favg/all/outputlength}{296.31111111111113}
\pgfkeyssetvalue{/flpy/func/mbfl/favg/all/time}{15774}
\pgfkeyssetvalue{/flpy/func/mbfl/favg/all/einspect}{95}
\pgfkeyssetvalue{/flpy/func/mbfl/favg/all/@1}{31}
\pgfkeyssetvalue{/flpy/func/mbfl/favg/all/@3}{51}
\pgfkeyssetvalue{/flpy/func/mbfl/favg/all/@5}{61}
\pgfkeyssetvalue{/flpy/func/mbfl/favg/all/@10}{67}
\pgfkeyssetvalue{/flpy/func/mbfl/favg/all/exam}{0.11566034435450817}
\pgfkeyssetvalue{/flpy/func/mbfl/favg/all/javaexam}{0.015038099010627524}
\pgfkeyssetvalue{/flpy/func/mbfl/favg/all/outputlength}{30.666666666666668}
\pgfkeyssetvalue{/flpy/stmt/sbfl/tarantula/cli/time}{30}
\pgfkeyssetvalue{/flpy/stmt/sbfl/tarantula/cli/einspect}{2356}
\pgfkeyssetvalue{/flpy/stmt/sbfl/tarantula/cli/@1}{9}
\pgfkeyssetvalue{/flpy/stmt/sbfl/tarantula/cli/@3}{42}
\pgfkeyssetvalue{/flpy/stmt/sbfl/tarantula/cli/@5}{60}
\pgfkeyssetvalue{/flpy/stmt/sbfl/tarantula/cli/@10}{74}
\pgfkeyssetvalue{/flpy/stmt/sbfl/tarantula/cli/exam}{0.05749241780411741}
\pgfkeyssetvalue{/flpy/stmt/sbfl/tarantula/cli/javaexam}{0.005648677113810365}
\pgfkeyssetvalue{/flpy/stmt/sbfl/tarantula/cli/outputlength}{687.1162790697674}
\pgfkeyssetvalue{/flpy/stmt/sbfl/tarantula/cli/distance}{5.1047232499089805}
\pgfkeyssetvalue{/flpy/stmt/sbfl/ochiai/cli/time}{30}
\pgfkeyssetvalue{/flpy/stmt/sbfl/ochiai/cli/einspect}{2356}
\pgfkeyssetvalue{/flpy/stmt/sbfl/ochiai/cli/@1}{9}
\pgfkeyssetvalue{/flpy/stmt/sbfl/ochiai/cli/@3}{42}
\pgfkeyssetvalue{/flpy/stmt/sbfl/ochiai/cli/@5}{60}
\pgfkeyssetvalue{/flpy/stmt/sbfl/ochiai/cli/@10}{74}
\pgfkeyssetvalue{/flpy/stmt/sbfl/ochiai/cli/exam}{0.05745867350106107}
\pgfkeyssetvalue{/flpy/stmt/sbfl/ochiai/cli/javaexam}{0.00561049277087819}
\pgfkeyssetvalue{/flpy/stmt/sbfl/ochiai/cli/outputlength}{687.1162790697674}
\pgfkeyssetvalue{/flpy/stmt/sbfl/ochiai/cli/distance}{5.3186060766626975}
\pgfkeyssetvalue{/flpy/stmt/sbfl/dstar/cli/time}{30}
\pgfkeyssetvalue{/flpy/stmt/sbfl/dstar/cli/einspect}{2356}
\pgfkeyssetvalue{/flpy/stmt/sbfl/dstar/cli/@1}{9}
\pgfkeyssetvalue{/flpy/stmt/sbfl/dstar/cli/@3}{42}
\pgfkeyssetvalue{/flpy/stmt/sbfl/dstar/cli/@5}{60}
\pgfkeyssetvalue{/flpy/stmt/sbfl/dstar/cli/@10}{74}
\pgfkeyssetvalue{/flpy/stmt/sbfl/dstar/cli/exam}{0.05745867350106107}
\pgfkeyssetvalue{/flpy/stmt/sbfl/dstar/cli/javaexam}{0.00561049277087819}
\pgfkeyssetvalue{/flpy/stmt/sbfl/dstar/cli/outputlength}{687.1162790697674}
\pgfkeyssetvalue{/flpy/stmt/sbfl/dstar/cli/distance}{5.3186060766626975}
\pgfkeyssetvalue{/flpy/stmt/mbfl/metallaxis/cli/time}{3770}
\pgfkeyssetvalue{/flpy/stmt/mbfl/metallaxis/cli/einspect}{2909}
\pgfkeyssetvalue{/flpy/stmt/mbfl/metallaxis/cli/@1}{9}
\pgfkeyssetvalue{/flpy/stmt/mbfl/metallaxis/cli/@3}{30}
\pgfkeyssetvalue{/flpy/stmt/mbfl/metallaxis/cli/@5}{35}
\pgfkeyssetvalue{/flpy/stmt/mbfl/metallaxis/cli/@10}{44}
\pgfkeyssetvalue{/flpy/stmt/mbfl/metallaxis/cli/exam}{0.17595006401716312}
\pgfkeyssetvalue{/flpy/stmt/mbfl/metallaxis/cli/javaexam}{0.0031892887239549244}
\pgfkeyssetvalue{/flpy/stmt/mbfl/metallaxis/cli/outputlength}{34.27906976744186}
\pgfkeyssetvalue{/flpy/stmt/mbfl/metallaxis/cli/distance}{22.788008401814235}
\pgfkeyssetvalue{/flpy/stmt/mbfl/muse/cli/time}{3770}
\pgfkeyssetvalue{/flpy/stmt/mbfl/muse/cli/einspect}{2912}
\pgfkeyssetvalue{/flpy/stmt/mbfl/muse/cli/@1}{9}
\pgfkeyssetvalue{/flpy/stmt/mbfl/muse/cli/@3}{35}
\pgfkeyssetvalue{/flpy/stmt/mbfl/muse/cli/@5}{42}
\pgfkeyssetvalue{/flpy/stmt/mbfl/muse/cli/@10}{47}
\pgfkeyssetvalue{/flpy/stmt/mbfl/muse/cli/exam}{0.17599488073180183}
\pgfkeyssetvalue{/flpy/stmt/mbfl/muse/cli/javaexam}{0.003266373473133597}
\pgfkeyssetvalue{/flpy/stmt/mbfl/muse/cli/outputlength}{34.27906976744186}
\pgfkeyssetvalue{/flpy/stmt/mbfl/muse/cli/distance}{5.668161380293332}
\pgfkeyssetvalue{/flpy/stmt/ps/favg/cli/time}{528}
\pgfkeyssetvalue{/flpy/stmt/ps/favg/cli/einspect}{8667}
\pgfkeyssetvalue{/flpy/stmt/ps/favg/cli/@1}{2}
\pgfkeyssetvalue{/flpy/stmt/ps/favg/cli/@3}{5}
\pgfkeyssetvalue{/flpy/stmt/ps/favg/cli/@5}{5}
\pgfkeyssetvalue{/flpy/stmt/ps/favg/cli/@10}{5}
\pgfkeyssetvalue{/flpy/stmt/ps/favg/cli/exam}{0.29720957961531863}
\pgfkeyssetvalue{/flpy/stmt/ps/favg/cli/javaexam}{0.0002078437110832702}
\pgfkeyssetvalue{/flpy/stmt/ps/favg/cli/outputlength}{0.3023255813953488}
\pgfkeyssetvalue{/flpy/stmt/ps/favg/cli/distance}{10.035731386409157}
\pgfkeyssetvalue{/flpy/stmt/st/favg/cli/time}{2}
\pgfkeyssetvalue{/flpy/stmt/st/favg/cli/einspect}{9124}
\pgfkeyssetvalue{/flpy/stmt/st/favg/cli/@1}{0}
\pgfkeyssetvalue{/flpy/stmt/st/favg/cli/@3}{9}
\pgfkeyssetvalue{/flpy/stmt/st/favg/cli/@5}{9}
\pgfkeyssetvalue{/flpy/stmt/st/favg/cli/@10}{14}
\pgfkeyssetvalue{/flpy/stmt/st/favg/cli/exam}{0.2675176621450724}
\pgfkeyssetvalue{/flpy/stmt/st/favg/cli/javaexam}{0.00837604832542256}
\pgfkeyssetvalue{/flpy/stmt/st/favg/cli/outputlength}{19.930232558139537}
\pgfkeyssetvalue{/flpy/stmt/st/favg/cli/distance}{8.503424399789266}
\pgfkeyssetvalue{/flpy/stmt/sbfl/favg/cli/time}{30}
\pgfkeyssetvalue{/flpy/stmt/sbfl/favg/cli/einspect}{2356}
\pgfkeyssetvalue{/flpy/stmt/sbfl/favg/cli/@1}{9}
\pgfkeyssetvalue{/flpy/stmt/sbfl/favg/cli/@3}{42}
\pgfkeyssetvalue{/flpy/stmt/sbfl/favg/cli/@5}{60}
\pgfkeyssetvalue{/flpy/stmt/sbfl/favg/cli/@10}{74}
\pgfkeyssetvalue{/flpy/stmt/sbfl/favg/cli/exam}{0.057469921602079845}
\pgfkeyssetvalue{/flpy/stmt/sbfl/favg/cli/javaexam}{0.005623220885188914}
\pgfkeyssetvalue{/flpy/stmt/sbfl/favg/cli/outputlength}{687.1162790697675}
\pgfkeyssetvalue{/flpy/stmt/sbfl/favg/cli/distance}{5.2473118010781254}
\pgfkeyssetvalue{/flpy/stmt/mbfl/favg/cli/time}{3770}
\pgfkeyssetvalue{/flpy/stmt/mbfl/favg/cli/einspect}{2910}
\pgfkeyssetvalue{/flpy/stmt/mbfl/favg/cli/@1}{9}
\pgfkeyssetvalue{/flpy/stmt/mbfl/favg/cli/@3}{33}
\pgfkeyssetvalue{/flpy/stmt/mbfl/favg/cli/@5}{38}
\pgfkeyssetvalue{/flpy/stmt/mbfl/favg/cli/@10}{45}
\pgfkeyssetvalue{/flpy/stmt/mbfl/favg/cli/exam}{0.1759724723744825}
\pgfkeyssetvalue{/flpy/stmt/mbfl/favg/cli/javaexam}{0.003227831098544261}
\pgfkeyssetvalue{/flpy/stmt/mbfl/favg/cli/outputlength}{34.27906976744186}
\pgfkeyssetvalue{/flpy/stmt/mbfl/favg/cli/distance}{14.228084891053783}
\pgfkeyssetvalue{/flpy/mod/sbfl/tarantula/all/time}{589}
\pgfkeyssetvalue{/flpy/mod/sbfl/tarantula/all/einspect}{2}
\pgfkeyssetvalue{/flpy/mod/sbfl/tarantula/all/@1}{59}
\pgfkeyssetvalue{/flpy/mod/sbfl/tarantula/all/@3}{84}
\pgfkeyssetvalue{/flpy/mod/sbfl/tarantula/all/@5}{91}
\pgfkeyssetvalue{/flpy/mod/sbfl/tarantula/all/@10}{98}
\pgfkeyssetvalue{/flpy/mod/sbfl/tarantula/all/exam}{0.042215430678112266}
\pgfkeyssetvalue{/flpy/mod/sbfl/tarantula/all/javaexam}{0.037501269388558946}
\pgfkeyssetvalue{/flpy/mod/sbfl/tarantula/all/outputlength}{20.918518518518518}
\pgfkeyssetvalue{/flpy/mod/sbfl/ochiai/all/time}{589}
\pgfkeyssetvalue{/flpy/mod/sbfl/ochiai/all/einspect}{2}
\pgfkeyssetvalue{/flpy/mod/sbfl/ochiai/all/@1}{61}
\pgfkeyssetvalue{/flpy/mod/sbfl/ochiai/all/@3}{87}
\pgfkeyssetvalue{/flpy/mod/sbfl/ochiai/all/@5}{93}
\pgfkeyssetvalue{/flpy/mod/sbfl/ochiai/all/@10}{98}
\pgfkeyssetvalue{/flpy/mod/sbfl/ochiai/all/exam}{0.04124790565790988}
\pgfkeyssetvalue{/flpy/mod/sbfl/ochiai/all/javaexam}{0.0365265240323849}
\pgfkeyssetvalue{/flpy/mod/sbfl/ochiai/all/outputlength}{20.918518518518518}
\pgfkeyssetvalue{/flpy/mod/sbfl/dstar/all/time}{589}
\pgfkeyssetvalue{/flpy/mod/sbfl/dstar/all/einspect}{2}
\pgfkeyssetvalue{/flpy/mod/sbfl/dstar/all/@1}{61}
\pgfkeyssetvalue{/flpy/mod/sbfl/dstar/all/@3}{87}
\pgfkeyssetvalue{/flpy/mod/sbfl/dstar/all/@5}{93}
\pgfkeyssetvalue{/flpy/mod/sbfl/dstar/all/@10}{98}
\pgfkeyssetvalue{/flpy/mod/sbfl/dstar/all/exam}{0.04126264935606302}
\pgfkeyssetvalue{/flpy/mod/sbfl/dstar/all/javaexam}{0.03654137775813619}
\pgfkeyssetvalue{/flpy/mod/sbfl/dstar/all/outputlength}{20.918518518518518}
\pgfkeyssetvalue{/flpy/mod/mbfl/metallaxis/all/time}{15774}
\pgfkeyssetvalue{/flpy/mod/mbfl/metallaxis/all/einspect}{6}
\pgfkeyssetvalue{/flpy/mod/mbfl/metallaxis/all/@1}{57}
\pgfkeyssetvalue{/flpy/mod/mbfl/metallaxis/all/@3}{82}
\pgfkeyssetvalue{/flpy/mod/mbfl/metallaxis/all/@5}{87}
\pgfkeyssetvalue{/flpy/mod/mbfl/metallaxis/all/@10}{92}
\pgfkeyssetvalue{/flpy/mod/mbfl/metallaxis/all/exam}{0.07248257070689533}
\pgfkeyssetvalue{/flpy/mod/mbfl/metallaxis/all/javaexam}{0.03660543125938902}
\pgfkeyssetvalue{/flpy/mod/mbfl/metallaxis/all/outputlength}{5.57037037037037}
\pgfkeyssetvalue{/flpy/mod/mbfl/muse/all/time}{15774}
\pgfkeyssetvalue{/flpy/mod/mbfl/muse/all/einspect}{7}
\pgfkeyssetvalue{/flpy/mod/mbfl/muse/all/@1}{47}
\pgfkeyssetvalue{/flpy/mod/mbfl/muse/all/@3}{77}
\pgfkeyssetvalue{/flpy/mod/mbfl/muse/all/@5}{85}
\pgfkeyssetvalue{/flpy/mod/mbfl/muse/all/@10}{87}
\pgfkeyssetvalue{/flpy/mod/mbfl/muse/all/exam}{0.07991533022926521}
\pgfkeyssetvalue{/flpy/mod/mbfl/muse/all/javaexam}{0.04463281154354846}
\pgfkeyssetvalue{/flpy/mod/mbfl/muse/all/outputlength}{5.57037037037037}
\pgfkeyssetvalue{/flpy/mod/ps/favg/all/time}{9751}
\pgfkeyssetvalue{/flpy/mod/ps/favg/all/einspect}{67}
\pgfkeyssetvalue{/flpy/mod/ps/favg/all/@1}{13}
\pgfkeyssetvalue{/flpy/mod/ps/favg/all/@3}{17}
\pgfkeyssetvalue{/flpy/mod/ps/favg/all/@5}{21}
\pgfkeyssetvalue{/flpy/mod/ps/favg/all/@10}{28}
\pgfkeyssetvalue{/flpy/mod/ps/favg/all/exam}{0.41634670834970977}
\pgfkeyssetvalue{/flpy/mod/ps/favg/all/javaexam}{0.023441892101774497}
\pgfkeyssetvalue{/flpy/mod/ps/favg/all/outputlength}{0.3925925925925926}
\pgfkeyssetvalue{/flpy/mod/st/favg/all/time}{2}
\pgfkeyssetvalue{/flpy/mod/st/favg/all/einspect}{61}
\pgfkeyssetvalue{/flpy/mod/st/favg/all/@1}{29}
\pgfkeyssetvalue{/flpy/mod/st/favg/all/@3}{33}
\pgfkeyssetvalue{/flpy/mod/st/favg/all/@5}{36}
\pgfkeyssetvalue{/flpy/mod/st/favg/all/@10}{41}
\pgfkeyssetvalue{/flpy/mod/st/favg/all/exam}{0.33632876193476285}
\pgfkeyssetvalue{/flpy/mod/st/favg/all/javaexam}{0.02838005830515854}
\pgfkeyssetvalue{/flpy/mod/st/favg/all/outputlength}{0.5851851851851851}
\pgfkeyssetvalue{/flpy/mod/sbfl/favg/all/time}{589}
\pgfkeyssetvalue{/flpy/mod/sbfl/favg/all/einspect}{2}
\pgfkeyssetvalue{/flpy/mod/sbfl/favg/all/@1}{60}
\pgfkeyssetvalue{/flpy/mod/sbfl/favg/all/@3}{86}
\pgfkeyssetvalue{/flpy/mod/sbfl/favg/all/@5}{92}
\pgfkeyssetvalue{/flpy/mod/sbfl/favg/all/@10}{98}
\pgfkeyssetvalue{/flpy/mod/sbfl/favg/all/exam}{0.041575328564028395}
\pgfkeyssetvalue{/flpy/mod/sbfl/favg/all/javaexam}{0.036856390393026676}
\pgfkeyssetvalue{/flpy/mod/sbfl/favg/all/outputlength}{20.918518518518518}
\pgfkeyssetvalue{/flpy/mod/mbfl/favg/all/time}{15774}
\pgfkeyssetvalue{/flpy/mod/mbfl/favg/all/einspect}{6}
\pgfkeyssetvalue{/flpy/mod/mbfl/favg/all/@1}{52}
\pgfkeyssetvalue{/flpy/mod/mbfl/favg/all/@3}{80}
\pgfkeyssetvalue{/flpy/mod/mbfl/favg/all/@5}{86}
\pgfkeyssetvalue{/flpy/mod/mbfl/favg/all/@10}{90}
\pgfkeyssetvalue{/flpy/mod/mbfl/favg/all/exam}{0.07619895046808027}
\pgfkeyssetvalue{/flpy/mod/mbfl/favg/all/javaexam}{0.04061912140146874}
\pgfkeyssetvalue{/flpy/mod/mbfl/favg/all/outputlength}{5.57037037037037}
\pgfkeyssetvalue{/flpy/stmt/sbfl/tarantula/pred/time}{1284}
\pgfkeyssetvalue{/flpy/stmt/sbfl/tarantula/pred/einspect}{377}
\pgfkeyssetvalue{/flpy/stmt/sbfl/tarantula/pred/@1}{12}
\pgfkeyssetvalue{/flpy/stmt/sbfl/tarantula/pred/@3}{23}
\pgfkeyssetvalue{/flpy/stmt/sbfl/tarantula/pred/@5}{38}
\pgfkeyssetvalue{/flpy/stmt/sbfl/tarantula/pred/@10}{50}
\pgfkeyssetvalue{/flpy/stmt/sbfl/tarantula/pred/exam}{0.015396784507531578}
\pgfkeyssetvalue{/flpy/stmt/sbfl/tarantula/pred/javaexam}{0.006521340740361081}
\pgfkeyssetvalue{/flpy/stmt/sbfl/tarantula/pred/outputlength}{3041.5384615384614}
\pgfkeyssetvalue{/flpy/stmt/sbfl/tarantula/pred/distance}{12.67360960552377}
\pgfkeyssetvalue{/flpy/stmt/sbfl/ochiai/pred/time}{1284}
\pgfkeyssetvalue{/flpy/stmt/sbfl/ochiai/pred/einspect}{372}
\pgfkeyssetvalue{/flpy/stmt/sbfl/ochiai/pred/@1}{12}
\pgfkeyssetvalue{/flpy/stmt/sbfl/ochiai/pred/@3}{23}
\pgfkeyssetvalue{/flpy/stmt/sbfl/ochiai/pred/@5}{38}
\pgfkeyssetvalue{/flpy/stmt/sbfl/ochiai/pred/@10}{50}
\pgfkeyssetvalue{/flpy/stmt/sbfl/ochiai/pred/exam}{0.015360611475504677}
\pgfkeyssetvalue{/flpy/stmt/sbfl/ochiai/pred/javaexam}{0.006483720787053104}
\pgfkeyssetvalue{/flpy/stmt/sbfl/ochiai/pred/outputlength}{3041.5384615384614}
\pgfkeyssetvalue{/flpy/stmt/sbfl/ochiai/pred/distance}{13.207893594627036}
\pgfkeyssetvalue{/flpy/stmt/sbfl/dstar/pred/time}{1284}
\pgfkeyssetvalue{/flpy/stmt/sbfl/dstar/pred/einspect}{371}
\pgfkeyssetvalue{/flpy/stmt/sbfl/dstar/pred/@1}{12}
\pgfkeyssetvalue{/flpy/stmt/sbfl/dstar/pred/@3}{23}
\pgfkeyssetvalue{/flpy/stmt/sbfl/dstar/pred/@5}{38}
\pgfkeyssetvalue{/flpy/stmt/sbfl/dstar/pred/@10}{50}
\pgfkeyssetvalue{/flpy/stmt/sbfl/dstar/pred/exam}{0.015352218336156572}
\pgfkeyssetvalue{/flpy/stmt/sbfl/dstar/pred/javaexam}{0.006474991922131078}
\pgfkeyssetvalue{/flpy/stmt/sbfl/dstar/pred/outputlength}{3041.5384615384614}
\pgfkeyssetvalue{/flpy/stmt/sbfl/dstar/pred/distance}{13.583468767178802}
\pgfkeyssetvalue{/flpy/stmt/mbfl/metallaxis/pred/time}{19671}
\pgfkeyssetvalue{/flpy/stmt/mbfl/metallaxis/pred/einspect}{1882}
\pgfkeyssetvalue{/flpy/stmt/mbfl/metallaxis/pred/@1}{12}
\pgfkeyssetvalue{/flpy/stmt/mbfl/metallaxis/pred/@3}{37}
\pgfkeyssetvalue{/flpy/stmt/mbfl/metallaxis/pred/@5}{44}
\pgfkeyssetvalue{/flpy/stmt/mbfl/metallaxis/pred/@10}{56}
\pgfkeyssetvalue{/flpy/stmt/mbfl/metallaxis/pred/exam}{0.028250016562632544}
\pgfkeyssetvalue{/flpy/stmt/mbfl/metallaxis/pred/javaexam}{0.003773762189448894}
\pgfkeyssetvalue{/flpy/stmt/mbfl/metallaxis/pred/outputlength}{146.46153846153845}
\pgfkeyssetvalue{/flpy/stmt/mbfl/metallaxis/pred/distance}{26.362795517508136}
\pgfkeyssetvalue{/flpy/stmt/mbfl/muse/pred/time}{19671}
\pgfkeyssetvalue{/flpy/stmt/mbfl/muse/pred/einspect}{1899}
\pgfkeyssetvalue{/flpy/stmt/mbfl/muse/pred/@1}{10}
\pgfkeyssetvalue{/flpy/stmt/mbfl/muse/pred/@3}{29}
\pgfkeyssetvalue{/flpy/stmt/mbfl/muse/pred/@5}{37}
\pgfkeyssetvalue{/flpy/stmt/mbfl/muse/pred/@10}{48}
\pgfkeyssetvalue{/flpy/stmt/mbfl/muse/pred/exam}{0.02715994157640326}
\pgfkeyssetvalue{/flpy/stmt/mbfl/muse/pred/javaexam}{0.0025141199831394965}
\pgfkeyssetvalue{/flpy/stmt/mbfl/muse/pred/outputlength}{146.46153846153845}
\pgfkeyssetvalue{/flpy/stmt/mbfl/muse/pred/distance}{32.61515162106886}
\pgfkeyssetvalue{/flpy/stmt/ps/favg/pred/time}{17287}
\pgfkeyssetvalue{/flpy/stmt/ps/favg/pred/einspect}{8425}
\pgfkeyssetvalue{/flpy/stmt/ps/favg/pred/@1}{8}
\pgfkeyssetvalue{/flpy/stmt/ps/favg/pred/@3}{13}
\pgfkeyssetvalue{/flpy/stmt/ps/favg/pred/@5}{17}
\pgfkeyssetvalue{/flpy/stmt/ps/favg/pred/@10}{17}
\pgfkeyssetvalue{/flpy/stmt/ps/favg/pred/exam}{0.14929941753331805}
\pgfkeyssetvalue{/flpy/stmt/ps/favg/pred/javaexam}{0.00010556862466765817}
\pgfkeyssetvalue{/flpy/stmt/ps/favg/pred/outputlength}{1.25}
\pgfkeyssetvalue{/flpy/stmt/ps/favg/pred/distance}{12.335695507867522}
\pgfkeyssetvalue{/flpy/stmt/st/favg/pred/time}{2}
\pgfkeyssetvalue{/flpy/stmt/st/favg/pred/einspect}{9194}
\pgfkeyssetvalue{/flpy/stmt/st/favg/pred/@1}{0}
\pgfkeyssetvalue{/flpy/stmt/st/favg/pred/@3}{2}
\pgfkeyssetvalue{/flpy/stmt/st/favg/pred/@5}{6}
\pgfkeyssetvalue{/flpy/stmt/st/favg/pred/@10}{17}
\pgfkeyssetvalue{/flpy/stmt/st/favg/pred/exam}{0.15410411274276103}
\pgfkeyssetvalue{/flpy/stmt/st/favg/pred/javaexam}{0.0007208885632961376}
\pgfkeyssetvalue{/flpy/stmt/st/favg/pred/outputlength}{47.15384615384615}
\pgfkeyssetvalue{/flpy/stmt/st/favg/pred/distance}{7.438756894266723}
\pgfkeyssetvalue{/flpy/stmt/sbfl/favg/pred/time}{1284}
\pgfkeyssetvalue{/flpy/stmt/sbfl/favg/pred/einspect}{374}
\pgfkeyssetvalue{/flpy/stmt/sbfl/favg/pred/@1}{12}
\pgfkeyssetvalue{/flpy/stmt/sbfl/favg/pred/@3}{23}
\pgfkeyssetvalue{/flpy/stmt/sbfl/favg/pred/@5}{38}
\pgfkeyssetvalue{/flpy/stmt/sbfl/favg/pred/@10}{50}
\pgfkeyssetvalue{/flpy/stmt/sbfl/favg/pred/exam}{0.015369871439730945}
\pgfkeyssetvalue{/flpy/stmt/sbfl/favg/pred/javaexam}{0.006493351149848421}
\pgfkeyssetvalue{/flpy/stmt/sbfl/favg/pred/outputlength}{3041.5384615384614}
\pgfkeyssetvalue{/flpy/stmt/sbfl/favg/pred/distance}{13.154990655776537}
\pgfkeyssetvalue{/flpy/stmt/mbfl/favg/pred/time}{19671}
\pgfkeyssetvalue{/flpy/stmt/mbfl/favg/pred/einspect}{1891}
\pgfkeyssetvalue{/flpy/stmt/mbfl/favg/pred/@1}{11}
\pgfkeyssetvalue{/flpy/stmt/mbfl/favg/pred/@3}{33}
\pgfkeyssetvalue{/flpy/stmt/mbfl/favg/pred/@5}{40}
\pgfkeyssetvalue{/flpy/stmt/mbfl/favg/pred/@10}{52}
\pgfkeyssetvalue{/flpy/stmt/mbfl/favg/pred/exam}{0.027704979069517903}
\pgfkeyssetvalue{/flpy/stmt/mbfl/favg/pred/javaexam}{0.0031439410862941952}
\pgfkeyssetvalue{/flpy/stmt/mbfl/favg/pred/outputlength}{146.46153846153845}
\pgfkeyssetvalue{/flpy/stmt/mbfl/favg/pred/distance}{29.488973569288497}
\pgfkeyssetvalue{/flpy/stmt/avfl/alfa/all/time}{-1}
\pgfkeyssetvalue{/flpy/stmt/avfl/alfa/all/einspect}{1575}
\pgfkeyssetvalue{/flpy/stmt/avfl/alfa/all/@1}{18}
\pgfkeyssetvalue{/flpy/stmt/avfl/alfa/all/@3}{36}
\pgfkeyssetvalue{/flpy/stmt/avfl/alfa/all/@5}{47}
\pgfkeyssetvalue{/flpy/stmt/avfl/alfa/all/@10}{59}
\pgfkeyssetvalue{/flpy/stmt/avfl/alfa/all/exam}{0.05128261837553908}
\pgfkeyssetvalue{/flpy/stmt/avfl/alfa/all/javaexam}{0.003302528846968437}
\pgfkeyssetvalue{/flpy/stmt/avfl/alfa/all/outputlength}{2548.362962962963}
\pgfkeyssetvalue{/flpy/stmt/avfl/sbst/all/time}{-1}
\pgfkeyssetvalue{/flpy/stmt/avfl/sbst/all/einspect}{1585}
\pgfkeyssetvalue{/flpy/stmt/avfl/sbst/all/@1}{12}
\pgfkeyssetvalue{/flpy/stmt/avfl/sbst/all/@3}{33}
\pgfkeyssetvalue{/flpy/stmt/avfl/sbst/all/@5}{44}
\pgfkeyssetvalue{/flpy/stmt/avfl/sbst/all/@10}{56}
\pgfkeyssetvalue{/flpy/stmt/avfl/sbst/all/exam}{0.051863927447538566}
\pgfkeyssetvalue{/flpy/stmt/avfl/sbst/all/javaexam}{0.0039583916263026715}
\pgfkeyssetvalue{/flpy/stmt/avfl/sbst/all/outputlength}{2548.3333333333335}
\pgfkeyssetvalue{/flpy/stmt/avfl/favg/all/time}{-1}
\pgfkeyssetvalue{/flpy/stmt/avfl/favg/all/einspect}{1580}
\pgfkeyssetvalue{/flpy/stmt/avfl/favg/all/@1}{15}
\pgfkeyssetvalue{/flpy/stmt/avfl/favg/all/@3}{35}
\pgfkeyssetvalue{/flpy/stmt/avfl/favg/all/@5}{46}
\pgfkeyssetvalue{/flpy/stmt/avfl/favg/all/@10}{57}
\pgfkeyssetvalue{/flpy/stmt/avfl/favg/all/exam}{0.05157327291153882}
\pgfkeyssetvalue{/flpy/stmt/avfl/favg/all/javaexam}{0.0036304602366355543}
\pgfkeyssetvalue{/flpy/stmt/avfl/favg/all/outputlength}{2548.3481481481485}
\pgfkeyssetvalue{/flpy/stmt/sbfl/tarantula/ds/time}{1726}
\pgfkeyssetvalue{/flpy/stmt/sbfl/tarantula/ds/einspect}{833}
\pgfkeyssetvalue{/flpy/stmt/sbfl/tarantula/ds/@1}{7}
\pgfkeyssetvalue{/flpy/stmt/sbfl/tarantula/ds/@3}{24}
\pgfkeyssetvalue{/flpy/stmt/sbfl/tarantula/ds/@5}{31}
\pgfkeyssetvalue{/flpy/stmt/sbfl/tarantula/ds/@10}{43}
\pgfkeyssetvalue{/flpy/stmt/sbfl/tarantula/ds/exam}{0.006802047086120662}
\pgfkeyssetvalue{/flpy/stmt/sbfl/tarantula/ds/javaexam}{0.001852224252099376}
\pgfkeyssetvalue{/flpy/stmt/sbfl/tarantula/ds/outputlength}{5775.357142857143}
\pgfkeyssetvalue{/flpy/stmt/sbfl/tarantula/ds/distance}{15.29586728769396}
\pgfkeyssetvalue{/flpy/stmt/sbfl/ochiai/ds/time}{1726}
\pgfkeyssetvalue{/flpy/stmt/sbfl/ochiai/ds/einspect}{825}
\pgfkeyssetvalue{/flpy/stmt/sbfl/ochiai/ds/@1}{7}
\pgfkeyssetvalue{/flpy/stmt/sbfl/ochiai/ds/@3}{24}
\pgfkeyssetvalue{/flpy/stmt/sbfl/ochiai/ds/@5}{31}
\pgfkeyssetvalue{/flpy/stmt/sbfl/ochiai/ds/@10}{43}
\pgfkeyssetvalue{/flpy/stmt/sbfl/ochiai/ds/exam}{0.006733464769448151}
\pgfkeyssetvalue{/flpy/stmt/sbfl/ochiai/ds/javaexam}{0.0017819691959958285}
\pgfkeyssetvalue{/flpy/stmt/sbfl/ochiai/ds/outputlength}{5775.357142857143}
\pgfkeyssetvalue{/flpy/stmt/sbfl/ochiai/ds/distance}{16.028889219116973}
\pgfkeyssetvalue{/flpy/stmt/sbfl/dstar/ds/time}{1726}
\pgfkeyssetvalue{/flpy/stmt/sbfl/dstar/ds/einspect}{824}
\pgfkeyssetvalue{/flpy/stmt/sbfl/dstar/ds/@1}{5}
\pgfkeyssetvalue{/flpy/stmt/sbfl/dstar/ds/@3}{21}
\pgfkeyssetvalue{/flpy/stmt/sbfl/dstar/ds/@5}{29}
\pgfkeyssetvalue{/flpy/stmt/sbfl/dstar/ds/@10}{43}
\pgfkeyssetvalue{/flpy/stmt/sbfl/dstar/ds/exam}{0.006728664395308205}
\pgfkeyssetvalue{/flpy/stmt/sbfl/dstar/ds/javaexam}{0.0017770517395597854}
\pgfkeyssetvalue{/flpy/stmt/sbfl/dstar/ds/outputlength}{5775.357142857143}
\pgfkeyssetvalue{/flpy/stmt/sbfl/dstar/ds/distance}{16.931973281172787}
\pgfkeyssetvalue{/flpy/stmt/mbfl/metallaxis/ds/time}{29799}
\pgfkeyssetvalue{/flpy/stmt/mbfl/metallaxis/ds/einspect}{14507}
\pgfkeyssetvalue{/flpy/stmt/mbfl/metallaxis/ds/@1}{7}
\pgfkeyssetvalue{/flpy/stmt/mbfl/metallaxis/ds/@3}{17}
\pgfkeyssetvalue{/flpy/stmt/mbfl/metallaxis/ds/@5}{21}
\pgfkeyssetvalue{/flpy/stmt/mbfl/metallaxis/ds/@10}{26}
\pgfkeyssetvalue{/flpy/stmt/mbfl/metallaxis/ds/exam}{0.1721599625176657}
\pgfkeyssetvalue{/flpy/stmt/mbfl/metallaxis/ds/javaexam}{0.0002519834799297213}
\pgfkeyssetvalue{/flpy/stmt/mbfl/metallaxis/ds/outputlength}{169.28571428571428}
\pgfkeyssetvalue{/flpy/stmt/mbfl/metallaxis/ds/distance}{37.81426957020117}
\pgfkeyssetvalue{/flpy/stmt/mbfl/muse/ds/time}{29799}
\pgfkeyssetvalue{/flpy/stmt/mbfl/muse/ds/einspect}{14530}
\pgfkeyssetvalue{/flpy/stmt/mbfl/muse/ds/@1}{0}
\pgfkeyssetvalue{/flpy/stmt/mbfl/muse/ds/@3}{7}
\pgfkeyssetvalue{/flpy/stmt/mbfl/muse/ds/@5}{17}
\pgfkeyssetvalue{/flpy/stmt/mbfl/muse/ds/@10}{21}
\pgfkeyssetvalue{/flpy/stmt/mbfl/muse/ds/exam}{0.17247685694574405}
\pgfkeyssetvalue{/flpy/stmt/mbfl/muse/ds/javaexam}{0.0008857723360864847}
\pgfkeyssetvalue{/flpy/stmt/mbfl/muse/ds/outputlength}{169.28571428571428}
\pgfkeyssetvalue{/flpy/stmt/mbfl/muse/ds/distance}{50.54031784669837}
\pgfkeyssetvalue{/flpy/stmt/ps/favg/ds/time}{15972}
\pgfkeyssetvalue{/flpy/stmt/ps/favg/ds/einspect}{22847}
\pgfkeyssetvalue{/flpy/stmt/ps/favg/ds/@1}{2}
\pgfkeyssetvalue{/flpy/stmt/ps/favg/ds/@3}{5}
\pgfkeyssetvalue{/flpy/stmt/ps/favg/ds/@5}{7}
\pgfkeyssetvalue{/flpy/stmt/ps/favg/ds/@10}{7}
\pgfkeyssetvalue{/flpy/stmt/ps/favg/ds/exam}{0.27861196417079476}
\pgfkeyssetvalue{/flpy/stmt/ps/favg/ds/javaexam}{2.3971379930046436e-05}
\pgfkeyssetvalue{/flpy/stmt/ps/favg/ds/outputlength}{1.0}
\pgfkeyssetvalue{/flpy/stmt/ps/favg/ds/distance}{18.320865667796674}
\pgfkeyssetvalue{/flpy/stmt/st/favg/ds/time}{2}
\pgfkeyssetvalue{/flpy/stmt/st/favg/ds/einspect}{15174}
\pgfkeyssetvalue{/flpy/stmt/st/favg/ds/@1}{0}
\pgfkeyssetvalue{/flpy/stmt/st/favg/ds/@3}{0}
\pgfkeyssetvalue{/flpy/stmt/st/favg/ds/@5}{0}
\pgfkeyssetvalue{/flpy/stmt/st/favg/ds/@10}{12}
\pgfkeyssetvalue{/flpy/stmt/st/favg/ds/exam}{0.17226288044301047}
\pgfkeyssetvalue{/flpy/stmt/st/favg/ds/javaexam}{0.0003469045961527693}
\pgfkeyssetvalue{/flpy/stmt/st/favg/ds/outputlength}{97.69047619047619}
\pgfkeyssetvalue{/flpy/stmt/st/favg/ds/distance}{8.173656907497048}
\pgfkeyssetvalue{/flpy/stmt/sbfl/favg/ds/time}{1726}
\pgfkeyssetvalue{/flpy/stmt/sbfl/favg/ds/einspect}{827}
\pgfkeyssetvalue{/flpy/stmt/sbfl/favg/ds/@1}{6}
\pgfkeyssetvalue{/flpy/stmt/sbfl/favg/ds/@3}{23}
\pgfkeyssetvalue{/flpy/stmt/sbfl/favg/ds/@5}{30}
\pgfkeyssetvalue{/flpy/stmt/sbfl/favg/ds/@10}{43}
\pgfkeyssetvalue{/flpy/stmt/sbfl/favg/ds/exam}{0.006754725416959006}
\pgfkeyssetvalue{/flpy/stmt/sbfl/favg/ds/javaexam}{0.0018037483958849966}
\pgfkeyssetvalue{/flpy/stmt/sbfl/favg/ds/outputlength}{5775.357142857142}
\pgfkeyssetvalue{/flpy/stmt/sbfl/favg/ds/distance}{16.085576595994574}
\pgfkeyssetvalue{/flpy/stmt/mbfl/favg/ds/time}{29799}
\pgfkeyssetvalue{/flpy/stmt/mbfl/favg/ds/einspect}{14519}
\pgfkeyssetvalue{/flpy/stmt/mbfl/favg/ds/@1}{4}
\pgfkeyssetvalue{/flpy/stmt/mbfl/favg/ds/@3}{12}
\pgfkeyssetvalue{/flpy/stmt/mbfl/favg/ds/@5}{19}
\pgfkeyssetvalue{/flpy/stmt/mbfl/favg/ds/@10}{24}
\pgfkeyssetvalue{/flpy/stmt/mbfl/favg/ds/exam}{0.17231840973170487}
\pgfkeyssetvalue{/flpy/stmt/mbfl/favg/ds/javaexam}{0.0005688779080081029}
\pgfkeyssetvalue{/flpy/stmt/mbfl/favg/ds/outputlength}{169.28571428571428}
\pgfkeyssetvalue{/flpy/stmt/mbfl/favg/ds/distance}{44.17729370844977}
\pgfkeyssetvalue{/flpy/stmt/sbfl/tarantula/all/time}{589}
\pgfkeyssetvalue{/flpy/stmt/sbfl/tarantula/all/einspect}{1586}
\pgfkeyssetvalue{/flpy/stmt/sbfl/tarantula/all/@1}{12}
\pgfkeyssetvalue{/flpy/stmt/sbfl/tarantula/all/@3}{30}
\pgfkeyssetvalue{/flpy/stmt/sbfl/tarantula/all/@5}{43}
\pgfkeyssetvalue{/flpy/stmt/sbfl/tarantula/all/@10}{54}
\pgfkeyssetvalue{/flpy/stmt/sbfl/tarantula/all/exam}{0.051662620778257194}
\pgfkeyssetvalue{/flpy/stmt/sbfl/tarantula/all/javaexam}{0.004189992725541162}
\pgfkeyssetvalue{/flpy/stmt/sbfl/tarantula/all/outputlength}{2521.348148148148}
\pgfkeyssetvalue{/flpy/stmt/sbfl/tarantula/all/distance}{10.310770590103361}
\pgfkeyssetvalue{/flpy/stmt/sbfl/ochiai/all/time}{589}
\pgfkeyssetvalue{/flpy/stmt/sbfl/ochiai/all/einspect}{1583}
\pgfkeyssetvalue{/flpy/stmt/sbfl/ochiai/all/@1}{12}
\pgfkeyssetvalue{/flpy/stmt/sbfl/ochiai/all/@3}{30}
\pgfkeyssetvalue{/flpy/stmt/sbfl/ochiai/all/@5}{43}
\pgfkeyssetvalue{/flpy/stmt/sbfl/ochiai/all/@10}{54}
\pgfkeyssetvalue{/flpy/stmt/sbfl/ochiai/all/exam}{0.05163232006860066}
\pgfkeyssetvalue{/flpy/stmt/sbfl/ochiai/all/javaexam}{0.004155904427177535}
\pgfkeyssetvalue{/flpy/stmt/sbfl/ochiai/all/outputlength}{2521.348148148148}
\pgfkeyssetvalue{/flpy/stmt/sbfl/ochiai/all/distance}{10.616494823874216}
\pgfkeyssetvalue{/flpy/stmt/sbfl/dstar/all/time}{589}
\pgfkeyssetvalue{/flpy/stmt/sbfl/dstar/all/einspect}{1583}
\pgfkeyssetvalue{/flpy/stmt/sbfl/dstar/all/@1}{11}
\pgfkeyssetvalue{/flpy/stmt/sbfl/dstar/all/@3}{30}
\pgfkeyssetvalue{/flpy/stmt/sbfl/dstar/all/@5}{42}
\pgfkeyssetvalue{/flpy/stmt/sbfl/dstar/all/@10}{54}
\pgfkeyssetvalue{/flpy/stmt/sbfl/dstar/all/exam}{0.051630826618868225}
\pgfkeyssetvalue{/flpy/stmt/sbfl/dstar/all/javaexam}{0.0041542242962285545}
\pgfkeyssetvalue{/flpy/stmt/sbfl/dstar/all/outputlength}{2521.348148148148}
\pgfkeyssetvalue{/flpy/stmt/sbfl/dstar/all/distance}{10.897454309847138}
\pgfkeyssetvalue{/flpy/stmt/mbfl/metallaxis/all/time}{15774}
\pgfkeyssetvalue{/flpy/stmt/mbfl/metallaxis/all/einspect}{6706}
\pgfkeyssetvalue{/flpy/stmt/mbfl/metallaxis/all/@1}{10}
\pgfkeyssetvalue{/flpy/stmt/mbfl/metallaxis/all/@3}{25}
\pgfkeyssetvalue{/flpy/stmt/mbfl/metallaxis/all/@5}{30}
\pgfkeyssetvalue{/flpy/stmt/mbfl/metallaxis/all/@10}{37}
\pgfkeyssetvalue{/flpy/stmt/mbfl/metallaxis/all/exam}{0.17256415741554368}
\pgfkeyssetvalue{/flpy/stmt/mbfl/metallaxis/all/javaexam}{0.003470449924030685}
\pgfkeyssetvalue{/flpy/stmt/mbfl/metallaxis/all/outputlength}{113.8962962962963}
\pgfkeyssetvalue{/flpy/stmt/mbfl/metallaxis/all/distance}{32.12070167204071}
\pgfkeyssetvalue{/flpy/stmt/mbfl/muse/all/time}{15774}
\pgfkeyssetvalue{/flpy/stmt/mbfl/muse/all/einspect}{6714}
\pgfkeyssetvalue{/flpy/stmt/mbfl/muse/all/@1}{6}
\pgfkeyssetvalue{/flpy/stmt/mbfl/muse/all/@3}{19}
\pgfkeyssetvalue{/flpy/stmt/mbfl/muse/all/@5}{25}
\pgfkeyssetvalue{/flpy/stmt/mbfl/muse/all/@10}{32}
\pgfkeyssetvalue{/flpy/stmt/mbfl/muse/all/exam}{0.17194447287313444}
\pgfkeyssetvalue{/flpy/stmt/mbfl/muse/all/javaexam}{0.0023244579620409615}
\pgfkeyssetvalue{/flpy/stmt/mbfl/muse/all/outputlength}{113.8962962962963}
\pgfkeyssetvalue{/flpy/stmt/mbfl/muse/all/distance}{22.90044297241492}
\pgfkeyssetvalue{/flpy/stmt/ps/favg/all/time}{9751}
\pgfkeyssetvalue{/flpy/stmt/ps/favg/all/einspect}{11945}
\pgfkeyssetvalue{/flpy/stmt/ps/favg/all/@1}{3}
\pgfkeyssetvalue{/flpy/stmt/ps/favg/all/@3}{5}
\pgfkeyssetvalue{/flpy/stmt/ps/favg/all/@5}{7}
\pgfkeyssetvalue{/flpy/stmt/ps/favg/all/@10}{7}
\pgfkeyssetvalue{/flpy/stmt/ps/favg/all/exam}{0.2787497218988966}
\pgfkeyssetvalue{/flpy/stmt/ps/favg/all/javaexam}{0.00010556862466765817}
\pgfkeyssetvalue{/flpy/stmt/ps/favg/all/outputlength}{1.0296296296296297}
\pgfkeyssetvalue{/flpy/stmt/ps/favg/all/distance}{17.116777563553907}
\pgfkeyssetvalue{/flpy/stmt/st/favg/all/time}{2}
\pgfkeyssetvalue{/flpy/stmt/st/favg/all/einspect}{9810}
\pgfkeyssetvalue{/flpy/stmt/st/favg/all/@1}{0}
\pgfkeyssetvalue{/flpy/stmt/st/favg/all/@3}{4}
\pgfkeyssetvalue{/flpy/stmt/st/favg/all/@5}{6}
\pgfkeyssetvalue{/flpy/stmt/st/favg/all/@10}{13}
\pgfkeyssetvalue{/flpy/stmt/st/favg/all/exam}{0.2241291127567302}
\pgfkeyssetvalue{/flpy/stmt/st/favg/all/javaexam}{0.002402072293431907}
\pgfkeyssetvalue{/flpy/stmt/st/favg/all/outputlength}{42.85925925925926}
\pgfkeyssetvalue{/flpy/stmt/st/favg/all/distance}{8.243865648880641}
\pgfkeyssetvalue{/flpy/stmt/sbfl/favg/all/time}{589}
\pgfkeyssetvalue{/flpy/stmt/sbfl/favg/all/einspect}{1584}
\pgfkeyssetvalue{/flpy/stmt/sbfl/favg/all/@1}{12}
\pgfkeyssetvalue{/flpy/stmt/sbfl/favg/all/@3}{30}
\pgfkeyssetvalue{/flpy/stmt/sbfl/favg/all/@5}{43}
\pgfkeyssetvalue{/flpy/stmt/sbfl/favg/all/@10}{54}
\pgfkeyssetvalue{/flpy/stmt/sbfl/favg/all/exam}{0.05164192248857536}
\pgfkeyssetvalue{/flpy/stmt/sbfl/favg/all/javaexam}{0.0041667071496490835}
\pgfkeyssetvalue{/flpy/stmt/sbfl/favg/all/outputlength}{2521.348148148148}
\pgfkeyssetvalue{/flpy/stmt/sbfl/favg/all/distance}{10.608239907941572}
\pgfkeyssetvalue{/flpy/stmt/mbfl/favg/all/time}{15774}
\pgfkeyssetvalue{/flpy/stmt/mbfl/favg/all/einspect}{6710}
\pgfkeyssetvalue{/flpy/stmt/mbfl/favg/all/@1}{8}
\pgfkeyssetvalue{/flpy/stmt/mbfl/favg/all/@3}{22}
\pgfkeyssetvalue{/flpy/stmt/mbfl/favg/all/@5}{27}
\pgfkeyssetvalue{/flpy/stmt/mbfl/favg/all/@10}{34}
\pgfkeyssetvalue{/flpy/stmt/mbfl/favg/all/exam}{0.17225431514433906}
\pgfkeyssetvalue{/flpy/stmt/mbfl/favg/all/javaexam}{0.0028974539430358234}
\pgfkeyssetvalue{/flpy/stmt/mbfl/favg/all/outputlength}{113.8962962962963}
\pgfkeyssetvalue{/flpy/stmt/mbfl/favg/all/distance}{27.510572322227816}
\pgfkeyssetvalue{/flpy/stmt/sbfl/tarantula/web/time}{231}
\pgfkeyssetvalue{/flpy/stmt/sbfl/tarantula/web/einspect}{770}
\pgfkeyssetvalue{/flpy/stmt/sbfl/tarantula/web/@1}{15}
\pgfkeyssetvalue{/flpy/stmt/sbfl/tarantula/web/@3}{15}
\pgfkeyssetvalue{/flpy/stmt/sbfl/tarantula/web/@5}{40}
\pgfkeyssetvalue{/flpy/stmt/sbfl/tarantula/web/@10}{45}
\pgfkeyssetvalue{/flpy/stmt/sbfl/tarantula/web/exam}{0.12359384557041267}
\pgfkeyssetvalue{/flpy/stmt/sbfl/tarantula/web/javaexam}{0.0048632250404120865}
\pgfkeyssetvalue{/flpy/stmt/sbfl/tarantula/web/outputlength}{1266.35}
\pgfkeyssetvalue{/flpy/stmt/sbfl/tarantula/web/distance}{12.720402450413863}
\pgfkeyssetvalue{/flpy/stmt/sbfl/ochiai/web/time}{231}
\pgfkeyssetvalue{/flpy/stmt/sbfl/ochiai/web/einspect}{770}
\pgfkeyssetvalue{/flpy/stmt/sbfl/ochiai/web/@1}{15}
\pgfkeyssetvalue{/flpy/stmt/sbfl/ochiai/web/@3}{15}
\pgfkeyssetvalue{/flpy/stmt/sbfl/ochiai/web/@5}{40}
\pgfkeyssetvalue{/flpy/stmt/sbfl/ochiai/web/@10}{45}
\pgfkeyssetvalue{/flpy/stmt/sbfl/ochiai/web/exam}{0.1236051424524732}
\pgfkeyssetvalue{/flpy/stmt/sbfl/ochiai/web/javaexam}{0.004879363443355731}
\pgfkeyssetvalue{/flpy/stmt/sbfl/ochiai/web/outputlength}{1266.35}
\pgfkeyssetvalue{/flpy/stmt/sbfl/ochiai/web/distance}{12.770402450413863}
\pgfkeyssetvalue{/flpy/stmt/sbfl/dstar/web/time}{231}
\pgfkeyssetvalue{/flpy/stmt/sbfl/dstar/web/einspect}{770}
\pgfkeyssetvalue{/flpy/stmt/sbfl/dstar/web/@1}{15}
\pgfkeyssetvalue{/flpy/stmt/sbfl/dstar/web/@3}{15}
\pgfkeyssetvalue{/flpy/stmt/sbfl/dstar/web/@5}{40}
\pgfkeyssetvalue{/flpy/stmt/sbfl/dstar/web/@10}{45}
\pgfkeyssetvalue{/flpy/stmt/sbfl/dstar/web/exam}{0.1236051424524732}
\pgfkeyssetvalue{/flpy/stmt/sbfl/dstar/web/javaexam}{0.004879363443355731}
\pgfkeyssetvalue{/flpy/stmt/sbfl/dstar/web/outputlength}{1266.35}
\pgfkeyssetvalue{/flpy/stmt/sbfl/dstar/web/distance}{12.770402450413863}
\pgfkeyssetvalue{/flpy/stmt/mbfl/metallaxis/web/time}{7753}
\pgfkeyssetvalue{/flpy/stmt/mbfl/metallaxis/web/einspect}{1469}
\pgfkeyssetvalue{/flpy/stmt/mbfl/metallaxis/web/@1}{10}
\pgfkeyssetvalue{/flpy/stmt/mbfl/metallaxis/web/@3}{20}
\pgfkeyssetvalue{/flpy/stmt/mbfl/metallaxis/web/@5}{20}
\pgfkeyssetvalue{/flpy/stmt/mbfl/metallaxis/web/@10}{25}
\pgfkeyssetvalue{/flpy/stmt/mbfl/metallaxis/web/exam}{0.21270208692968975}
\pgfkeyssetvalue{/flpy/stmt/mbfl/metallaxis/web/javaexam}{0.004927144409854644}
\pgfkeyssetvalue{/flpy/stmt/mbfl/metallaxis/web/outputlength}{98.65}
\pgfkeyssetvalue{/flpy/stmt/mbfl/metallaxis/web/distance}{39.80167049619566}
\pgfkeyssetvalue{/flpy/stmt/mbfl/muse/web/time}{7753}
\pgfkeyssetvalue{/flpy/stmt/mbfl/muse/web/einspect}{1460}
\pgfkeyssetvalue{/flpy/stmt/mbfl/muse/web/@1}{5}
\pgfkeyssetvalue{/flpy/stmt/mbfl/muse/web/@3}{15}
\pgfkeyssetvalue{/flpy/stmt/mbfl/muse/web/@5}{20}
\pgfkeyssetvalue{/flpy/stmt/mbfl/muse/web/@10}{25}
\pgfkeyssetvalue{/flpy/stmt/mbfl/muse/web/exam}{0.21214506830792698}
\pgfkeyssetvalue{/flpy/stmt/mbfl/muse/web/javaexam}{0.0035345978554476515}
\pgfkeyssetvalue{/flpy/stmt/mbfl/muse/web/outputlength}{98.65}
\pgfkeyssetvalue{/flpy/stmt/mbfl/muse/web/distance}{15.337936969627435}
\pgfkeyssetvalue{/flpy/stmt/ps/favg/web/time}{828}
\pgfkeyssetvalue{/flpy/stmt/ps/favg/web/einspect}{2362}
\pgfkeyssetvalue{/flpy/stmt/ps/favg/web/@1}{5}
\pgfkeyssetvalue{/flpy/stmt/ps/favg/web/@3}{5}
\pgfkeyssetvalue{/flpy/stmt/ps/favg/web/@5}{5}
\pgfkeyssetvalue{/flpy/stmt/ps/favg/web/@10}{5}
\pgfkeyssetvalue{/flpy/stmt/ps/favg/web/exam}{0.274473269502389}
\pgfkeyssetvalue{/flpy/stmt/ps/favg/web/javaexam}{0.00022716946842344388}
\pgfkeyssetvalue{/flpy/stmt/ps/favg/web/outputlength}{1.05}
\pgfkeyssetvalue{/flpy/stmt/ps/favg/web/distance}{20.308287272828476}
\pgfkeyssetvalue{/flpy/stmt/st/favg/web/time}{1}
\pgfkeyssetvalue{/flpy/stmt/st/favg/web/einspect}{2319}
\pgfkeyssetvalue{/flpy/stmt/st/favg/web/@1}{0}
\pgfkeyssetvalue{/flpy/stmt/st/favg/web/@3}{5}
\pgfkeyssetvalue{/flpy/stmt/st/favg/web/@5}{5}
\pgfkeyssetvalue{/flpy/stmt/st/favg/web/@10}{15}
\pgfkeyssetvalue{/flpy/stmt/st/favg/web/exam}{0.2631058856251216}
\pgfkeyssetvalue{/flpy/stmt/st/favg/web/javaexam}{0.0014071383509049152}
\pgfkeyssetvalue{/flpy/stmt/st/favg/web/outputlength}{22.65}
\pgfkeyssetvalue{/flpy/stmt/st/favg/web/distance}{8.70625}
\pgfkeyssetvalue{/flpy/stmt/sbfl/favg/web/time}{231}
\pgfkeyssetvalue{/flpy/stmt/sbfl/favg/web/einspect}{770}
\pgfkeyssetvalue{/flpy/stmt/sbfl/favg/web/@1}{15}
\pgfkeyssetvalue{/flpy/stmt/sbfl/favg/web/@3}{15}
\pgfkeyssetvalue{/flpy/stmt/sbfl/favg/web/@5}{40}
\pgfkeyssetvalue{/flpy/stmt/sbfl/favg/web/@10}{45}
\pgfkeyssetvalue{/flpy/stmt/sbfl/favg/web/exam}{0.1236013768251197}
\pgfkeyssetvalue{/flpy/stmt/sbfl/favg/web/javaexam}{0.00487398397570785}
\pgfkeyssetvalue{/flpy/stmt/sbfl/favg/web/outputlength}{1266.35}
\pgfkeyssetvalue{/flpy/stmt/sbfl/favg/web/distance}{12.753735783747196}
\pgfkeyssetvalue{/flpy/stmt/mbfl/favg/web/time}{7753}
\pgfkeyssetvalue{/flpy/stmt/mbfl/favg/web/einspect}{1465}
\pgfkeyssetvalue{/flpy/stmt/mbfl/favg/web/@1}{8}
\pgfkeyssetvalue{/flpy/stmt/mbfl/favg/web/@3}{18}
\pgfkeyssetvalue{/flpy/stmt/mbfl/favg/web/@5}{20}
\pgfkeyssetvalue{/flpy/stmt/mbfl/favg/web/@10}{25}
\pgfkeyssetvalue{/flpy/stmt/mbfl/favg/web/exam}{0.21242357761880837}
\pgfkeyssetvalue{/flpy/stmt/mbfl/favg/web/javaexam}{0.004230871132651148}
\pgfkeyssetvalue{/flpy/stmt/mbfl/favg/web/outputlength}{98.65}
\pgfkeyssetvalue{/flpy/stmt/mbfl/favg/web/distance}{27.56980373291155}
\pgfkeyssetvalue{/flpy/stmt/sbfl/tarantula/crash/time}{890}
\pgfkeyssetvalue{/flpy/stmt/sbfl/tarantula/crash/einspect}{903}
\pgfkeyssetvalue{/flpy/stmt/sbfl/tarantula/crash/@1}{14}
\pgfkeyssetvalue{/flpy/stmt/sbfl/tarantula/crash/@3}{31}
\pgfkeyssetvalue{/flpy/stmt/sbfl/tarantula/crash/@5}{43}
\pgfkeyssetvalue{/flpy/stmt/sbfl/tarantula/crash/@10}{53}
\pgfkeyssetvalue{/flpy/stmt/sbfl/tarantula/crash/exam}{0.04662088093730834}
\pgfkeyssetvalue{/flpy/stmt/sbfl/tarantula/crash/javaexam}{0.00254115899205032}
\pgfkeyssetvalue{/flpy/stmt/sbfl/tarantula/crash/outputlength}{3147.469387755102}
\pgfkeyssetvalue{/flpy/stmt/sbfl/tarantula/crash/distance}{11.688373677354202}
\pgfkeyssetvalue{/flpy/stmt/sbfl/ochiai/crash/time}{890}
\pgfkeyssetvalue{/flpy/stmt/sbfl/ochiai/crash/einspect}{895}
\pgfkeyssetvalue{/flpy/stmt/sbfl/ochiai/crash/@1}{14}
\pgfkeyssetvalue{/flpy/stmt/sbfl/ochiai/crash/@3}{31}
\pgfkeyssetvalue{/flpy/stmt/sbfl/ochiai/crash/@5}{43}
\pgfkeyssetvalue{/flpy/stmt/sbfl/ochiai/crash/@10}{53}
\pgfkeyssetvalue{/flpy/stmt/sbfl/ochiai/crash/exam}{0.04656670706671579}
\pgfkeyssetvalue{/flpy/stmt/sbfl/ochiai/crash/javaexam}{0.002480828999799537}
\pgfkeyssetvalue{/flpy/stmt/sbfl/ochiai/crash/outputlength}{3147.469387755102}
\pgfkeyssetvalue{/flpy/stmt/sbfl/ochiai/crash/distance}{12.337086353267805}
\pgfkeyssetvalue{/flpy/stmt/sbfl/dstar/crash/time}{890}
\pgfkeyssetvalue{/flpy/stmt/sbfl/dstar/crash/einspect}{894}
\pgfkeyssetvalue{/flpy/stmt/sbfl/dstar/crash/@1}{14}
\pgfkeyssetvalue{/flpy/stmt/sbfl/dstar/crash/@3}{31}
\pgfkeyssetvalue{/flpy/stmt/sbfl/dstar/crash/@5}{43}
\pgfkeyssetvalue{/flpy/stmt/sbfl/dstar/crash/@10}{53}
\pgfkeyssetvalue{/flpy/stmt/sbfl/dstar/crash/exam}{0.04655780006169332}
\pgfkeyssetvalue{/flpy/stmt/sbfl/dstar/crash/javaexam}{0.0024709098351154164}
\pgfkeyssetvalue{/flpy/stmt/sbfl/dstar/crash/outputlength}{3147.469387755102}
\pgfkeyssetvalue{/flpy/stmt/sbfl/dstar/crash/distance}{12.735655924139065}
\pgfkeyssetvalue{/flpy/stmt/mbfl/metallaxis/crash/time}{18278}
\pgfkeyssetvalue{/flpy/stmt/mbfl/metallaxis/crash/einspect}{7793}
\pgfkeyssetvalue{/flpy/stmt/mbfl/metallaxis/crash/@1}{12}
\pgfkeyssetvalue{/flpy/stmt/mbfl/metallaxis/crash/@3}{29}
\pgfkeyssetvalue{/flpy/stmt/mbfl/metallaxis/crash/@5}{33}
\pgfkeyssetvalue{/flpy/stmt/mbfl/metallaxis/crash/@10}{37}
\pgfkeyssetvalue{/flpy/stmt/mbfl/metallaxis/crash/exam}{0.16728461318059687}
\pgfkeyssetvalue{/flpy/stmt/mbfl/metallaxis/crash/javaexam}{0.0008556603018041637}
\pgfkeyssetvalue{/flpy/stmt/mbfl/metallaxis/crash/outputlength}{104.40816326530613}
\pgfkeyssetvalue{/flpy/stmt/mbfl/metallaxis/crash/distance}{42.69853431266601}
\pgfkeyssetvalue{/flpy/stmt/mbfl/muse/crash/time}{18278}
\pgfkeyssetvalue{/flpy/stmt/mbfl/muse/crash/einspect}{7818}
\pgfkeyssetvalue{/flpy/stmt/mbfl/muse/crash/@1}{2}
\pgfkeyssetvalue{/flpy/stmt/mbfl/muse/crash/@3}{14}
\pgfkeyssetvalue{/flpy/stmt/mbfl/muse/crash/@5}{20}
\pgfkeyssetvalue{/flpy/stmt/mbfl/muse/crash/@10}{31}
\pgfkeyssetvalue{/flpy/stmt/mbfl/muse/crash/exam}{0.1682339412771017}
\pgfkeyssetvalue{/flpy/stmt/mbfl/muse/crash/javaexam}{0.0026447786375248315}
\pgfkeyssetvalue{/flpy/stmt/mbfl/muse/crash/outputlength}{104.40816326530613}
\pgfkeyssetvalue{/flpy/stmt/mbfl/muse/crash/distance}{43.09180646796559}
\pgfkeyssetvalue{/flpy/stmt/ps/favg/crash/time}{11419}
\pgfkeyssetvalue{/flpy/stmt/ps/favg/crash/einspect}{15607}
\pgfkeyssetvalue{/flpy/stmt/ps/favg/crash/@1}{0}
\pgfkeyssetvalue{/flpy/stmt/ps/favg/crash/@3}{0}
\pgfkeyssetvalue{/flpy/stmt/ps/favg/crash/@5}{0}
\pgfkeyssetvalue{/flpy/stmt/ps/favg/crash/@10}{0}
\pgfkeyssetvalue{/flpy/stmt/ps/favg/crash/exam}{0.29940246505546325}
\pgfkeyssetvalue{/flpy/stmt/ps/favg/crash/javaexam}{None}
\pgfkeyssetvalue{/flpy/stmt/ps/favg/crash/outputlength}{0.30612244897959184}
\pgfkeyssetvalue{/flpy/stmt/ps/favg/crash/distance}{13.835314811282768}
\pgfkeyssetvalue{/flpy/stmt/st/favg/crash/time}{2}
\pgfkeyssetvalue{/flpy/stmt/st/favg/crash/einspect}{5273}
\pgfkeyssetvalue{/flpy/stmt/st/favg/crash/@1}{0}
\pgfkeyssetvalue{/flpy/stmt/st/favg/crash/@3}{10}
\pgfkeyssetvalue{/flpy/stmt/st/favg/crash/@5}{16}
\pgfkeyssetvalue{/flpy/stmt/st/favg/crash/@10}{37}
\pgfkeyssetvalue{/flpy/stmt/st/favg/crash/exam}{0.09286682401942466}
\pgfkeyssetvalue{/flpy/stmt/st/favg/crash/javaexam}{0.002402072293431907}
\pgfkeyssetvalue{/flpy/stmt/st/favg/crash/outputlength}{118.08163265306122}
\pgfkeyssetvalue{/flpy/stmt/st/favg/crash/distance}{5.161670665283396}
\pgfkeyssetvalue{/flpy/stmt/sbfl/favg/crash/time}{890}
\pgfkeyssetvalue{/flpy/stmt/sbfl/favg/crash/einspect}{897}
\pgfkeyssetvalue{/flpy/stmt/sbfl/favg/crash/@1}{14}
\pgfkeyssetvalue{/flpy/stmt/sbfl/favg/crash/@3}{31}
\pgfkeyssetvalue{/flpy/stmt/sbfl/favg/crash/@5}{43}
\pgfkeyssetvalue{/flpy/stmt/sbfl/favg/crash/@10}{53}
\pgfkeyssetvalue{/flpy/stmt/sbfl/favg/crash/exam}{0.04658179602190582}
\pgfkeyssetvalue{/flpy/stmt/sbfl/favg/crash/javaexam}{0.0024976326089884244}
\pgfkeyssetvalue{/flpy/stmt/sbfl/favg/crash/outputlength}{3147.469387755102}
\pgfkeyssetvalue{/flpy/stmt/sbfl/favg/crash/distance}{12.253705318253692}
\pgfkeyssetvalue{/flpy/stmt/mbfl/favg/crash/time}{18278}
\pgfkeyssetvalue{/flpy/stmt/mbfl/favg/crash/einspect}{7806}
\pgfkeyssetvalue{/flpy/stmt/mbfl/favg/crash/@1}{7}
\pgfkeyssetvalue{/flpy/stmt/mbfl/favg/crash/@3}{21}
\pgfkeyssetvalue{/flpy/stmt/mbfl/favg/crash/@5}{27}
\pgfkeyssetvalue{/flpy/stmt/mbfl/favg/crash/@10}{34}
\pgfkeyssetvalue{/flpy/stmt/mbfl/favg/crash/exam}{0.16775927722884929}
\pgfkeyssetvalue{/flpy/stmt/mbfl/favg/crash/javaexam}{0.0017502194696644977}
\pgfkeyssetvalue{/flpy/stmt/mbfl/favg/crash/outputlength}{104.40816326530613}
\pgfkeyssetvalue{/flpy/stmt/mbfl/favg/crash/distance}{42.8951703903158}
\pgfkeyssetvalue{/flpy/stmt/sbfl/tarantula/dev/time}{38}
\pgfkeyssetvalue{/flpy/stmt/sbfl/tarantula/dev/einspect}{2081}
\pgfkeyssetvalue{/flpy/stmt/sbfl/tarantula/dev/@1}{20}
\pgfkeyssetvalue{/flpy/stmt/sbfl/tarantula/dev/@3}{33}
\pgfkeyssetvalue{/flpy/stmt/sbfl/tarantula/dev/@5}{37}
\pgfkeyssetvalue{/flpy/stmt/sbfl/tarantula/dev/@10}{47}
\pgfkeyssetvalue{/flpy/stmt/sbfl/tarantula/dev/exam}{0.05815723168207835}
\pgfkeyssetvalue{/flpy/stmt/sbfl/tarantula/dev/javaexam}{0.005337890808826007}
\pgfkeyssetvalue{/flpy/stmt/sbfl/tarantula/dev/outputlength}{1431.4666666666667}
\pgfkeyssetvalue{/flpy/stmt/sbfl/tarantula/dev/distance}{9.187215160881484}
\pgfkeyssetvalue{/flpy/stmt/sbfl/ochiai/dev/time}{38}
\pgfkeyssetvalue{/flpy/stmt/sbfl/ochiai/dev/einspect}{2081}
\pgfkeyssetvalue{/flpy/stmt/sbfl/ochiai/dev/@1}{20}
\pgfkeyssetvalue{/flpy/stmt/sbfl/ochiai/dev/@3}{33}
\pgfkeyssetvalue{/flpy/stmt/sbfl/ochiai/dev/@5}{37}
\pgfkeyssetvalue{/flpy/stmt/sbfl/ochiai/dev/@10}{47}
\pgfkeyssetvalue{/flpy/stmt/sbfl/ochiai/dev/exam}{0.05815772931163908}
\pgfkeyssetvalue{/flpy/stmt/sbfl/ochiai/dev/javaexam}{0.005338443730560148}
\pgfkeyssetvalue{/flpy/stmt/sbfl/ochiai/dev/outputlength}{1431.4666666666667}
\pgfkeyssetvalue{/flpy/stmt/sbfl/ochiai/dev/distance}{9.196844790511113}
\pgfkeyssetvalue{/flpy/stmt/sbfl/dstar/dev/time}{38}
\pgfkeyssetvalue{/flpy/stmt/sbfl/dstar/dev/einspect}{2081}
\pgfkeyssetvalue{/flpy/stmt/sbfl/dstar/dev/@1}{20}
\pgfkeyssetvalue{/flpy/stmt/sbfl/dstar/dev/@3}{33}
\pgfkeyssetvalue{/flpy/stmt/sbfl/dstar/dev/@5}{37}
\pgfkeyssetvalue{/flpy/stmt/sbfl/dstar/dev/@10}{47}
\pgfkeyssetvalue{/flpy/stmt/sbfl/dstar/dev/exam}{0.05815772931163908}
\pgfkeyssetvalue{/flpy/stmt/sbfl/dstar/dev/javaexam}{0.005338443730560148}
\pgfkeyssetvalue{/flpy/stmt/sbfl/dstar/dev/outputlength}{1431.4666666666667}
\pgfkeyssetvalue{/flpy/stmt/sbfl/dstar/dev/distance}{9.196844790511113}
\pgfkeyssetvalue{/flpy/stmt/mbfl/metallaxis/dev/time}{18694}
\pgfkeyssetvalue{/flpy/stmt/mbfl/metallaxis/dev/einspect}{4718}
\pgfkeyssetvalue{/flpy/stmt/mbfl/metallaxis/dev/@1}{13}
\pgfkeyssetvalue{/flpy/stmt/mbfl/metallaxis/dev/@3}{33}
\pgfkeyssetvalue{/flpy/stmt/mbfl/metallaxis/dev/@5}{40}
\pgfkeyssetvalue{/flpy/stmt/mbfl/metallaxis/dev/@10}{50}
\pgfkeyssetvalue{/flpy/stmt/mbfl/metallaxis/dev/exam}{0.14151827780082132}
\pgfkeyssetvalue{/flpy/stmt/mbfl/metallaxis/dev/javaexam}{0.006784306210421347}
\pgfkeyssetvalue{/flpy/stmt/mbfl/metallaxis/dev/outputlength}{160.63333333333333}
\pgfkeyssetvalue{/flpy/stmt/mbfl/metallaxis/dev/distance}{32.40592108583738}
\pgfkeyssetvalue{/flpy/stmt/mbfl/muse/dev/time}{18694}
\pgfkeyssetvalue{/flpy/stmt/mbfl/muse/dev/einspect}{4721}
\pgfkeyssetvalue{/flpy/stmt/mbfl/muse/dev/@1}{10}
\pgfkeyssetvalue{/flpy/stmt/mbfl/muse/dev/@3}{17}
\pgfkeyssetvalue{/flpy/stmt/mbfl/muse/dev/@5}{17}
\pgfkeyssetvalue{/flpy/stmt/mbfl/muse/dev/@10}{30}
\pgfkeyssetvalue{/flpy/stmt/mbfl/muse/dev/exam}{0.13859315361752955}
\pgfkeyssetvalue{/flpy/stmt/mbfl/muse/dev/javaexam}{0.002165689078908047}
\pgfkeyssetvalue{/flpy/stmt/mbfl/muse/dev/outputlength}{160.63333333333333}
\pgfkeyssetvalue{/flpy/stmt/mbfl/muse/dev/distance}{13.945892432317322}
\pgfkeyssetvalue{/flpy/stmt/ps/favg/dev/time}{20210}
\pgfkeyssetvalue{/flpy/stmt/ps/favg/dev/einspect}{7768}
\pgfkeyssetvalue{/flpy/stmt/ps/favg/dev/@1}{3}
\pgfkeyssetvalue{/flpy/stmt/ps/favg/dev/@3}{7}
\pgfkeyssetvalue{/flpy/stmt/ps/favg/dev/@5}{10}
\pgfkeyssetvalue{/flpy/stmt/ps/favg/dev/@10}{10}
\pgfkeyssetvalue{/flpy/stmt/ps/favg/dev/exam}{0.25533442158903913}
\pgfkeyssetvalue{/flpy/stmt/ps/favg/dev/javaexam}{7.84488638762666e-05}
\pgfkeyssetvalue{/flpy/stmt/ps/favg/dev/outputlength}{2.1}
\pgfkeyssetvalue{/flpy/stmt/ps/favg/dev/distance}{23.4528805986718}
\pgfkeyssetvalue{/flpy/stmt/st/favg/dev/time}{1}
\pgfkeyssetvalue{/flpy/stmt/st/favg/dev/einspect}{8279}
\pgfkeyssetvalue{/flpy/stmt/st/favg/dev/@1}{0}
\pgfkeyssetvalue{/flpy/stmt/st/favg/dev/@3}{0}
\pgfkeyssetvalue{/flpy/stmt/st/favg/dev/@5}{10}
\pgfkeyssetvalue{/flpy/stmt/st/favg/dev/@10}{13}
\pgfkeyssetvalue{/flpy/stmt/st/favg/dev/exam}{0.20856706862705324}
\pgfkeyssetvalue{/flpy/stmt/st/favg/dev/javaexam}{0.00278045955163677}
\pgfkeyssetvalue{/flpy/stmt/st/favg/dev/outputlength}{12.433333333333334}
\pgfkeyssetvalue{/flpy/stmt/st/favg/dev/distance}{7.661867443102397}
\pgfkeyssetvalue{/flpy/stmt/sbfl/favg/dev/time}{38}
\pgfkeyssetvalue{/flpy/stmt/sbfl/favg/dev/einspect}{2081}
\pgfkeyssetvalue{/flpy/stmt/sbfl/favg/dev/@1}{20}
\pgfkeyssetvalue{/flpy/stmt/sbfl/favg/dev/@3}{33}
\pgfkeyssetvalue{/flpy/stmt/sbfl/favg/dev/@5}{37}
\pgfkeyssetvalue{/flpy/stmt/sbfl/favg/dev/@10}{47}
\pgfkeyssetvalue{/flpy/stmt/sbfl/favg/dev/exam}{0.05815756343511883}
\pgfkeyssetvalue{/flpy/stmt/sbfl/favg/dev/javaexam}{0.005338259423315434}
\pgfkeyssetvalue{/flpy/stmt/sbfl/favg/dev/outputlength}{1431.4666666666665}
\pgfkeyssetvalue{/flpy/stmt/sbfl/favg/dev/distance}{9.193634913967903}
\pgfkeyssetvalue{/flpy/stmt/mbfl/favg/dev/time}{18694}
\pgfkeyssetvalue{/flpy/stmt/mbfl/favg/dev/einspect}{4720}
\pgfkeyssetvalue{/flpy/stmt/mbfl/favg/dev/@1}{12}
\pgfkeyssetvalue{/flpy/stmt/mbfl/favg/dev/@3}{25}
\pgfkeyssetvalue{/flpy/stmt/mbfl/favg/dev/@5}{28}
\pgfkeyssetvalue{/flpy/stmt/mbfl/favg/dev/@10}{40}
\pgfkeyssetvalue{/flpy/stmt/mbfl/favg/dev/exam}{0.14005571570917544}
\pgfkeyssetvalue{/flpy/stmt/mbfl/favg/dev/javaexam}{0.004474997644664697}
\pgfkeyssetvalue{/flpy/stmt/mbfl/favg/dev/outputlength}{160.63333333333333}
\pgfkeyssetvalue{/flpy/stmt/mbfl/favg/dev/distance}{23.17590675907735}
\pgfkeyssetvalue{/flpy/stmt/sbfl/tarantula/mut/time}{521}
\pgfkeyssetvalue{/flpy/stmt/sbfl/tarantula/mut/einspect}{526}
\pgfkeyssetvalue{/flpy/stmt/sbfl/tarantula/mut/@1}{12}
\pgfkeyssetvalue{/flpy/stmt/sbfl/tarantula/mut/@3}{35}
\pgfkeyssetvalue{/flpy/stmt/sbfl/tarantula/mut/@5}{50}
\pgfkeyssetvalue{/flpy/stmt/sbfl/tarantula/mut/@10}{57}
\pgfkeyssetvalue{/flpy/stmt/sbfl/tarantula/mut/exam}{0.008854186103602725}
\pgfkeyssetvalue{/flpy/stmt/sbfl/tarantula/mut/javaexam}{0.004213188676947862}
\pgfkeyssetvalue{/flpy/stmt/sbfl/tarantula/mut/outputlength}{2396.175675675676}
\pgfkeyssetvalue{/flpy/stmt/sbfl/tarantula/mut/distance}{9.56423176802314}
\pgfkeyssetvalue{/flpy/stmt/sbfl/ochiai/mut/time}{521}
\pgfkeyssetvalue{/flpy/stmt/sbfl/ochiai/mut/einspect}{522}
\pgfkeyssetvalue{/flpy/stmt/sbfl/ochiai/mut/@1}{12}
\pgfkeyssetvalue{/flpy/stmt/sbfl/ochiai/mut/@3}{35}
\pgfkeyssetvalue{/flpy/stmt/sbfl/ochiai/mut/@5}{50}
\pgfkeyssetvalue{/flpy/stmt/sbfl/ochiai/mut/@10}{57}
\pgfkeyssetvalue{/flpy/stmt/sbfl/ochiai/mut/exam}{0.00882554716841561}
\pgfkeyssetvalue{/flpy/stmt/sbfl/ochiai/mut/javaexam}{0.0041841574275801015}
\pgfkeyssetvalue{/flpy/stmt/sbfl/ochiai/mut/outputlength}{2396.175675675676}
\pgfkeyssetvalue{/flpy/stmt/sbfl/ochiai/mut/distance}{9.939073970576185}
\pgfkeyssetvalue{/flpy/stmt/sbfl/dstar/mut/time}{521}
\pgfkeyssetvalue{/flpy/stmt/sbfl/dstar/mut/einspect}{522}
\pgfkeyssetvalue{/flpy/stmt/sbfl/dstar/mut/@1}{11}
\pgfkeyssetvalue{/flpy/stmt/sbfl/dstar/mut/@3}{34}
\pgfkeyssetvalue{/flpy/stmt/sbfl/dstar/mut/@5}{49}
\pgfkeyssetvalue{/flpy/stmt/sbfl/dstar/mut/@10}{57}
\pgfkeyssetvalue{/flpy/stmt/sbfl/dstar/mut/exam}{0.008822822631741587}
\pgfkeyssetvalue{/flpy/stmt/sbfl/dstar/mut/javaexam}{0.004181395568485886}
\pgfkeyssetvalue{/flpy/stmt/sbfl/dstar/mut/outputlength}{2396.175675675676}
\pgfkeyssetvalue{/flpy/stmt/sbfl/dstar/mut/distance}{10.45163519498624}
\pgfkeyssetvalue{/flpy/stmt/mbfl/metallaxis/mut/time}{17744}
\pgfkeyssetvalue{/flpy/stmt/mbfl/metallaxis/mut/einspect}{482}
\pgfkeyssetvalue{/flpy/stmt/mbfl/metallaxis/mut/@1}{18}
\pgfkeyssetvalue{/flpy/stmt/mbfl/metallaxis/mut/@3}{46}
\pgfkeyssetvalue{/flpy/stmt/mbfl/metallaxis/mut/@5}{54}
\pgfkeyssetvalue{/flpy/stmt/mbfl/metallaxis/mut/@10}{68}
\pgfkeyssetvalue{/flpy/stmt/mbfl/metallaxis/mut/exam}{0.007928819574336011}
\pgfkeyssetvalue{/flpy/stmt/mbfl/metallaxis/mut/javaexam}{0.003470449924030685}
\pgfkeyssetvalue{/flpy/stmt/mbfl/metallaxis/mut/outputlength}{138.6891891891892}
\pgfkeyssetvalue{/flpy/stmt/mbfl/metallaxis/mut/distance}{18.397460537354096}
\pgfkeyssetvalue{/flpy/stmt/mbfl/muse/mut/time}{17744}
\pgfkeyssetvalue{/flpy/stmt/mbfl/muse/mut/einspect}{496}
\pgfkeyssetvalue{/flpy/stmt/mbfl/muse/mut/@1}{11}
\pgfkeyssetvalue{/flpy/stmt/mbfl/muse/mut/@3}{35}
\pgfkeyssetvalue{/flpy/stmt/mbfl/muse/mut/@5}{46}
\pgfkeyssetvalue{/flpy/stmt/mbfl/muse/mut/@10}{58}
\pgfkeyssetvalue{/flpy/stmt/mbfl/muse/mut/exam}{0.006798313990211014}
\pgfkeyssetvalue{/flpy/stmt/mbfl/muse/mut/javaexam}{0.0023244579620409615}
\pgfkeyssetvalue{/flpy/stmt/mbfl/muse/mut/outputlength}{138.6891891891892}
\pgfkeyssetvalue{/flpy/stmt/mbfl/muse/mut/distance}{27.049774881803454}
\pgfkeyssetvalue{/flpy/stmt/ps/favg/mut/time}{12932}
\pgfkeyssetvalue{/flpy/stmt/ps/favg/mut/einspect}{10081}
\pgfkeyssetvalue{/flpy/stmt/ps/favg/mut/@1}{5}
\pgfkeyssetvalue{/flpy/stmt/ps/favg/mut/@3}{9}
\pgfkeyssetvalue{/flpy/stmt/ps/favg/mut/@5}{12}
\pgfkeyssetvalue{/flpy/stmt/ps/favg/mut/@10}{12}
\pgfkeyssetvalue{/flpy/stmt/ps/favg/mut/exam}{0.2043890182994061}
\pgfkeyssetvalue{/flpy/stmt/ps/favg/mut/javaexam}{0.00010556862466765817}
\pgfkeyssetvalue{/flpy/stmt/ps/favg/mut/outputlength}{1.135135135135135}
\pgfkeyssetvalue{/flpy/stmt/ps/favg/mut/distance}{12.612974953312067}
\pgfkeyssetvalue{/flpy/stmt/st/favg/mut/time}{2}
\pgfkeyssetvalue{/flpy/stmt/st/favg/mut/einspect}{9304}
\pgfkeyssetvalue{/flpy/stmt/st/favg/mut/@1}{0}
\pgfkeyssetvalue{/flpy/stmt/st/favg/mut/@3}{4}
\pgfkeyssetvalue{/flpy/stmt/st/favg/mut/@5}{5}
\pgfkeyssetvalue{/flpy/stmt/st/favg/mut/@10}{19}
\pgfkeyssetvalue{/flpy/stmt/st/favg/mut/exam}{0.17048376746971602}
\pgfkeyssetvalue{/flpy/stmt/st/favg/mut/javaexam}{0.0007175107088746468}
\pgfkeyssetvalue{/flpy/stmt/st/favg/mut/outputlength}{35.270270270270274}
\pgfkeyssetvalue{/flpy/stmt/st/favg/mut/distance}{8.629261601376617}
\pgfkeyssetvalue{/flpy/stmt/sbfl/favg/mut/time}{521}
\pgfkeyssetvalue{/flpy/stmt/sbfl/favg/mut/einspect}{524}
\pgfkeyssetvalue{/flpy/stmt/sbfl/favg/mut/@1}{12}
\pgfkeyssetvalue{/flpy/stmt/sbfl/favg/mut/@3}{35}
\pgfkeyssetvalue{/flpy/stmt/sbfl/favg/mut/@5}{50}
\pgfkeyssetvalue{/flpy/stmt/sbfl/favg/mut/@10}{57}
\pgfkeyssetvalue{/flpy/stmt/sbfl/favg/mut/exam}{0.008834185301253307}
\pgfkeyssetvalue{/flpy/stmt/sbfl/favg/mut/javaexam}{0.004192913891004616}
\pgfkeyssetvalue{/flpy/stmt/sbfl/favg/mut/outputlength}{2396.175675675676}
\pgfkeyssetvalue{/flpy/stmt/sbfl/favg/mut/distance}{9.984980311195187}
\pgfkeyssetvalue{/flpy/stmt/mbfl/favg/mut/time}{17744}
\pgfkeyssetvalue{/flpy/stmt/mbfl/favg/mut/einspect}{489}
\pgfkeyssetvalue{/flpy/stmt/mbfl/favg/mut/@1}{14}
\pgfkeyssetvalue{/flpy/stmt/mbfl/favg/mut/@3}{41}
\pgfkeyssetvalue{/flpy/stmt/mbfl/favg/mut/@5}{50}
\pgfkeyssetvalue{/flpy/stmt/mbfl/favg/mut/@10}{63}
\pgfkeyssetvalue{/flpy/stmt/mbfl/favg/mut/exam}{0.007363566782273513}
\pgfkeyssetvalue{/flpy/stmt/mbfl/favg/mut/javaexam}{0.0028974539430358234}
\pgfkeyssetvalue{/flpy/stmt/mbfl/favg/mut/outputlength}{138.6891891891892}
\pgfkeyssetvalue{/flpy/stmt/mbfl/favg/mut/distance}{22.723617709578775}
\pgfkeyssetvalue{/flpy/mod/avfl/alfa/all/time}{-1}
\pgfkeyssetvalue{/flpy/mod/avfl/alfa/all/einspect}{2}
\pgfkeyssetvalue{/flpy/mod/avfl/alfa/all/@1}{70}
\pgfkeyssetvalue{/flpy/mod/avfl/alfa/all/@3}{89}
\pgfkeyssetvalue{/flpy/mod/avfl/alfa/all/@5}{93}
\pgfkeyssetvalue{/flpy/mod/avfl/alfa/all/@10}{99}
\pgfkeyssetvalue{/flpy/mod/avfl/alfa/all/exam}{0.03858549430871499}
\pgfkeyssetvalue{/flpy/mod/avfl/alfa/all/javaexam}{0.03384424394177811}
\pgfkeyssetvalue{/flpy/mod/avfl/alfa/all/outputlength}{20.918518518518518}
\pgfkeyssetvalue{/flpy/mod/avfl/sbst/all/time}{-1}
\pgfkeyssetvalue{/flpy/mod/avfl/sbst/all/einspect}{2}
\pgfkeyssetvalue{/flpy/mod/avfl/sbst/all/@1}{64}
\pgfkeyssetvalue{/flpy/mod/avfl/sbst/all/@3}{87}
\pgfkeyssetvalue{/flpy/mod/avfl/sbst/all/@5}{93}
\pgfkeyssetvalue{/flpy/mod/avfl/sbst/all/@10}{98}
\pgfkeyssetvalue{/flpy/mod/avfl/sbst/all/exam}{0.04098960206373947}
\pgfkeyssetvalue{/flpy/mod/avfl/sbst/all/javaexam}{0.03626629279945203}
\pgfkeyssetvalue{/flpy/mod/avfl/sbst/all/outputlength}{20.918518518518518}
\pgfkeyssetvalue{/flpy/mod/avfl/favg/all/time}{-1}
\pgfkeyssetvalue{/flpy/mod/avfl/favg/all/einspect}{2}
\pgfkeyssetvalue{/flpy/mod/avfl/favg/all/@1}{67}
\pgfkeyssetvalue{/flpy/mod/avfl/favg/all/@3}{88}
\pgfkeyssetvalue{/flpy/mod/avfl/favg/all/@5}{93}
\pgfkeyssetvalue{/flpy/mod/avfl/favg/all/@10}{98}
\pgfkeyssetvalue{/flpy/mod/avfl/favg/all/exam}{0.039787548186227234}
\pgfkeyssetvalue{/flpy/mod/avfl/favg/all/javaexam}{0.03505526837061507}
\pgfkeyssetvalue{/flpy/mod/avfl/favg/all/outputlength}{20.918518518518518}
\pgfkeyssetvalue{/flpy/stmt/sbfl/ochiai/all/@1/java}{4}
\pgfkeyssetvalue{/flpy/stmt/sbfl/ochiai/all/@3/java}{23}
\pgfkeyssetvalue{/flpy/stmt/sbfl/ochiai/all/@5/java}{31}
\pgfkeyssetvalue{/flpy/stmt/sbfl/ochiai/all/@10/java}{44}
\pgfkeyssetvalue{/flpy/stmt/sbfl/ochiai/all/javaexam/java}{0.033}
\pgfkeyssetvalue{/flpy/stmt/sbfl/dstar/all/@1/java}{5}
\pgfkeyssetvalue{/flpy/stmt/sbfl/dstar/all/@3/java}{24}
\pgfkeyssetvalue{/flpy/stmt/sbfl/dstar/all/@5/java}{31}
\pgfkeyssetvalue{/flpy/stmt/sbfl/dstar/all/@10/java}{43}
\pgfkeyssetvalue{/flpy/stmt/sbfl/dstar/all/javaexam/java}{0.033}
\pgfkeyssetvalue{/flpy/stmt/mbfl/metallaxis/all/@1/java}{6}
\pgfkeyssetvalue{/flpy/stmt/mbfl/metallaxis/all/@3/java}{22}
\pgfkeyssetvalue{/flpy/stmt/mbfl/metallaxis/all/@5/java}{29}
\pgfkeyssetvalue{/flpy/stmt/mbfl/metallaxis/all/@10/java}{36}
\pgfkeyssetvalue{/flpy/stmt/mbfl/metallaxis/all/javaexam/java}{0.118}
\pgfkeyssetvalue{/flpy/stmt/mbfl/muse/all/@1/java}{7}
\pgfkeyssetvalue{/flpy/stmt/mbfl/muse/all/@3/java}{12}
\pgfkeyssetvalue{/flpy/stmt/mbfl/muse/all/@5/java}{16}
\pgfkeyssetvalue{/flpy/stmt/mbfl/muse/all/@10/java}{19}
\pgfkeyssetvalue{/flpy/stmt/mbfl/muse/all/javaexam/java}{0.304}
\pgfkeyssetvalue{/flpy/stmt/ps/favg/all/@1/java}{1}
\pgfkeyssetvalue{/flpy/stmt/ps/favg/all/@3/java}{4}
\pgfkeyssetvalue{/flpy/stmt/ps/favg/all/@5/java}{6}
\pgfkeyssetvalue{/flpy/stmt/ps/favg/all/@10/java}{6}
\pgfkeyssetvalue{/flpy/stmt/ps/favg/all/javaexam/java}{0.331}
\pgfkeyssetvalue{/flpy/stmt/st/favg/all/@1/java}{6}
\pgfkeyssetvalue{/flpy/stmt/st/favg/all/@3/java}{9}
\pgfkeyssetvalue{/flpy/stmt/st/favg/all/@5/java}{11}
\pgfkeyssetvalue{/flpy/stmt/st/favg/all/@10/java}{11}
\pgfkeyssetvalue{/flpy/stmt/st/favg/all/javaexam/java}{0.311}
\pgfkeyssetvalue{/flpy/stmt/comb/sbst/all/time}{591}
\pgfkeyssetvalue{/flpy/stmt/comb/sbst/all/@10}{56}
\pgfkeyssetvalue{/flpy/stmt/comb/sbst/all/javaexam}{0.003933670244659881}
\pgfkeyssetvalue{/flpy/stmt/comb/sbst/all/einspect}{1584}
\pgfkeyssetvalue{/flpy/stmt/comb/sbst/all/exam}{0.051843632951264776}
\pgfkeyssetvalue{/flpy/stmt/comb/sbst/all/@1}{12}
\pgfkeyssetvalue{/flpy/stmt/comb/sbst/all/@5}{41}
\pgfkeyssetvalue{/flpy/stmt/comb/sbst/all/@3}{32}
\pgfkeyssetvalue{/flpy/mod/comb/alfa/all/@10}{99}
\pgfkeyssetvalue{/flpy/mod/comb/alfa/all/javaexam}{0.033990342493138946}
\pgfkeyssetvalue{/flpy/mod/comb/alfa/all/einspect}{2}
\pgfkeyssetvalue{/flpy/mod/comb/alfa/all/exam}{0.03873051064858431}
\pgfkeyssetvalue{/flpy/mod/comb/alfa/all/@1}{70}
\pgfkeyssetvalue{/flpy/mod/comb/alfa/all/@5}{93}
\pgfkeyssetvalue{/flpy/mod/comb/alfa/all/@3}{89}
\pgfkeyssetvalue{/flpy/func/comb/sbst/all/@10}{79}
\pgfkeyssetvalue{/flpy/func/comb/sbst/all/javaexam}{0.015366033282319778}
\pgfkeyssetvalue{/flpy/func/comb/sbst/all/einspect}{67}
\pgfkeyssetvalue{/flpy/func/comb/sbst/all/exam}{0.07423039067992367}
\pgfkeyssetvalue{/flpy/func/comb/sbst/all/@1}{44}
\pgfkeyssetvalue{/flpy/func/comb/sbst/all/@5}{73}
\pgfkeyssetvalue{/flpy/func/comb/sbst/all/@3}{64}
\pgfkeyssetvalue{/flpy/func/comb/alfa/all/@10}{84}
\pgfkeyssetvalue{/flpy/func/comb/alfa/all/javaexam}{0.012833193698378412}
\pgfkeyssetvalue{/flpy/func/comb/alfa/all/einspect}{66}
\pgfkeyssetvalue{/flpy/func/comb/alfa/all/exam}{0.07188516884294091}
\pgfkeyssetvalue{/flpy/func/comb/alfa/all/@1}{53}
\pgfkeyssetvalue{/flpy/func/comb/alfa/all/@5}{77}
\pgfkeyssetvalue{/flpy/func/comb/alfa/all/@3}{70}
\pgfkeyssetvalue{/flpy/stmt/comb/alfa/all/@10/java}{52}
\pgfkeyssetvalue{/flpy/stmt/comb/alfa/all/javaexam/java}{0.018620908434973688}
\pgfkeyssetvalue{/flpy/stmt/comb/alfa/all/@1/java}{19}
\pgfkeyssetvalue{/flpy/stmt/comb/alfa/all/@5/java}{42}
\pgfkeyssetvalue{/flpy/stmt/comb/alfa/all/@3/java}{33}
\pgfkeyssetvalue{/flpy/stmt/comb/sbst/all/@10/java}{40}
\pgfkeyssetvalue{/flpy/stmt/comb/sbst/all/javaexam/java}{0.02646145339076598}
\pgfkeyssetvalue{/flpy/stmt/comb/sbst/all/@1/java}{10}
\pgfkeyssetvalue{/flpy/stmt/comb/sbst/all/@5/java}{30}
\pgfkeyssetvalue{/flpy/stmt/comb/sbst/all/@3/java}{23}
\pgfkeyssetvalue{/flpy/mod/comb/sbst/all/@10}{98}
\pgfkeyssetvalue{/flpy/mod/comb/sbst/all/javaexam}{0.03620025132540632}
\pgfkeyssetvalue{/flpy/mod/comb/sbst/all/einspect}{2}
\pgfkeyssetvalue{/flpy/mod/comb/sbst/all/exam}{0.040924049785797854}
\pgfkeyssetvalue{/flpy/mod/comb/sbst/all/@1}{64}
\pgfkeyssetvalue{/flpy/mod/comb/sbst/all/@5}{93}
\pgfkeyssetvalue{/flpy/mod/comb/sbst/all/@3}{87}
\pgfkeyssetvalue{/flpy/stmt/comb/alfa/all/time}{26116}
\pgfkeyssetvalue{/flpy/stmt/comb/alfa/all/@10}{60}
\pgfkeyssetvalue{/flpy/stmt/comb/alfa/all/javaexam}{0.0033443617631956917}
\pgfkeyssetvalue{/flpy/stmt/comb/alfa/all/einspect}{1580}
\pgfkeyssetvalue{/flpy/stmt/comb/alfa/all/exam}{0.05131980318996325}
\pgfkeyssetvalue{/flpy/stmt/comb/alfa/all/@1}{20}
\pgfkeyssetvalue{/flpy/stmt/comb/alfa/all/@5}{49}
\pgfkeyssetvalue{/flpy/stmt/comb/alfa/all/@3}{39}
\pgfkeyssetvalue{/flpy/stmt/comb/favg/all/@10}{58}
\pgfkeyssetvalue{/flpy/stmt/comb/favg/all/javaexam}{0.003639016003927786}
\pgfkeyssetvalue{/flpy/stmt/comb/favg/all/einspect}{1582}
\pgfkeyssetvalue{/flpy/stmt/comb/favg/all/exam}{0.05158171807061401}
\pgfkeyssetvalue{/flpy/stmt/comb/favg/all/@1}{16}
\pgfkeyssetvalue{/flpy/stmt/comb/favg/all/@5}{45}
\pgfkeyssetvalue{/flpy/stmt/comb/favg/all/@3}{35}
\pgfkeyssetvalue{/flpy/mod/comb/favg/all/@10}{98}
\pgfkeyssetvalue{/flpy/mod/comb/favg/all/javaexam}{0.035095296909272634}
\pgfkeyssetvalue{/flpy/mod/comb/favg/all/einspect}{2}
\pgfkeyssetvalue{/flpy/mod/comb/favg/all/exam}{0.03982728021719108}
\pgfkeyssetvalue{/flpy/mod/comb/favg/all/@1}{67}
\pgfkeyssetvalue{/flpy/mod/comb/favg/all/@5}{93}
\pgfkeyssetvalue{/flpy/mod/comb/favg/all/@3}{88}
\pgfkeyssetvalue{/flpy/func/comb/favg/all/@10}{81}
\pgfkeyssetvalue{/flpy/func/comb/favg/all/javaexam}{0.014099613490349095}
\pgfkeyssetvalue{/flpy/func/comb/favg/all/einspect}{66}
\pgfkeyssetvalue{/flpy/func/comb/favg/all/exam}{0.0730577797614323}
\pgfkeyssetvalue{/flpy/func/comb/favg/all/@1}{49}
\pgfkeyssetvalue{/flpy/func/comb/favg/all/@5}{75}
\pgfkeyssetvalue{/flpy/func/comb/favg/all/@3}{67}
\pgfkeyssetvalue{/flpy/stmt/comb/favg/all/@10/java}{46}
\pgfkeyssetvalue{/flpy/stmt/comb/favg/all/javaexam/java}{0.022541180912869836}
\pgfkeyssetvalue{/flpy/stmt/comb/favg/all/@1/java}{14}
\pgfkeyssetvalue{/flpy/stmt/comb/favg/all/@5/java}{36}
\pgfkeyssetvalue{/flpy/stmt/comb/favg/all/@3/java}{28}
\pgfkeyssetvalue{/flpy/stmt/comb/alfa/all/outputlength}{2548.362962962963}
\pgfkeyssetvalue{/flpy/stmt/comb/sbst/all/outputlength}{2548.362962962963}
\pgfkeyssetvalue{/flpy/stmt/comb/favg/all/outputlength}{2548.362962962963}
\pgfkeyssetvalue{/flpy/func/comb/alfa/all/outputlength}{296.34814814814814}
\pgfkeyssetvalue{/flpy/func/comb/sbst/all/outputlength}{296.34814814814814}
\pgfkeyssetvalue{/flpy/func/comb/favg/all/outputlength}{296.34814814814814}
\pgfkeyssetvalue{/flpy/mod/comb/alfa/all/outputlength}{20.918518518518518}
\pgfkeyssetvalue{/flpy/mod/comb/sbst/all/outputlength}{20.918518518518518}
\pgfkeyssetvalue{/flpy/mod/comb/favg/all/outputlength}{20.918518518518518}
\pgfkeyssetvalue{/flpy/stmt/alfafl/favg/all/einspect}{1577.5}
\pgfkeyssetvalue{/flpy/stmt/alfafl/favg/all/@1}{19.0}
\pgfkeyssetvalue{/flpy/stmt/alfafl/favg/all/@3}{37.5}
\pgfkeyssetvalue{/flpy/stmt/alfafl/favg/all/@5}{48.0}
\pgfkeyssetvalue{/flpy/stmt/alfafl/favg/all/@10}{59.5}
\pgfkeyssetvalue{/flpy/stmt/alfafl/favg/all/javaexam}{0.003323445305082064}
\pgfkeyssetvalue{/flpy/stmt/alfafl/favg/all/outputlength}{2548.362962962963}
\pgfkeyssetvalue{/flpy/stmt/sbstfl/favg/all/einspect}{1584.5}
\pgfkeyssetvalue{/flpy/stmt/sbstfl/favg/all/@1}{12.0}
\pgfkeyssetvalue{/flpy/stmt/sbstfl/favg/all/@3}{32.5}
\pgfkeyssetvalue{/flpy/stmt/sbstfl/favg/all/@5}{42.5}
\pgfkeyssetvalue{/flpy/stmt/sbstfl/favg/all/@10}{56.0}
\pgfkeyssetvalue{/flpy/stmt/sbstfl/favg/all/javaexam}{0.003946030935481276}
\pgfkeyssetvalue{/flpy/stmt/sbstfl/favg/all/outputlength}{2548.3481481481485}
\pgfkeyssetvalue{/flpy/func/alfafl/favg/all/einspect}{66.0}
\pgfkeyssetvalue{/flpy/func/alfafl/favg/all/@1}{53.0}
\pgfkeyssetvalue{/flpy/func/alfafl/favg/all/@3}{70.5}
\pgfkeyssetvalue{/flpy/func/alfafl/favg/all/@5}{76.5}
\pgfkeyssetvalue{/flpy/func/alfafl/favg/all/@10}{84.0}
\pgfkeyssetvalue{/flpy/func/alfafl/favg/all/javaexam}{0.012919761855964022}
\pgfkeyssetvalue{/flpy/func/alfafl/favg/all/outputlength}{296.34814814814814}
\pgfkeyssetvalue{/flpy/func/sbstfl/favg/all/einspect}{66.5}
\pgfkeyssetvalue{/flpy/func/sbstfl/favg/all/@1}{44.0}
\pgfkeyssetvalue{/flpy/func/sbstfl/favg/all/@3}{64.0}
\pgfkeyssetvalue{/flpy/func/sbstfl/favg/all/@5}{73.0}
\pgfkeyssetvalue{/flpy/func/sbstfl/favg/all/@10}{79.0}
\pgfkeyssetvalue{/flpy/func/sbstfl/favg/all/javaexam}{0.015317222329690433}
\pgfkeyssetvalue{/flpy/func/sbstfl/favg/all/outputlength}{296.34814814814814}
\pgfkeyssetvalue{/flpy/mod/alfafl/favg/all/einspect}{2.0}
\pgfkeyssetvalue{/flpy/mod/alfafl/favg/all/@1}{70.0}
\pgfkeyssetvalue{/flpy/mod/alfafl/favg/all/@3}{89.0}
\pgfkeyssetvalue{/flpy/mod/alfafl/favg/all/@5}{93.0}
\pgfkeyssetvalue{/flpy/mod/alfafl/favg/all/@10}{99.0}
\pgfkeyssetvalue{/flpy/mod/alfafl/favg/all/javaexam}{0.033917293217458525}
\pgfkeyssetvalue{/flpy/mod/alfafl/favg/all/outputlength}{20.918518518518518}
\pgfkeyssetvalue{/flpy/mod/sbstfl/favg/all/einspect}{2.0}
\pgfkeyssetvalue{/flpy/mod/sbstfl/favg/all/@1}{64.0}
\pgfkeyssetvalue{/flpy/mod/sbstfl/favg/all/@3}{87.0}
\pgfkeyssetvalue{/flpy/mod/sbstfl/favg/all/@5}{93.0}
\pgfkeyssetvalue{/flpy/mod/sbstfl/favg/all/@10}{98.0}
\pgfkeyssetvalue{/flpy/mod/sbstfl/favg/all/javaexam}{0.03623327206242918}
\pgfkeyssetvalue{/flpy/mod/sbstfl/favg/all/outputlength}{20.918518518518518}
\pgfkeyssetvalue{/flpy/test/all/max/n}{1036}
\pgfkeyssetvalue{/flpy/test/all/max/p}{100.0}
\pgfkeyssetvalue{/flpy/test/all/mean/n}{86.64444444444445}
\pgfkeyssetvalue{/flpy/test/all/mean/p}{4.5918127002072735}
\pgfkeyssetvalue{/flpy/test/all/median/n}{53}
\pgfkeyssetvalue{/flpy/test/all/median/p}{8.938740293356341}
\pgfkeyssetvalue{/flpy/test/all/min/n}{1}
\pgfkeyssetvalue{/flpy/test/all/min/p}{0.008731336767659128}
\pgfkeyssetvalue{/flpy/test/black/max/n}{129}
\pgfkeyssetvalue{/flpy/test/black/max/p}{93.47826086956522}
\pgfkeyssetvalue{/flpy/test/black/mean/n}{83.3076923076923}
\pgfkeyssetvalue{/flpy/test/black/mean/p}{91.23841617523168}
\pgfkeyssetvalue{/flpy/test/black/median/n}{91}
\pgfkeyssetvalue{/flpy/test/black/median/p}{91.0}
\pgfkeyssetvalue{/flpy/test/black/min/n}{16}
\pgfkeyssetvalue{/flpy/test/black/min/p}{88.46153846153845}
\pgfkeyssetvalue{/flpy/test/cli/max/n}{126}
\pgfkeyssetvalue{/flpy/test/cli/max/p}{100.0}
\pgfkeyssetvalue{/flpy/test/cli/mean/n}{40.55813953488372}
\pgfkeyssetvalue{/flpy/test/cli/mean/p}{17.859703020993344}
\pgfkeyssetvalue{/flpy/test/cli/median/n}{17}
\pgfkeyssetvalue{/flpy/test/cli/median/p}{15.151515151515152}
\pgfkeyssetvalue{/flpy/test/cli/min/n}{1}
\pgfkeyssetvalue{/flpy/test/cli/min/p}{0.6122448979591837}
\pgfkeyssetvalue{/flpy/test/cookiecutter/max/n}{60}
\pgfkeyssetvalue{/flpy/test/cookiecutter/max/p}{28.037383177570092}
\pgfkeyssetvalue{/flpy/test/cookiecutter/mean/n}{39.75}
\pgfkeyssetvalue{/flpy/test/cookiecutter/mean/p}{20.866141732283463}
\pgfkeyssetvalue{/flpy/test/cookiecutter/median/n}{44.0}
\pgfkeyssetvalue{/flpy/test/cookiecutter/median/p}{26.30627608825283}
\pgfkeyssetvalue{/flpy/test/cookiecutter/min/n}{11}
\pgfkeyssetvalue{/flpy/test/cookiecutter/min/p}{5.213270142180095}
\pgfkeyssetvalue{/flpy/test/dev/max/n}{198}
\pgfkeyssetvalue{/flpy/test/dev/max/p}{93.47826086956522}
\pgfkeyssetvalue{/flpy/test/dev/mean/n}{80.76666666666667}
\pgfkeyssetvalue{/flpy/test/dev/mean/p}{19.916159789577513}
\pgfkeyssetvalue{/flpy/test/dev/median/n}{80.0}
\pgfkeyssetvalue{/flpy/test/dev/median/p}{27.584583061653262}
\pgfkeyssetvalue{/flpy/test/dev/min/n}{2}
\pgfkeyssetvalue{/flpy/test/dev/min/p}{0.2242152466367713}
\pgfkeyssetvalue{/flpy/test/ds/max/n}{1036}
\pgfkeyssetvalue{/flpy/test/ds/max/p}{51.61290322580645}
\pgfkeyssetvalue{/flpy/test/ds/mean/n}{112.80952380952381}
\pgfkeyssetvalue{/flpy/test/ds/mean/p}{2.13487854442557}
\pgfkeyssetvalue{/flpy/test/ds/median/n}{67.5}
\pgfkeyssetvalue{/flpy/test/ds/median/p}{3.169367738415564}
\pgfkeyssetvalue{/flpy/test/ds/min/n}{1}
\pgfkeyssetvalue{/flpy/test/ds/min/p}{0.008731336767659128}
\pgfkeyssetvalue{/flpy/test/fastapi/max/n}{282}
\pgfkeyssetvalue{/flpy/test/fastapi/max/p}{49.91150442477876}
\pgfkeyssetvalue{/flpy/test/fastapi/mean/n}{37.61538461538461}
\pgfkeyssetvalue{/flpy/test/fastapi/mean/p}{8.801295896328295}
\pgfkeyssetvalue{/flpy/test/fastapi/median/n}{5}
\pgfkeyssetvalue{/flpy/test/fastapi/median/p}{1.5424164524421593}
\pgfkeyssetvalue{/flpy/test/fastapi/min/n}{1}
\pgfkeyssetvalue{/flpy/test/fastapi/min/p}{0.2506265664160401}
\pgfkeyssetvalue{/flpy/test/httpie/max/n}{17}
\pgfkeyssetvalue{/flpy/test/httpie/max/p}{100.0}
\pgfkeyssetvalue{/flpy/test/httpie/mean/n}{8.0}
\pgfkeyssetvalue{/flpy/test/httpie/mean/p}{7.191011235955057}
\pgfkeyssetvalue{/flpy/test/httpie/median/n}{7.0}
\pgfkeyssetvalue{/flpy/test/httpie/median/p}{4.658247968915577}
\pgfkeyssetvalue{/flpy/test/httpie/min/n}{1}
\pgfkeyssetvalue{/flpy/test/httpie/min/p}{0.7874015748031495}
\pgfkeyssetvalue{/flpy/test/keras/max/n}{288}
\pgfkeyssetvalue{/flpy/test/keras/max/p}{51.61290322580645}
\pgfkeyssetvalue{/flpy/test/keras/mean/n}{76.55555555555556}
\pgfkeyssetvalue{/flpy/test/keras/mean/p}{13.945956886954761}
\pgfkeyssetvalue{/flpy/test/keras/median/n}{58.0}
\pgfkeyssetvalue{/flpy/test/keras/median/p}{10.51003654604828}
\pgfkeyssetvalue{/flpy/test/keras/min/n}{18}
\pgfkeyssetvalue{/flpy/test/keras/min/p}{3.0100334448160537}
\pgfkeyssetvalue{/flpy/test/luigi/max/n}{198}
\pgfkeyssetvalue{/flpy/test/luigi/max/p}{33.22147651006711}
\pgfkeyssetvalue{/flpy/test/luigi/mean/n}{90.84615384615384}
\pgfkeyssetvalue{/flpy/test/luigi/mean/p}{11.559166095722814}
\pgfkeyssetvalue{/flpy/test/luigi/median/n}{91}
\pgfkeyssetvalue{/flpy/test/luigi/median/p}{11.303744798890431}
\pgfkeyssetvalue{/flpy/test/luigi/min/n}{2}
\pgfkeyssetvalue{/flpy/test/luigi/min/p}{0.2242152466367713}
\pgfkeyssetvalue{/flpy/test/pandas/max/n}{1036}
\pgfkeyssetvalue{/flpy/test/pandas/max/p}{8.938740293356341}
\pgfkeyssetvalue{/flpy/test/pandas/mean/n}{159.83333333333334}
\pgfkeyssetvalue{/flpy/test/pandas/mean/p}{1.3935915134781662}
\pgfkeyssetvalue{/flpy/test/pandas/median/n}{91.5}
\pgfkeyssetvalue{/flpy/test/pandas/median/p}{0.7735426408340571}
\pgfkeyssetvalue{/flpy/test/pandas/min/n}{1}
\pgfkeyssetvalue{/flpy/test/pandas/min/p}{0.008731336767659128}
\pgfkeyssetvalue{/flpy/test/sanic/max/n}{298}
\pgfkeyssetvalue{/flpy/test/sanic/max/p}{64.22413793103449}
\pgfkeyssetvalue{/flpy/test/sanic/mean/n}{220.33333333333334}
\pgfkeyssetvalue{/flpy/test/sanic/mean/p}{47.34957020057307}
\pgfkeyssetvalue{/flpy/test/sanic/median/n}{265}
\pgfkeyssetvalue{/flpy/test/sanic/median/p}{56.50319829424307}
\pgfkeyssetvalue{/flpy/test/sanic/min/n}{98}
\pgfkeyssetvalue{/flpy/test/sanic/min/p}{21.166306695464364}
\pgfkeyssetvalue{/flpy/test/spacy/max/n}{152}
\pgfkeyssetvalue{/flpy/test/spacy/max/p}{16.890380313199106}
\pgfkeyssetvalue{/flpy/test/spacy/mean/n}{80.5}
\pgfkeyssetvalue{/flpy/test/spacy/mean/p}{8.614232209737828}
\pgfkeyssetvalue{/flpy/test/spacy/median/n}{75.0}
\pgfkeyssetvalue{/flpy/test/spacy/median/p}{7.8189486028731885}
\pgfkeyssetvalue{/flpy/test/spacy/min/n}{13}
\pgfkeyssetvalue{/flpy/test/spacy/min/p}{1.3713080168776373}
\pgfkeyssetvalue{/flpy/test/thefuck/max/n}{18}
\pgfkeyssetvalue{/flpy/test/thefuck/max/p}{5.426356589147287}
\pgfkeyssetvalue{/flpy/test/thefuck/mean/n}{7.375}
\pgfkeyssetvalue{/flpy/test/thefuck/mean/p}{1.953965888392118}
\pgfkeyssetvalue{/flpy/test/thefuck/median/n}{5.0}
\pgfkeyssetvalue{/flpy/test/thefuck/median/p}{1.7137534627375657}
\pgfkeyssetvalue{/flpy/test/thefuck/min/n}{3}
\pgfkeyssetvalue{/flpy/test/thefuck/min/p}{0.6122448979591837}
\pgfkeyssetvalue{/flpy/test/tornado/max/n}{787}
\pgfkeyssetvalue{/flpy/test/tornado/max/p}{76.7056530214425}
\pgfkeyssetvalue{/flpy/test/tornado/mean/n}{410.5}
\pgfkeyssetvalue{/flpy/test/tornado/mean/p}{41.88775510204082}
\pgfkeyssetvalue{/flpy/test/tornado/median/n}{411.5}
\pgfkeyssetvalue{/flpy/test/tornado/median/p}{40.481153747979235}
\pgfkeyssetvalue{/flpy/test/tornado/min/n}{32}
\pgfkeyssetvalue{/flpy/test/tornado/min/p}{3.361344537815126}
\pgfkeyssetvalue{/flpy/test/tqdm/max/n}{77}
\pgfkeyssetvalue{/flpy/test/tqdm/max/p}{100.0}
\pgfkeyssetvalue{/flpy/test/tqdm/mean/n}{47.857142857142854}
\pgfkeyssetvalue{/flpy/test/tqdm/mean/p}{82.1078431372549}
\pgfkeyssetvalue{/flpy/test/tqdm/median/n}{63}
\pgfkeyssetvalue{/flpy/test/tqdm/median/p}{95.23809523809523}
\pgfkeyssetvalue{/flpy/test/tqdm/min/n}{1}
\pgfkeyssetvalue{/flpy/test/tqdm/min/p}{1.6666666666666667}
\pgfkeyssetvalue{/flpy/test/web/max/n}{787}
\pgfkeyssetvalue{/flpy/test/web/max/p}{76.7056530214425}
\pgfkeyssetvalue{/flpy/test/web/mean/n}{139.6}
\pgfkeyssetvalue{/flpy/test/web/mean/p}{25.680647534952172}
\pgfkeyssetvalue{/flpy/test/web/median/n}{32.0}
\pgfkeyssetvalue{/flpy/test/web/median/p}{4.458177002869124}
\pgfkeyssetvalue{/flpy/test/web/min/n}{1}
\pgfkeyssetvalue{/flpy/test/web/min/p}{0.2506265664160401}
\pgfkeyssetvalue{/flpy/test/youtube-dl/max/n}{126}
\pgfkeyssetvalue{/flpy/test/youtube-dl/max/p}{55.75221238938053}
\pgfkeyssetvalue{/flpy/test/youtube-dl/mean/n}{78.6875}
\pgfkeyssetvalue{/flpy/test/youtube-dl/mean/p}{43.82178907065785}
\pgfkeyssetvalue{/flpy/test/youtube-dl/median/n}{89.5}
\pgfkeyssetvalue{/flpy/test/youtube-dl/median/p}{51.03860539674058}
\pgfkeyssetvalue{/flpy/test/youtube-dl/min/n}{15}
\pgfkeyssetvalue{/flpy/test/youtube-dl/min/p}{14.285714285714285} %
\pgfkeyssetvalue{/flpy/num extra tests threshold}{50}
\pgfkeyssetvalue{/flpy/max extra tests}{642}
\pgfkeyssetvalue{/flpy/zero transitive delta}{0.481481481481481}
\pgfkeyssetvalue{/flpy/10percent transitive delta}{0.696296296296296}
\pgfkeyssetvalue{/flpy/perc more than threshold transitive delta}{0.296296296296296}
\pgfkeyssetvalue{/flpy/num more than threshold transitive delta}{40}
\pgfkeyssetvalue{/flpy/projects more than threshold transitive delta}{7}
\pgfkeyssetvalue{/flpy/extra tests new experiments}{35}
\pgfkeyssetvalue{/flpy/extra experiments no change}{20}
\pgfkeyssetvalue{/flpy/extra experiments top-1 change}{0}
\pgfkeyssetvalue{/flpy/extra experiments top-3 change}{0}
\pgfkeyssetvalue{/flpy/extra experiments top-5 change}{0}
\pgfkeyssetvalue{/flpy/extra experiments top-10 change}{-1}
\pgfkeyssetvalue{/flpy/extra experiments top-10 change einspect nontransitive}{13.3333333333333}
\pgfkeyssetvalue{/flpy/extra experiments top-10 change einspect transitive}{9.33333333333333}
\pgfkeyssetvalue{/flpy/discussion/python/java/e_inspect/dstar/pval}{2.79690539361965e-06}
\pgfkeyssetvalue{/flpy/discussion/python/java/e_inspect/dstar/cohend}{-0.273513849984438}
\pgfkeyssetvalue{/flpy/discussion/python/java/e_inspect/ochiai/pval}{1.44410933105894e-06}
\pgfkeyssetvalue{/flpy/discussion/python/java/e_inspect/ochiai/cohend}{-0.281315489158626}
\pgfkeyssetvalue{/flpy/discussion/python/java/exam/dstar/pval}{4.80227415794797e-28}
\pgfkeyssetvalue{/flpy/discussion/python/java/exam/dstar/cohend}{-0.641103848946986}
\pgfkeyssetvalue{/flpy/discussion/python/java/exam/ochiai/pval}{5.0343356159017e-28}
\pgfkeyssetvalue{/flpy/discussion/python/java/exam/ochiai/cohend}{-0.640854860462704}
\pgfkeyssetvalue{/flpy/discussion/python/java/@5/dstar/pval}{0.0203315101206638}
\pgfkeyssetvalue{/flpy/discussion/python/java/@5/dstar/cohend}{0.111297852474323}
\pgfkeyssetvalue{/flpy/discussion/python/java/@5/ochiai/pval}{0.0134696328365303}
\pgfkeyssetvalue{/flpy/discussion/python/java/@5/ochiai/cohend}{0.11870525988173}
\pgfkeyssetvalue{/flpy/discussion/python/java/@10/dstar/pval}{0.0346377109968765}
\pgfkeyssetvalue{/flpy/discussion/python/java/@10/dstar/cohend}{0.106567071272954}
\pgfkeyssetvalue{/flpy/discussion/python/java/@10/ochiai/pval}{0.0397409905158073}
\pgfkeyssetvalue{/flpy/discussion/python/java/@10/ochiai/cohend}{0.103765950824774}
\pgfkeyssetvalue{/flpy/discussion/dstar/ochiai/e_inspect/pval}{0.361310428526179}
\pgfkeyssetvalue{/flpy/discussion/dstar/ochiai/e_inspect/cohend}{0.00674897119341565}
\pgfkeyssetvalue{/flpy/discussion/dstar/tarantula/e_inspect/pval}{0.0865712192909329}
\pgfkeyssetvalue{/flpy/discussion/dstar/tarantula/e_inspect/cohend}{-0.00126200274348421}
\pgfkeyssetvalue{/flpy/discussion/ochiai/dstar/e_inspect/pval}{0.361310428526179}
\pgfkeyssetvalue{/flpy/discussion/ochiai/dstar/e_inspect/cohend}{-0.00674897119341563}
\pgfkeyssetvalue{/flpy/discussion/ochiai/tarantula/e_inspect/pval}{0.0407602539105694}
\pgfkeyssetvalue{/flpy/discussion/ochiai/tarantula/e_inspect/cohend}{-0.00801097393689985}
\pgfkeyssetvalue{/flpy/discussion/tarantula/dstar/e_inspect/pval}{0.0865712192909329}
\pgfkeyssetvalue{/flpy/discussion/tarantula/dstar/e_inspect/cohend}{0.00126200274348423}
\pgfkeyssetvalue{/flpy/discussion/tarantula/ochiai/e_inspect/pval}{0.0407602539105694}
\pgfkeyssetvalue{/flpy/discussion/tarantula/ochiai/e_inspect/cohend}{0.00801097393689988}
\pgfkeyssetvalue{/flpy/discussion/dstar/ochiai/exam/pval}{0.583882420770365}
\pgfkeyssetvalue{/flpy/discussion/dstar/ochiai/exam/cohend}{0.00241426611796983}
\pgfkeyssetvalue{/flpy/discussion/dstar/tarantula/exam/pval}{0.0934924836898357}
\pgfkeyssetvalue{/flpy/discussion/dstar/tarantula/exam/cohend}{-0.00949245541838134}
\pgfkeyssetvalue{/flpy/discussion/ochiai/dstar/exam/pval}{0.583882420770365}
\pgfkeyssetvalue{/flpy/discussion/ochiai/dstar/exam/cohend}{-0.00241426611796982}
\pgfkeyssetvalue{/flpy/discussion/ochiai/tarantula/exam/pval}{0.0559296803251122}
\pgfkeyssetvalue{/flpy/discussion/ochiai/tarantula/exam/cohend}{-0.0119067215363512}
\pgfkeyssetvalue{/flpy/discussion/tarantula/dstar/exam/pval}{0.0934924836898357}
\pgfkeyssetvalue{/flpy/discussion/tarantula/dstar/exam/cohend}{0.00949245541838135}
\pgfkeyssetvalue{/flpy/discussion/tarantula/ochiai/exam/pval}{0.0559296803251122}
\pgfkeyssetvalue{/flpy/discussion/tarantula/ochiai/exam/cohend}{0.0119067215363512}
\pgfkeyssetvalue{/flpy/discussion/python/java/e_inspect/dstar/esmagnitude}{s}
\pgfkeyssetvalue{/flpy/discussion/python/e_inspect/dstar/mean}{85.1922751322751}
\pgfkeyssetvalue{/flpy/discussion/java/e_inspect/dstar/mean}{177.061174756479}
\pgfkeyssetvalue{/flpy/discussion/python/java/e_inspect/ochiai/esmagnitude}{s}
\pgfkeyssetvalue{/flpy/discussion/python/e_inspect/ochiai/mean}{85.5330158730159}
\pgfkeyssetvalue{/flpy/discussion/java/e_inspect/ochiai/mean}{177.985777931082}
\pgfkeyssetvalue{/flpy/discussion/python/java/exam/dstar/esmagnitude}{l}
\pgfkeyssetvalue{/flpy/discussion/python/exam/dstar/mean}{-0.107418467292241}
\pgfkeyssetvalue{/flpy/discussion/java/exam/dstar/mean}{0.000987470636658941}
\pgfkeyssetvalue{/flpy/discussion/python/java/exam/ochiai/esmagnitude}{l}
\pgfkeyssetvalue{/flpy/discussion/python/exam/ochiai/mean}{-0.107416973842509}
\pgfkeyssetvalue{/flpy/discussion/java/exam/ochiai/mean}{0.000935287809059463}
\pgfkeyssetvalue{/flpy/discussion/python/java/@5/dstar/esmagnitude}{n}
\pgfkeyssetvalue{/flpy/discussion/python/@5/dstar/mean}{0.422222222222222}
\pgfkeyssetvalue{/flpy/discussion/java/@5/dstar/mean}{0.310924369747899}
\pgfkeyssetvalue{/flpy/discussion/python/java/@5/ochiai/esmagnitude}{n}
\pgfkeyssetvalue{/flpy/discussion/python/@5/ochiai/mean}{0.42962962962963}
\pgfkeyssetvalue{/flpy/discussion/java/@5/ochiai/mean}{0.310924369747899}
\pgfkeyssetvalue{/flpy/discussion/python/java/@10/dstar/esmagnitude}{n}
\pgfkeyssetvalue{/flpy/discussion/python/@10/dstar/mean}{0.540740740740741}
\pgfkeyssetvalue{/flpy/discussion/java/@10/dstar/mean}{0.434173669467787}
\pgfkeyssetvalue{/flpy/discussion/python/java/@10/ochiai/esmagnitude}{n}
\pgfkeyssetvalue{/flpy/discussion/python/@10/ochiai/mean}{0.540740740740741}
\pgfkeyssetvalue{/flpy/discussion/java/@10/ochiai/mean}{0.436974789915966}
\pgfkeyssetvalue{/flpy/discussion/dstar/ochiai/e_inspect/esmagnitude}{n}
\pgfkeyssetvalue{/flpy/discussion/dstar/tarantula/e_inspect/esmagnitude}{n}
\pgfkeyssetvalue{/flpy/discussion/ochiai/dstar/e_inspect/esmagnitude}{n}
\pgfkeyssetvalue{/flpy/discussion/ochiai/tarantula/e_inspect/esmagnitude}{n}
\pgfkeyssetvalue{/flpy/discussion/tarantula/dstar/e_inspect/esmagnitude}{n}
\pgfkeyssetvalue{/flpy/discussion/tarantula/ochiai/e_inspect/esmagnitude}{n}
\pgfkeyssetvalue{/flpy/discussion/dstar/ochiai/exam/esmagnitude}{n}
\pgfkeyssetvalue{/flpy/discussion/dstar/tarantula/exam/esmagnitude}{n}
\pgfkeyssetvalue{/flpy/discussion/ochiai/dstar/exam/esmagnitude}{n}
\pgfkeyssetvalue{/flpy/discussion/ochiai/tarantula/exam/esmagnitude}{n}
\pgfkeyssetvalue{/flpy/discussion/tarantula/dstar/exam/esmagnitude}{n}
\pgfkeyssetvalue{/flpy/discussion/tarantula/ochiai/exam/esmagnitude}{n}
\pgfkeyssetvalue{/flpy/statement/einspect/MBFL:PS/corrTau}{0.0748108683990909}
\pgfkeyssetvalue{/flpy/statement/einspect/MBFL:PS/WilcoxonP}{7.44970812013263e-20}
\pgfkeyssetvalue{/flpy/statement/einspect/MBFL:PS/CliffD}{-0.697997256515775}
\pgfkeyssetvalue{/flpy/statement/einspect/MBFL:ST/corrTau}{-0.00551600728116301}
\pgfkeyssetvalue{/flpy/statement/einspect/MBFL:ST/WilcoxonP}{4.69632453473045e-14}
\pgfkeyssetvalue{/flpy/statement/einspect/MBFL:ST/CliffD}{-0.453772290809328}
\pgfkeyssetvalue{/flpy/statement/einspect/MBFL:SBFL/corrTau}{0.540395240400838}
\pgfkeyssetvalue{/flpy/statement/einspect/MBFL:SBFL/WilcoxonP}{0.00520931038798318}
\pgfkeyssetvalue{/flpy/statement/einspect/MBFL:SBFL/CliffD}{0.176186556927298}
\pgfkeyssetvalue{/flpy/statement/einspect/PS:ST/corrTau}{0.146485617820881}
\pgfkeyssetvalue{/flpy/statement/einspect/PS:ST/WilcoxonP}{0.0116745886019255}
\pgfkeyssetvalue{/flpy/statement/einspect/PS:ST/CliffD}{0.183539094650206}
\pgfkeyssetvalue{/flpy/statement/einspect/PS:SBFL/corrTau}{0.0570253731815212}
\pgfkeyssetvalue{/flpy/statement/einspect/PS:SBFL/WilcoxonP}{9.17718332925928e-21}
\pgfkeyssetvalue{/flpy/statement/einspect/PS:SBFL/CliffD}{0.783758573388203}
\pgfkeyssetvalue{/flpy/statement/einspect/ST:SBFL/corrTau}{0.0667513688819089}
\pgfkeyssetvalue{/flpy/statement/einspect/ST:SBFL/WilcoxonP}{3.57426094089792e-17}
\pgfkeyssetvalue{/flpy/statement/einspect/ST:SBFL/CliffD}{0.562194787379973}
\pgfkeyssetvalue{/flpy/statement/time/MBFL:PS/corrTau}{0.561967938087341}
\pgfkeyssetvalue{/flpy/statement/time/MBFL:PS/WilcoxonP}{5.76724770171956e-08}
\pgfkeyssetvalue{/flpy/statement/time/MBFL:PS/CliffD}{0.336954732510288}
\pgfkeyssetvalue{/flpy/statement/time/MBFL:ST/corrTau}{0.179878385848535}
\pgfkeyssetvalue{/flpy/statement/time/MBFL:ST/WilcoxonP}{6.78946425811962e-24}
\pgfkeyssetvalue{/flpy/statement/time/MBFL:ST/CliffD}{0.998353909465021}
\pgfkeyssetvalue{/flpy/statement/time/MBFL:SBFL/corrTau}{0.592482034273079}
\pgfkeyssetvalue{/flpy/statement/time/MBFL:SBFL/WilcoxonP}{7.54381658530857e-23}
\pgfkeyssetvalue{/flpy/statement/time/MBFL:SBFL/CliffD}{0.766913580246914}
\pgfkeyssetvalue{/flpy/statement/time/PS:ST/corrTau}{0.085461580983969}
\pgfkeyssetvalue{/flpy/statement/time/PS:ST/WilcoxonP}{7.42416508740563e-24}
\pgfkeyssetvalue{/flpy/statement/time/PS:ST/CliffD}{0.96323731138546}
\pgfkeyssetvalue{/flpy/statement/time/PS:SBFL/corrTau}{0.324267551133223}
\pgfkeyssetvalue{/flpy/statement/time/PS:SBFL/WilcoxonP}{1.96521841180797e-14}
\pgfkeyssetvalue{/flpy/statement/time/PS:SBFL/CliffD}{0.485432098765432}
\pgfkeyssetvalue{/flpy/statement/time/ST:SBFL/corrTau}{0.200884466556108}
\pgfkeyssetvalue{/flpy/statement/time/ST:SBFL/WilcoxonP}{8.48796405900451e-24}
\pgfkeyssetvalue{/flpy/statement/time/ST:SBFL/CliffD}{-0.864362139917696}
\pgfkeyssetvalue{/flpy/statement-intra-SBFL/einspect/DStar:Ochiai/corrTau}{0.993881655156452}
\pgfkeyssetvalue{/flpy/statement-intra-SBFL/einspect/DStar:Ochiai/WilcoxonP}{1}
\pgfkeyssetvalue{/flpy/statement-intra-SBFL/einspect/DStar:Ochiai/CliffD}{0.00219478737997256}
\pgfkeyssetvalue{/flpy/statement-intra-SBFL/einspect/DStar:Tarantula/corrTau}{0.936853579376171}
\pgfkeyssetvalue{/flpy/statement-intra-SBFL/einspect/DStar:Tarantula/WilcoxonP}{0.0908887852457271}
\pgfkeyssetvalue{/flpy/statement-intra-SBFL/einspect/DStar:Tarantula/CliffD}{-0.0205761316872428}
\pgfkeyssetvalue{/flpy/statement-intra-SBFL/einspect/Ochiai:Tarantula/corrTau}{0.943224617816058}
\pgfkeyssetvalue{/flpy/statement-intra-SBFL/einspect/Ochiai:Tarantula/WilcoxonP}{0.0584752615656529}
\pgfkeyssetvalue{/flpy/statement-intra-SBFL/einspect/Ochiai:Tarantula/CliffD}{-0.0227709190672154}
\pgfkeyssetvalue{/flpy/statement-intra-MBFL/einspect/Metallaxis:Muse/corrTau}{0.615555642276624}
\pgfkeyssetvalue{/flpy/statement-intra-MBFL/einspect/Metallaxis:Muse/WilcoxonP}{0.00366191684658065}
\pgfkeyssetvalue{/flpy/statement-intra-MBFL/einspect/Metallaxis:Muse/CliffD}{-0.10803840877915}
\pgfkeyssetvalue{/flpy/m6/mutability/mbfl/significant}{0.87}
\pgfkeyssetvalue{/flpy/m2/einspect/family/lower 50:MBFL}{-2.5262775}
\pgfkeyssetvalue{/flpy/m2/einspect/family/lower 50:PS}{1.117685}
\pgfkeyssetvalue{/flpy/m2/einspect/family/lower 50:ST}{1.2182375}
\pgfkeyssetvalue{/flpy/m2/einspect/family/lower 50:SBFL}{-2.88013}
\pgfkeyssetvalue{/flpy/m2/einspect/family/lower 70:MBFL}{-2.7709055}
\pgfkeyssetvalue{/flpy/m2/einspect/family/lower 70:PS}{0.94775795}
\pgfkeyssetvalue{/flpy/m2/einspect/family/lower 70:ST}{1.0616545}
\pgfkeyssetvalue{/flpy/m2/einspect/family/lower 70:SBFL}{-3.1258475}
\pgfkeyssetvalue{/flpy/m2/einspect/family/lower 90:MBFL}{-3.2180915}
\pgfkeyssetvalue{/flpy/m2/einspect/family/lower 90:PS}{0.6775959}
\pgfkeyssetvalue{/flpy/m2/einspect/family/lower 90:ST}{0.7788652}
\pgfkeyssetvalue{/flpy/m2/einspect/family/lower 90:SBFL}{-3.5455925}
\pgfkeyssetvalue{/flpy/m2/einspect/family/lower 95:MBFL}{-3.51031975}
\pgfkeyssetvalue{/flpy/m2/einspect/family/lower 95:PS}{0.535925525}
\pgfkeyssetvalue{/flpy/m2/einspect/family/lower 95:ST}{0.644767625}
\pgfkeyssetvalue{/flpy/m2/einspect/family/lower 95:SBFL}{-3.76600675}
\pgfkeyssetvalue{/flpy/m2/einspect/family/lower 99:MBFL}{-4.0641862}
\pgfkeyssetvalue{/flpy/m2/einspect/family/lower 99:PS}{0.232960895}
\pgfkeyssetvalue{/flpy/m2/einspect/family/lower 99:ST}{0.33158403}
\pgfkeyssetvalue{/flpy/m2/einspect/family/lower 99:SBFL}{-4.2131177}
\pgfkeyssetvalue{/flpy/m2/einspect/family/upper 99:MBFL}{-0.702209474999999}
\pgfkeyssetvalue{/flpy/m2/einspect/family/upper 99:PS}{2.59205895}
\pgfkeyssetvalue{/flpy/m2/einspect/family/upper 99:ST}{2.70007475}
\pgfkeyssetvalue{/flpy/m2/einspect/family/upper 99:SBFL}{-0.978632644999998}
\pgfkeyssetvalue{/flpy/m2/einspect/family/upper 95:MBFL}{-1.0036095}
\pgfkeyssetvalue{/flpy/m2/einspect/family/upper 95:PS}{2.30851975}
\pgfkeyssetvalue{/flpy/m2/einspect/family/upper 95:ST}{2.4006195}
\pgfkeyssetvalue{/flpy/m2/einspect/family/upper 95:SBFL}{-1.33387875}
\pgfkeyssetvalue{/flpy/m2/einspect/family/upper 90:MBFL}{-1.1621255}
\pgfkeyssetvalue{/flpy/m2/einspect/family/upper 90:PS}{2.153348}
\pgfkeyssetvalue{/flpy/m2/einspect/family/upper 90:ST}{2.257362}
\pgfkeyssetvalue{/flpy/m2/einspect/family/upper 90:SBFL}{-1.514786}
\pgfkeyssetvalue{/flpy/m2/einspect/family/upper 70:MBFL}{-1.4888745}
\pgfkeyssetvalue{/flpy/m2/einspect/family/upper 70:PS}{1.862479}
\pgfkeyssetvalue{/flpy/m2/einspect/family/upper 70:ST}{1.977887}
\pgfkeyssetvalue{/flpy/m2/einspect/family/upper 70:SBFL}{-1.837451}
\pgfkeyssetvalue{/flpy/m2/einspect/family/upper 50:MBFL}{-1.6824225}
\pgfkeyssetvalue{/flpy/m2/einspect/family/upper 50:PS}{1.703535}
\pgfkeyssetvalue{/flpy/m2/einspect/family/upper 50:ST}{1.810125}
\pgfkeyssetvalue{/flpy/m2/einspect/family/upper 50:SBFL}{-2.0364775}
\pgfkeyssetvalue{/flpy/m2/einspect/family/MBFL}{-2.1329160825}
\pgfkeyssetvalue{/flpy/m2/einspect/family/PS}{1.411442171875}
\pgfkeyssetvalue{/flpy/m2/einspect/family/ST}{1.5187468519}
\pgfkeyssetvalue{/flpy/m2/einspect/family/SBFL}{-2.4731523275}
\pgfkeyssetvalue{/flpy/m2/einspect/category/lower 50:CL}{0.8576515}
\pgfkeyssetvalue{/flpy/m2/einspect/category/lower 50:DEV}{-1.5181075}
\pgfkeyssetvalue{/flpy/m2/einspect/category/lower 50:DS}{0.1622345}
\pgfkeyssetvalue{/flpy/m2/einspect/category/lower 50:WEB}{-1.3309425}
\pgfkeyssetvalue{/flpy/m2/einspect/category/lower 70:CL}{0.73787805}
\pgfkeyssetvalue{/flpy/m2/einspect/category/lower 70:DEV}{-1.6876235}
\pgfkeyssetvalue{/flpy/m2/einspect/category/lower 70:DS}{0.04688615}
\pgfkeyssetvalue{/flpy/m2/einspect/category/lower 70:WEB}{-1.5082515}
\pgfkeyssetvalue{/flpy/m2/einspect/category/lower 90:CL}{0.53094055}
\pgfkeyssetvalue{/flpy/m2/einspect/category/lower 90:DEV}{-2.0290295}
\pgfkeyssetvalue{/flpy/m2/einspect/category/lower 90:DS}{-0.1804682}
\pgfkeyssetvalue{/flpy/m2/einspect/category/lower 90:WEB}{-1.894603}
\pgfkeyssetvalue{/flpy/m2/einspect/category/lower 95:CL}{0.40753025}
\pgfkeyssetvalue{/flpy/m2/einspect/category/lower 95:DEV}{-2.22066825}
\pgfkeyssetvalue{/flpy/m2/einspect/category/lower 95:DS}{-0.31008255}
\pgfkeyssetvalue{/flpy/m2/einspect/category/lower 95:WEB}{-2.123565}
\pgfkeyssetvalue{/flpy/m2/einspect/category/lower 99:CL}{0.17324546}
\pgfkeyssetvalue{/flpy/m2/einspect/category/lower 99:DEV}{-2.64810525}
\pgfkeyssetvalue{/flpy/m2/einspect/category/lower 99:DS}{-0.55104856}
\pgfkeyssetvalue{/flpy/m2/einspect/category/lower 99:WEB}{-2.5519435}
\pgfkeyssetvalue{/flpy/m2/einspect/category/upper 99:CL}{2.07250355}
\pgfkeyssetvalue{/flpy/m2/einspect/category/upper 99:DEV}{-0.177008155}
\pgfkeyssetvalue{/flpy/m2/einspect/category/upper 99:DS}{1.3620486}
\pgfkeyssetvalue{/flpy/m2/einspect/category/upper 99:WEB}{0.0438689545000003}
\pgfkeyssetvalue{/flpy/m2/einspect/category/upper 95:CL}{1.7999425}
\pgfkeyssetvalue{/flpy/m2/einspect/category/upper 95:DEV}{-0.40564255}
\pgfkeyssetvalue{/flpy/m2/einspect/category/upper 95:DS}{1.12326375}
\pgfkeyssetvalue{/flpy/m2/einspect/category/upper 95:WEB}{-0.193428325}
\pgfkeyssetvalue{/flpy/m2/einspect/category/upper 90:CL}{1.6730295}
\pgfkeyssetvalue{/flpy/m2/einspect/category/upper 90:DEV}{-0.52403775}
\pgfkeyssetvalue{/flpy/m2/einspect/category/upper 90:DS}{0.99301}
\pgfkeyssetvalue{/flpy/m2/einspect/category/upper 90:WEB}{-0.31669545}
\pgfkeyssetvalue{/flpy/m2/einspect/category/upper 70:CL}{1.4488815}
\pgfkeyssetvalue{/flpy/m2/einspect/category/upper 70:DEV}{-0.7515989}
\pgfkeyssetvalue{/flpy/m2/einspect/category/upper 70:DS}{0.7566167}
\pgfkeyssetvalue{/flpy/m2/einspect/category/upper 70:WEB}{-0.5624215}
\pgfkeyssetvalue{/flpy/m2/einspect/category/upper 50:CL}{1.3150325}
\pgfkeyssetvalue{/flpy/m2/einspect/category/upper 50:DEV}{-0.9015535}
\pgfkeyssetvalue{/flpy/m2/einspect/category/upper 50:DS}{0.621022}
\pgfkeyssetvalue{/flpy/m2/einspect/category/upper 50:WEB}{-0.70125675}
\pgfkeyssetvalue{/flpy/m2/einspect/category/CL}{1.089910848175}
\pgfkeyssetvalue{/flpy/m2/einspect/category/DEV}{-1.225378347225}
\pgfkeyssetvalue{/flpy/m2/einspect/category/DS}{0.393016353284575}
\pgfkeyssetvalue{/flpy/m2/einspect/category/WEB}{-1.04013070633775}
\pgfkeyssetvalue{/flpy/m2/time/family/lower 50:MBFL}{1.5663375}
\pgfkeyssetvalue{/flpy/m2/time/family/lower 50:PS}{0.87735675}
\pgfkeyssetvalue{/flpy/m2/time/family/lower 50:ST}{-2.946875}
\pgfkeyssetvalue{/flpy/m2/time/family/lower 50:SBFL}{-2.48673}
\pgfkeyssetvalue{/flpy/m2/time/family/lower 70:MBFL}{1.407346}
\pgfkeyssetvalue{/flpy/m2/time/family/lower 70:PS}{0.7103954}
\pgfkeyssetvalue{/flpy/m2/time/family/lower 70:ST}{-3.2094265}
\pgfkeyssetvalue{/flpy/m2/time/family/lower 70:SBFL}{-2.733209}
\pgfkeyssetvalue{/flpy/m2/time/family/lower 90:MBFL}{1.1276775}
\pgfkeyssetvalue{/flpy/m2/time/family/lower 90:PS}{0.425159}
\pgfkeyssetvalue{/flpy/m2/time/family/lower 90:ST}{-3.6871695}
\pgfkeyssetvalue{/flpy/m2/time/family/lower 90:SBFL}{-3.182601}
\pgfkeyssetvalue{/flpy/m2/time/family/lower 95:MBFL}{0.967736825}
\pgfkeyssetvalue{/flpy/m2/time/family/lower 95:PS}{0.245108975}
\pgfkeyssetvalue{/flpy/m2/time/family/lower 95:ST}{-3.92562525}
\pgfkeyssetvalue{/flpy/m2/time/family/lower 95:SBFL}{-3.45412175}
\pgfkeyssetvalue{/flpy/m2/time/family/lower 99:MBFL}{0.626437305}
\pgfkeyssetvalue{/flpy/m2/time/family/lower 99:PS}{-0.0473359034999997}
\pgfkeyssetvalue{/flpy/m2/time/family/lower 99:ST}{-4.3522941}
\pgfkeyssetvalue{/flpy/m2/time/family/lower 99:SBFL}{-4.0203258}
\pgfkeyssetvalue{/flpy/m2/time/family/upper 99:MBFL}{2.97825215}
\pgfkeyssetvalue{/flpy/m2/time/family/upper 99:PS}{2.31325605}
\pgfkeyssetvalue{/flpy/m2/time/family/upper 99:ST}{-1.0736534}
\pgfkeyssetvalue{/flpy/m2/time/family/upper 99:SBFL}{-0.523117094999996}
\pgfkeyssetvalue{/flpy/m2/time/family/upper 95:MBFL}{2.70623925}
\pgfkeyssetvalue{/flpy/m2/time/family/upper 95:PS}{2.0606435}
\pgfkeyssetvalue{/flpy/m2/time/family/upper 95:ST}{-1.43820975}
\pgfkeyssetvalue{/flpy/m2/time/family/upper 95:SBFL}{-0.86405605}
\pgfkeyssetvalue{/flpy/m2/time/family/upper 90:MBFL}{2.578099}
\pgfkeyssetvalue{/flpy/m2/time/family/upper 90:PS}{1.9228}
\pgfkeyssetvalue{/flpy/m2/time/family/upper 90:ST}{-1.5988785}
\pgfkeyssetvalue{/flpy/m2/time/family/upper 90:SBFL}{-1.037878}
\pgfkeyssetvalue{/flpy/m2/time/family/upper 70:MBFL}{2.3357665}
\pgfkeyssetvalue{/flpy/m2/time/family/upper 70:PS}{1.6489455}
\pgfkeyssetvalue{/flpy/m2/time/family/upper 70:ST}{-1.9309575}
\pgfkeyssetvalue{/flpy/m2/time/family/upper 70:SBFL}{-1.40563}
\pgfkeyssetvalue{/flpy/m2/time/family/upper 50:MBFL}{2.17211}
\pgfkeyssetvalue{/flpy/m2/time/family/upper 50:PS}{1.4788225}
\pgfkeyssetvalue{/flpy/m2/time/family/upper 50:ST}{-2.12484}
\pgfkeyssetvalue{/flpy/m2/time/family/upper 50:SBFL}{-1.63189}
\pgfkeyssetvalue{/flpy/m2/time/family/MBFL}{1.8625292902}
\pgfkeyssetvalue{/flpy/m2/time/family/PS}{1.1746280953275}
\pgfkeyssetvalue{/flpy/m2/time/family/ST}{-2.564445435}
\pgfkeyssetvalue{/flpy/m2/time/family/SBFL}{-2.06971550775}
\pgfkeyssetvalue{/flpy/m2/time/category/lower 50:CL}{-0.8226835}
\pgfkeyssetvalue{/flpy/m2/time/category/lower 50:DEV}{0.035917725}
\pgfkeyssetvalue{/flpy/m2/time/category/lower 50:DS}{0.22997525}
\pgfkeyssetvalue{/flpy/m2/time/category/lower 50:WEB}{-0.23261775}
\pgfkeyssetvalue{/flpy/m2/time/category/lower 70:CL}{-0.91694215}
\pgfkeyssetvalue{/flpy/m2/time/category/lower 70:DEV}{-0.0503907649999998}
\pgfkeyssetvalue{/flpy/m2/time/category/lower 70:DS}{0.15723225}
\pgfkeyssetvalue{/flpy/m2/time/category/lower 70:WEB}{-0.32312575}
\pgfkeyssetvalue{/flpy/m2/time/category/lower 90:CL}{-1.097481}
\pgfkeyssetvalue{/flpy/m2/time/category/lower 90:DEV}{-0.19879025}
\pgfkeyssetvalue{/flpy/m2/time/category/lower 90:DS}{0.0213224}
\pgfkeyssetvalue{/flpy/m2/time/category/lower 90:WEB}{-0.4800993}
\pgfkeyssetvalue{/flpy/m2/time/category/lower 95:CL}{-1.193378}
\pgfkeyssetvalue{/flpy/m2/time/category/lower 95:DEV}{-0.2838526}
\pgfkeyssetvalue{/flpy/m2/time/category/lower 95:DS}{-0.0632587875}
\pgfkeyssetvalue{/flpy/m2/time/category/lower 95:WEB}{-0.578146275}
\pgfkeyssetvalue{/flpy/m2/time/category/lower 99:CL}{-1.46282195}
\pgfkeyssetvalue{/flpy/m2/time/category/lower 99:DEV}{-0.487885135}
\pgfkeyssetvalue{/flpy/m2/time/category/lower 99:DS}{-0.22869648}
\pgfkeyssetvalue{/flpy/m2/time/category/lower 99:WEB}{-0.7566351}
\pgfkeyssetvalue{/flpy/m2/time/category/upper 99:CL}{-0.110575348999999}
\pgfkeyssetvalue{/flpy/m2/time/category/upper 99:DEV}{0.738491820000001}
\pgfkeyssetvalue{/flpy/m2/time/category/upper 99:DS}{0.903944345}
\pgfkeyssetvalue{/flpy/m2/time/category/upper 99:WEB}{0.51016309}
\pgfkeyssetvalue{/flpy/m2/time/category/upper 95:CL}{-0.215460325}
\pgfkeyssetvalue{/flpy/m2/time/category/upper 95:DEV}{0.599098075}
\pgfkeyssetvalue{/flpy/m2/time/category/upper 95:DS}{0.769458525}
\pgfkeyssetvalue{/flpy/m2/time/category/upper 95:WEB}{0.3571841}
\pgfkeyssetvalue{/flpy/m2/time/category/upper 90:CL}{-0.2763427}
\pgfkeyssetvalue{/flpy/m2/time/category/upper 90:DEV}{0.52126265}
\pgfkeyssetvalue{/flpy/m2/time/category/upper 90:DS}{0.71423625}
\pgfkeyssetvalue{/flpy/m2/time/category/upper 90:WEB}{0.282513849999999}
\pgfkeyssetvalue{/flpy/m2/time/category/upper 70:CL}{-0.4095806}
\pgfkeyssetvalue{/flpy/m2/time/category/upper 70:DEV}{0.39238235}
\pgfkeyssetvalue{/flpy/m2/time/category/upper 70:DS}{0.5725315}
\pgfkeyssetvalue{/flpy/m2/time/category/upper 70:WEB}{0.1460154}
\pgfkeyssetvalue{/flpy/m2/time/category/upper 50:CL}{-0.494305}
\pgfkeyssetvalue{/flpy/m2/time/category/upper 50:DEV}{0.31617325}
\pgfkeyssetvalue{/flpy/m2/time/category/upper 50:DS}{0.49809975}
\pgfkeyssetvalue{/flpy/m2/time/category/upper 50:WEB}{0.068308625}
\pgfkeyssetvalue{/flpy/m2/time/category/CL}{-0.663901802125}
\pgfkeyssetvalue{/flpy/m2/time/category/DEV}{0.172121719109}
\pgfkeyssetvalue{/flpy/m2/time/category/DS}{0.3640750835075}
\pgfkeyssetvalue{/flpy/m2/time/category/WEB}{-0.08639932358645}
\pgfkeyssetvalue{/flpy/m6/einspect/logmutability/lower 50:logmutability MBFL}{-0.7547445}
\pgfkeyssetvalue{/flpy/m6/einspect/logmutability/lower 50:logmutability PS}{0.3759405}
\pgfkeyssetvalue{/flpy/m6/einspect/logmutability/lower 50:logmutability ST}{0.26624775}
\pgfkeyssetvalue{/flpy/m6/einspect/logmutability/lower 50:logmutability SBFL}{-0.515623}
\pgfkeyssetvalue{/flpy/m6/einspect/logmutability/lower 70:logmutability MBFL}{-0.8841669}
\pgfkeyssetvalue{/flpy/m6/einspect/logmutability/lower 70:logmutability PS}{0.25446255}
\pgfkeyssetvalue{/flpy/m6/einspect/logmutability/lower 70:logmutability ST}{0.147502}
\pgfkeyssetvalue{/flpy/m6/einspect/logmutability/lower 70:logmutability SBFL}{-0.76394705}
\pgfkeyssetvalue{/flpy/m6/einspect/logmutability/lower 80:logmutability MBFL}{-0.9720413}
\pgfkeyssetvalue{/flpy/m6/einspect/logmutability/lower 80:logmutability PS}{0.1822384}
\pgfkeyssetvalue{/flpy/m6/einspect/logmutability/lower 80:logmutability ST}{0.07889756}
\pgfkeyssetvalue{/flpy/m6/einspect/logmutability/lower 80:logmutability SBFL}{-0.9411668}
\pgfkeyssetvalue{/flpy/m6/einspect/logmutability/lower 87:logmutability MBFL}{-1.05968925}
\pgfkeyssetvalue{/flpy/m6/einspect/logmutability/lower 87:logmutability PS}{0.115193315}
\pgfkeyssetvalue{/flpy/m6/einspect/logmutability/lower 87:logmutability ST}{0.0086214319}
\pgfkeyssetvalue{/flpy/m6/einspect/logmutability/lower 87:logmutability SBFL}{-1.11161255}
\pgfkeyssetvalue{/flpy/m6/einspect/logmutability/lower 90:logmutability MBFL}{-1.1205065}
\pgfkeyssetvalue{/flpy/m6/einspect/logmutability/lower 90:logmutability PS}{0.085546755}
\pgfkeyssetvalue{/flpy/m6/einspect/logmutability/lower 90:logmutability ST}{-0.027511595}
\pgfkeyssetvalue{/flpy/m6/einspect/logmutability/lower 90:logmutability SBFL}{-1.225794}
\pgfkeyssetvalue{/flpy/m6/einspect/logmutability/lower 95:logmutability MBFL}{-1.28001225}
\pgfkeyssetvalue{/flpy/m6/einspect/logmutability/lower 95:logmutability PS}{-0.00431204149999994}
\pgfkeyssetvalue{/flpy/m6/einspect/logmutability/lower 95:logmutability ST}{-0.130285675}
\pgfkeyssetvalue{/flpy/m6/einspect/logmutability/lower 95:logmutability SBFL}{-1.45534775}
\pgfkeyssetvalue{/flpy/m6/einspect/logmutability/lower 99:logmutability MBFL}{-1.4713408}
\pgfkeyssetvalue{/flpy/m6/einspect/logmutability/lower 99:logmutability PS}{-0.178970675}
\pgfkeyssetvalue{/flpy/m6/einspect/logmutability/lower 99:logmutability ST}{-0.261238665}
\pgfkeyssetvalue{/flpy/m6/einspect/logmutability/lower 99:logmutability SBFL}{-1.8345544}
\pgfkeyssetvalue{/flpy/m6/einspect/logmutability/upper 99:logmutability MBFL}{0.25625661}
\pgfkeyssetvalue{/flpy/m6/einspect/logmutability/upper 99:logmutability PS}{1.56417885}
\pgfkeyssetvalue{/flpy/m6/einspect/logmutability/upper 99:logmutability ST}{1.4698448}
\pgfkeyssetvalue{/flpy/m6/einspect/logmutability/upper 99:logmutability SBFL}{1.3301171}
\pgfkeyssetvalue{/flpy/m6/einspect/logmutability/upper 95:logmutability MBFL}{0.1087532}
\pgfkeyssetvalue{/flpy/m6/einspect/logmutability/upper 95:logmutability PS}{1.3620565}
\pgfkeyssetvalue{/flpy/m6/einspect/logmutability/upper 95:logmutability ST}{1.25861325}
\pgfkeyssetvalue{/flpy/m6/einspect/logmutability/upper 95:logmutability SBFL}{0.992993675}
\pgfkeyssetvalue{/flpy/m6/einspect/logmutability/upper 90:logmutability MBFL}{0.0151696599999999}
\pgfkeyssetvalue{/flpy/m6/einspect/logmutability/upper 90:logmutability PS}{1.221132}
\pgfkeyssetvalue{/flpy/m6/einspect/logmutability/upper 90:logmutability ST}{1.1073075}
\pgfkeyssetvalue{/flpy/m6/einspect/logmutability/upper 90:logmutability SBFL}{0.84015345}
\pgfkeyssetvalue{/flpy/m6/einspect/logmutability/upper 87:logmutability MBFL}{-0.0229480944999999}
\pgfkeyssetvalue{/flpy/m6/einspect/logmutability/upper 87:logmutability PS}{1.16346775}
\pgfkeyssetvalue{/flpy/m6/einspect/logmutability/upper 87:logmutability ST}{1.0500677}
\pgfkeyssetvalue{/flpy/m6/einspect/logmutability/upper 87:logmutability SBFL}{0.76403437}
\pgfkeyssetvalue{/flpy/m6/einspect/logmutability/upper 80:logmutability MBFL}{-0.0932349600000002}
\pgfkeyssetvalue{/flpy/m6/einspect/logmutability/upper 80:logmutability PS}{1.068632}
\pgfkeyssetvalue{/flpy/m6/einspect/logmutability/upper 80:logmutability ST}{0.955611}
\pgfkeyssetvalue{/flpy/m6/einspect/logmutability/upper 80:logmutability SBFL}{0.6521943}
\pgfkeyssetvalue{/flpy/m6/einspect/logmutability/upper 70:logmutability MBFL}{-0.16356845}
\pgfkeyssetvalue{/flpy/m6/einspect/logmutability/upper 70:logmutability PS}{0.9829318}
\pgfkeyssetvalue{/flpy/m6/einspect/logmutability/upper 70:logmutability ST}{0.87455715}
\pgfkeyssetvalue{/flpy/m6/einspect/logmutability/upper 70:logmutability SBFL}{0.52767515}
\pgfkeyssetvalue{/flpy/m6/einspect/logmutability/upper 50:logmutability MBFL}{-0.27993625}
\pgfkeyssetvalue{/flpy/m6/einspect/logmutability/upper 50:logmutability PS}{0.85509625}
\pgfkeyssetvalue{/flpy/m6/einspect/logmutability/upper 50:logmutability ST}{0.749353}
\pgfkeyssetvalue{/flpy/m6/einspect/logmutability/upper 50:logmutability SBFL}{0.331809}
\pgfkeyssetvalue{/flpy/m6/einspect/logmutability/logmutability MBFL}{-0.5291964366535}
\pgfkeyssetvalue{/flpy/m6/einspect/logmutability/logmutability PS}{0.625550630091}
\pgfkeyssetvalue{/flpy/m6/einspect/logmutability/logmutability ST}{0.51708127530175}
\pgfkeyssetvalue{/flpy/m6/einspect/logmutability/logmutability SBFL}{-0.109275289464}
\pgfkeyssetvalue{/flpy/m6/einspect/crashing/lower 50:crashing MBFL}{-1.5439225}
\pgfkeyssetvalue{/flpy/m6/einspect/crashing/lower 50:crashing PS}{0.6979465}
\pgfkeyssetvalue{/flpy/m6/einspect/crashing/lower 50:crashing ST}{-2.229575}
\pgfkeyssetvalue{/flpy/m6/einspect/crashing/lower 50:crashing SBFL}{-0.557907}
\pgfkeyssetvalue{/flpy/m6/einspect/crashing/lower 70:crashing MBFL}{-1.743097}
\pgfkeyssetvalue{/flpy/m6/einspect/crashing/lower 70:crashing PS}{0.50589745}
\pgfkeyssetvalue{/flpy/m6/einspect/crashing/lower 70:crashing ST}{-2.479337}
\pgfkeyssetvalue{/flpy/m6/einspect/crashing/lower 70:crashing SBFL}{-0.87455075}
\pgfkeyssetvalue{/flpy/m6/einspect/crashing/lower 80:crashing MBFL}{-1.874443}
\pgfkeyssetvalue{/flpy/m6/einspect/crashing/lower 80:crashing PS}{0.3868135}
\pgfkeyssetvalue{/flpy/m6/einspect/crashing/lower 80:crashing ST}{-2.653464}
\pgfkeyssetvalue{/flpy/m6/einspect/crashing/lower 80:crashing SBFL}{-1.070707}
\pgfkeyssetvalue{/flpy/m6/einspect/crashing/lower 87:crashing MBFL}{-2.00333665}
\pgfkeyssetvalue{/flpy/m6/einspect/crashing/lower 87:crashing PS}{0.248499565}
\pgfkeyssetvalue{/flpy/m6/einspect/crashing/lower 87:crashing ST}{-2.8226749}
\pgfkeyssetvalue{/flpy/m6/einspect/crashing/lower 87:crashing SBFL}{-1.3077794}
\pgfkeyssetvalue{/flpy/m6/einspect/crashing/lower 90:crashing MBFL}{-2.074271}
\pgfkeyssetvalue{/flpy/m6/einspect/crashing/lower 90:crashing PS}{0.1867658}
\pgfkeyssetvalue{/flpy/m6/einspect/crashing/lower 90:crashing ST}{-2.9160775}
\pgfkeyssetvalue{/flpy/m6/einspect/crashing/lower 90:crashing SBFL}{-1.4007575}
\pgfkeyssetvalue{/flpy/m6/einspect/crashing/lower 95:crashing MBFL}{-2.25423675}
\pgfkeyssetvalue{/flpy/m6/einspect/crashing/lower 95:crashing PS}{0.0457672750000002}
\pgfkeyssetvalue{/flpy/m6/einspect/crashing/lower 95:crashing ST}{-3.157148}
\pgfkeyssetvalue{/flpy/m6/einspect/crashing/lower 95:crashing SBFL}{-1.6832555}
\pgfkeyssetvalue{/flpy/m6/einspect/crashing/lower 99:crashing MBFL}{-2.5844665}
\pgfkeyssetvalue{/flpy/m6/einspect/crashing/lower 99:crashing PS}{-0.22599418}
\pgfkeyssetvalue{/flpy/m6/einspect/crashing/lower 99:crashing ST}{-3.72883035}
\pgfkeyssetvalue{/flpy/m6/einspect/crashing/lower 99:crashing SBFL}{-2.2694496}
\pgfkeyssetvalue{/flpy/m6/einspect/crashing/upper 99:crashing MBFL}{0.107556140000001}
\pgfkeyssetvalue{/flpy/m6/einspect/crashing/upper 99:crashing PS}{2.48447375}
\pgfkeyssetvalue{/flpy/m6/einspect/crashing/upper 99:crashing ST}{-0.172546394999999}
\pgfkeyssetvalue{/flpy/m6/einspect/crashing/upper 99:crashing SBFL}{2.1516732}
\pgfkeyssetvalue{/flpy/m6/einspect/crashing/upper 95:crashing MBFL}{-0.190584025}
\pgfkeyssetvalue{/flpy/m6/einspect/crashing/upper 95:crashing PS}{2.14954175}
\pgfkeyssetvalue{/flpy/m6/einspect/crashing/upper 95:crashing ST}{-0.4905315}
\pgfkeyssetvalue{/flpy/m6/einspect/crashing/upper 95:crashing SBFL}{1.60724175}
\pgfkeyssetvalue{/flpy/m6/einspect/crashing/upper 90:crashing MBFL}{-0.343026500000002}
\pgfkeyssetvalue{/flpy/m6/einspect/crashing/upper 90:crashing PS}{1.952465}
\pgfkeyssetvalue{/flpy/m6/einspect/crashing/upper 90:crashing ST}{-0.7176368}
\pgfkeyssetvalue{/flpy/m6/einspect/crashing/upper 90:crashing SBFL}{1.340869}
\pgfkeyssetvalue{/flpy/m6/einspect/crashing/upper 87:crashing MBFL}{-0.399429795}
\pgfkeyssetvalue{/flpy/m6/einspect/crashing/upper 87:crashing PS}{1.8871615}
\pgfkeyssetvalue{/flpy/m6/einspect/crashing/upper 87:crashing ST}{-0.81324367}
\pgfkeyssetvalue{/flpy/m6/einspect/crashing/upper 87:crashing SBFL}{1.25926105}
\pgfkeyssetvalue{/flpy/m6/einspect/crashing/upper 80:crashing MBFL}{-0.5205948}
\pgfkeyssetvalue{/flpy/m6/einspect/crashing/upper 80:crashing PS}{1.768569}
\pgfkeyssetvalue{/flpy/m6/einspect/crashing/upper 80:crashing ST}{-0.953928}
\pgfkeyssetvalue{/flpy/m6/einspect/crashing/upper 80:crashing SBFL}{1.075584}
\pgfkeyssetvalue{/flpy/m6/einspect/crashing/upper 70:crashing MBFL}{-0.6415296}
\pgfkeyssetvalue{/flpy/m6/einspect/crashing/upper 70:crashing PS}{1.627476}
\pgfkeyssetvalue{/flpy/m6/einspect/crashing/upper 70:crashing ST}{-1.105512}
\pgfkeyssetvalue{/flpy/m6/einspect/crashing/upper 70:crashing SBFL}{0.8747363}
\pgfkeyssetvalue{/flpy/m6/einspect/crashing/upper 50:crashing MBFL}{-0.83476175}
\pgfkeyssetvalue{/flpy/m6/einspect/crashing/upper 50:crashing PS}{1.41554}
\pgfkeyssetvalue{/flpy/m6/einspect/crashing/upper 50:crashing ST}{-1.3252025}
\pgfkeyssetvalue{/flpy/m6/einspect/crashing/upper 50:crashing SBFL}{0.581589}
\pgfkeyssetvalue{/flpy/m6/einspect/crashing/crashing MBFL}{-1.1932547987975}
\pgfkeyssetvalue{/flpy/m6/einspect/crashing/crashing PS}{1.0644355637525}
\pgfkeyssetvalue{/flpy/m6/einspect/crashing/crashing ST}{-1.7871074044125}
\pgfkeyssetvalue{/flpy/m6/einspect/crashing/crashing SBFL}{0.00502130608705}
\pgfkeyssetvalue{/flpy/m6/einspect/predicate/lower 50:predicate MBFL}{-0.59165975}
\pgfkeyssetvalue{/flpy/m6/einspect/predicate/lower 50:predicate PS}{-0.286226}
\pgfkeyssetvalue{/flpy/m6/einspect/predicate/lower 50:predicate ST}{0.23657575}
\pgfkeyssetvalue{/flpy/m6/einspect/predicate/lower 50:predicate SBFL}{-0.637463}
\pgfkeyssetvalue{/flpy/m6/einspect/predicate/lower 70:predicate MBFL}{-0.7619661}
\pgfkeyssetvalue{/flpy/m6/einspect/predicate/lower 70:predicate PS}{-0.4756684}
\pgfkeyssetvalue{/flpy/m6/einspect/predicate/lower 70:predicate ST}{0.054673455}
\pgfkeyssetvalue{/flpy/m6/einspect/predicate/lower 70:predicate SBFL}{-0.95892545}
\pgfkeyssetvalue{/flpy/m6/einspect/predicate/lower 80:predicate MBFL}{-0.9020018}
\pgfkeyssetvalue{/flpy/m6/einspect/predicate/lower 80:predicate PS}{-0.5995323}
\pgfkeyssetvalue{/flpy/m6/einspect/predicate/lower 80:predicate ST}{-0.06710679}
\pgfkeyssetvalue{/flpy/m6/einspect/predicate/lower 80:predicate SBFL}{-1.167584}
\pgfkeyssetvalue{/flpy/m6/einspect/predicate/lower 87:predicate MBFL}{-1.01068645}
\pgfkeyssetvalue{/flpy/m6/einspect/predicate/lower 87:predicate PS}{-0.71890176}
\pgfkeyssetvalue{/flpy/m6/einspect/predicate/lower 87:predicate ST}{-0.183570605}
\pgfkeyssetvalue{/flpy/m6/einspect/predicate/lower 87:predicate SBFL}{-1.3752713}
\pgfkeyssetvalue{/flpy/m6/einspect/predicate/lower 90:predicate MBFL}{-1.071021}
\pgfkeyssetvalue{/flpy/m6/einspect/predicate/lower 90:predicate PS}{-0.777294}
\pgfkeyssetvalue{/flpy/m6/einspect/predicate/lower 90:predicate ST}{-0.2229248}
\pgfkeyssetvalue{/flpy/m6/einspect/predicate/lower 90:predicate SBFL}{-1.4714315}
\pgfkeyssetvalue{/flpy/m6/einspect/predicate/lower 95:predicate MBFL}{-1.21743375}
\pgfkeyssetvalue{/flpy/m6/einspect/predicate/lower 95:predicate PS}{-0.935833725}
\pgfkeyssetvalue{/flpy/m6/einspect/predicate/lower 95:predicate ST}{-0.375249225}
\pgfkeyssetvalue{/flpy/m6/einspect/predicate/lower 95:predicate SBFL}{-1.776968}
\pgfkeyssetvalue{/flpy/m6/einspect/predicate/lower 99:predicate MBFL}{-1.54477605}
\pgfkeyssetvalue{/flpy/m6/einspect/predicate/lower 99:predicate PS}{-1.1746152}
\pgfkeyssetvalue{/flpy/m6/einspect/predicate/lower 99:predicate ST}{-0.651481505}
\pgfkeyssetvalue{/flpy/m6/einspect/predicate/lower 99:predicate SBFL}{-2.28115135}
\pgfkeyssetvalue{/flpy/m6/einspect/predicate/upper 99:predicate MBFL}{0.973989965000001}
\pgfkeyssetvalue{/flpy/m6/einspect/predicate/upper 99:predicate PS}{1.32411835}
\pgfkeyssetvalue{/flpy/m6/einspect/predicate/upper 99:predicate ST}{1.8931751}
\pgfkeyssetvalue{/flpy/m6/einspect/predicate/upper 99:predicate SBFL}{2.06602095}
\pgfkeyssetvalue{/flpy/m6/einspect/predicate/upper 95:predicate MBFL}{0.694470375000001}
\pgfkeyssetvalue{/flpy/m6/einspect/predicate/upper 95:predicate PS}{1.03394625}
\pgfkeyssetvalue{/flpy/m6/einspect/predicate/upper 95:predicate ST}{1.5518615}
\pgfkeyssetvalue{/flpy/m6/einspect/predicate/upper 95:predicate SBFL}{1.53690225}
\pgfkeyssetvalue{/flpy/m6/einspect/predicate/upper 90:predicate MBFL}{0.55932135}
\pgfkeyssetvalue{/flpy/m6/einspect/predicate/upper 90:predicate PS}{0.8917026}
\pgfkeyssetvalue{/flpy/m6/einspect/predicate/upper 90:predicate ST}{1.399975}
\pgfkeyssetvalue{/flpy/m6/einspect/predicate/upper 90:predicate SBFL}{1.302487}
\pgfkeyssetvalue{/flpy/m6/einspect/predicate/upper 87:predicate MBFL}{0.493542565}
\pgfkeyssetvalue{/flpy/m6/einspect/predicate/upper 87:predicate PS}{0.83245287}
\pgfkeyssetvalue{/flpy/m6/einspect/predicate/upper 87:predicate ST}{1.3270633}
\pgfkeyssetvalue{/flpy/m6/einspect/predicate/upper 87:predicate SBFL}{1.19595145}
\pgfkeyssetvalue{/flpy/m6/einspect/predicate/upper 80:predicate MBFL}{0.3862723}
\pgfkeyssetvalue{/flpy/m6/einspect/predicate/upper 80:predicate PS}{0.7108285}
\pgfkeyssetvalue{/flpy/m6/einspect/predicate/upper 80:predicate ST}{1.221673}
\pgfkeyssetvalue{/flpy/m6/einspect/predicate/upper 80:predicate SBFL}{0.9985885}
\pgfkeyssetvalue{/flpy/m6/einspect/predicate/upper 70:predicate MBFL}{0.2497229}
\pgfkeyssetvalue{/flpy/m6/einspect/predicate/upper 70:predicate PS}{0.5776847}
\pgfkeyssetvalue{/flpy/m6/einspect/predicate/upper 70:predicate ST}{1.105351}
\pgfkeyssetvalue{/flpy/m6/einspect/predicate/upper 70:predicate SBFL}{0.81038855}
\pgfkeyssetvalue{/flpy/m6/einspect/predicate/upper 50:predicate MBFL}{0.0721588}
\pgfkeyssetvalue{/flpy/m6/einspect/predicate/upper 50:predicate PS}{0.39488575}
\pgfkeyssetvalue{/flpy/m6/einspect/predicate/upper 50:predicate ST}{0.919414}
\pgfkeyssetvalue{/flpy/m6/einspect/predicate/upper 50:predicate SBFL}{0.49630225}
\pgfkeyssetvalue{/flpy/m6/einspect/predicate/predicate MBFL}{-0.2579525699835}
\pgfkeyssetvalue{/flpy/m6/einspect/predicate/predicate PS}{0.055023161026325}
\pgfkeyssetvalue{/flpy/m6/einspect/predicate/predicate ST}{0.58027280242025}
\pgfkeyssetvalue{/flpy/m6/einspect/predicate/predicate SBFL}{-0.07751487353475}
\pgfkeyssetvalue{/flpy/deltas/m2/einspect/category/CL/CL:coef.delta}{0}
\pgfkeyssetvalue{/flpy/deltas/m2/einspect/category/CL/CL:outcome.delta}{0}
\pgfkeyssetvalue{/flpy/deltas/m2/einspect/category/CL/CL:ratio.delta}{1}
\pgfkeyssetvalue{/flpy/deltas/m2/einspect/category/DEV/CL:coef.delta}{-2.3152891954}
\pgfkeyssetvalue{/flpy/deltas/m2/einspect/category/DEV/CL:outcome.delta}{-159.299830699841}
\pgfkeyssetvalue{/flpy/deltas/m2/einspect/category/DEV/CL:ratio.delta}{0.0987376254057489}
\pgfkeyssetvalue{/flpy/deltas/m2/einspect/category/DS/CL:coef.delta}{-0.696894494890425}
\pgfkeyssetvalue{/flpy/deltas/m2/einspect/category/DS/CL:outcome.delta}{-88.7064991797041}
\pgfkeyssetvalue{/flpy/deltas/m2/einspect/category/DS/CL:ratio.delta}{0.498129849044942}
\pgfkeyssetvalue{/flpy/deltas/m2/einspect/category/WEB/CL:coef.delta}{-2.13004155451275}
\pgfkeyssetvalue{/flpy/deltas/m2/einspect/category/WEB/CL:outcome.delta}{-155.748049074124}
\pgfkeyssetvalue{/flpy/deltas/m2/einspect/category/WEB/CL:ratio.delta}{0.118832355729168}
\pgfkeyssetvalue{/flpy/deltas/m2/einspect/category/CL/DEV:coef.delta}{2.3152891954}
\pgfkeyssetvalue{/flpy/deltas/m2/einspect/category/CL/DEV:outcome.delta}{159.299830699841}
\pgfkeyssetvalue{/flpy/deltas/m2/einspect/category/CL/DEV:ratio.delta}{10.1278514233114}
\pgfkeyssetvalue{/flpy/deltas/m2/einspect/category/DEV/DEV:coef.delta}{0}
\pgfkeyssetvalue{/flpy/deltas/m2/einspect/category/DEV/DEV:outcome.delta}{0}
\pgfkeyssetvalue{/flpy/deltas/m2/einspect/category/DEV/DEV:ratio.delta}{1}
\pgfkeyssetvalue{/flpy/deltas/m2/einspect/category/DS/DEV:coef.delta}{1.61839470050957}
\pgfkeyssetvalue{/flpy/deltas/m2/einspect/category/DS/DEV:outcome.delta}{70.5933315201374}
\pgfkeyssetvalue{/flpy/deltas/m2/einspect/category/DS/DEV:ratio.delta}{5.04498510064369}
\pgfkeyssetvalue{/flpy/deltas/m2/einspect/category/WEB/DEV:coef.delta}{0.18524764088725}
\pgfkeyssetvalue{/flpy/deltas/m2/einspect/category/WEB/DEV:outcome.delta}{3.55178162571792}
\pgfkeyssetvalue{/flpy/deltas/m2/einspect/category/WEB/DEV:ratio.delta}{1.2035164431071}
\pgfkeyssetvalue{/flpy/deltas/m2/einspect/category/CL/DS:coef.delta}{0.696894494890425}
\pgfkeyssetvalue{/flpy/deltas/m2/einspect/category/CL/DS:outcome.delta}{88.7064991797041}
\pgfkeyssetvalue{/flpy/deltas/m2/einspect/category/CL/DS:ratio.delta}{2.00750868858248}
\pgfkeyssetvalue{/flpy/deltas/m2/einspect/category/DEV/DS:coef.delta}{-1.61839470050957}
\pgfkeyssetvalue{/flpy/deltas/m2/einspect/category/DEV/DS:outcome.delta}{-70.5933315201374}
\pgfkeyssetvalue{/flpy/deltas/m2/einspect/category/DEV/DS:ratio.delta}{0.198216640892043}
\pgfkeyssetvalue{/flpy/deltas/m2/einspect/category/DS/DS:coef.delta}{0}
\pgfkeyssetvalue{/flpy/deltas/m2/einspect/category/DS/DS:outcome.delta}{0}
\pgfkeyssetvalue{/flpy/deltas/m2/einspect/category/DS/DS:ratio.delta}{1}
\pgfkeyssetvalue{/flpy/deltas/m2/einspect/category/WEB/DS:coef.delta}{-1.43314705962233}
\pgfkeyssetvalue{/flpy/deltas/m2/einspect/category/WEB/DS:outcome.delta}{-67.0415498944194}
\pgfkeyssetvalue{/flpy/deltas/m2/einspect/category/WEB/DS:ratio.delta}{0.23855698661103}
\pgfkeyssetvalue{/flpy/deltas/m2/einspect/category/CL/WEB:coef.delta}{2.13004155451275}
\pgfkeyssetvalue{/flpy/deltas/m2/einspect/category/CL/WEB:outcome.delta}{155.748049074124}
\pgfkeyssetvalue{/flpy/deltas/m2/einspect/category/CL/WEB:ratio.delta}{8.41521649439574}
\pgfkeyssetvalue{/flpy/deltas/m2/einspect/category/DEV/WEB:coef.delta}{-0.18524764088725}
\pgfkeyssetvalue{/flpy/deltas/m2/einspect/category/DEV/WEB:outcome.delta}{-3.55178162571792}
\pgfkeyssetvalue{/flpy/deltas/m2/einspect/category/DEV/WEB:ratio.delta}{0.830898493931926}
\pgfkeyssetvalue{/flpy/deltas/m2/einspect/category/DS/WEB:coef.delta}{1.43314705962233}
\pgfkeyssetvalue{/flpy/deltas/m2/einspect/category/DS/WEB:outcome.delta}{67.0415498944194}
\pgfkeyssetvalue{/flpy/deltas/m2/einspect/category/DS/WEB:ratio.delta}{4.19187052203385}
\pgfkeyssetvalue{/flpy/deltas/m2/einspect/category/WEB/WEB:coef.delta}{0}
\pgfkeyssetvalue{/flpy/deltas/m2/einspect/category/WEB/WEB:outcome.delta}{0}
\pgfkeyssetvalue{/flpy/deltas/m2/einspect/category/WEB/WEB:ratio.delta}{1}
\pgfkeyssetvalue{/flpy/deltas/m2/einspect/family/MBFL/MBFL:coef.delta}{0}
\pgfkeyssetvalue{/flpy/deltas/m2/einspect/family/MBFL/MBFL:outcome.delta}{0}
\pgfkeyssetvalue{/flpy/deltas/m2/einspect/family/MBFL/MBFL:ratio.delta}{1}
\pgfkeyssetvalue{/flpy/deltas/m2/einspect/family/PS/MBFL:coef.delta}{3.544358254375}
\pgfkeyssetvalue{/flpy/deltas/m2/einspect/family/PS/MBFL:outcome.delta}{236.740767910714}
\pgfkeyssetvalue{/flpy/deltas/m2/einspect/family/PS/MBFL:ratio.delta}{34.6174626071753}
\pgfkeyssetvalue{/flpy/deltas/m2/einspect/family/SBFL/MBFL:coef.delta}{-0.340236245}
\pgfkeyssetvalue{/flpy/deltas/m2/einspect/family/SBFL/MBFL:outcome.delta}{-2.03095396348916}
\pgfkeyssetvalue{/flpy/deltas/m2/einspect/family/SBFL/MBFL:ratio.delta}{0.711602190443701}
\pgfkeyssetvalue{/flpy/deltas/m2/einspect/family/ST/MBFL:coef.delta}{3.6516629344}
\pgfkeyssetvalue{/flpy/deltas/m2/einspect/family/ST/MBFL:outcome.delta}{264.354891888454}
\pgfkeyssetvalue{/flpy/deltas/m2/einspect/family/ST/MBFL:ratio.delta}{38.5387001222087}
\pgfkeyssetvalue{/flpy/deltas/m2/einspect/family/MBFL/PS:coef.delta}{-3.544358254375}
\pgfkeyssetvalue{/flpy/deltas/m2/einspect/family/MBFL/PS:outcome.delta}{-236.740767910714}
\pgfkeyssetvalue{/flpy/deltas/m2/einspect/family/MBFL/PS:ratio.delta}{0.0288871547677422}
\pgfkeyssetvalue{/flpy/deltas/m2/einspect/family/PS/PS:coef.delta}{0}
\pgfkeyssetvalue{/flpy/deltas/m2/einspect/family/PS/PS:outcome.delta}{0}
\pgfkeyssetvalue{/flpy/deltas/m2/einspect/family/PS/PS:ratio.delta}{1}
\pgfkeyssetvalue{/flpy/deltas/m2/einspect/family/SBFL/PS:coef.delta}{-3.884594499375}
\pgfkeyssetvalue{/flpy/deltas/m2/einspect/family/SBFL/PS:outcome.delta}{-238.771721874203}
\pgfkeyssetvalue{/flpy/deltas/m2/einspect/family/SBFL/PS:ratio.delta}{0.0205561626084115}
\pgfkeyssetvalue{/flpy/deltas/m2/einspect/family/ST/PS:coef.delta}{0.107304680025}
\pgfkeyssetvalue{/flpy/deltas/m2/einspect/family/ST/PS:outcome.delta}{27.6141239777402}
\pgfkeyssetvalue{/flpy/deltas/m2/einspect/family/ST/PS:ratio.delta}{1.11327339497785}
\pgfkeyssetvalue{/flpy/deltas/m2/einspect/family/MBFL/SBFL:coef.delta}{0.340236245}
\pgfkeyssetvalue{/flpy/deltas/m2/einspect/family/MBFL/SBFL:outcome.delta}{2.03095396348916}
\pgfkeyssetvalue{/flpy/deltas/m2/einspect/family/MBFL/SBFL:ratio.delta}{1.40527954161647}
\pgfkeyssetvalue{/flpy/deltas/m2/einspect/family/PS/SBFL:coef.delta}{3.884594499375}
\pgfkeyssetvalue{/flpy/deltas/m2/einspect/family/PS/SBFL:outcome.delta}{238.771721874203}
\pgfkeyssetvalue{/flpy/deltas/m2/einspect/family/PS/SBFL:ratio.delta}{48.6472119845366}
\pgfkeyssetvalue{/flpy/deltas/m2/einspect/family/SBFL/SBFL:coef.delta}{0}
\pgfkeyssetvalue{/flpy/deltas/m2/einspect/family/SBFL/SBFL:outcome.delta}{0}
\pgfkeyssetvalue{/flpy/deltas/m2/einspect/family/SBFL/SBFL:ratio.delta}{1}
\pgfkeyssetvalue{/flpy/deltas/m2/einspect/family/ST/SBFL:coef.delta}{3.9918991794}
\pgfkeyssetvalue{/flpy/deltas/m2/einspect/family/ST/SBFL:outcome.delta}{266.385845851943}
\pgfkeyssetvalue{/flpy/deltas/m2/einspect/family/ST/SBFL:ratio.delta}{54.1576468422321}
\pgfkeyssetvalue{/flpy/deltas/m2/einspect/family/MBFL/ST:coef.delta}{-3.6516629344}
\pgfkeyssetvalue{/flpy/deltas/m2/einspect/family/MBFL/ST:outcome.delta}{-264.354891888454}
\pgfkeyssetvalue{/flpy/deltas/m2/einspect/family/MBFL/ST:ratio.delta}{0.0259479431539968}
\pgfkeyssetvalue{/flpy/deltas/m2/einspect/family/PS/ST:coef.delta}{-0.107304680025}
\pgfkeyssetvalue{/flpy/deltas/m2/einspect/family/PS/ST:outcome.delta}{-27.6141239777402}
\pgfkeyssetvalue{/flpy/deltas/m2/einspect/family/PS/ST:ratio.delta}{0.898251951866593}
\pgfkeyssetvalue{/flpy/deltas/m2/einspect/family/SBFL/ST:coef.delta}{-3.9918991794}
\pgfkeyssetvalue{/flpy/deltas/m2/einspect/family/SBFL/ST:outcome.delta}{-266.385845851943}
\pgfkeyssetvalue{/flpy/deltas/m2/einspect/family/SBFL/ST:ratio.delta}{0.0184646131858927}
\pgfkeyssetvalue{/flpy/deltas/m2/einspect/family/ST/ST:coef.delta}{0}
\pgfkeyssetvalue{/flpy/deltas/m2/einspect/family/ST/ST:outcome.delta}{0}
\pgfkeyssetvalue{/flpy/deltas/m2/einspect/family/ST/ST:ratio.delta}{1}
\pgfkeyssetvalue{/flpy/deltas/m2/time/category/CL/CL:coef.delta}{0}
\pgfkeyssetvalue{/flpy/deltas/m2/time/category/CL/CL:outcome.delta}{0}
\pgfkeyssetvalue{/flpy/deltas/m2/time/category/CL/CL:ratio.delta}{1}
\pgfkeyssetvalue{/flpy/deltas/m2/time/category/DEV/CL:coef.delta}{0.836023521234}
\pgfkeyssetvalue{/flpy/deltas/m2/time/category/DEV/CL:outcome.delta}{4404.28925690882}
\pgfkeyssetvalue{/flpy/deltas/m2/time/category/DEV/CL:ratio.delta}{2.30717428207451}
\pgfkeyssetvalue{/flpy/deltas/m2/time/category/DS/CL:coef.delta}{1.0279768856325}
\pgfkeyssetvalue{/flpy/deltas/m2/time/category/DS/CL:outcome.delta}{11118.0965964186}
\pgfkeyssetvalue{/flpy/deltas/m2/time/category/DS/CL:ratio.delta}{2.79540468667046}
\pgfkeyssetvalue{/flpy/deltas/m2/time/category/WEB/CL:coef.delta}{0.57750247853855}
\pgfkeyssetvalue{/flpy/deltas/m2/time/category/WEB/CL:outcome.delta}{1444.1427878494}
\pgfkeyssetvalue{/flpy/deltas/m2/time/category/WEB/CL:ratio.delta}{1.78158332711254}
\pgfkeyssetvalue{/flpy/deltas/m2/time/category/CL/DEV:coef.delta}{-0.836023521234}
\pgfkeyssetvalue{/flpy/deltas/m2/time/category/CL/DEV:outcome.delta}{-4404.28925690882}
\pgfkeyssetvalue{/flpy/deltas/m2/time/category/CL/DEV:ratio.delta}{0.433430628873361}
\pgfkeyssetvalue{/flpy/deltas/m2/time/category/DEV/DEV:coef.delta}{0}
\pgfkeyssetvalue{/flpy/deltas/m2/time/category/DEV/DEV:outcome.delta}{0}
\pgfkeyssetvalue{/flpy/deltas/m2/time/category/DEV/DEV:ratio.delta}{1}
\pgfkeyssetvalue{/flpy/deltas/m2/time/category/DS/DEV:coef.delta}{0.1919533643985}
\pgfkeyssetvalue{/flpy/deltas/m2/time/category/DS/DEV:outcome.delta}{6713.80733950981}
\pgfkeyssetvalue{/flpy/deltas/m2/time/category/DS/DEV:ratio.delta}{1.21161401129912}
\pgfkeyssetvalue{/flpy/deltas/m2/time/category/WEB/DEV:coef.delta}{-0.25852104269545}
\pgfkeyssetvalue{/flpy/deltas/m2/time/category/WEB/DEV:outcome.delta}{-2960.14646905942}
\pgfkeyssetvalue{/flpy/deltas/m2/time/category/WEB/DEV:ratio.delta}{0.772192781860685}
\pgfkeyssetvalue{/flpy/deltas/m2/time/category/CL/DS:coef.delta}{-1.0279768856325}
\pgfkeyssetvalue{/flpy/deltas/m2/time/category/CL/DS:outcome.delta}{-11118.0965964186}
\pgfkeyssetvalue{/flpy/deltas/m2/time/category/CL/DS:ratio.delta}{0.357729957586597}
\pgfkeyssetvalue{/flpy/deltas/m2/time/category/DEV/DS:coef.delta}{-0.1919533643985}
\pgfkeyssetvalue{/flpy/deltas/m2/time/category/DEV/DS:outcome.delta}{-6713.80733950981}
\pgfkeyssetvalue{/flpy/deltas/m2/time/category/DEV/DS:ratio.delta}{0.825345358071402}
\pgfkeyssetvalue{/flpy/deltas/m2/time/category/DS/DS:coef.delta}{0}
\pgfkeyssetvalue{/flpy/deltas/m2/time/category/DS/DS:outcome.delta}{0}
\pgfkeyssetvalue{/flpy/deltas/m2/time/category/DS/DS:ratio.delta}{1}
\pgfkeyssetvalue{/flpy/deltas/m2/time/category/WEB/DS:coef.delta}{-0.45047440709395}
\pgfkeyssetvalue{/flpy/deltas/m2/time/category/WEB/DS:outcome.delta}{-9673.95380856923}
\pgfkeyssetvalue{/flpy/deltas/m2/time/category/WEB/DS:ratio.delta}{0.637325728044959}
\pgfkeyssetvalue{/flpy/deltas/m2/time/category/CL/WEB:coef.delta}{-0.57750247853855}
\pgfkeyssetvalue{/flpy/deltas/m2/time/category/CL/WEB:outcome.delta}{-1444.1427878494}
\pgfkeyssetvalue{/flpy/deltas/m2/time/category/CL/WEB:ratio.delta}{0.561298472421565}
\pgfkeyssetvalue{/flpy/deltas/m2/time/category/DEV/WEB:coef.delta}{0.25852104269545}
\pgfkeyssetvalue{/flpy/deltas/m2/time/category/DEV/WEB:outcome.delta}{2960.14646905942}
\pgfkeyssetvalue{/flpy/deltas/m2/time/category/DEV/WEB:ratio.delta}{1.29501340013874}
\pgfkeyssetvalue{/flpy/deltas/m2/time/category/DS/WEB:coef.delta}{0.45047440709395}
\pgfkeyssetvalue{/flpy/deltas/m2/time/category/DS/WEB:outcome.delta}{9673.95380856923}
\pgfkeyssetvalue{/flpy/deltas/m2/time/category/DS/WEB:ratio.delta}{1.56905638042822}
\pgfkeyssetvalue{/flpy/deltas/m2/time/category/WEB/WEB:coef.delta}{0}
\pgfkeyssetvalue{/flpy/deltas/m2/time/category/WEB/WEB:outcome.delta}{0}
\pgfkeyssetvalue{/flpy/deltas/m2/time/category/WEB/WEB:ratio.delta}{1}
\pgfkeyssetvalue{/flpy/deltas/m2/time/family/MBFL/MBFL:coef.delta}{0}
\pgfkeyssetvalue{/flpy/deltas/m2/time/family/MBFL/MBFL:outcome.delta}{0}
\pgfkeyssetvalue{/flpy/deltas/m2/time/family/MBFL/MBFL:ratio.delta}{1}
\pgfkeyssetvalue{/flpy/deltas/m2/time/family/PS/MBFL:coef.delta}{-0.6879011948725}
\pgfkeyssetvalue{/flpy/deltas/m2/time/family/PS/MBFL:outcome.delta}{-341812618709.654}
\pgfkeyssetvalue{/flpy/deltas/m2/time/family/PS/MBFL:ratio.delta}{0.50262988498193}
\pgfkeyssetvalue{/flpy/deltas/m2/time/family/SBFL/MBFL:coef.delta}{-3.93224479795}
\pgfkeyssetvalue{/flpy/deltas/m2/time/family/SBFL/MBFL:outcome.delta}{-341818237478.079}
\pgfkeyssetvalue{/flpy/deltas/m2/time/family/SBFL/MBFL:ratio.delta}{0.0195996259335646}
\pgfkeyssetvalue{/flpy/deltas/m2/time/family/ST/MBFL:coef.delta}{-4.4269747252}
\pgfkeyssetvalue{/flpy/deltas/m2/time/family/ST/MBFL:outcome.delta}{-341818237497.854}
\pgfkeyssetvalue{/flpy/deltas/m2/time/family/ST/MBFL:ratio.delta}{0.0119505888555931}
\pgfkeyssetvalue{/flpy/deltas/m2/time/family/MBFL/PS:coef.delta}{0.6879011948725}
\pgfkeyssetvalue{/flpy/deltas/m2/time/family/MBFL/PS:outcome.delta}{341812618709.654}
\pgfkeyssetvalue{/flpy/deltas/m2/time/family/MBFL/PS:ratio.delta}{1.98953550093018}
\pgfkeyssetvalue{/flpy/deltas/m2/time/family/PS/PS:coef.delta}{0}
\pgfkeyssetvalue{/flpy/deltas/m2/time/family/PS/PS:outcome.delta}{0}
\pgfkeyssetvalue{/flpy/deltas/m2/time/family/PS/PS:ratio.delta}{1}
\pgfkeyssetvalue{/flpy/deltas/m2/time/family/SBFL/PS:coef.delta}{-3.2443436030775}
\pgfkeyssetvalue{/flpy/deltas/m2/time/family/SBFL/PS:outcome.delta}{-5618768.42415353}
\pgfkeyssetvalue{/flpy/deltas/m2/time/family/SBFL/PS:ratio.delta}{0.0389941515997786}
\pgfkeyssetvalue{/flpy/deltas/m2/time/family/ST/PS:coef.delta}{-3.7390735303275}
\pgfkeyssetvalue{/flpy/deltas/m2/time/family/ST/PS:outcome.delta}{-5618788.19925456}
\pgfkeyssetvalue{/flpy/deltas/m2/time/family/ST/PS:ratio.delta}{0.023776120785223}
\pgfkeyssetvalue{/flpy/deltas/m2/time/family/MBFL/SBFL:coef.delta}{3.93224479795}
\pgfkeyssetvalue{/flpy/deltas/m2/time/family/MBFL/SBFL:outcome.delta}{341818237478.079}
\pgfkeyssetvalue{/flpy/deltas/m2/time/family/MBFL/SBFL:ratio.delta}{51.0213819074723}
\pgfkeyssetvalue{/flpy/deltas/m2/time/family/PS/SBFL:coef.delta}{3.2443436030775}
\pgfkeyssetvalue{/flpy/deltas/m2/time/family/PS/SBFL:outcome.delta}{5618768.42415353}
\pgfkeyssetvalue{/flpy/deltas/m2/time/family/PS/SBFL:ratio.delta}{25.6448713197719}
\pgfkeyssetvalue{/flpy/deltas/m2/time/family/SBFL/SBFL:coef.delta}{0}
\pgfkeyssetvalue{/flpy/deltas/m2/time/family/SBFL/SBFL:outcome.delta}{0}
\pgfkeyssetvalue{/flpy/deltas/m2/time/family/SBFL/SBFL:ratio.delta}{1}
\pgfkeyssetvalue{/flpy/deltas/m2/time/family/ST/SBFL:coef.delta}{-0.49472992725}
\pgfkeyssetvalue{/flpy/deltas/m2/time/family/ST/SBFL:outcome.delta}{-19.7751010302638}
\pgfkeyssetvalue{/flpy/deltas/m2/time/family/ST/SBFL:ratio.delta}{0.609735558020399}
\pgfkeyssetvalue{/flpy/deltas/m2/time/family/MBFL/ST:coef.delta}{4.4269747252}
\pgfkeyssetvalue{/flpy/deltas/m2/time/family/MBFL/ST:outcome.delta}{341818237497.854}
\pgfkeyssetvalue{/flpy/deltas/m2/time/family/MBFL/ST:ratio.delta}{83.6778850049703}
\pgfkeyssetvalue{/flpy/deltas/m2/time/family/PS/ST:coef.delta}{3.7390735303275}
\pgfkeyssetvalue{/flpy/deltas/m2/time/family/PS/ST:outcome.delta}{5618788.19925456}
\pgfkeyssetvalue{/flpy/deltas/m2/time/family/PS/ST:ratio.delta}{42.0590057155794}
\pgfkeyssetvalue{/flpy/deltas/m2/time/family/SBFL/ST:coef.delta}{0.49472992725}
\pgfkeyssetvalue{/flpy/deltas/m2/time/family/SBFL/ST:outcome.delta}{19.7751010302638}
\pgfkeyssetvalue{/flpy/deltas/m2/time/family/SBFL/ST:ratio.delta}{1.64005524500926}
\pgfkeyssetvalue{/flpy/deltas/m2/time/family/ST/ST:coef.delta}{0}
\pgfkeyssetvalue{/flpy/deltas/m2/time/family/ST/ST:outcome.delta}{0}
\pgfkeyssetvalue{/flpy/deltas/m2/time/family/ST/ST:ratio.delta}{1}
\pgfkeyssetvalue{/flpy/deltas/m3/einspect/family/MBFL/MBFL:coef.delta}{0}
\pgfkeyssetvalue{/flpy/deltas/m3/einspect/family/MBFL/MBFL:outcome.delta}{0}
\pgfkeyssetvalue{/flpy/deltas/m3/einspect/family/MBFL/MBFL:ratio.delta}{1}
\pgfkeyssetvalue{/flpy/deltas/m3/einspect/family/PS/MBFL:coef.delta}{1.26812964375}
\pgfkeyssetvalue{/flpy/deltas/m3/einspect/family/PS/MBFL:outcome.delta}{17.691932576341}
\pgfkeyssetvalue{/flpy/deltas/m3/einspect/family/PS/MBFL:ratio.delta}{3.55419872397402}
\pgfkeyssetvalue{/flpy/deltas/m3/einspect/family/SBFL/MBFL:coef.delta}{-2.0025586375}
\pgfkeyssetvalue{/flpy/deltas/m3/einspect/family/SBFL/MBFL:outcome.delta}{-5.9915887305987}
\pgfkeyssetvalue{/flpy/deltas/m3/einspect/family/SBFL/MBFL:ratio.delta}{0.134989451923003}
\pgfkeyssetvalue{/flpy/deltas/m3/einspect/family/ST/MBFL:coef.delta}{-0.444935924999999}
\pgfkeyssetvalue{/flpy/deltas/m3/einspect/family/ST/MBFL:outcome.delta}{-2.48758491702068}
\pgfkeyssetvalue{/flpy/deltas/m3/einspect/family/ST/MBFL:ratio.delta}{0.64086533816332}
\pgfkeyssetvalue{/flpy/deltas/m3/einspect/family/MBFL/PS:coef.delta}{-1.26812964375}
\pgfkeyssetvalue{/flpy/deltas/m3/einspect/family/MBFL/PS:outcome.delta}{-17.691932576341}
\pgfkeyssetvalue{/flpy/deltas/m3/einspect/family/MBFL/PS:ratio.delta}{0.281357368470911}
\pgfkeyssetvalue{/flpy/deltas/m3/einspect/family/PS/PS:coef.delta}{0}
\pgfkeyssetvalue{/flpy/deltas/m3/einspect/family/PS/PS:outcome.delta}{0}
\pgfkeyssetvalue{/flpy/deltas/m3/einspect/family/PS/PS:ratio.delta}{1}
\pgfkeyssetvalue{/flpy/deltas/m3/einspect/family/SBFL/PS:coef.delta}{-3.27068828125}
\pgfkeyssetvalue{/flpy/deltas/m3/einspect/family/SBFL/PS:outcome.delta}{-23.6835213069397}
\pgfkeyssetvalue{/flpy/deltas/m3/einspect/family/SBFL/PS:ratio.delta}{0.0379802769643868}
\pgfkeyssetvalue{/flpy/deltas/m3/einspect/family/ST/PS:coef.delta}{-1.71306556875}
\pgfkeyssetvalue{/flpy/deltas/m3/einspect/family/ST/PS:outcome.delta}{-20.1795174933617}
\pgfkeyssetvalue{/flpy/deltas/m3/einspect/family/ST/PS:ratio.delta}{0.180312185089852}
\pgfkeyssetvalue{/flpy/deltas/m3/einspect/family/MBFL/SBFL:coef.delta}{2.0025586375}
\pgfkeyssetvalue{/flpy/deltas/m3/einspect/family/MBFL/SBFL:outcome.delta}{5.9915887305987}
\pgfkeyssetvalue{/flpy/deltas/m3/einspect/family/MBFL/SBFL:ratio.delta}{7.40798622228936}
\pgfkeyssetvalue{/flpy/deltas/m3/einspect/family/PS/SBFL:coef.delta}{3.27068828125}
\pgfkeyssetvalue{/flpy/deltas/m3/einspect/family/PS/SBFL:outcome.delta}{23.6835213069397}
\pgfkeyssetvalue{/flpy/deltas/m3/einspect/family/PS/SBFL:ratio.delta}{26.3294551784779}
\pgfkeyssetvalue{/flpy/deltas/m3/einspect/family/SBFL/SBFL:coef.delta}{0}
\pgfkeyssetvalue{/flpy/deltas/m3/einspect/family/SBFL/SBFL:outcome.delta}{0}
\pgfkeyssetvalue{/flpy/deltas/m3/einspect/family/SBFL/SBFL:ratio.delta}{1}
\pgfkeyssetvalue{/flpy/deltas/m3/einspect/family/ST/SBFL:coef.delta}{1.5576227125}
\pgfkeyssetvalue{/flpy/deltas/m3/einspect/family/ST/SBFL:outcome.delta}{3.50400381357802}
\pgfkeyssetvalue{/flpy/deltas/m3/einspect/family/ST/SBFL:ratio.delta}{4.74752159545668}
\pgfkeyssetvalue{/flpy/deltas/m3/einspect/family/MBFL/ST:coef.delta}{0.444935924999999}
\pgfkeyssetvalue{/flpy/deltas/m3/einspect/family/MBFL/ST:outcome.delta}{2.48758491702068}
\pgfkeyssetvalue{/flpy/deltas/m3/einspect/family/MBFL/ST:ratio.delta}{1.56039021062668}
\pgfkeyssetvalue{/flpy/deltas/m3/einspect/family/PS/ST:coef.delta}{1.71306556875}
\pgfkeyssetvalue{/flpy/deltas/m3/einspect/family/PS/ST:outcome.delta}{20.1795174933617}
\pgfkeyssetvalue{/flpy/deltas/m3/einspect/family/PS/ST:ratio.delta}{5.54593689551089}
\pgfkeyssetvalue{/flpy/deltas/m3/einspect/family/SBFL/ST:coef.delta}{-1.5576227125}
\pgfkeyssetvalue{/flpy/deltas/m3/einspect/family/SBFL/ST:outcome.delta}{-3.50400381357802}
\pgfkeyssetvalue{/flpy/deltas/m3/einspect/family/SBFL/ST:ratio.delta}{0.210636219318515}
\pgfkeyssetvalue{/flpy/deltas/m3/einspect/family/ST/ST:coef.delta}{0}
\pgfkeyssetvalue{/flpy/deltas/m3/einspect/family/ST/ST:outcome.delta}{0}
\pgfkeyssetvalue{/flpy/deltas/m3/einspect/family/ST/ST:ratio.delta}{1}
\pgfkeyssetvalue{/flpy/deltas/m4/einspect/crashing/crashing MBFL/crashing MBFL:coef.delta}{0}
\pgfkeyssetvalue{/flpy/deltas/m4/einspect/crashing/crashing MBFL/crashing MBFL:outcome.delta}{0}
\pgfkeyssetvalue{/flpy/deltas/m4/einspect/crashing/crashing MBFL/crashing MBFL:ratio.delta}{1}
\pgfkeyssetvalue{/flpy/deltas/m4/einspect/crashing/crashing PS/crashing MBFL:coef.delta}{2.15683540262}
\pgfkeyssetvalue{/flpy/deltas/m4/einspect/crashing/crashing PS/crashing MBFL:outcome.delta}{143.705228253692}
\pgfkeyssetvalue{/flpy/deltas/m4/einspect/crashing/crashing PS/crashing MBFL:ratio.delta}{8.64374037250212}
\pgfkeyssetvalue{/flpy/deltas/m4/einspect/crashing/crashing SBFL/crashing MBFL:coef.delta}{1.1803868772335}
\pgfkeyssetvalue{/flpy/deltas/m4/einspect/crashing/crashing SBFL/crashing MBFL:outcome.delta}{42.4067681429403}
\pgfkeyssetvalue{/flpy/deltas/m4/einspect/crashing/crashing SBFL/crashing MBFL:ratio.delta}{3.25563348975706}
\pgfkeyssetvalue{/flpy/deltas/m4/einspect/crashing/crashing ST/crashing MBFL:coef.delta}{-0.67235787205}
\pgfkeyssetvalue{/flpy/deltas/m4/einspect/crashing/crashing ST/crashing MBFL:outcome.delta}{-9.20272133254542}
\pgfkeyssetvalue{/flpy/deltas/m4/einspect/crashing/crashing ST/crashing MBFL:ratio.delta}{0.510503455753343}
\pgfkeyssetvalue{/flpy/deltas/m4/einspect/crashing/crashing MBFL/crashing PS:coef.delta}{-2.15683540262}
\pgfkeyssetvalue{/flpy/deltas/m4/einspect/crashing/crashing MBFL/crashing PS:outcome.delta}{-143.705228253692}
\pgfkeyssetvalue{/flpy/deltas/m4/einspect/crashing/crashing MBFL/crashing PS:ratio.delta}{0.11569065669548}
\pgfkeyssetvalue{/flpy/deltas/m4/einspect/crashing/crashing PS/crashing PS:coef.delta}{0}
\pgfkeyssetvalue{/flpy/deltas/m4/einspect/crashing/crashing PS/crashing PS:outcome.delta}{0}
\pgfkeyssetvalue{/flpy/deltas/m4/einspect/crashing/crashing PS/crashing PS:ratio.delta}{1}
\pgfkeyssetvalue{/flpy/deltas/m4/einspect/crashing/crashing SBFL/crashing PS:coef.delta}{-0.9764485253865}
\pgfkeyssetvalue{/flpy/deltas/m4/einspect/crashing/crashing SBFL/crashing PS:outcome.delta}{-101.298460110752}
\pgfkeyssetvalue{/flpy/deltas/m4/einspect/crashing/crashing SBFL/crashing PS:ratio.delta}{0.37664637638979}
\pgfkeyssetvalue{/flpy/deltas/m4/einspect/crashing/crashing ST/crashing PS:coef.delta}{-2.82919327467}
\pgfkeyssetvalue{/flpy/deltas/m4/einspect/crashing/crashing ST/crashing PS:outcome.delta}{-152.907949586238}
\pgfkeyssetvalue{/flpy/deltas/m4/einspect/crashing/crashing ST/crashing PS:ratio.delta}{0.059060480041416}
\pgfkeyssetvalue{/flpy/deltas/m4/einspect/crashing/crashing MBFL/crashing SBFL:coef.delta}{-1.1803868772335}
\pgfkeyssetvalue{/flpy/deltas/m4/einspect/crashing/crashing MBFL/crashing SBFL:outcome.delta}{-42.4067681429403}
\pgfkeyssetvalue{/flpy/deltas/m4/einspect/crashing/crashing MBFL/crashing SBFL:ratio.delta}{0.307159882445681}
\pgfkeyssetvalue{/flpy/deltas/m4/einspect/crashing/crashing PS/crashing SBFL:coef.delta}{0.9764485253865}
\pgfkeyssetvalue{/flpy/deltas/m4/einspect/crashing/crashing PS/crashing SBFL:outcome.delta}{101.298460110752}
\pgfkeyssetvalue{/flpy/deltas/m4/einspect/crashing/crashing PS/crashing SBFL:ratio.delta}{2.65501027670874}
\pgfkeyssetvalue{/flpy/deltas/m4/einspect/crashing/crashing SBFL/crashing SBFL:coef.delta}{0}
\pgfkeyssetvalue{/flpy/deltas/m4/einspect/crashing/crashing SBFL/crashing SBFL:outcome.delta}{0}
\pgfkeyssetvalue{/flpy/deltas/m4/einspect/crashing/crashing SBFL/crashing SBFL:ratio.delta}{1}
\pgfkeyssetvalue{/flpy/deltas/m4/einspect/crashing/crashing ST/crashing SBFL:coef.delta}{-1.8527447492835}
\pgfkeyssetvalue{/flpy/deltas/m4/einspect/crashing/crashing ST/crashing SBFL:outcome.delta}{-51.6094894754857}
\pgfkeyssetvalue{/flpy/deltas/m4/einspect/crashing/crashing ST/crashing SBFL:ratio.delta}{0.15680618145731}
\pgfkeyssetvalue{/flpy/deltas/m4/einspect/crashing/crashing MBFL/crashing ST:coef.delta}{0.67235787205}
\pgfkeyssetvalue{/flpy/deltas/m4/einspect/crashing/crashing MBFL/crashing ST:outcome.delta}{9.20272133254542}
\pgfkeyssetvalue{/flpy/deltas/m4/einspect/crashing/crashing MBFL/crashing ST:ratio.delta}{1.95885059881586}
\pgfkeyssetvalue{/flpy/deltas/m4/einspect/crashing/crashing PS/crashing ST:coef.delta}{2.82919327467}
\pgfkeyssetvalue{/flpy/deltas/m4/einspect/crashing/crashing PS/crashing ST:outcome.delta}{152.907949586238}
\pgfkeyssetvalue{/flpy/deltas/m4/einspect/crashing/crashing PS/crashing ST:ratio.delta}{16.9317960046846}
\pgfkeyssetvalue{/flpy/deltas/m4/einspect/crashing/crashing SBFL/crashing ST:coef.delta}{1.8527447492835}
\pgfkeyssetvalue{/flpy/deltas/m4/einspect/crashing/crashing SBFL/crashing ST:outcome.delta}{51.6094894754857}
\pgfkeyssetvalue{/flpy/deltas/m4/einspect/crashing/crashing SBFL/crashing ST:ratio.delta}{6.37729961093558}
\pgfkeyssetvalue{/flpy/deltas/m4/einspect/crashing/crashing ST/crashing ST:coef.delta}{0}
\pgfkeyssetvalue{/flpy/deltas/m4/einspect/crashing/crashing ST/crashing ST:outcome.delta}{0}
\pgfkeyssetvalue{/flpy/deltas/m4/einspect/crashing/crashing ST/crashing ST:ratio.delta}{1}
\pgfkeyssetvalue{/flpy/deltas/m4/einspect/mutable/ismutable MBFL/ismutable MBFL:coef.delta}{0}
\pgfkeyssetvalue{/flpy/deltas/m4/einspect/mutable/ismutable MBFL/ismutable MBFL:outcome.delta}{0}
\pgfkeyssetvalue{/flpy/deltas/m4/einspect/mutable/ismutable MBFL/ismutable MBFL:ratio.delta}{1}
\pgfkeyssetvalue{/flpy/deltas/m4/einspect/mutable/ismutable PS/ismutable MBFL:coef.delta}{1.2555643154823}
\pgfkeyssetvalue{/flpy/deltas/m4/einspect/mutable/ismutable PS/ismutable MBFL:outcome.delta}{116.316100292253}
\pgfkeyssetvalue{/flpy/deltas/m4/einspect/mutable/ismutable PS/ismutable MBFL:ratio.delta}{3.50981846049072}
\pgfkeyssetvalue{/flpy/deltas/m4/einspect/mutable/ismutable SBFL/ismutable MBFL:coef.delta}{-0.04124731340195}
\pgfkeyssetvalue{/flpy/deltas/m4/einspect/mutable/ismutable SBFL/ismutable MBFL:outcome.delta}{-1.87269582786142}
\pgfkeyssetvalue{/flpy/deltas/m4/einspect/mutable/ismutable SBFL/ismutable MBFL:ratio.delta}{0.959591780691417}
\pgfkeyssetvalue{/flpy/deltas/m4/einspect/mutable/ismutable ST/ismutable MBFL:coef.delta}{1.1185544808608}
\pgfkeyssetvalue{/flpy/deltas/m4/einspect/mutable/ismutable ST/ismutable MBFL:outcome.delta}{95.4893148666807}
\pgfkeyssetvalue{/flpy/deltas/m4/einspect/mutable/ismutable ST/ismutable MBFL:ratio.delta}{3.0604270983866}
\pgfkeyssetvalue{/flpy/deltas/m4/einspect/mutable/ismutable MBFL/ismutable PS:coef.delta}{-1.2555643154823}
\pgfkeyssetvalue{/flpy/deltas/m4/einspect/mutable/ismutable MBFL/ismutable PS:outcome.delta}{-116.316100292253}
\pgfkeyssetvalue{/flpy/deltas/m4/einspect/mutable/ismutable MBFL/ismutable PS:ratio.delta}{0.284915020892615}
\pgfkeyssetvalue{/flpy/deltas/m4/einspect/mutable/ismutable PS/ismutable PS:coef.delta}{0}
\pgfkeyssetvalue{/flpy/deltas/m4/einspect/mutable/ismutable PS/ismutable PS:outcome.delta}{0}
\pgfkeyssetvalue{/flpy/deltas/m4/einspect/mutable/ismutable PS/ismutable PS:ratio.delta}{1}
\pgfkeyssetvalue{/flpy/deltas/m4/einspect/mutable/ismutable SBFL/ismutable PS:coef.delta}{-1.29681162888425}
\pgfkeyssetvalue{/flpy/deltas/m4/einspect/mutable/ismutable SBFL/ismutable PS:outcome.delta}{-118.188796120114}
\pgfkeyssetvalue{/flpy/deltas/m4/einspect/mutable/ismutable SBFL/ismutable PS:ratio.delta}{0.273402112244077}
\pgfkeyssetvalue{/flpy/deltas/m4/einspect/mutable/ismutable ST/ismutable PS:coef.delta}{-0.1370098346215}
\pgfkeyssetvalue{/flpy/deltas/m4/einspect/mutable/ismutable ST/ismutable PS:outcome.delta}{-20.8267854255721}
\pgfkeyssetvalue{/flpy/deltas/m4/einspect/mutable/ismutable ST/ismutable PS:ratio.delta}{0.871961650677144}
\pgfkeyssetvalue{/flpy/deltas/m4/einspect/mutable/ismutable MBFL/ismutable SBFL:coef.delta}{0.04124731340195}
\pgfkeyssetvalue{/flpy/deltas/m4/einspect/mutable/ismutable MBFL/ismutable SBFL:outcome.delta}{1.87269582786142}
\pgfkeyssetvalue{/flpy/deltas/m4/einspect/mutable/ismutable MBFL/ismutable SBFL:ratio.delta}{1.04210980139854}
\pgfkeyssetvalue{/flpy/deltas/m4/einspect/mutable/ismutable PS/ismutable SBFL:coef.delta}{1.29681162888425}
\pgfkeyssetvalue{/flpy/deltas/m4/einspect/mutable/ismutable PS/ismutable SBFL:outcome.delta}{118.188796120114}
\pgfkeyssetvalue{/flpy/deltas/m4/einspect/mutable/ismutable PS/ismutable SBFL:ratio.delta}{3.6576162188069}
\pgfkeyssetvalue{/flpy/deltas/m4/einspect/mutable/ismutable SBFL/ismutable SBFL:coef.delta}{0}
\pgfkeyssetvalue{/flpy/deltas/m4/einspect/mutable/ismutable SBFL/ismutable SBFL:outcome.delta}{0}
\pgfkeyssetvalue{/flpy/deltas/m4/einspect/mutable/ismutable SBFL/ismutable SBFL:ratio.delta}{1}
\pgfkeyssetvalue{/flpy/deltas/m4/einspect/mutable/ismutable ST/ismutable SBFL:coef.delta}{1.15980179426275}
\pgfkeyssetvalue{/flpy/deltas/m4/einspect/mutable/ismutable ST/ismutable SBFL:outcome.delta}{97.3620106945421}
\pgfkeyssetvalue{/flpy/deltas/m4/einspect/mutable/ismutable ST/ismutable SBFL:ratio.delta}{3.18930107569436}
\pgfkeyssetvalue{/flpy/deltas/m4/einspect/mutable/ismutable MBFL/ismutable ST:coef.delta}{-1.1185544808608}
\pgfkeyssetvalue{/flpy/deltas/m4/einspect/mutable/ismutable MBFL/ismutable ST:outcome.delta}{-95.4893148666807}
\pgfkeyssetvalue{/flpy/deltas/m4/einspect/mutable/ismutable MBFL/ismutable ST:ratio.delta}{0.326751779360201}
\pgfkeyssetvalue{/flpy/deltas/m4/einspect/mutable/ismutable PS/ismutable ST:coef.delta}{0.1370098346215}
\pgfkeyssetvalue{/flpy/deltas/m4/einspect/mutable/ismutable PS/ismutable ST:outcome.delta}{20.8267854255721}
\pgfkeyssetvalue{/flpy/deltas/m4/einspect/mutable/ismutable PS/ismutable ST:ratio.delta}{1.14683942719663}
\pgfkeyssetvalue{/flpy/deltas/m4/einspect/mutable/ismutable SBFL/ismutable ST:coef.delta}{-1.15980179426275}
\pgfkeyssetvalue{/flpy/deltas/m4/einspect/mutable/ismutable SBFL/ismutable ST:outcome.delta}{-97.3620106945421}
\pgfkeyssetvalue{/flpy/deltas/m4/einspect/mutable/ismutable SBFL/ismutable ST:ratio.delta}{0.313548321800345}
\pgfkeyssetvalue{/flpy/deltas/m4/einspect/mutable/ismutable ST/ismutable ST:coef.delta}{0}
\pgfkeyssetvalue{/flpy/deltas/m4/einspect/mutable/ismutable ST/ismutable ST:outcome.delta}{0}
\pgfkeyssetvalue{/flpy/deltas/m4/einspect/mutable/ismutable ST/ismutable ST:ratio.delta}{1}
\pgfkeyssetvalue{/flpy/deltas/m4/einspect/predicate/predicate MBFL/predicate MBFL:coef.delta}{0}
\pgfkeyssetvalue{/flpy/deltas/m4/einspect/predicate/predicate MBFL/predicate MBFL:outcome.delta}{0}
\pgfkeyssetvalue{/flpy/deltas/m4/einspect/predicate/predicate MBFL/predicate MBFL:ratio.delta}{1}
\pgfkeyssetvalue{/flpy/deltas/m4/einspect/predicate/predicate PS/predicate MBFL:coef.delta}{0.39294543644445}
\pgfkeyssetvalue{/flpy/deltas/m4/einspect/predicate/predicate PS/predicate MBFL:outcome.delta}{18.0006348893976}
\pgfkeyssetvalue{/flpy/deltas/m4/einspect/predicate/predicate PS/predicate MBFL:ratio.delta}{1.48133756007987}
\pgfkeyssetvalue{/flpy/deltas/m4/einspect/predicate/predicate SBFL/predicate MBFL:coef.delta}{0.43850001253675}
\pgfkeyssetvalue{/flpy/deltas/m4/einspect/predicate/predicate SBFL/predicate MBFL:outcome.delta}{20.5826199020843}
\pgfkeyssetvalue{/flpy/deltas/m4/einspect/predicate/predicate SBFL/predicate MBFL:ratio.delta}{1.55037992304284}
\pgfkeyssetvalue{/flpy/deltas/m4/einspect/predicate/predicate ST/predicate MBFL:coef.delta}{0.94674469269725}
\pgfkeyssetvalue{/flpy/deltas/m4/einspect/predicate/predicate ST/predicate MBFL:outcome.delta}{58.9866924010772}
\pgfkeyssetvalue{/flpy/deltas/m4/einspect/predicate/predicate ST/predicate MBFL:ratio.delta}{2.57730606534059}
\pgfkeyssetvalue{/flpy/deltas/m4/einspect/predicate/predicate MBFL/predicate PS:coef.delta}{-0.39294543644445}
\pgfkeyssetvalue{/flpy/deltas/m4/einspect/predicate/predicate MBFL/predicate PS:outcome.delta}{-18.0006348893976}
\pgfkeyssetvalue{/flpy/deltas/m4/einspect/predicate/predicate MBFL/predicate PS:ratio.delta}{0.67506558055956}
\pgfkeyssetvalue{/flpy/deltas/m4/einspect/predicate/predicate PS/predicate PS:coef.delta}{0}
\pgfkeyssetvalue{/flpy/deltas/m4/einspect/predicate/predicate PS/predicate PS:outcome.delta}{0}
\pgfkeyssetvalue{/flpy/deltas/m4/einspect/predicate/predicate PS/predicate PS:ratio.delta}{1}
\pgfkeyssetvalue{/flpy/deltas/m4/einspect/predicate/predicate SBFL/predicate PS:coef.delta}{0.0455545760923}
\pgfkeyssetvalue{/flpy/deltas/m4/einspect/predicate/predicate SBFL/predicate PS:outcome.delta}{2.58198501268673}
\pgfkeyssetvalue{/flpy/deltas/m4/einspect/predicate/predicate SBFL/predicate PS:ratio.delta}{1.0466081228368}
\pgfkeyssetvalue{/flpy/deltas/m4/einspect/predicate/predicate ST/predicate PS:coef.delta}{0.5537992562528}
\pgfkeyssetvalue{/flpy/deltas/m4/einspect/predicate/predicate ST/predicate PS:outcome.delta}{40.9860575116796}
\pgfkeyssetvalue{/flpy/deltas/m4/einspect/predicate/predicate ST/predicate PS:ratio.delta}{1.73985061527882}
\pgfkeyssetvalue{/flpy/deltas/m4/einspect/predicate/predicate MBFL/predicate SBFL:coef.delta}{-0.43850001253675}
\pgfkeyssetvalue{/flpy/deltas/m4/einspect/predicate/predicate MBFL/predicate SBFL:outcome.delta}{-20.5826199020843}
\pgfkeyssetvalue{/flpy/deltas/m4/einspect/predicate/predicate MBFL/predicate SBFL:ratio.delta}{0.645003192531902}
\pgfkeyssetvalue{/flpy/deltas/m4/einspect/predicate/predicate PS/predicate SBFL:coef.delta}{-0.0455545760923}
\pgfkeyssetvalue{/flpy/deltas/m4/einspect/predicate/predicate PS/predicate SBFL:outcome.delta}{-2.58198501268673}
\pgfkeyssetvalue{/flpy/deltas/m4/einspect/predicate/predicate PS/predicate SBFL:ratio.delta}{0.955467455468936}
\pgfkeyssetvalue{/flpy/deltas/m4/einspect/predicate/predicate SBFL/predicate SBFL:coef.delta}{0}
\pgfkeyssetvalue{/flpy/deltas/m4/einspect/predicate/predicate SBFL/predicate SBFL:outcome.delta}{0}
\pgfkeyssetvalue{/flpy/deltas/m4/einspect/predicate/predicate SBFL/predicate SBFL:ratio.delta}{1}
\pgfkeyssetvalue{/flpy/deltas/m4/einspect/predicate/predicate ST/predicate SBFL:coef.delta}{0.5082446801605}
\pgfkeyssetvalue{/flpy/deltas/m4/einspect/predicate/predicate ST/predicate SBFL:outcome.delta}{38.4040724989929}
\pgfkeyssetvalue{/flpy/deltas/m4/einspect/predicate/predicate ST/predicate SBFL:ratio.delta}{1.66237064027652}
\pgfkeyssetvalue{/flpy/deltas/m4/einspect/predicate/predicate MBFL/predicate ST:coef.delta}{-0.94674469269725}
\pgfkeyssetvalue{/flpy/deltas/m4/einspect/predicate/predicate MBFL/predicate ST:outcome.delta}{-58.9866924010772}
\pgfkeyssetvalue{/flpy/deltas/m4/einspect/predicate/predicate MBFL/predicate ST:ratio.delta}{0.388002035710046}
\pgfkeyssetvalue{/flpy/deltas/m4/einspect/predicate/predicate PS/predicate ST:coef.delta}{-0.5537992562528}
\pgfkeyssetvalue{/flpy/deltas/m4/einspect/predicate/predicate PS/predicate ST:outcome.delta}{-40.9860575116796}
\pgfkeyssetvalue{/flpy/deltas/m4/einspect/predicate/predicate PS/predicate ST:ratio.delta}{0.574761988884743}
\pgfkeyssetvalue{/flpy/deltas/m4/einspect/predicate/predicate SBFL/predicate ST:coef.delta}{-0.5082446801605}
\pgfkeyssetvalue{/flpy/deltas/m4/einspect/predicate/predicate SBFL/predicate ST:outcome.delta}{-38.4040724989929}
\pgfkeyssetvalue{/flpy/deltas/m4/einspect/predicate/predicate SBFL/predicate ST:ratio.delta}{0.601550566264609}
\pgfkeyssetvalue{/flpy/deltas/m4/einspect/predicate/predicate ST/predicate ST:coef.delta}{0}
\pgfkeyssetvalue{/flpy/deltas/m4/einspect/predicate/predicate ST/predicate ST:outcome.delta}{0}
\pgfkeyssetvalue{/flpy/deltas/m4/einspect/predicate/predicate ST/predicate ST:ratio.delta}{1}
\pgfkeyssetvalue{/flpy/deltas/m6/einspect/category/CL/CL:coef.delta}{0}
\pgfkeyssetvalue{/flpy/deltas/m6/einspect/category/CL/CL:outcome.delta}{0}
\pgfkeyssetvalue{/flpy/deltas/m6/einspect/category/CL/CL:ratio.delta}{1}
\pgfkeyssetvalue{/flpy/deltas/m6/einspect/category/DEV/CL:coef.delta}{-2.46104017479625}
\pgfkeyssetvalue{/flpy/deltas/m6/einspect/category/DEV/CL:outcome.delta}{-127.292291497055}
\pgfkeyssetvalue{/flpy/deltas/m6/einspect/category/DEV/CL:ratio.delta}{0.085346129887339}
\pgfkeyssetvalue{/flpy/deltas/m6/einspect/category/DS/CL:coef.delta}{-0.31623790102325}
\pgfkeyssetvalue{/flpy/deltas/m6/einspect/category/DS/CL:outcome.delta}{-37.7309056784644}
\pgfkeyssetvalue{/flpy/deltas/m6/einspect/category/DS/CL:ratio.delta}{0.728886026808034}
\pgfkeyssetvalue{/flpy/deltas/m6/einspect/category/WEB/CL:coef.delta}{-2.33572496300125}
\pgfkeyssetvalue{/flpy/deltas/m6/einspect/category/WEB/CL:outcome.delta}{-125.706562599304}
\pgfkeyssetvalue{/flpy/deltas/m6/einspect/category/WEB/CL:ratio.delta}{0.0967403239427568}
\pgfkeyssetvalue{/flpy/deltas/m6/einspect/category/CL/DEV:coef.delta}{2.46104017479625}
\pgfkeyssetvalue{/flpy/deltas/m6/einspect/category/CL/DEV:outcome.delta}{127.292291497055}
\pgfkeyssetvalue{/flpy/deltas/m6/einspect/category/CL/DEV:ratio.delta}{11.7169929242257}
\pgfkeyssetvalue{/flpy/deltas/m6/einspect/category/DEV/DEV:coef.delta}{0}
\pgfkeyssetvalue{/flpy/deltas/m6/einspect/category/DEV/DEV:outcome.delta}{0}
\pgfkeyssetvalue{/flpy/deltas/m6/einspect/category/DEV/DEV:ratio.delta}{1}
\pgfkeyssetvalue{/flpy/deltas/m6/einspect/category/DS/DEV:coef.delta}{2.144802273773}
\pgfkeyssetvalue{/flpy/deltas/m6/einspect/category/DS/DEV:outcome.delta}{89.5613858185908}
\pgfkeyssetvalue{/flpy/deltas/m6/einspect/category/DS/DEV:ratio.delta}{8.54035241867673}
\pgfkeyssetvalue{/flpy/deltas/m6/einspect/category/WEB/DEV:coef.delta}{0.125315211795}
\pgfkeyssetvalue{/flpy/deltas/m6/einspect/category/WEB/DEV:outcome.delta}{1.58572889775118}
\pgfkeyssetvalue{/flpy/deltas/m6/einspect/category/WEB/DEV:ratio.delta}{1.13350569112459}
\pgfkeyssetvalue{/flpy/deltas/m6/einspect/category/CL/DS:coef.delta}{0.31623790102325}
\pgfkeyssetvalue{/flpy/deltas/m6/einspect/category/CL/DS:outcome.delta}{37.7309056784644}
\pgfkeyssetvalue{/flpy/deltas/m6/einspect/category/CL/DS:ratio.delta}{1.37195660668546}
\pgfkeyssetvalue{/flpy/deltas/m6/einspect/category/DEV/DS:coef.delta}{-2.144802273773}
\pgfkeyssetvalue{/flpy/deltas/m6/einspect/category/DEV/DS:outcome.delta}{-89.5613858185908}
\pgfkeyssetvalue{/flpy/deltas/m6/einspect/category/DEV/DS:ratio.delta}{0.11709118675397}
\pgfkeyssetvalue{/flpy/deltas/m6/einspect/category/DS/DS:coef.delta}{0}
\pgfkeyssetvalue{/flpy/deltas/m6/einspect/category/DS/DS:outcome.delta}{0}
\pgfkeyssetvalue{/flpy/deltas/m6/einspect/category/DS/DS:ratio.delta}{1}
\pgfkeyssetvalue{/flpy/deltas/m6/einspect/category/WEB/DS:coef.delta}{-2.019487061978}
\pgfkeyssetvalue{/flpy/deltas/m6/einspect/category/WEB/DS:outcome.delta}{-87.9756569208396}
\pgfkeyssetvalue{/flpy/deltas/m6/einspect/category/WEB/DS:ratio.delta}{0.132723526566157}
\pgfkeyssetvalue{/flpy/deltas/m6/einspect/category/CL/WEB:coef.delta}{2.33572496300125}
\pgfkeyssetvalue{/flpy/deltas/m6/einspect/category/CL/WEB:outcome.delta}{125.706562599304}
\pgfkeyssetvalue{/flpy/deltas/m6/einspect/category/CL/WEB:ratio.delta}{10.3369511207314}
\pgfkeyssetvalue{/flpy/deltas/m6/einspect/category/DEV/WEB:coef.delta}{-0.125315211795}
\pgfkeyssetvalue{/flpy/deltas/m6/einspect/category/DEV/WEB:outcome.delta}{-1.58572889775118}
\pgfkeyssetvalue{/flpy/deltas/m6/einspect/category/DEV/WEB:ratio.delta}{0.882218772989017}
\pgfkeyssetvalue{/flpy/deltas/m6/einspect/category/DS/WEB:coef.delta}{2.019487061978}
\pgfkeyssetvalue{/flpy/deltas/m6/einspect/category/DS/WEB:outcome.delta}{87.9756569208396}
\pgfkeyssetvalue{/flpy/deltas/m6/einspect/category/DS/WEB:ratio.delta}{7.53445923169877}
\pgfkeyssetvalue{/flpy/deltas/m6/einspect/category/WEB/WEB:coef.delta}{0}
\pgfkeyssetvalue{/flpy/deltas/m6/einspect/category/WEB/WEB:outcome.delta}{0}
\pgfkeyssetvalue{/flpy/deltas/m6/einspect/category/WEB/WEB:ratio.delta}{1}
\pgfkeyssetvalue{/flpy/deltas/m6/einspect/crashing/crashing MBFL/crashing MBFL:coef.delta}{0}
\pgfkeyssetvalue{/flpy/deltas/m6/einspect/crashing/crashing MBFL/crashing MBFL:outcome.delta}{0}
\pgfkeyssetvalue{/flpy/deltas/m6/einspect/crashing/crashing MBFL/crashing MBFL:ratio.delta}{1}
\pgfkeyssetvalue{/flpy/deltas/m6/einspect/crashing/crashing PS/crashing MBFL:coef.delta}{2.25769036255}
\pgfkeyssetvalue{/flpy/deltas/m6/einspect/crashing/crashing PS/crashing MBFL:outcome.delta}{154.28417324165}
\pgfkeyssetvalue{/flpy/deltas/m6/einspect/crashing/crashing PS/crashing MBFL:ratio.delta}{9.56098124560721}
\pgfkeyssetvalue{/flpy/deltas/m6/einspect/crashing/crashing SBFL/crashing MBFL:coef.delta}{1.19827610488455}
\pgfkeyssetvalue{/flpy/deltas/m6/einspect/crashing/crashing SBFL/crashing MBFL:outcome.delta}{41.7095915736318}
\pgfkeyssetvalue{/flpy/deltas/m6/einspect/crashing/crashing SBFL/crashing MBFL:ratio.delta}{3.31439831994265}
\pgfkeyssetvalue{/flpy/deltas/m6/einspect/crashing/crashing ST/crashing MBFL:coef.delta}{-0.593852605615}
\pgfkeyssetvalue{/flpy/deltas/m6/einspect/crashing/crashing ST/crashing MBFL:outcome.delta}{-8.07023171987616}
\pgfkeyssetvalue{/flpy/deltas/m6/einspect/crashing/crashing ST/crashing MBFL:ratio.delta}{0.552195788867669}
\pgfkeyssetvalue{/flpy/deltas/m6/einspect/crashing/crashing MBFL/crashing PS:coef.delta}{-2.25769036255}
\pgfkeyssetvalue{/flpy/deltas/m6/einspect/crashing/crashing MBFL/crashing PS:outcome.delta}{-154.28417324165}
\pgfkeyssetvalue{/flpy/deltas/m6/einspect/crashing/crashing MBFL/crashing PS:ratio.delta}{0.104591775081606}
\pgfkeyssetvalue{/flpy/deltas/m6/einspect/crashing/crashing PS/crashing PS:coef.delta}{0}
\pgfkeyssetvalue{/flpy/deltas/m6/einspect/crashing/crashing PS/crashing PS:outcome.delta}{0}
\pgfkeyssetvalue{/flpy/deltas/m6/einspect/crashing/crashing PS/crashing PS:ratio.delta}{1}
\pgfkeyssetvalue{/flpy/deltas/m6/einspect/crashing/crashing SBFL/crashing PS:coef.delta}{-1.05941425766545}
\pgfkeyssetvalue{/flpy/deltas/m6/einspect/crashing/crashing SBFL/crashing PS:outcome.delta}{-112.574581668018}
\pgfkeyssetvalue{/flpy/deltas/m6/einspect/crashing/crashing SBFL/crashing PS:ratio.delta}{0.346658803610294}
\pgfkeyssetvalue{/flpy/deltas/m6/einspect/crashing/crashing ST/crashing PS:coef.delta}{-2.851542968165}
\pgfkeyssetvalue{/flpy/deltas/m6/einspect/crashing/crashing ST/crashing PS:outcome.delta}{-162.354404961526}
\pgfkeyssetvalue{/flpy/deltas/m6/einspect/crashing/crashing ST/crashing PS:ratio.delta}{0.0577551377502572}
\pgfkeyssetvalue{/flpy/deltas/m6/einspect/crashing/crashing MBFL/crashing SBFL:coef.delta}{-1.19827610488455}
\pgfkeyssetvalue{/flpy/deltas/m6/einspect/crashing/crashing MBFL/crashing SBFL:outcome.delta}{-41.7095915736318}
\pgfkeyssetvalue{/flpy/deltas/m6/einspect/crashing/crashing MBFL/crashing SBFL:ratio.delta}{0.301713886946848}
\pgfkeyssetvalue{/flpy/deltas/m6/einspect/crashing/crashing PS/crashing SBFL:coef.delta}{1.05941425766545}
\pgfkeyssetvalue{/flpy/deltas/m6/einspect/crashing/crashing PS/crashing SBFL:outcome.delta}{112.574581668018}
\pgfkeyssetvalue{/flpy/deltas/m6/einspect/crashing/crashing PS/crashing SBFL:ratio.delta}{2.88468081463807}
\pgfkeyssetvalue{/flpy/deltas/m6/einspect/crashing/crashing SBFL/crashing SBFL:coef.delta}{0}
\pgfkeyssetvalue{/flpy/deltas/m6/einspect/crashing/crashing SBFL/crashing SBFL:outcome.delta}{0}
\pgfkeyssetvalue{/flpy/deltas/m6/einspect/crashing/crashing SBFL/crashing SBFL:ratio.delta}{1}
\pgfkeyssetvalue{/flpy/deltas/m6/einspect/crashing/crashing ST/crashing SBFL:coef.delta}{-1.79212871049955}
\pgfkeyssetvalue{/flpy/deltas/m6/einspect/crashing/crashing ST/crashing SBFL:outcome.delta}{-49.779823293508}
\pgfkeyssetvalue{/flpy/deltas/m6/einspect/crashing/crashing ST/crashing SBFL:ratio.delta}{0.166605137814946}
\pgfkeyssetvalue{/flpy/deltas/m6/einspect/crashing/crashing MBFL/crashing ST:coef.delta}{0.593852605615}
\pgfkeyssetvalue{/flpy/deltas/m6/einspect/crashing/crashing MBFL/crashing ST:outcome.delta}{8.07023171987616}
\pgfkeyssetvalue{/flpy/deltas/m6/einspect/crashing/crashing MBFL/crashing ST:ratio.delta}{1.81095187641796}
\pgfkeyssetvalue{/flpy/deltas/m6/einspect/crashing/crashing PS/crashing ST:coef.delta}{2.851542968165}
\pgfkeyssetvalue{/flpy/deltas/m6/einspect/crashing/crashing PS/crashing ST:outcome.delta}{162.354404961526}
\pgfkeyssetvalue{/flpy/deltas/m6/einspect/crashing/crashing PS/crashing ST:ratio.delta}{17.3144769271293}
\pgfkeyssetvalue{/flpy/deltas/m6/einspect/crashing/crashing SBFL/crashing ST:coef.delta}{1.79212871049955}
\pgfkeyssetvalue{/flpy/deltas/m6/einspect/crashing/crashing SBFL/crashing ST:outcome.delta}{49.779823293508}
\pgfkeyssetvalue{/flpy/deltas/m6/einspect/crashing/crashing SBFL/crashing ST:ratio.delta}{6.00221585669667}
\pgfkeyssetvalue{/flpy/deltas/m6/einspect/crashing/crashing ST/crashing ST:coef.delta}{0}
\pgfkeyssetvalue{/flpy/deltas/m6/einspect/crashing/crashing ST/crashing ST:outcome.delta}{0}
\pgfkeyssetvalue{/flpy/deltas/m6/einspect/crashing/crashing ST/crashing ST:ratio.delta}{1}
\pgfkeyssetvalue{/flpy/deltas/m6/einspect/family/MBFL/MBFL:coef.delta}{0}
\pgfkeyssetvalue{/flpy/deltas/m6/einspect/family/MBFL/MBFL:outcome.delta}{0}
\pgfkeyssetvalue{/flpy/deltas/m6/einspect/family/MBFL/MBFL:ratio.delta}{1}
\pgfkeyssetvalue{/flpy/deltas/m6/einspect/family/PS/MBFL:coef.delta}{1.00344464824718}
\pgfkeyssetvalue{/flpy/deltas/m6/einspect/family/PS/MBFL:outcome.delta}{48.7130426307368}
\pgfkeyssetvalue{/flpy/deltas/m6/einspect/family/PS/MBFL:ratio.delta}{2.72766149875258}
\pgfkeyssetvalue{/flpy/deltas/m6/einspect/family/SBFL/MBFL:coef.delta}{0.07872333934125}
\pgfkeyssetvalue{/flpy/deltas/m6/einspect/family/SBFL/MBFL:outcome.delta}{2.30938748896576}
\pgfkeyssetvalue{/flpy/deltas/m6/einspect/family/SBFL/MBFL:ratio.delta}{1.08190496012241}
\pgfkeyssetvalue{/flpy/deltas/m6/einspect/family/ST/MBFL:coef.delta}{1.0592698107795}
\pgfkeyssetvalue{/flpy/deltas/m6/einspect/family/ST/MBFL:outcome.delta}{53.1286021962154}
\pgfkeyssetvalue{/flpy/deltas/m6/einspect/family/ST/MBFL:ratio.delta}{2.88426416157028}
\pgfkeyssetvalue{/flpy/deltas/m6/einspect/family/MBFL/PS:coef.delta}{-1.00344464824718}
\pgfkeyssetvalue{/flpy/deltas/m6/einspect/family/MBFL/PS:outcome.delta}{-48.7130426307368}
\pgfkeyssetvalue{/flpy/deltas/m6/einspect/family/MBFL/PS:ratio.delta}{0.366614405950784}
\pgfkeyssetvalue{/flpy/deltas/m6/einspect/family/PS/PS:coef.delta}{0}
\pgfkeyssetvalue{/flpy/deltas/m6/einspect/family/PS/PS:outcome.delta}{0}
\pgfkeyssetvalue{/flpy/deltas/m6/einspect/family/PS/PS:ratio.delta}{1}
\pgfkeyssetvalue{/flpy/deltas/m6/einspect/family/SBFL/PS:coef.delta}{-0.924721308905925}
\pgfkeyssetvalue{/flpy/deltas/m6/einspect/family/SBFL/PS:outcome.delta}{-46.403655141771}
\pgfkeyssetvalue{/flpy/deltas/m6/einspect/family/SBFL/PS:ratio.delta}{0.396641944250486}
\pgfkeyssetvalue{/flpy/deltas/m6/einspect/family/ST/PS:coef.delta}{0.055825162532325}
\pgfkeyssetvalue{/flpy/deltas/m6/einspect/family/ST/PS:outcome.delta}{4.41555956547859}
\pgfkeyssetvalue{/flpy/deltas/m6/einspect/family/ST/PS:ratio.delta}{1.05741279219922}
\pgfkeyssetvalue{/flpy/deltas/m6/einspect/family/MBFL/SBFL:coef.delta}{-0.07872333934125}
\pgfkeyssetvalue{/flpy/deltas/m6/einspect/family/MBFL/SBFL:outcome.delta}{-2.30938748896576}
\pgfkeyssetvalue{/flpy/deltas/m6/einspect/family/MBFL/SBFL:ratio.delta}{0.924295605306082}
\pgfkeyssetvalue{/flpy/deltas/m6/einspect/family/PS/SBFL:coef.delta}{0.924721308905925}
\pgfkeyssetvalue{/flpy/deltas/m6/einspect/family/PS/SBFL:outcome.delta}{46.403655141771}
\pgfkeyssetvalue{/flpy/deltas/m6/einspect/family/PS/SBFL:ratio.delta}{2.52116553605961}
\pgfkeyssetvalue{/flpy/deltas/m6/einspect/family/SBFL/SBFL:coef.delta}{0}
\pgfkeyssetvalue{/flpy/deltas/m6/einspect/family/SBFL/SBFL:outcome.delta}{0}
\pgfkeyssetvalue{/flpy/deltas/m6/einspect/family/SBFL/SBFL:ratio.delta}{1}
\pgfkeyssetvalue{/flpy/deltas/m6/einspect/family/ST/SBFL:coef.delta}{0.98054647143825}
\pgfkeyssetvalue{/flpy/deltas/m6/einspect/family/ST/SBFL:outcome.delta}{50.8192147072496}
\pgfkeyssetvalue{/flpy/deltas/m6/einspect/family/ST/SBFL:ratio.delta}{2.66591268908124}
\pgfkeyssetvalue{/flpy/deltas/m6/einspect/family/MBFL/ST:coef.delta}{-1.0592698107795}
\pgfkeyssetvalue{/flpy/deltas/m6/einspect/family/MBFL/ST:outcome.delta}{-53.1286021962154}
\pgfkeyssetvalue{/flpy/deltas/m6/einspect/family/MBFL/ST:ratio.delta}{0.346708881011638}
\pgfkeyssetvalue{/flpy/deltas/m6/einspect/family/PS/ST:coef.delta}{-0.055825162532325}
\pgfkeyssetvalue{/flpy/deltas/m6/einspect/family/PS/ST:outcome.delta}{-4.41555956547859}
\pgfkeyssetvalue{/flpy/deltas/m6/einspect/family/PS/ST:ratio.delta}{0.945704466011032}
\pgfkeyssetvalue{/flpy/deltas/m6/einspect/family/SBFL/ST:coef.delta}{-0.98054647143825}
\pgfkeyssetvalue{/flpy/deltas/m6/einspect/family/SBFL/ST:outcome.delta}{-50.8192147072496}
\pgfkeyssetvalue{/flpy/deltas/m6/einspect/family/SBFL/ST:ratio.delta}{0.375106058084983}
\pgfkeyssetvalue{/flpy/deltas/m6/einspect/family/ST/ST:coef.delta}{0}
\pgfkeyssetvalue{/flpy/deltas/m6/einspect/family/ST/ST:outcome.delta}{0}
\pgfkeyssetvalue{/flpy/deltas/m6/einspect/family/ST/ST:ratio.delta}{1}
\pgfkeyssetvalue{/flpy/deltas/m6/einspect/logmutability/logmutability MBFL/logmutability MBFL:coef.delta}{0}
\pgfkeyssetvalue{/flpy/deltas/m6/einspect/logmutability/logmutability MBFL/logmutability MBFL:outcome.delta}{0}
\pgfkeyssetvalue{/flpy/deltas/m6/einspect/logmutability/logmutability MBFL/logmutability MBFL:ratio.delta}{1}
\pgfkeyssetvalue{/flpy/deltas/m6/einspect/logmutability/logmutability PS/logmutability MBFL:coef.delta}{1.1547470667445}
\pgfkeyssetvalue{/flpy/deltas/m6/einspect/logmutability/logmutability PS/logmutability MBFL:outcome.delta}{76.0849129074274}
\pgfkeyssetvalue{/flpy/deltas/m6/einspect/logmutability/logmutability PS/logmutability MBFL:ratio.delta}{3.17322070297245}
\pgfkeyssetvalue{/flpy/deltas/m6/einspect/logmutability/logmutability SBFL/logmutability MBFL:coef.delta}{0.4199211471895}
\pgfkeyssetvalue{/flpy/deltas/m6/einspect/logmutability/logmutability SBFL/logmutability MBFL:outcome.delta}{18.2697822004949}
\pgfkeyssetvalue{/flpy/deltas/m6/einspect/logmutability/logmutability SBFL/logmutability MBFL:ratio.delta}{1.52184154940398}
\pgfkeyssetvalue{/flpy/deltas/m6/einspect/logmutability/logmutability ST/logmutability MBFL:coef.delta}{1.04627771195525}
\pgfkeyssetvalue{/flpy/deltas/m6/einspect/logmutability/logmutability ST/logmutability MBFL:outcome.delta}{64.665044105586}
\pgfkeyssetvalue{/flpy/deltas/m6/einspect/logmutability/logmutability ST/logmutability MBFL:ratio.delta}{2.84703388991987}
\pgfkeyssetvalue{/flpy/deltas/m6/einspect/logmutability/logmutability MBFL/logmutability PS:coef.delta}{-1.1547470667445}
\pgfkeyssetvalue{/flpy/deltas/m6/einspect/logmutability/logmutability MBFL/logmutability PS:outcome.delta}{-76.0849129074274}
\pgfkeyssetvalue{/flpy/deltas/m6/einspect/logmutability/logmutability MBFL/logmutability PS:ratio.delta}{0.315137235510681}
\pgfkeyssetvalue{/flpy/deltas/m6/einspect/logmutability/logmutability PS/logmutability PS:coef.delta}{0}
\pgfkeyssetvalue{/flpy/deltas/m6/einspect/logmutability/logmutability PS/logmutability PS:outcome.delta}{0}
\pgfkeyssetvalue{/flpy/deltas/m6/einspect/logmutability/logmutability PS/logmutability PS:ratio.delta}{1}
\pgfkeyssetvalue{/flpy/deltas/m6/einspect/logmutability/logmutability SBFL/logmutability PS:coef.delta}{-0.734825919555}
\pgfkeyssetvalue{/flpy/deltas/m6/einspect/logmutability/logmutability SBFL/logmutability PS:outcome.delta}{-57.8151307069324}
\pgfkeyssetvalue{/flpy/deltas/m6/einspect/logmutability/logmutability SBFL/logmutability PS:ratio.delta}{0.479588938764461}
\pgfkeyssetvalue{/flpy/deltas/m6/einspect/logmutability/logmutability ST/logmutability PS:coef.delta}{-0.10846935478925}
\pgfkeyssetvalue{/flpy/deltas/m6/einspect/logmutability/logmutability ST/logmutability PS:outcome.delta}{-11.4198688018414}
\pgfkeyssetvalue{/flpy/deltas/m6/einspect/logmutability/logmutability ST/logmutability PS:ratio.delta}{0.897206389474571}
\pgfkeyssetvalue{/flpy/deltas/m6/einspect/logmutability/logmutability MBFL/logmutability SBFL:coef.delta}{-0.4199211471895}
\pgfkeyssetvalue{/flpy/deltas/m6/einspect/logmutability/logmutability MBFL/logmutability SBFL:outcome.delta}{-18.2697822004949}
\pgfkeyssetvalue{/flpy/deltas/m6/einspect/logmutability/logmutability MBFL/logmutability SBFL:ratio.delta}{0.657098631846165}
\pgfkeyssetvalue{/flpy/deltas/m6/einspect/logmutability/logmutability PS/logmutability SBFL:coef.delta}{0.734825919555}
\pgfkeyssetvalue{/flpy/deltas/m6/einspect/logmutability/logmutability PS/logmutability SBFL:outcome.delta}{57.8151307069324}
\pgfkeyssetvalue{/flpy/deltas/m6/einspect/logmutability/logmutability PS/logmutability SBFL:ratio.delta}{2.08511898246912}
\pgfkeyssetvalue{/flpy/deltas/m6/einspect/logmutability/logmutability SBFL/logmutability SBFL:coef.delta}{0}
\pgfkeyssetvalue{/flpy/deltas/m6/einspect/logmutability/logmutability SBFL/logmutability SBFL:outcome.delta}{0}
\pgfkeyssetvalue{/flpy/deltas/m6/einspect/logmutability/logmutability SBFL/logmutability SBFL:ratio.delta}{1}
\pgfkeyssetvalue{/flpy/deltas/m6/einspect/logmutability/logmutability ST/logmutability SBFL:coef.delta}{0.62635656476575}
\pgfkeyssetvalue{/flpy/deltas/m6/einspect/logmutability/logmutability ST/logmutability SBFL:outcome.delta}{46.395261905091}
\pgfkeyssetvalue{/flpy/deltas/m6/einspect/logmutability/logmutability ST/logmutability SBFL:ratio.delta}{1.87078207388601}
\pgfkeyssetvalue{/flpy/deltas/m6/einspect/logmutability/logmutability MBFL/logmutability ST:coef.delta}{-1.04627771195525}
\pgfkeyssetvalue{/flpy/deltas/m6/einspect/logmutability/logmutability MBFL/logmutability ST:outcome.delta}{-64.665044105586}
\pgfkeyssetvalue{/flpy/deltas/m6/einspect/logmutability/logmutability MBFL/logmutability ST:ratio.delta}{0.351242745490516}
\pgfkeyssetvalue{/flpy/deltas/m6/einspect/logmutability/logmutability PS/logmutability ST:coef.delta}{0.10846935478925}
\pgfkeyssetvalue{/flpy/deltas/m6/einspect/logmutability/logmutability PS/logmutability ST:outcome.delta}{11.4198688018414}
\pgfkeyssetvalue{/flpy/deltas/m6/einspect/logmutability/logmutability PS/logmutability ST:ratio.delta}{1.11457075175939}
\pgfkeyssetvalue{/flpy/deltas/m6/einspect/logmutability/logmutability SBFL/logmutability ST:coef.delta}{-0.62635656476575}
\pgfkeyssetvalue{/flpy/deltas/m6/einspect/logmutability/logmutability SBFL/logmutability ST:outcome.delta}{-46.395261905091}
\pgfkeyssetvalue{/flpy/deltas/m6/einspect/logmutability/logmutability SBFL/logmutability ST:ratio.delta}{0.534535804014193}
\pgfkeyssetvalue{/flpy/deltas/m6/einspect/logmutability/logmutability ST/logmutability ST:coef.delta}{0}
\pgfkeyssetvalue{/flpy/deltas/m6/einspect/logmutability/logmutability ST/logmutability ST:outcome.delta}{0}
\pgfkeyssetvalue{/flpy/deltas/m6/einspect/logmutability/logmutability ST/logmutability ST:ratio.delta}{1}
\pgfkeyssetvalue{/flpy/deltas/m6/einspect/predicate/predicate MBFL/predicate MBFL:coef.delta}{0}
\pgfkeyssetvalue{/flpy/deltas/m6/einspect/predicate/predicate MBFL/predicate MBFL:outcome.delta}{0}
\pgfkeyssetvalue{/flpy/deltas/m6/einspect/predicate/predicate MBFL/predicate MBFL:ratio.delta}{1}
\pgfkeyssetvalue{/flpy/deltas/m6/einspect/predicate/predicate PS/predicate MBFL:coef.delta}{0.312975731009825}
\pgfkeyssetvalue{/flpy/deltas/m6/einspect/predicate/predicate PS/predicate MBFL:outcome.delta}{16.8747752096196}
\pgfkeyssetvalue{/flpy/deltas/m6/einspect/predicate/predicate PS/predicate MBFL:ratio.delta}{1.36748834306373}
\pgfkeyssetvalue{/flpy/deltas/m6/einspect/predicate/predicate SBFL/predicate MBFL:coef.delta}{0.18043769644875}
\pgfkeyssetvalue{/flpy/deltas/m6/einspect/predicate/predicate SBFL/predicate MBFL:outcome.delta}{9.08013367503642}
\pgfkeyssetvalue{/flpy/deltas/m6/einspect/predicate/predicate SBFL/predicate MBFL:ratio.delta}{1.19774149560713}
\pgfkeyssetvalue{/flpy/deltas/m6/einspect/predicate/predicate ST/predicate MBFL:coef.delta}{0.83822537240375}
\pgfkeyssetvalue{/flpy/deltas/m6/einspect/predicate/predicate ST/predicate MBFL:outcome.delta}{60.257942312032}
\pgfkeyssetvalue{/flpy/deltas/m6/einspect/predicate/predicate ST/predicate MBFL:ratio.delta}{2.31225993333523}
\pgfkeyssetvalue{/flpy/deltas/m6/einspect/predicate/predicate MBFL/predicate PS:coef.delta}{-0.312975731009825}
\pgfkeyssetvalue{/flpy/deltas/m6/einspect/predicate/predicate MBFL/predicate PS:outcome.delta}{-16.8747752096196}
\pgfkeyssetvalue{/flpy/deltas/m6/einspect/predicate/predicate MBFL/predicate PS:ratio.delta}{0.731267659481175}
\pgfkeyssetvalue{/flpy/deltas/m6/einspect/predicate/predicate PS/predicate PS:coef.delta}{0}
\pgfkeyssetvalue{/flpy/deltas/m6/einspect/predicate/predicate PS/predicate PS:outcome.delta}{0}
\pgfkeyssetvalue{/flpy/deltas/m6/einspect/predicate/predicate PS/predicate PS:ratio.delta}{1}
\pgfkeyssetvalue{/flpy/deltas/m6/einspect/predicate/predicate SBFL/predicate PS:coef.delta}{-0.132538034561075}
\pgfkeyssetvalue{/flpy/deltas/m6/einspect/predicate/predicate SBFL/predicate PS:outcome.delta}{-7.79464153458316}
\pgfkeyssetvalue{/flpy/deltas/m6/einspect/predicate/predicate SBFL/predicate PS:ratio.delta}{0.875869620156105}
\pgfkeyssetvalue{/flpy/deltas/m6/einspect/predicate/predicate ST/predicate PS:coef.delta}{0.525249641393925}
\pgfkeyssetvalue{/flpy/deltas/m6/einspect/predicate/predicate ST/predicate PS:outcome.delta}{43.3831671024124}
\pgfkeyssetvalue{/flpy/deltas/m6/einspect/predicate/predicate ST/predicate PS:ratio.delta}{1.69088090956215}
\pgfkeyssetvalue{/flpy/deltas/m6/einspect/predicate/predicate MBFL/predicate SBFL:coef.delta}{-0.18043769644875}
\pgfkeyssetvalue{/flpy/deltas/m6/einspect/predicate/predicate MBFL/predicate SBFL:outcome.delta}{-9.08013367503642}
\pgfkeyssetvalue{/flpy/deltas/m6/einspect/predicate/predicate MBFL/predicate SBFL:ratio.delta}{0.834904696604092}
\pgfkeyssetvalue{/flpy/deltas/m6/einspect/predicate/predicate PS/predicate SBFL:coef.delta}{0.132538034561075}
\pgfkeyssetvalue{/flpy/deltas/m6/einspect/predicate/predicate PS/predicate SBFL:outcome.delta}{7.79464153458316}
\pgfkeyssetvalue{/flpy/deltas/m6/einspect/predicate/predicate PS/predicate SBFL:ratio.delta}{1.14172244017525}
\pgfkeyssetvalue{/flpy/deltas/m6/einspect/predicate/predicate SBFL/predicate SBFL:coef.delta}{0}
\pgfkeyssetvalue{/flpy/deltas/m6/einspect/predicate/predicate SBFL/predicate SBFL:outcome.delta}{0}
\pgfkeyssetvalue{/flpy/deltas/m6/einspect/predicate/predicate SBFL/predicate SBFL:ratio.delta}{1}
\pgfkeyssetvalue{/flpy/deltas/m6/einspect/predicate/predicate ST/predicate SBFL:coef.delta}{0.657787675955}
\pgfkeyssetvalue{/flpy/deltas/m6/einspect/predicate/predicate ST/predicate SBFL:outcome.delta}{51.1778086369955}
\pgfkeyssetvalue{/flpy/deltas/m6/einspect/predicate/predicate ST/predicate SBFL:ratio.delta}{1.93051667811105}
\pgfkeyssetvalue{/flpy/deltas/m6/einspect/predicate/predicate MBFL/predicate ST:coef.delta}{-0.83822537240375}
\pgfkeyssetvalue{/flpy/deltas/m6/einspect/predicate/predicate MBFL/predicate ST:outcome.delta}{-60.257942312032}
\pgfkeyssetvalue{/flpy/deltas/m6/einspect/predicate/predicate MBFL/predicate ST:ratio.delta}{0.4324773290335}
\pgfkeyssetvalue{/flpy/deltas/m6/einspect/predicate/predicate PS/predicate ST:coef.delta}{-0.525249641393925}
\pgfkeyssetvalue{/flpy/deltas/m6/einspect/predicate/predicate PS/predicate ST:outcome.delta}{-43.3831671024124}
\pgfkeyssetvalue{/flpy/deltas/m6/einspect/predicate/predicate PS/predicate ST:ratio.delta}{0.591407706092646}
\pgfkeyssetvalue{/flpy/deltas/m6/einspect/predicate/predicate SBFL/predicate ST:coef.delta}{-0.657787675955}
\pgfkeyssetvalue{/flpy/deltas/m6/einspect/predicate/predicate SBFL/predicate ST:outcome.delta}{-51.1778086369955}
\pgfkeyssetvalue{/flpy/deltas/m6/einspect/predicate/predicate SBFL/predicate ST:ratio.delta}{0.517996042892759}
\pgfkeyssetvalue{/flpy/deltas/m6/einspect/predicate/predicate ST/predicate ST:coef.delta}{0}
\pgfkeyssetvalue{/flpy/deltas/m6/einspect/predicate/predicate ST/predicate ST:outcome.delta}{0}
\pgfkeyssetvalue{/flpy/deltas/m6/einspect/predicate/predicate ST/predicate ST:ratio.delta}{1}
\pgfkeyssetvalue{}{}
 
\pgfkeyssetvalue{/flpy/number of families}{4}
\pgfkeyssetvalue{/flpy/number of categories}{4}
\pgfkeyssetvalue{/flpy/number of project subjects}{13}
\pgfkeyssetvalue{/flpy/number of techniques}{7}
\pgfkeyssetvalue{/flpy/number of granularities}{3}
\pgfkeyssetvalue{/flpy/number of subjects}{135}
\pgfkeyssetvalue{/flpy/number of experiments}{945}
\pgfkeyssetvalue{/flpy/number of CLI}{43}
\pgfkeyssetvalue{/flpy/number of DEV}{30}
\pgfkeyssetvalue{/flpy/number of DS}{42}
\pgfkeyssetvalue{/flpy/number of WEB}{20}

\pgfkeyssetvalue{/flpy/crashing bugs}{49}
\pgfkeyssetvalue{/flpy/predicate bugs}{52}
\pgfkeyssetvalue{/flpy/mutable bugs}{74}
\pgfkeyssetvalue{/flpy/other bugs}{34}

\pgfkeyssetvalue{/flpy/critpred/number of subjects}{29}
\pgfkeyssetvalue{/flpy/critpred/SBFL/@1p}{12.64367816091954}
\pgfkeyssetvalue{/flpy/critpred/SBFL/@3p}{16.091954022988507}
\pgfkeyssetvalue{/flpy/critpred/SBFL/@5p}{19.54022988505747}
\pgfkeyssetvalue{/flpy/critpred/SBFL/@10p}{31.034482758620687}
\pgfkeyssetvalue{/flpy/critpred/PS/@1p}{13.793103448275861}
\pgfkeyssetvalue{/flpy/critpred/PS/@3p}{24.137931034482758}
\pgfkeyssetvalue{/flpy/critpred/PS/@5p}{31.03448275862069}
\pgfkeyssetvalue{/flpy/critpred/PS/@10p}{31.03448275862069}
\pgfkeyssetvalue{/flpy/critpred/MBFL/@1p}{5.172413793103448}
\pgfkeyssetvalue{/flpy/critpred/MBFL/@3p}{17.241379310344826}
\pgfkeyssetvalue{/flpy/critpred/MBFL/@5p}{24.137931034482758}
\pgfkeyssetvalue{/flpy/critpred/MBFL/@10p}{31.034482758620687}

\newcommand{\reducedstrut}{%
  \vrule width 0pt height .9\ht\strutbox depth .9\dp\strutbox\relax%
}

\newcommand{\best}[2][bcol!50]{%
  \begingroup
  \setlength{\fboxsep}{0pt}%
  \colorbox{#1}{\reducedstrut#2\/}%
  \endgroup
}

\usepackage{mathtools}

\usepackage{amsmath,amssymb}
\DeclareMathOperator*{\mean}{mean}

\DeclareDocumentCommand{\spellout}{m O{13}}{%
  \IfStrEq{#1}{ }{\NAN}{%
    \IfInteger{#1}{%
      \expandafter\ifnum#1<#2{\numberstringnum{#1}}\else#1\fi%
    }{\NAN}%
}}

\DeclareDocumentCommand{\n}{t. t: o m o O{} t| t!}{%
  \begingroup%
  \pgfkeys{/pgf/fpu=true}%
  \IfBooleanTF{#2}{%
    \pgfkeyssetvalue{/tmp/value}{#4}%
    \pgfkeyssetvalue{/tmp/found}{found}%
  }{%
    \pgfkeysifdefined{\keyfamily#4}{%
      \pgfkeyssetvalue{/tmp/value}{\pgfkeysvalueof{\keyfamily#4}}%
      \pgfkeyssetvalue{/tmp/found}{found}%
    }{}%
  }%
  \pgfkeysifdefined{/tmp/found}{%
    \IfNoValueF{#5}{%
      \pgfkeyssetvalue{/tmp/multiplier}{#5}%
      \pgfmathparse{\pgfkeysvalueof{/tmp/multiplier} * \pgfkeysvalueof{/tmp/value}}%
      \pgfkeyslet{/tmp/value}\pgfmathresult%
    }%
    \IfBooleanTF{#1}{%
      \IfBooleanTF{#8}{%
        \spellout{\pgfkeysvalueof{/tmp/value}}}{%
        \pgfkeysvalueof{/tmp/value}}%
    }{%
      \IfNoValueTF{#3}{%
        \pgfmathprintnumber%
        [set thousands separator={\,},int detect,#6]%
        {\pgfkeysvalueof{/tmp/value}}%
      }{%
        {\pgfmathprintnumber%
          [precision=#3,fixed,zerofill,set thousands separator={\,},#6]%
          {\pgfkeysvalueof{/tmp/value}}}%
      }%
    }%
    \IfBooleanT{#7}{{\smaller[1.2]\%}}%
  }{%
    \NAN%
  }%
  \pgfkeys{/pgf/fpu=false}%
  \endgroup%
}

\usepackage{boxedminipage}
{\smallskip
	\noindent
	\let\emph=\textbf
	\begin{boxedminipage}{\columnwidth}\begin{center}\em}%
     {\end{center}\end{boxedminipage}%
	\smallskip
 }

\usepackage{tcolorbox}

\newcounter{nfinding}[subsection]
\newcounter{totalfindings}

\makeatletter
\renewcommand{\p@nfinding}{\the\value{subsection}.}
\makeatother

\ExplSyntaxOn
\DeclareDocumentCommand{\finding}{s o m}{%
  \IfBooleanF{#1}{\refstepcounter{nfinding}}%
  \stepcounter{totalfindings}%
  \begin{tcolorbox}[
    boxsep=0pt,
    left=6pt,
    right=6pt,
    top=4pt,
    bottom=4pt,
    colframe=white,
    colback=bcol!20!white,
    boxrule=0.3pt, %
    ]
    \small%
  \IfBooleanTF{#1}%
  {\IfNoValueTF{#2}%
    {\tl_set:Nn \l_header{Finding}}%
    {\tl_set:Nn \l_header{Finding~#2}}}%
  {\IfNoValueTF{#2}%
    {\tl_set:Nn \l_header{Finding~\the\value{subsection}.\the\value{nfinding}}}%
    {\tl_set:Nn \l_header{Finding~#2.\thenfinding}}}%
  \begin{description}[style=nextline,
    leftmargin=!,
    labelwidth=\widthof{\textbf{\tl_use:N \l_header :}}]%
  \item[\textbf{\tl_use:N \l_header :}] #3
  \end{description}%
\end{tcolorbox}
}
\ExplSyntaxOff

\newcounter{comparisonfinding}
\newcommand{\newrow}{%
  \refstepcounter{comparisonfinding}\the\value{comparisonfinding}%
}

\usepackage{booktabs}
\usepackage{relsize}
\usepackage[skip=1pt]{subcaption}  %

\graphicspath{{./figs/}}
\DeclareGraphicsExtensions{.pdf,.png}
\usepackage{amsmath}
\usepackage{array}

\usepackage{listings}

\let\origthelstnumber\thelstnumber
\makeatletter
\newcommand*\SuppressNumber{%
  \lst@AddToHook{OnNewLine}{%
    \let\thelstnumber\relax%
     \advance\c@lstnumber-\@ne\relax%
    }%
}

\newcommand*\ReactivateNumber[1]{%
  \setcounter{lstnumber}{\numexpr#1-1\relax}
  \lst@AddToHook{OnNewLine}{%
   \let\thelstnumber\origthelstnumber%
   \refstepcounter{lstnumber}
  }%
}

\makeatother

    \newcommand{\Py}[1]{\mbox{\lstinline[basicstyle=\ttfamily,language=Python]|#1|}}

\newcommand{\fp}{{\smaller \textsc{FauxPy}}\xspace}
\newcommand{\tech}[1]{\ensuremath{\mathsf{#1}}}
\newcommand{\tr}[1]{\ensuremath{\textsl{tr}}}
\newcommand{\project}[1]{\textsf{#1}\xspace}

\newcommand{\category}[1]{\textsc{#1}\xspace}
\newcommand{\dev}{\category{dev}}
\newcommand{\ds}{\category{ds}}
\newcommand{\cli}{\category{cl}}
\newcommand{\web}{\category{web}}

\newcommand{\muse}{{Muse}\xspace}

\usepackage{multicol,multirow}

\usepackage{xurl}

\usepackage{savesym}
\savesymbol{endnote}

\usepackage{enotez}

\restoresymbol{enotez}{endnote}

\setenotez{list-name={URL References}}
\setenotez{counter-format=arabic}
\setenotez{mark-cs={\formatEndNoteMark}}
\newcommand{\formatEndNoteMark}[1]{\textsuperscript{\{\color{blue}{\textsf{#1}}\}}}

\DeclareDocumentCommand{\projURL}{O{} m}
{\enotezendnote{\IfNoValueF{#1}{#1\xspace}\url{#2}}}

\newcommand{\bip}{{\smaller[0.5]{\textsc{BugsInPy}}}\xspace}

\DeclareRobustCommand{\var}[1]{\ensuremath{\textnormal{\textsl{#1}}}}

\DeclareRobustCommand{\dist}[1]{\ensuremath{\textsf{#1}}}

\DeclareDocumentCommand{\einspect}{s o o}%
{\ensuremath{\IfBooleanTF{#1}{\widetilde{\mathcal{I}}}{\mathcal{I}}\IfNoValueF{#2}{_{#2}}\IfNoValueF{#3}{(#3)}}\xspace}

\newcommand{\einspectso}{\ensuremath{E_{\mathrm{inspect}}}\xspace}

\newcommand{\exam}{\ensuremath{\mathcal{E}}\xspace}

\DeclareDocumentCommand{\outputlen}{m o}%
{\ensuremath{|#1{\IfNoValueF{#2}{_{#2}}}|}\xspace}

\newcommand{\combine}{\textsf{\smaller[0.5]CombineFL}\xspace}
\newcommand{\average}{\textsf{\smaller[0.5]AvgFL}\xspace}

\newcommand{\timecost}{\ensuremath{T}\xspace}

\newcommand{\truth}{\ensuremath{\mathcal{F}}}
\newcommand{\locs}[1][n]{\ensuremath{\langle \ell_1, s_1 \rangle\, \ldots\, \langle \ell_{#1}, s_{#1} \rangle}}
\newcommand{\metric}[1]{\ensuremath{\mathsf{#1}}}

\title{An Empirical Study of Fault Localization in Python Programs}

\author{Mohammad Rezaalipour \and
        Carlo A.\ Furia}

\institute{M. Rezaalipour \at
              Software Institute, Università della Svizzera italiana, Lugano, Switzerland \\
              \email{rezaam@usi.ch}           %
           \and
           C. A.\ Furia \at
              Software Institute, Università della Svizzera italiana, Lugano, Switzerland \\
              \url{https://bugcounting.net/}
}

\date{}

\makeatletter
\let\Title\@title
\makeatother

\usepackage{silence}
\usepackage{hyperref}

\hypersetup{
  pdftitle={\Title{}},
  pdfauthor={},
  pdfsubject={},
  pdfkeywords={},
  pdfcreator={pdfTeX}
}

\DeclareDocumentCommand{\tabref}{s O{f} m}{%
  \IfBooleanTF{#1}{\cite[#2~#3]{Zou:2021}}{#2~\ref{#3}}
}

\usepackage[normalem]{ulem}

\begin{document}

\maketitle


\begin{abstract}
  Despite its massive popularity as a programming language,
  especially in novel domains like data science programs,
  there is comparatively little research about fault localization
  that targets Python.
  Even though it is plausible that several findings about programming
  languages like C/C++ and Java---the most common choices for fault localization research---carry over to other languages, whether
  the dynamic nature of Python and how the language is used in practice
  affect the capabilities of classic fault localization approaches
  remain open questions to investigate.

  This paper is the first multi-family large-scale empirical study of fault localization
  on real-world Python programs and faults.
  Using Zou et al.'s recent large-scale empirical study
  of fault localization in Java~\cite{Zou:2021}
  as the basis of our study,
  we investigated the
  effectiveness (i.e., localization accuracy),
  efficiency (i.e., runtime performance),
  and other features (e.g., different entity granularities)
  of \n.{number of techniques}!
  well-known fault-localization techniques
  in \n.{number of families}! families
  (spectrum-based, mutation-based, predicate switching, and stack-trace based)
  on \n.{number of subjects} faults
  from \n.{number of project subjects}! open-source Python projects
  from the \bip curated collection~\cite{Widyasari:2020}.
  
  The results replicate for Python several results known about Java,
  and shed light on whether Python's peculiarities affect the capabilities
  of fault localization.
  The replication package that accompanies this paper includes
  detailed data about our experiments,
  as well as the tool \fp that we implemented to conduct the study.

\end{abstract}

\section{Introduction}
\label{sec:introduction}

It is commonplace that debugging
is an activity that takes up a disproportionate amount of
time and resources in software development~\cite{CodeComplete}.
This also explains the popularity of \emph{fault localization}
as a research subject in software engineering:
identifying locations in a program's source code
that are implicated in some observed failures
(such as crashes or other kinds of runtime errors)
is a key step of debugging.
This paper contributes to the empirical knowledge about
the capabilities of fault localization techniques,
targeting the Python programming language.

Despite the massive amount of work on fault localization
(see \autoref{sec:related-work})
and the popularity of the Python programming language,%
\footnote{TIOBE language popularity index: \url{https://www.tiobe.com/tiobe-index/}}%
\footnote{Popularity of Programming Language Index: \url{https://pypl.github.io/PYPL.html}}
most empirical studies of fault localization target
languages like Java or C.
This leaves open the question of whether
Python's characteristics%
---such as the fact that it is dynamically typed,
or that it is dominant in certain application domains
such as data science---%
affect the capabilities of classic fault localization
techniques---developed and tested primarily on
different kinds of languages and programs.

This paper fills this knowledge gap:
to our knowledge,
it is the first multi-family large-scale empirical study of fault
localization in real-world Python programs.
The starting point is Zou et al.'s recent extensive study~\cite{Zou:2021}
of fault localization for Java.
This paper's main contribution is a differentiated conceptual
replication~\cite{DBLP:conf/laser/JuzgadoG10}
of Zou et al.'s study,
sharing several of its features:
\begin{enumerate*}
\item it experiments
  with several different families (spectrum-based, mutation-based, predicate switching, and stack-trace-based) of fault localization techniques;
\item it targets a large number
  of faults in real-world projects
  (\n.{number of subjects} faults in \n.{number of project subjects} projects) ;
\item it studies
  fault localization effectiveness at different granularities (statement, function, and module);
\item it considers
  combinations of complementary fault localization techniques.
\end{enumerate*}
The fundamental novelty of our replication is that it targets
the Python programming language;
furthermore, 
\begin{enumerate*}
\item it analyzes fault localization effectiveness of different kinds of faults
  and different categories of projects;
\item it estimates the contributions of different fault localization features
  by means of regression statistical models;
\item it compares its main findings for Python to Zou et al.'s~\cite{Zou:2021} for Java.
\end{enumerate*}

The main \emph{findings} of our Python fault localization study are as follows:

\begin{enumerate}[label=\arabic*.]
\item Spectrum-based fault localization techniques are the most effective, followed by mutation-based fault localization techniques.
\item Predicate switching and stack-trace fault localization
  are considerably less effective, but they can work well on small sets of faults that match their characteristics.
\item Stack-trace is by far the fastest fault localization technique,
  predicate switching and mutation-based fault localization techniques are
  the most time consuming.
\item Bugs in data-science related projects tend to be harder to localize
  than those in other categories of projects.
\item Combining fault localization techniques boosts their effectiveness
  with only a modest hit on efficiency.
\item The main findings about relative effectiveness
  still hold at all granularity levels.
\item Most of Zou et al.~\cite{Zou:2021}'s findings about fault localization
  in Java carry over to Python.
\end{enumerate}

A practical challenge to carry out a large-scale fault localization study
of Python projects 
was that, at the time of writing,
there were no open-source tools that support a variety
of fault localization techniques for this programming language.
Thus, to perform this study,
we implemented \fp:
a fault-localization tool for Python that
supports \n.{number of techniques}! fault localization techniques
in \n.{number of families}! families,
is highly configurable,
and works with the most common Python unit testing frameworks
(such as Pytest and Unittest).
The present paper is not a paper about \fp, 
which we plan to present in detail in a separate publication.
Nevertheless, we briefly discuss the key features of \fp,
and make the tool available as part of this paper's replication package%
---which also includes all the detailed experimental artifacts and data
that support further independent analysis and replicability.

The rest of the paper is organized as follows.
\autoref{sec:preliminaries} presents the fault localization techniques
that fall within the scope of the empirical study,
and outlines \fp's features.
\autoref{sec:related-work} summarizes the most relevant
related work in fault localization,
demonstrating how Python is underrepresented in this area.
\autoref{sec:design} presents in detail the paper's research questions,
and the experimental methodology that we followed to answer them.
\autoref{sec:experimental-results} details the experimental results
for each investigated research question,
and presents any limitations and threats to the validity of the findings.
\autoref{sec:conclusions} concludes with a high-level discussion of the
main results, and of possible avenues for future work.

\paragraph{Replication package.}
For reproducibility,
all experimental artifacts of this paper's empirical study,
and the implementation of the \fp tool, are available:
  \begin{center}
    \url{https://doi.org/10.6084/m9.figshare.23254688}
  \end{center}

\section{Fault Localization and \fp}
\label{sec:preliminaries}

Fault localization techniques~\cite{zeller:whyfail,Wong:2016}
relate program failures
(such as crashes or assertion violations)
to faulty locations in the program's source code
that are responsible for the failures.
Concretely, a fault localization technique $L$
assigns a \emph{suspiciousness score} $L_T(e)$
to any program entity $e$%
---usually, a location, function, or module---%
given test inputs $T$
that trigger a failure in the program.
The suspiciousness score $L_T(e)$
should be higher the more likely $e$
is the location of a fault that is ultimately responsible
for the failure.
Thus, a list of all program entities $e_1, e_2, \ldots$
ordered by decreasing suspiciousness score
$L_T(e_1) \geq L_T(e_2) \geq \ldots$
is fault localization technique $L$'s overall output.

Let $T = P \cup F$ be a set of tests
partitioned into passing $P$ and failing $F$,
such that $F \neq \emptyset$%
---there is at least one failing test---%
and all failing tests originate in the same fault.
Tests $T$ and a program $p$
are thus the target of a single fault localization run.
Then, fault localization techniques differ
in what kind of information they extract from $T$ and $p$ 
to compute suspiciousness scores.
A fault localization \emph{family}
is a group of techniques that combine
the same kind of information according to different formulas.
Sections~\ref{sec:preliminaries:sbfl}--\ref{sec:preliminaries:st}
describe \n.{number of families}! common FL families
that comprise a total of \n.{number of techniques}!
FL families.
As \autoref{sec:preliminaries:granularities}
further explains,
a FL technique's \emph{granularity} denotes
the kind of program entities it analyzes for suspiciousness%
---from individual program locations to functions or files/modules.
Some FL techniques are only defined for a certain granularity level,
whereas others can be applied to different granularities.

While FL techniques are usually applicable
to any programming language,
we could not find any
comprehensive implementation
of the most common fault localization techniques
for Python at the time of writing.
Therefore, we implemented \fp%
---an automated fault localization tool
for Python implementing several widely used
techniques---%
and used it to perform the empirical study
described in the rest of the paper.
\autoref{sec:fp-implementation}
outlines \fp's main features and some details
of its implementation.

\subsection{Spectrum-Based Fault Localization}
\label{sec:preliminaries:sbfl}

Techniques
in the spectrum-based fault localization (SBFL) family
compute suspiciousness scores based on
a program's spectra~\cite{Reps:1997}---in other words,
its concrete execution traces.
The key heuristics of SBFL techniques is that
a program entity's suspiciousness is higher
the more often the entity is covered (reached) by failing tests
and the less often it is covered by passing tests.
The various techniques in the SBFL family
differ in
what formula they use to assign suspiciousness scores
based on an entity's coverage in passing and failing tests.

\begin{figure}[!tb]
  \centering
  \begin{align}
    \label{eq:tarantula}
    \tech{Tarantula}_T(e) &=
       \frac{F^+(e)/|F|}
       {F^+(e)/|F| + P^+(e)/|P|}
    \\
    \label{eq:ochiai}
    \tech{Ochiai}_T(e) &=
       \frac{F^+(e)}
       {\sqrt{|F| \times (F^+(e) + P^+(e))}}
    \\
    \label{eq:dstar}
    \tech{DStar}_T(e) &= \frac{(F^+(e))^{2}}{P^+(e) + F^-(e)}
  \end{align}
  \caption{SBFL formulas to compute the suspiciousness score of an entity $e$ given tests $T = P \cup F$ partitioned into passing $P$ and failing $F$. All formulas compute a score that is higher the more failing tests $F^+(e)$ cover $e$, and lower the more passing tests $P^+(e)$ cover $e$---capturing the basic heuristics of SBFL.}
  \label{fig:sfbl-formulas}
\end{figure}

Given tests $T = P \cup F$ as above, and a program entity $e$:
\begin{enumerate*}
\item $P^+(e)$ is the number of passing tests that cover $e$;
\item $P^-(e)$ is the number of passing tests that do not cover $e$;
\item $F^+(e)$ is the number of failing tests that cover $e$;
\item and
  $F^-(e)$ is the number of failing tests that do not cover $e$.
\end{enumerate*}
\autoref{fig:sfbl-formulas}
shows how 
Tarantula~\cite{Jones:2005},
Ochiai~\cite{Abreu:2007}, and
DStar~\cite{Wong:2014}%
---three widely used SBFL techniques~\cite{Pearson:2017}---%
compute suspiciousness scores
given this coverage information.
DStar's formula \eqref{eq:dstar},
in particular,
takes the second power of the numerator,
as recommended by other empirical studies~\cite{Zou:2021,Wong:2014}.\footnote{
  Suspiciousness score formulas are typically expressed as \emph{ratios};
  when a denominator is zero, this leads to an undefined score.
  There are different strategies to account for suspiciousness scores
  in these degenerate cases~\cite{Sarhan:2022}.
  In this paper, we implicitly add a small constant $\epsilon = 0.1$ to the
  denominator of every suspiciousness score formula.
  When the denominator is not zero, adding $\epsilon$ is practically irrelevant;
  when the denominator is zero and the numerator is also zero,
  adding $\epsilon$ gives a very low suspiciousness of zero
  (which reflects that the entity is hardly covered by any tests);
  when the denominator is zero and the numerator is positive,
  adding $\epsilon$ gives a large suspiciousness
  (which reflects that the entity is only covered by failing tests).
}

\subsection{Mutation-Based Fault Localization}
\label{sec:preliminaries:mbfl}

Techniques
in the mutation-based fault localization (MBFL) family
compute suspiciousness scores based on
mutation analysis~\cite{Jia:2011},
which 
generates many \emph{mutants} of a program $p$
by applying random transformations to it 
(for example, change a comparison operator $<$ to $\leq$ in
an expression).
A mutant $m$ of $p$ is thus a variant
of $p$ whose behavior differs from $p$'s 
at, or after, the location where $m$ differs from $p$.
The key idea of mutation analysis is
to collect information 
about $p$'s runtime behavior
based on how it differs from its mutants'.
Accordingly, when a test $t$ behaves differently on $p$ than on $m$
(for example, $p$ passes $t$ but $m$ fails it),
we say that $t$ \emph{kills} $m$.

\begin{figure}[!tb]
  \centering
  \begin{align}
    \label{eq:metallaxis}
    \tech{Metallaxis}_T(m) &=
       \frac{F_{\sim}^k(m)}
       {\sqrt{|F| \times (F_{\sim}^k(m) + P^k(m))}}
    & \tech{Metallaxis}_T(e) &= \max_{\substack{m \in M \\ m \text{ mutates } e}} \tech{Metallaxis}_T(m)
    \\
    \label{eq:muse}
    \tech{\muse}_T(m) &= \frac{F^k(m) - P^k(m) \times \sum_{n \in M} F^k(n) / \sum_{n \in M} P^k(n)}{|F|}
    & \tech{\muse}_T(e) &= \mean_{\substack{m \in M \\ m \text{ mutates } e}} \tech{\muse}_T(m)
  \end{align}
  \caption{MBFL formulas to compute the suspiciousness score of a mutant $m$ given tests $T = P \cup F$ partitioned into passing $P$ and failing $F$. All formulas compute a score that is higher the more failing tests $F^k(m)$ kill $m$, and lower the more passing tests $P^k(m)$ kill $m$---capturing the basic heuristics of mutation analysis. On the right, MBFL formulas to compute the suspiciousness score of a program entity $e$ by aggregating the suspiciousness score of all mutants $m \in M$ that modified $e$ in the original program.}
  \label{fig:mfbl-formulas}
\end{figure}

To perform fault localization on a program $p$,
MBFL techniques first generate
a large number of mutants $M = \{ m_1, m_2, \ldots \}$ of $p$
by systematically applying each mutation operator
to each statement in $p$ that is executed in any failing test $F$.
Then, given tests $T = P \cup F$ as above, and a mutant $m \in M$:
\begin{enumerate*}
\item $P^k(m)$ is the number of tests that $p$ passes but $m$ fails (that is, the tests in $P$ that \emph{kill} $m$);
\item 
  $F^k(m)$ is the number of tests that $p$ fails but $m$ passes (that is, the tests in $F$ that \emph{kill} $m$);
\item and
  $F^k_{\sim}(m)$ is the number of tests that $p$ fails and behave differently on $m$, either because they pass on $m$ or because they still fail but lead to a different stack trace (this is a \emph{weaker} notion of tests that \emph{kill} $m$~\cite{Papadakis:2015}).
\end{enumerate*}
\autoref{fig:mfbl-formulas}
shows how 
Metallaxis~\cite{Papadakis:2015} 
and \muse~\cite{Moon:2014}%
---two widely used MBFL techniques---%
compute suspiciousness scores
of each \emph{mutant} in $M$.

Metallaxis's formula \eqref{eq:metallaxis}
is formally equivalent to
Ochiai's---except that it is computed for each mutant
and measures \emph{killing} tests
instead of \emph{covering} tests.
In \muse's formula \eqref{eq:muse}, $\sum_{n \in M} F^k(n)$ is the total number
of failing tests in $F$ that kill any mutant in $M$,
and $\sum_{n \in M} P^k(n)$ is the total number
of passing tests in $P$ that kill any mutant in $M$
(these are called $\mathit{f2p}$ and $\mathit{p2f}$ in \muse's paper~\cite{Moon:2014}).

Finally, MBFL computes a suspiciousness score for
a program entity $e$ by aggregating the
suspiciousness scores of all mutants that modified
$e$ in the original program $p$;
when this is the case, we say that a mutant $m$ \emph{mutates} $e$.
The right-hand side of \autoref{fig:mfbl-formulas}
shows Metallaxis's and \muse's
suspiciousness formulas for entities:
Metallaxis \eqref{eq:metallaxis}
takes the largest (maximum) mutant score,
whereas \muse \eqref{eq:muse}
takes the average (mean) of the mutant scores.

\subsection{Predicate Switching}
\label{sec:preliminaries:ps}

The predicate switching (PS)~\cite{Zhang:2006}
fault localization technique
identifies \emph{critical predicates}:
branching conditions
(such as those of \Py{if} and \Py{while} statements)
that are related to a program's failure.
PS's key idea is that if forcibly changing
a predicate's value turns a failing test into passing one,
the predicate's location is a suspicious program entity.

For each failing test $t \in F$, 
PS starts from
$t$'s execution trace
(the sequence of all statements executed by $t$),
and finds $t$'s subsequence $b^t_1\,b^t_2\,\ldots$
of \emph{branching} statements.
Then, by instrumenting the program $p$ under analysis,
it generates, for each branching statement $b^t_k$,
a new execution of $t$
where the \emph{predicate} (branching condition)
$c^t_k$ evaluated by statement $b^t_k$
is forcibly \emph{switched} (negated) at runtime
(that is, the new execution takes the \emph{other} branch at $b^t_k$).
If switching predicate $c^t_k$
makes the test execution pass,
then $c^t_k$ is a \emph{critical predicate}.
Finally, PS assigns a (positive) suspiciousness score
to all critical predicates in all tests $F$:
$\tech{PS}_F(c_k^t)$
is higher,
the fewer critical predicates are evaluated between
$c_k^t$ and the failure location
when executing $t \in F$~\cite{Zou:2021}.\footnote{%
  The actual value of the suspiciousness score is immaterial,
  as long as the resulting ranking is consistent with this criterion. In \fp's current implementation,
  $\tech{PS}_F(c_k^t) = 1/10^d$, where $d$ is the number
  of critical predicates \emph{other than} $c_k^t$ evaluated
  after $c_k^t$ in $t$.
}
For example, the most suspicious program entity $e$
will be the location of the last critical predicate
evaluated before any test failure.

PS has some distinctive features compared to other FL techniques.
First, it only uses failing tests for its dynamic analysis;
any passing tests $P$ are ignored.
Second, the only program entities it can report
as suspicious are locations of predicates;
thus, it usually reports a shorter list of suspicious locations
than SBFL and MBFL techniques.
Third, while MBFL mutates program code,
PS dynamically mutates individual \emph{program executions}.
For example, suppose that a loop \Py{while $\;c$: $B$}
executes its body $B$ twice---and hence, 
the loop condition $c$
is evaluated three times---in a failing test.
Then, PS will generate three variants of this test execution:
\begin{enumerate*}
\item one where the loop body never executes
  ($c$ is false the first time it is evaluated);
\item one where the loop body executes once
  ($c$ is false the second time it is evaluated);
\item one where the loop body executes three or more times
  ($c$ is true the third time it is evaluated).
\end{enumerate*}

\subsection{Stack Trace Fault Localization}
\label{sec:preliminaries:st}

When a program execution fails with a crash
(for example, an uncaught exception),
the language runtime
usually prints its stack trace
(the chain of methods active when the crash occurred)
as debugging information to the user.
In fact,
it is known that 
stack trace information helps developers
debug failing programs~\cite{Bettenburg:2008};
and a bug is more likely to be fixed if it is close
to the top of a stack trace~\cite{Schroter:2010}.
Based on these empirical findings,
Zou et al.~\cite{Zou:2021}
proposed the stack trace fault localization technique (ST),
which uses the simple heuristics of
assigning suspiciousness based on how close
a program entity is to the top of a stack trace.

Concretely, given a failing test $t \in F$,
its \emph{stack trace}
is a sequence $f_1\,f_2\,\ldots$ of
the stack frames of all functions
that were executing when $t$ terminated with a failure,
listed in reverse order of execution;
thus, $f_1$ is the most recently called function,
which was directly called by $f_2$, and so on.
ST assigns a (positive) suspiciousness score
to any program entity $e$
that belongs to any function $f_k$
in $t$'s stack trace:
$\tech{ST}_t(e) = 1/k$,
so that
$e$'s suspiciousness is higher,
the closer to the failure
$e$'s function was called.\footnote{%
  As in PS, the actual value of the suspiciousness score
  is immaterial, as long as the resulting ranking is consistent with this criterion.
}
In particular, the most suspicious program entities
will be all those in the function $f_1$
called in the top stack frame.
Then, the overall suspiciousness score of $e$
is the maximum in all failing tests $F$:
$\tech{ST}_F(e) = \max_{t \in F} \tech{ST}_t(e)$.

\subsection{Granularities}
\label{sec:preliminaries:granularities}

Fault localization \emph{granularity}
refers to the kinds of program entity
that a FL technique ranks.
The most widely studied granularity
is \emph{statement-level},
where each statement in a program
may receive a different suspiciousness score~\cite{Pearson:2017,Wong:2014}.
However,
coarser granularities have also been considered,
such as \emph{function-level}
(also called method-level)~\cite{Le:2016,Xuan:2014}
and \emph{module-level}
(also called file-level)~\cite{Saha:2013,Zhou:2012}.

In practice,
implementations of FL techniques
that support different levels of granularity
focus on the finest granularity
(usually, statement-level granularity),
whose information they use to perform FL
at coarser granularities.
Namely, the suspiciousness of a function
is the maximum suspiciousness of any statements in its definition;
and the suspiciousness of a module
is the maximum suspiciousness of any functions belonging to it.\footnote{
  Other approaches aggregate the suspiciousness scores
  of finer-granularity entities by average or by minimum.
  We take the maximum for consistency with Zou et al.~\cite{Zou:2021}.
}

\subsection{\fp: Features and Implementation}
\label{sec:fp-implementation}

Despite its popularity as a programming language,
we could not find off-the-shelf implementations
of fault localization techniques for Python
at the time of writing~\cite{Sarhan:2022}.
The only exception is CharmFL~\cite{Sarhan:2021}%
---a plugin for the PyCharm IDE---%
which only implements SBFL techniques.
Therefore,
to conduct an extensive empirical study of FL in Python,
we implemented \fp:
a fault localization tool for Python programs.

\fp supports all \n.{number of techniques}! FL
techniques described in Sections~\ref{sec:preliminaries:sbfl}--\ref{sec:preliminaries:st};
it can localize faults at the level of 
statements, functions, or modules  (\autoref{sec:preliminaries:granularities}).
To make \fp a flexible and extensible tool,
easy to use with a variety of other commonly used
Python development tools,
we implemented it as a stand-alone command-line tool
that works with tests 
in the formats supported by 
Pytest, Unittest, and Hypothesis~\cite{MacIver:2019}%
---three popular Python testing frameworks.

While running,
\fp stores intermediate analysis data
in an SQLite database;
upon completing a FL localization run,
it returns to the user a human-readable summary%
---including suspiciousness scores and ranking of program entities.
The database improves performance
(for example by caching intermediate results)
but also facilitates \emph{incremental} analyses%
---for example, where we provide different batches of tests
in different runs.

\fp's implementation uses Coverage.py~\cite{Batchelder:2023}%
---a popular code-coverage measurement library---%
to collect the execution traces needed for SBFL and MBFL.
It also uses the state-of-the-art
mutation-testing framework Cosmic Ray~\cite{CosmicRay:2023}
to generate mutants for MBFL;
since Cosmic Ray is easily configurable to use some or all
of its mutation operators%
---or even to add new user-defined mutation operators---%
\fp's MBFL implementation is also fully configurable.
To implement PS in \fp,
we developed an instrumentation library
that can selectively change the runtime value of predicates
in different runs as required by the PS technique.
The implementation of \fp is available as open-source
(see this paper's replication package).

\section{Related Work}
\label{sec:related-work}

Fault localization has been an intensely researched topic
for over two decades,
whose popularity does not seem to wane~\cite{Wong:2016}.
This section summarizes a selection of studies
that are directly relevant for the paper;
Wong's recent survey~\cite{Wong:2016} provides a broader summary for
interested readers.

\paragraph{Spectrum-based fault localization.}
The Tarantula SBFL technique~\cite{Jones:2005}
was one of the earliest, most influential FL techniques,
also thanks to its empirical evaluation showing it
is more effective than other competing techniques~\cite{Renieres:2003, Cleve:2005}.
The Ochiai SBFL technique~\cite{Abreu:2007}
improved over Tarantula, and it often still is considered
the ``standard'' SBFL technique.

These earlier empirical studies~\cite{Jones:2005,Abreu:2007},
as well as other contemporary and later studies of FL~\cite{Papadakis:2015},
used the Siemens suite~\cite{Hutchins:94}:
a set of seven small C programs with seeded bugs.
Since then,
the scale and realism of FL empirical studies
has significantly improved over the years,
targeting real-world bugs affecting projects of realistic size.
For example,
Ochiai's effectiveness was confirmed~\cite{Le:2013}
on a collection of more realistic C and Java programs~\cite{Do:2005}.
When Wong et al.~\cite{Wong:2014} proposed DStar,
a new SBFL technique,
they demonstrated its capabilities in a sweeping comparison
involving 38 other SBFL techniques
(including the ``classic'' Tarantula and Ochiai).
In contrast,
numerous empirical results
about fault localization in Java based on experiments with artificial faults
were found not to hold
to experiments with real-world faults~\cite{Pearson:2017}
using the Defects4J curated collection~\cite{Just:2014}.

\paragraph{Mutation-based fault localization.}
With the introduction of novel fault localization families%
---most notably, MBFL---%
empirical comparison of techniques belonging to different families
became more common~\cite{Moon:2014,Papadakis:2015,Pearson:2017,Zou:2021}.
The \muse MBFL technique
was introduced to overcome a specific limitation of SBFL techniques:
the so-called ``tie set problem''.
This occurs when SBFL assigns the same suspiciousness score
to different program entities, simply because they belong to the same
simple control-flow block
(see \autoref{sec:preliminaries:sbfl} for details on how SBFL works).
Metallaxis-FL~\cite{Papadakis:2015}
(which we simply call ``Metallaxis'' in this paper)
is another take on MBFL that can improve over SBFL techniques.

The comparison between MBFL and SBFL is especially delicate
given how MBFL works.
As demonstrated by Pearson et al.~\cite{Pearson:2017},
MBFL's effectiveness crucially depends on whether
it is applied to bugs that are ``similar'' to
those introduced by its mutation operators.
This explains why the MBFL studies
targeting artificially seeded faults~\cite{Moon:2014,Papadakis:2015}
found MBFL to outperform SBFL;
whereas studies targeting real-world faults~\cite{Pearson:2017,Zou:2021}
found the opposite to be the case%
---a result also confirmed by the present paper
in \autoref{sec:experimental-results:rq1}.

\paragraph{Mutation testing.}
MBFL techniques rely on mutation testing to generate mutants
of a faulty program that may help locate the fault.
Therefore, the selection of mutation operators that are
used for mutation testing impacts the effectiveness of MBFL techniques.
Research in mutation testing has grown considerably
in the last decade,
developing a large variety of mutation operators
tailored to specific programming languages, applications, and faults~\cite{Papadakis:2019}.
Despite these recent developments,
the fundamental set of mutation operators introduced 
in Offut et al.'s seminal work~\cite{Offutt:1996}
remains the basis of basically every application to mutation testing.
These fundamental operators generate mutants by
modifying or removing arithmetic, logical, and relational operators,
as well as constants and variables in a program,
and hence are widely applicable and domain-agnostic.
Notably, the Cosmic Ray~\cite{CosmicRay:2023}
Python mutation testing framework (used in our implementation of \fp),
the two other popular Python mutation testing frameworks
MutPy~\cite{Derezinska:2014}
and \project{mutmut},%
\footnote{\url{https://mutmut.readthedocs.io}}
as well as the popular Java mutation testing
frameworks Pitest\footnote{\url{https://pitest.org}},
MuJava~\cite{Ma:2005}
and
Major~\cite{Just:2014:major}
(the latter used in Zou et al.'s MBFL experiments~\cite{Zou:2021})
all offer Offut et al.'s fundamental operators.
This helps make experiments with mutation testing techniques
meaningfully comparable.

\paragraph{Empirical comparisons.}
This paper's study design is based on Zou et al.'s empirical comparison
of fault localization on Java programs~\cite{Zou:2021}.
We chose their study because it is fairly recent (it was published in 2021),
it is comprehensive (it targets 11 fault localization techniques in seven families, as well as combinations of some of these techniques),
and it targets realistic programs and faults
(357 bugs in five projects from the Defects4J curated collection).

Ours is a differentiated conceptual
replication~\cite{DBLP:conf/laser/JuzgadoG10}
of Zou et al.'s study~\cite{Zou:2021}.
We target a comparable number of subjects
(\n.{number of subjects}! \bip~\cite{Widyasari:2020} bugs
vs.~357 Defects4J~\cite{Just:2014} bugs)
from a wide selection of projects
(\n.{number of project subjects}! real-world Python projects
vs.~\spellout{5} real-world Java projects).
We study \cite{Zou:2021}'s
four main fault localization families SBFL, MBFL, PS, and ST,
but we
exclude three other families that
featured in their study:
DS (dynamic slicing~\cite{Hammacher:2008}), IRBFL (Information retrieval-based fault localization~\cite{Zhou:2012}),
and HBFL (history-based fault localization~\cite{Rahman:2011}).
IRBFL and HBFL 
were shown to be scarcely effective by Zou et al.~\cite{Zhou:2012},
and rely on different kinds of artifacts that
may not always be available when dynamically analyzing a program
as done by the other ``mainstream'' fault localization techniques.
Namely, IRBFL analyzes bug reports, which may not be available
for all bugs;
HBFL mines commit histories of programs.
In contrast, our study only includes techniques that
solely rely on \emph{tests} to perform fault localization;
this help make a comparison between techniques consistent.
Finally, we excluded DS for practical reasons:
implementing it
requires accurate data- and control-dependency static analyses~\cite{zeller:whyfail}.
These are available in languages like Java
through widely used frameworks like Soot~\cite{Soot,Soot-Bodden};
in contrast, Python currently offers few mature static analysis tools
(e.g, Scalpel~\cite{Li:2022}),
none with the features required to implement DS.
Unfortunately,
dynamic slicing has been implemented for Python in the past~\cite{ds-python}
but no implementation is publicly available;
and building it from scratch is outside the present paper's scope.

\paragraph{Python fault localization.}
Despite Python's popularity as a programming language,
the vast majority of fault localization empirical studies target
other languages---mostly C, C++, and Java.
To our knowledge,
CharmFL~\cite{Szatmari:2022,Sarhan:2021}
is the only available implementation of fault localization techniques
for Python; the tool is limited to SBFL techniques.
We could not find any realistic-size empirical study
of fault localization using Python programs
comparing techniques of different families.
This gap in both the availability of tools~\cite{Sarhan:2022}
and the empirical knowledge about fault localization in Python
motivated the present work.

Note that numerous recent empirical studies
looked into fault localization for deep-learning models
implemented in Python~\cite{Eniser:2019,Guo:2020,Zhang:2020,Zhang:2021,Schoop:2021,Wardat:2021}.
This is a very different problem, using very different techniques,
than ``classic'' program-based fault localization,
which is the topic of our paper.

\paragraph{Deep learning-based fault localization.}
Deep learning models have recently been applied to
the software fault localization problem.
The key idea of techniques such as
DeepFL~\cite{Li:2019}, GRACE~\cite{Lou:2021}, and DEEPRL4FL~\cite{Li:2021}
is to train a deep learning model to identify suspicious
locations, giving it as input coverage information,
as well as other encoded information about
the source code of the faulty programs
(such as the data and control-flow dependencies).
While these approaches are promising,
we could not include them in our empirical study
since they do not have the same level of maturity as the other ``classic''
FL techniques we considered.
First, DeepFL and GRACE only work at function-level granularity,
whereas the bulk of FL research targets statement-level granularity.
Second, there are no reference implementations of techniques
such as DEEPRL4FL that we can use for our experiments.\footnote{
  The replication package of DEEPRL4FL~\cite{Li:2021} is not
  available at the time of writing.
}
Third, the performance of a deep learning-based technique
usually depends on the training set.
Fourth, training a deep learning model is usually a time consuming process;
how to account for this overhead when comparing efficiency
is tricky.

Nevertheless, our empirical study does feature
one FL technique that is based on machine learning:
\combine, which is Zou et al.'s application of learning to rank to fault localization~\cite{Zou:2021}.
The same paper also discusses how \combine
outperforms other state-of-the-art machine learning-based
fault localization techniques
such as MULTRIC~\cite{Li:2017}, Savant~\cite{Le:2016}, TraPT~\cite{Li:2017}, and FLUCCS~\cite{Sohn:2017}.
Therefore, \combine is a valid representative
of the capabilities of pre-deep learning
machine learning FL techniques.

\paragraph{Python vs.\ Java SBFL comparison.}
To our knowledge,
Widyasari et al.'s
recent empirical study of spectrum-based
fault localization~\cite{Widyasari:2022}
is the only currently available
large-scale study targeting real-world Python projects.
Like our work,
they use the bugs
in the \bip curated collection as experimental subjects~\cite{Widyasari:2020};
and they compare their results to those obtained by others
for Java~\cite{Pearson:2017}.
Besides these high-level similarities, the scopes of
our study and Widyasari et al.'s are fundamentally different:
\begin{enumerate*}
\item We are especially interested in comparing fault localization techniques
  in different \emph{families};
  they consider exclusively five \emph{spectrum}-based techniques,
  and drill down into the relative performance of these techniques.
\item Accordingly, we consider orthogonal categorization of bugs:
  we classify bugs (see \autoref{sec:design:faults-kinds})
  according to characteristics that match the capabilities
  of different fault-localization families (e.g.,
  stack-trace fault localization works for bugs that result in a crash);
  they classify bugs according to syntactic characteristics
  (e.g., multi-line vs.\ single-line patch).
\item Most important, even though both our paper and Widyasari et al.'s
  compare Python to Java, the framing of our comparisons
  is quite different:
  in \autoref{sec:experimental-results:rq6},
  we compare our findings about fault localization in Python
  to Zou et al.~\cite{Zou:2021}'s findings about fault localization in Java;
  for example, we confirm that
  SBFL techniques are generally more effective than MBFL techniques
  in Python, as they were found to be in Java.
  In contrast, Widyasari et al.\ directly compare
  various SBFL effectiveness metrics they collected on Python programs
  against the same metrics Pearson et al.~\cite{Pearson:2017}
  collected on Java programs;
  for example, Widyasari et al.\ report that the percentage of bugs in \bip
  that their implementation of the
  Ochiai SBFL technique correctly localized within the top-5
  positions is considerably lower than the percentage of bugs in
  Defects4J that Pearson et al.'s implementation of
  the Ochiai SBFL technique correctly localized within the top-5.
\end{enumerate*}

It is also important to note
that there are several technical differences
between ours and Widyasari et al.'s methodology.
First, we handle ties between suspiciousness scores
by computing the \einspectso rank
(described in~\autoref{sec:design:classic-metrics});
whereas they use average rank
(as well as other effectiveness metrics).
Even though we also take our subjects from \bip,
we carefully selected a subset of bugs that are
fully analyzable on our infrastructure
with all fault localization techniques we consider
(\autoref{sec:design:subjects}, \autoref{sec:design:experimental-setup});
whereas they use all \bip available bugs.
The selection of subjects
is likely to impact the value of \emph{some} metrics more than others (see~\autoref{sec:design:classic-metrics});
for example, the exam score
is undefined for bugs that a fault localization technique cannot localize,
whereas the top-$k$ counts are lower the more faults cannot be localized.
These and numerous other differences make
our results and Widyasari et al.'s incomparable and mostly complementary.
A replication of their comparison following our methodology
is an interesting direction for future work, 
but clearly outside the present paper's scope.
In \autoref{sec:future-work-widyasari}
we present some additional data,
and outline a few directions for future work
that are directly inspired by Widyasari et al.'s study~\cite{Widyasari:2022}.

\section{Experimental Design}
\label{sec:design}

Our experiments assess and compare the effectiveness and efficiency
of the \n.{number of techniques}! FL
techniques described in \autoref{sec:preliminaries},
as well as of their combinations,
on real-world Python programs and faults.
To this end, we target the following research questions:

\begin{description}
\item[RQ1.] How \emph{effective} are the fault localization
  techniques?
  \\
  RQ1 compares fault localization techniques according to how
  accurately they identify program entities that are responsible
  for a fault.

\item[RQ2.] How \emph{efficient} are the fault localization
  techniques?
  \\
  RQ2 compares fault localization techniques according to their
  running time.

\item[RQ3.] Do fault localization techniques behave differently 
  on \emph{different} faults?
  \\
  RQ3 investigates whether the fault localization techniques'
  effectiveness and efficiency depend on which kinds of faults and programs
  it analyzes.

\item[RQ4.] Does \emph{combining} fault localization techniques
  improve their effectiveness?
  \\
  RQ4 studies whether combining the information of different fault
  localization techniques for the same faults improves the effectiveness
  compared to applying each technique in isolation.

\item[RQ5.] How does program entity \emph{granularity} impact fault
  localization effectiveness?
  \\
  RQ5 analyzes the relation between effectiveness and granularity:
  does the relative effectiveness of fault localization techniques
  change as they target coarser-grained program entities?

\item[RQ6.] Are fault localization techniques as effective on Python
  programs as they are on \emph{Java} programs?
  \\
  RQ6 compares our overall results to Zou et al.~\cite{Zou:2021}'s,
  exploring similarities and differences between Java and Python
  programs.
\end{description}

\subsection{Subjects}
\label{sec:design:subjects}

To have a representative collection of realistic Python bugs,%
\footnote{Henceforth, we use the terms ``bug'' and ``fault'' as synonyms.}
we used \bip~\cite{Widyasari:2020},
a curated dataset of real bugs collected from real-world Python projects,
with all the information needed to reproduce the bugs in controlled experiments.
\autoref{tab:bugsinpy-projects} overviews \bip's
501 bugs from 17 projects.

\paragraph{Project category.}
\label{sec:project-category-classification}
Columns \textsc{category}
in \autoref{tab:bugsinpy-projects} and \autoref{tab:bugsinpy-selected-projects}
partition all \bip projects into four non-overlapping categories:

\begin{description}
\item[Command line (\cli)] projects consist of tools mainly used through their command line interface.

\item[Development (\dev)] projects offer libraries and utilities useful to software developers.

\item[Data science (\ds)] projects consist of machine learning and numerical computation frameworks.

\item[Web (\web)] projects offer libraries and utilities useful for web development.
\end{description}
We classified the projects according to their description
in their respective repositories,
as well as how they are presented in \bip.
Like any classification, the boundaries between categories
may be somewhat fuzzy, but the main focus of most projects
is quite obvious
(such as \ds for \project{keras} and \project{pandas}, or \cli for \project{youtube-dl}).

\paragraph{Unique bugs.}
Each bug $b = \langle p_b^-, p_b^+, F_b, P_b\rangle$ in \bip
consists of:
\begin{enumerate*}
\item a \emph{faulty} version $p_b^-$ of the project,
  such that tests in $F_b$ all fail on it (all due to the same root cause);
\item a \emph{fixed} version $p_b^+$ of the project,
  such that all tests in $F_b \cup P_b$ pass on it;
\item a collection of \emph{failing} $F_b$ and \emph{passing} $P_b$ tests,
  such that tests in $P_b$ pass
  on both the faulty $p_b^-$ and fixed $p_b^+$ versions of the project,
  whereas tests in $F_b$
  fail on the faulty $p_b^-$ version and pass on the fixed $p_b^+$ version
  of the project.
\end{enumerate*}

\paragraph{Bug selection.}
Despite \bip's careful curation,
several of its bugs cannot be reproduced because
their dependencies are missing or no longer available;
this is a well-known problem that plagues reproducibility
of experiments involving Python programs~\cite{Mukherjee:2021}.
In order to identify which \bip bugs were reproducible
at the time of our experiments on our infrastructure,
we took the following steps for each bug $b$:

\begin{enumerate}
\item Using \bip's scripts,
  we generated and executed the faulty $p_b^-$
  version and checked that tests in $F_b$ fail whereas tests in $P_b$ pass on it;
  and we generated and executed the fixed $p_b^+$
  version and checked that all tests in $F_b \cup P_b$ pass on it.
  Out of all of \bip's bugs, 120 failed this step;
  we did not include them in our experiments.
  
\item Python projects often have two sets of dependencies (\emph{requirements}):
  one for users and one for developers; both are needed to run fault localization
  experiments, which require to instrument the project code.
  Another 39 bugs in \bip miss some development dependencies;
  we did not include them in our experiments.

\item Two bugs resulted in an empty ground truth (\autoref{sec:design:ground-truth}):
  essentially, there is no way of localizing the fault in $p_b^-$;
  we did not include these bugs in our experiments.
\end{enumerate}
This resulted in $501 - 120 - 39 - 2 =$ 340 bugs in 13 projects
(all but \project{ansible}, \project{matplotlib}, \project{PySnooper}, and \project{scrapy})
that we could reproduce in our experiments.

However, this is still an impractically large number:
just \emph{reproducing} each of these bugs in \bip
takes nearly a full week of running time,
and each FL experiment may require to rerun the same tests several times
(hundreds of times in the case of MBFL).
Thus, we first discarded 27 bugs that each take more than 48 hours to reproduce.
We estimate that including these 27 bugs in the experiments
would have taken over 14 CPU-months just for the MBFL experiments%
---not counting other FL techniques, nor the time for setup
and dealing with unexpected failures.

Running all the fault localization experiments
for each of the remaining $313 = 340 - 27$ bugs
takes approximately eleven CPU-hours,
for a total of nearly five CPU-months.
We selected \n.{number of subjects}! bugs out of the 313
using stratified random sampling with the four project categories as the ``strata'',
picking:
\n.{number of CLI}! bugs in category \cli,
\n.{number of DEV}! bugs in category \dev,
\n.{number of DS}! bugs in category \ds,
and \n.{number of WEB}! bugs in category \web.
This gives us a still sizable, balanced, and representative\footnote{%
For example, this sample size is sufficient to 
estimate a ratio with up to 5.5\% error
and 90\% probability with the most conservative (i.e., 50\%)
a priori assumption~\cite{ss-estimate}.}
sample of all bugs in \bip,
which we could exhaustively analyze in around two CPU-months worth of experiments.
In all, we used this selection of \n.{number of subjects}! bugs
as our empirical study's subjects.
\autoref{tab:bugsinpy-selected-projects}
gives some details about the selected projects and their bugs.

As a side comment,
note that
our experiments with \bip were generally more time consuming than
Zou et al.'s experiments with Defects4J.
For example, the average per-bug running time of MBFL in our experiments
(\n{stmt/mbfl/metallaxis/all/time} seconds in \autoref{table:efficiency-family-bug-type})
was 3.3 times larger than in Zou et al.'s (4800 seconds in~\cite[Table~9]{Zou:2021}).
Even more strikingly,
running all fault localization experiments on the 357 Defects4J
bugs took less than one CPU-month;\footnote{
  The sum of column \textsc{average} in~\cite[Table~9]{Zou:2021} multiplied by 357 gives $2.04$ million seconds or $0.79$ months.
}
in contrast,
running MBFL on just 27 ``time consuming'' bugs in \bip takes over 14 CPU-months.
This difference may be partly due to the different characteristics of
projects in Defects4J vs.\ \bip,
and partly to the dynamic nature of Python
(which is run by an interpreter).

\begin{table}[!tb]

  \small
  \setlength{\tabcolsep}{2.5pt}
    \centering
    \begin{tabular}{lrrrrrrrrll}
    \toprule
    \multicolumn{1}{c}{\textsc{project}} & 
    \multicolumn{1}{c}{\textsc{kLOC}} & 
    \multicolumn{1}{c}{\textsc{|f|}} &
    \multicolumn{1}{c}{\textsc{|m|}} &
    \multicolumn{1}{c}{\textsc{bugs}} & 
    \multicolumn{1}{c}{\textsc{subjects}} & 
    \multicolumn{1}{c}{\textsc{tests}} & 
    \multicolumn{1}{c}{\textsc{test kLOC}} & 
    \multicolumn{1}{c}{\textsc{category}} & 
    \multicolumn{1}{c}{\textsc{description}} \\
    
    \midrule

\project{ansible} &
82.6  &
\numprint{3713} &
493 &
18   &
0        &
\numprint{1830}  &
103.1     &
\dev        &
IT automation platform \\

\project{black} & 
93.5     & 
421 &
27 &
23   & 
13       & 
153    & 
6.8       & 
\dev     & 
Code formatter \\

\project{cookiecutter} & 
1.6 & 
62 &
18 &
4    & 
4        & 
218    & 
4.1       &
\dev     &
Developer tool \\

\project{fastapi} & 
4.7   & 
160 &
40 &
16   & 
13       & 
595    & 
16.8      & 
\web      & 
Web framework for building APIs \\

\project{httpie} & 
3.5    &
197 &
34 &
5    & 
4        & 
217    & 
2.4       & 
\cli      & 
Command-line HTTP client \\

\project{keras} & 
6.7   & 
150 &
119 &
45   & 
18       & 
616    & 
13.6      & 
\ds       & 
Deep learning API \\

\project{luigi} & 
22.0   & 
\numprint{2004} &
120 &
33   & 
13       & 
\numprint{1508}   & 
21.2      & 
\dev     & 
Pipelines of batch jobs management tool \\

\project{matplotlib} & 
99.6  & 
\numprint{5526} &
147 &
30   & 
0        & 
\numprint{2484}   & 
34.9      &
\ds        & 
Plotting library \\

\project{pandas} & 
128.0  & 
\numprint{5466} &
234 &
169  & 
18       & 
\numprint{12226}  & 
200.9     & 
\ds      & 
Data analysis toolkit \\

\project{PySnooper} & 
0.7    & 
60 &
7 &
3    & 
0        & 
49     & 
3.9       & 
\dev        & 
Debugging tool \\

\project{sanic} & 
7.3   & 
462 &
61 &
5    & 
3        & 
466    & 
8.3       & 
\web      & 
Web server and web framework \\

\project{scrapy} & 
15.7   & 
\numprint{1509} &
179 &
40   & 
0        & 
\numprint{1572}   & 
24.5      &
\web        &
Web crawling and web scraping framework \\

\project{spaCy} & 
97.2    & 
852 &
415 &
10   & 
6        & 
\numprint{986}   & 
13.4 & 
\ds       & 
Natural language processing library \\

\project{thefuck} & 
4.7   & 
604 &
203 &
32   & 
16       & 
\numprint{614}   & 
7.3       & 
\cli      & 
Console command tool \\

\project{tornado} & 
17.9   & 
\numprint{1124} &
35 &
16   & 
4        & 
\numprint{926}   & 
13.1      & 
\web      & 
Web server \\

\project{tqdm} & 
3.3    & 
200 &
28 &
9    & 
7        & 
120     & 
2.7       &
\cli      &
Progress bar for Python and CLI \\

\project{youtube-dl} & 
125.0  & 
\numprint{3078} &
818 &
43   & 
16       & 
\numprint{237}   & 
5.1       & 
\cli      & 
Video downloader \\

\cmidrule{2-8}
\multicolumn{1}{r}{\textbf{total}} &
714.0 &
\numprint{25588} & 
\numprint{2978} & 
501 &
135 &
\numprint{24817} &
482.1 &
& \\
 \bottomrule
    \end{tabular}
    \caption{Overview of projects in \bip. %
      For each \textsc{project}, the table reports
      the project's overall size in \textsc{kLOC}
      (thousands of non-empty non-comment lines of code, excluding tests),
      the number $\textsc{|F|}$ of functions (excluding test functions),
      the number $\textsc{|m|}$ of modules (excluding test modules),
      the number of \textsc{bugs} included in \bip,
      how many we selected as \textsc{subjects} for our experiments,
      the corresponding number of \textsc{tests} (i.e., test functions),
      their size in kLOC (\textsc{test kLOC},
      thousands of non-empty non-comment lines of test code),
      the \textsc{category} the project belongs to
      (\cli: command line;
       \dev: development tools;
       \ds: data science;
       \web: web tools),
       and a brief \textsc{description} of the project.
       Consistently with what done by the authors of \bip~\cite{Widyasari:2020},
       the project statistics reported here refer to the \emph{latest} version of
       the projects on 2020-06-19.}
    \label{tab:bugsinpy-projects}
\end{table}

\begin{table}[!tb]

\small
\setlength{\tabcolsep}{8pt}
\centering
\begin{tabular}{llrrrrrr}
\toprule
\multicolumn{1}{c}{\textsc{category}} & 
\multicolumn{1}{c}{\textsc{project}} & 
\multicolumn{2}{c}{\textsc{bugs (subjects)}} & 
\multicolumn{2}{c}{\textsc{tests}} &
\multicolumn{2}{c}{\textsc{ground truth}}
\\
\cmidrule(lr){3-4}
\cmidrule(lr){5-6}
\cmidrule(lr){7-8}
& & 
\multicolumn{1}{c}{$C$} & \multicolumn{1}{r}{$P$} &
\multicolumn{1}{c}{$C$} & \multicolumn{1}{r}{$P$} &
\multicolumn{1}{c}{$C$} & \multicolumn{1}{r}{$P$}
\\
      
\midrule
\multirow{4}{*}{\cli}
& \project{httpie}
& \multirow{4}{*}{43} 
& 4
& \multirow{4}{*}{1188} 
& 217
& \multirow{4}{*}{139} 
& 12
\\ & \project{thefuck} 
& 
& 16 
& 
& 614 
& 
& 55
\\ & \project{tqdm} 
& 
& 7 
& 
& 120 
& 
& 22
\\ & \project{youtube-dl} 
& 
& 16 
& 
& 237 
& 
& 50
\\

\midrule
\multirow{3}{*}{\dev}
& \project{black}
& \multirow{3}{*}{30} 
& 13
& \multirow{3}{*}{1879} 
& 153 
& \multirow{3}{*}{300} 
& 208
\\ & \project{cookiecutter} 
& 
& 4 
& 
& 218 
& 
& 19
\\ & \project{luigi} 
& 
& 13 
& 
& 1508 
& 
& 73
\\

\midrule
\multirow{3}{*}{\ds}
& \project{keras}
& \multirow{3}{*}{42} 
& 18
& \multirow{3}{*}{13828} 
& 616 
& \multirow{3}{*}{186} 
& 111
\\ & \project{pandas} 
& 
& 18 
& 
& 12226 
& 
& 64
\\ & \project{spaCy} 
& 
& 6 
& 
& 986 
& 
& 11
\\

\midrule
\multirow{3}{*}{\web}
& \project{fastapi}
& \multirow{3}{*}{20} 
& 13
& \multirow{3}{*}{1987} 
& 595 
& \multirow{3}{*}{174} 
& 156
\\ & \project{sanic} 
& 
& 3 
& 
& 466 
& 
& 6
\\ & \project{tornado} 
& 
& 4 
& 
& 926 
& 
& 12
\\
      
\midrule
\multicolumn{2}{r}{\textbf{total}} 
& 135 
& 135 
& 18882 
& 18882 
& 799 
& 799
\\
\bottomrule
\end{tabular}
\caption{Selected \bip bugs used in the paper's experiments. %
The \textsc{project}s are grouped by \textsc{category};
the table reports---for each project individually
(column $P$),
as well as for
all projects in the category (column $C$)---the number of \textsc{bugs}
selected as \textsc{subjects} for our experiments,
the corresponding number of \textsc{tests} (i.e., test functions),
and the total number of program locations that
make up the \textsc{ground truth}
(described in \autoref{sec:design:ground-truth}).
}
\label{tab:bugsinpy-selected-projects}
\end{table}

\subsection{Faulty Locations: Ground Truth}
\label{sec:design:ground-truth}

A fault localization technique's
effectiveness
measures how accurately
the technique's list of suspicious entities
matches the actual fault locations in a program%
---fault localization's \emph{ground truth}.
It is customary to use programmer-written patches
as ground truth~\cite{Zou:2021, Pearson:2017}:
the program locations modified by the patches
that fix a certain bug correspond to
the bug's actual fault locations.

Concretely, here is how to determine
the ground truth of
a bug $b = \langle p_b^-, p_b^+, F_b, P_b\rangle$ in \bip.
The programmer-written fix $p_b^+$
consists of a series
of \emph{edits}
to the faulty program $p_b^-$.
Each edit can be of three kinds:
\begin{enumerate*}
\item \emph{add}, which inserts into $p_b^+$ a new program location;
\item \emph{remove}, which deletes a program location in $p_b^-$;
\item \emph{modify}, which takes a program location in $p_b^-$
  and changes parts of it, without changing its location, in $p_b^+$.
\end{enumerate*}
Take, for instance,
the program in \autoref{code:ex-fixed},
which modifies the program in \autoref{code:ex-buggy};
the edited program includes
\spellout{2} adds (lines~\ref{l:gt:add1}, \ref{l:gt:add2}),
\spellout{1} remove (line~\ref{l:gt:remove}),
and \spellout{1} modify (line~\ref{l:gt:modify}).

Bug $b$'s \emph{ground truth} $\truth(b)$
is a set of locations in $p_b^-$ that are
affected by the edits, determined as follows.
First of all, ignore any blank or comment lines,
since these do not affect a program's behavior and hence
cannot be responsible for a fault.
Then, finding the ground truth locations
corresponding to removes and modifies is
straightforward:
a location $\ell$ that is removed or modified in $p_b^+$
exists by definition also in $p_b^-$,
and hence it is part of the ground truth.
In \autoref{fig:ground-truth},
line~\ref{l:gt:4} is modified
and line~\ref{l:gt:8} is removed
by the edit that transforms \autoref{code:ex-buggy} into \autoref{code:ex-fixed};
thus \ref{l:gt:4} and \ref{l:gt:8} are part of
the example's ground truth.

Finding the ground truth locations
corresponding to \emph{adds}
is more involved~\cite{Sarhan:2022},
because a location $\ell$ that is added to $p_b^+$
does not exist in $p_b^-$:
$b$ is a fault of omission~\cite{Pearson:2017}.\footnote{%
  In \bip, 41\% of all fixes include at least one add edit.
}
A common solution~\cite{Zou:2021, Pearson:2017}
is to take as ground truth
the location in $p_b^-$ that immediately \emph{follows}
$\ell$.
In \autoref{fig:ground-truth},
line~\ref{l:gt:2} corresponds to
the first non-blank line that follows
the assignment statement 
that is added at line~\ref{l:gt:add1} in \autoref{code:ex-fixed};
thus \ref{l:gt:2} is part of
the example's ground truth.
However, an add at $\ell$ is actually a modification
between two other locations;
therefore, the location that immediately \emph{precedes}
$\ell$ should also be part of the ground truth,
since it identifies the same insertion location.
In \autoref{fig:ground-truth},
line~\ref{l:gt:1} precedes
the assignment statement
that is added at line~\ref{l:gt:add1} in \autoref{code:ex-buggy};
thus \ref{l:gt:1} is also part of
the example's ground truth.

A location's \emph{scope}
poses a final complication to determine the ground truth
of adds.
Consider line~\ref{l:gt:add2},
added in \autoref{code:ex-fixed}
at the very end of function \Py{foo}'s body.
The (non-blank, non-comment)
location that follows it in \autoref{code:ex-buggy}
is line~\ref{l:gt:7};
however, line~\ref{l:gt:7}
marks the beginning of another function \Py{bar}'s
definition.
Function \Py{bar} cannot be the location of
a fault in \Py{foo},
since the two functions are independent%
---in fact, the fact that \Py{bar}'s declaration
follows \Py{foo}'s is immaterial.
Therefore, we only include a location in the ground truth
if it is within the same \emph{scope} as the
location $\ell$ that has been added.
If $\ell$ is part of a function body (including methods),
its scope is the function declaration;
if $\ell$ is part of a class outside any function
(e.g., an attribute),
its scope is the class declaration;
and otherwise $\ell$'s scope is the module it belongs to.
In \autoref{fig:ground-truth},
both lines~\ref{l:gt:1} and \ref{l:gt:2}
are within the same module as the added statement
at line~\ref{l:gt:add1} in \autoref{code:ex-buggy}.
In contrast,
line~\ref{l:gt:7} is within a different scope
than the added statement
at line~\ref{l:gt:add2} in \autoref{code:ex-buggy}.
Therefore, lines~\ref{l:gt:1}, \ref{l:gt:2}, and \ref{l:gt:6}
are part of the ground truth,
but not line~\ref{l:gt:7}.

\begin{figure}[!bt]
  \centering
  \begin{subfigure}[t]{0.4\linewidth}
\begin{lstlisting}[language=Python]
(*\best[acol!50]{a = 3}*) (*\label{l:gt:1}*)




(*\best[acol!50]{c = 5}*) (*\label{l:gt:2}*)

# Function foo
def foo(y):(*\label{l:gt:3}*)
  (*\best[ccol!50]{if y > 3:}*)(*\label{l:gt:4}*)
    a = y (*\label{l:gt:5}*)
  (*\best[acol!50]{y = y * ~2}*) (*\label{l:gt:6}*)


# Function bar
def bar(z):(*\label{l:gt:7}*)
   (*\best[bcol!50]{z = z + 2}*) (*\label{l:gt:8}*)
   return z (*\label{l:gt:9}*)
\end{lstlisting}
    \caption{Faulty program version.
      Lines with colored background are the ground truth locations.
      Extra blank lines are added for readability.}
    \label{code:ex-buggy}
  \end{subfigure}
  \hspace{16mm}
  \begin{subfigure}[t]{0.4\linewidth}
\begin{lstlisting}[language=Python,firstnumber=last]
a = 3

# Global variable b
(*\best[acol]{\textcolor{white}{\uline{b = None}}}*)         (*\textcolor{acol}{\# add}*)(*\label{l:gt:add1}*)

c = 5

# Function foo
def foo(y):
  if y > (*\best[ccol]{\textcolor{white}{\uwave{100}}}*):    (*\textcolor{ccol}{\# modify}*)(*\label{l:gt:modify}*)
    a = y
  y = y * 2
  (*\best[acol]{\textcolor{white}{\uline{a = y}}}*)          (*\textcolor{acol}{\# add}*)(*\label{l:gt:add2}*)

# Function bar
def bar(z):
  (*\best[bcol]{\textcolor{white}{\sout{z = z + 2}}}*)      (*\textcolor{bcol}{\# remove}*)(*\label{l:gt:remove}*)
  return z
\end{lstlisting}
    \caption{Fixed program version, which edits \autoref{code:ex-buggy}'s program with two \best[acol]{\textcolor{white}{\uline{adds}}},
      one \best[ccol]{\textcolor{white}{\uwave{modify}}},
      and one \best[bcol]{\textcolor{white}{\sout{remove}}}.
    }
    \label{code:ex-fixed}
  \end{subfigure}
  \caption{An example of program edit, and the corresponding ground truth faulty locations.}
  \label{fig:ground-truth}
\end{figure}

Our definition of ground truth refines
that used in related work~\cite{Zou:2021, Pearson:2017}
by including the location that precedes an add,
and by considering only locations within scope.
We found that this definition better captures
the programmer's intent and their corrective impact on
a program's behavior.

How to best characterize bugs of omissions
(fixed by an add)
in fault localization remains an open issue~\cite{Sarhan:2022}.
Pearson et al.'s study~\cite{Pearson:2017}
proposed the first viable solution: including the location following an add.
Zou et al.~\cite{Zou:2021} followed the same approach,
and hence we also include the location following an add
in our ground truth computation.
We also noticed that, by also including the location preceding an add,
and by taking scope into account,
our ground truth computation becomes more comprehensive;
in particular, it also works for statements added at the very end of a file%
---a location that has no following lines.
While our approach is usually more precise,
it is not necessarily the preferable alternative in all cases.
Consider again, for instance,
the add at line~\ref{l:gt:add2} in \autoref{fig:ground-truth};
if we ignored the scope (and the preceding statement), only line~\ref{l:gt:7} would be included in its ground truth.
If this fault localization information were consumed
by a developer, it could still be useful and actionable
even if it reports a line outside the scope of the actual add location:
the developer would use the location as a starting point
for their inspection of the nearby code;
and they may prefer a smaller, if slightly imprecise, ground truth
to a larger, redundant one.
However, this paper's focus is strictly evaluating
the effectiveness of FL techniques as rigorously as possible%
---for which our stricter ground truth computation is more appropriate.

\begin{table}[!bt]

  \small
  \setlength{\tabcolsep}{8pt}
  \centering
  
\begin{tabular}{l *{10}{r}}
  \toprule
  & \multicolumn{10}{c}{\textsc{program entity} $\ell$}
  \\
  \cmidrule(lr){2-11}
  & $\ell_1$ & $\ell_2$ & $\ell_3$ & $\ell_4$ & $\ell_5$ & $\ell_6$ & $\ell_7$ & $\ell_8$ & $\ell_9$ & $\ell_{10}$ \\
  \midrule
  suspiciousness score $s$ of $\ell$
  & 10 & 7 & 4 & 4 & 4 & 3 & 3 & 2 & 2 & 2  \\
  $\ell \in \truth(b)$?
  &  & \nay & & \nay &  & & & \nay & \nay &  \\
  \cmidrule(lr){2-11}
  $\textsf{start}(\ell)$
  & 1 & 2 & 3 & 3 & 3 & 6 & 6 & 8 & 8 & 8  \\
  $\textsf{ties}(\ell)$
  & 1 & 1 & 3 & 3 & 3 & 2 & 2 & 3 & 3 & 3  \\
  $\textsf{faulty}(\ell)$
  & 0 & 1 & 1 & 1 & 1 & 0 & 0 & 2 & 2 & 2  \\
  \cmidrule(lr){2-11}
  $\einspect[b][\ell, \locs]$
  & 1.0 & 2.0 & 4.0 & 4.0 & 4.0 & 6.0 & 6.0 & 8.3 & 8.3 & 8.3 \\
  \bottomrule
\end{tabular}
\caption{An example of calculating the \einspectso metric $\einspect[b][\ell, \locs]$ for a list of 10 suspicious locations $\ell_1, \ldots, \ell_{10}$ ordered by their decreasing suspiciousness scores $s_1, \ldots, s_{10}$ . For each location $\ell$, the table reports its suspiciousness score $s$, and whether $\ell$ is a faulty location $\ell \in \truth(b)$;
based on this ranking of locations, it also shows the lowest rank $\mathsf{start}(\ell)$ of the first location whose score is equal to $\ell$'s, the number $\mathsf{ties}(\ell)$ of locations whose score is equal to $\ell$'s, the number of faulty locations among these, and the corresponding \einspectso value $\einspect[b][\ell, L]$---computed according to \eqref{eq:e-inspect}.}
\label{tab:einspect-example}
\end{table}

\begin{figure}[!bt]
\centering
\begin{tikzpicture}[ultra thick]
\coordinate (L) at (0,0);
\coordinate (R) at (10mm,0);
\coordinate (B) at ($(L)!0.5!(R)+(0,-10mm)$);
\draw[dcol] let \p1 = ($(R)-(L)$), \n2 = {veclen(\x1,\y1)}
     in (L) circle (\n2) node [label={[label distance=8mm]170:predicate}] {};
\draw[ecol] let \p1 = ($(R)-(L)$), \n2 = {veclen(\x1,\y1)}
      in (R) circle (\n2) node [label={[label distance=8mm]10:mutable}] {};
\draw[fcol] let \p1 = ($(R)-(L)$), \n2 = {veclen(\x1,\y1)}
      in (B) circle (\n2) node [label={[label distance=8mm]-30:crashing}] {};
\node at ($(L)+(-12pt,7pt)$) {4};
\node at ($(R)+(12pt,7pt)$) {19};
\node at ($(B)+(0pt,-10pt)$) {20};
\node at ($(L)!0.5!(R)+(0,10pt)$) {29};
\node (X) at ($(L)!0.5!(R)+(0,-10pt)$) {16};
\node at ($(X)+(-18pt,-8pt)$) {3};
\node at ($(X)+(18pt,-8pt)$) {10};
\node at ($(X)+(10mm,18mm)$) {34};
\draw[gcol] (X) ellipse (35mm and 25mm) node [label={[label distance=27mm]40:all bugs}] {};
\end{tikzpicture}
\caption{Classification of the \n.{number of subjects} \bip bugs used in our experiments into three categories.}
\label{fig:fault-classification}
\end{figure}

\subsection{Classification of Faults}
\label{sec:design:faults-kinds}

\paragraph{Bug kind.}
The information used by each fault localization technique 
naturally captures the behavior of different \emph{kinds} of faults.
Stack trace fault localization analyzes the call stack
after a program terminates with a crash;
predicate switching targets branching conditions
as program entities to perform fault localization;
and MBFL crucially relies on the analysis of
mutants to track suspicious locations.

Correspondingly, we classify a bug $b = \langle p_b^-, p_b^+, F_b, P_b \rangle$ as:

\begin{description}
\item[Crashing] bug if any failing test in $F_b$ terminates abruptly
  with an unexpected uncaught exception.
\item[Predicate] bug if any faulty entity in the ground truth $\truth(b)$
  includes a branching predicate (such as an \Py{if} or \Py{while} condition).
\item[Mutable] bug if any of the mutants generated by MBFL's mutation operators
  mutates any locations in the ground truth $\truth(b)$.
  Precisely, a bug $b$'s \emph{mutability} is the percentage of all mutants
  of $p_b^-$ that mutate locations in $\truth(b)$; and $b$ is mutable if
  its mutability is greater than zero.
\end{description}
The notion of crashing and predicate bugs is from Zou et al.~\cite{Zou:2021}.

We introduced the notion of mutable bug to
try to capture scenarios where MBFL techniques have a fighting chance
to correctly localize bugs.
Since MBFL uses mutant analysis for fault localization,
its capabilities depend on the mutation operators
that are used to generate the mutants.
Therefore, the notion of mutable bugs is somewhat
dependent on the applied mutation operators.\footnote{
  In this sense, ``mutable'' is a qualitatively different attribute
  than ``crashing'' and ``predicate''. Whether a bug $b$ is ``crashing''
  exclusively depends on the failing tests that trigger the bug;
  whether $b$ is a ``predicate'' bug
  depends on the branching syntactic structure of $b$'s program
  and how it relates to $b$.
  In contrast, whether $b$ is a ``mutable'' bug
  depends on the mutation operators used to analyze $b$,
  and on whether they can change the program so as to
  effectively affect $b$'s buggy behavior.
}
Our implementation of \fp
uses the standard operators offered by the popular
Python mutation testing framework Cosmic Ray~\cite{CosmicRay:2023}.
As we discussed in \autoref{sec:related-work},
Cosmic Ray features a set of mutation operators
that are largely similar to several other general-purpose
mutation testing frameworks---all based on Offut et al.'s
well known work~\cite{Offutt:1996}.
These strong similarities between
the mutation operators offered by most widely used
mutation testing frameworks
suggest that our definition of ``mutable bug''
is not strongly dependent on the specific mutation testing framework
that is used.
Correspondingly, bugs that we classify as ``mutable''
are likely to remain amenable to localization with MBFL
provided one uses (at least) this standard set of core mutation operators.
Conversely, we expect that
devising new, specialized mutation operators %
may extend the number of bugs that we can classify as ``mutable'', 
and hence that are more likely to be amenable to localization with MBFL
techniques.

\autoref{fig:fault-classification} shows the kind
of the \n.{number of subjects} \bip bugs we used in the experiments,
consisting of
\n.{crashing bugs} crashing bugs, \n.{predicate bugs} predicate bugs, \n.{mutable bugs} mutable bugs,
and \n.{other bugs} bugs that do not belong to any of these categories.

\paragraph{Project category.}
Another, orthogonal classification of bugs is according to the
project \emph{category} they belong to.
We classify a bug $b$ as a \cli, \dev, \ds, or \web bug
according to the category of project
(\autoref{tab:bugsinpy-selected-projects})
$b$ belongs to.

\subsection{Ranking Program Entities}
\label{sec:design:ranking-einspect}

Running a fault localization technique $L$
on a bug $b$
returns a list of program entities $\ell_1, \ell_2, \ldots$,
sorted by their decreasing suspiciousness scores $s_1 \geq s_2 \geq \ldots$.
The programmer (or, more realistically, a tool~\cite{parnin-orso-11,Gazzola:2019})
will go through the entities in this order
until a faulty entity (that is an $\ell \in \truth(b)$ that matches $b$'s ground truth)
is found.
In this idealized process, the earlier a faulty entity appears in the list,
the less time the programmer will spend going through the list,
the more effective fault localization technique $L$ is on bug $b$.
Thus, a program entity's \emph{rank} in the sorted list of suspicious entities
is a key measure of fault localization effectiveness.

Computing a program entity $\ell$'s rank is trivial if there are no
\emph{ties} between scores.
For example, consider \autoref{tab:einspect-example}'s first two program entities
$\ell_1$ and $\ell_2$, with suspiciousness scores $s_1 = 10$ and $s_2 = 7$.
Obviously, $\ell_1$'s rank is $1$ and $\ell_2$'s is $2$;
since $\ell_2$ is faulty ($\ell_2 \in \truth(b)$),
its rank is also a measure of how many entities will need to be inspected
in the aforementioned debugging process.

When several program entities tie the same suspiciousness score,
their relative order in a ranking is immaterial~\cite{Debroy:2010}.
Thus, it is a common practice to give all of them the same \emph{average}
rank~\cite{Sarhan:2022,Steimann:2013}, capturing an average-case number of program entities inspected
while going through the fault localization output list.
For example, consider \autoref{tab:einspect-example}'s first five program entities
$\ell_1, \ldots, \ell_5$; $\ell_3$, $\ell_4$, and $\ell_5$
all have the same suspiciousness score $s = 4$.
Thus, they all have the same average rank $4 = (3 + 4 + 5) / 3$,
which is a proxy of how many entities will need to be inspected
if $\ell_4$ were faulty but $\ell_2$ were not.

Capturing the ``average number of inspected entities''
is trickier still if more than one entity is faulty among a bunch of tied entities.
Consider now all of \autoref{tab:einspect-example}'s ten program entities;
entities $\ell_8$, $\ell_9$, and $\ell_{10}$
all have the suspiciousness score $s = 2$;
$\ell_8$ and $\ell_9$ are faulty, whereas $\ell_{10}$ is not.
Their average rank $9 = (8 + 9 + 10) / 3$
overestimates the number of entities to be inspected
(assuming now that these are the only faulty entities in the output),
since two entities out of three are faulty,
and hence it is more likely that the faulty entity will appear
before rank 9.

To properly account for such scenarios,
Zou et al.~\cite{Zou:2021} introduced the \einspectso
metric, which ranks a program entity $\ell$ within
a list $\locs$ of
program entities $\ell_1, \ldots, \ell_n$ with suspiciousness scores $s_1 \geq \ldots \geq s_n$
as:
\begin{equation}
  \einspect[b][\ell, \locs]
  \quad=\quad
  \metric{start}(\ell) + \sum_{k = 1}^{\metric{ties}(\ell) - \metric{faulty}(\ell)}
  k \:\frac
  {\binom{\metric{ties}(\ell) - k - 1}{\metric{faulty}(\ell) - 1}}
  {\binom{\metric{ties}(\ell)}{\metric{faulty}(\ell)}}
  \label{eq:e-inspect}
\end{equation}
In \eqref{eq:e-inspect},
$\metric{start}(\ell)$ is the position $k$ of the first entity among those
with the same score as $\ell$'s;
$\metric{ties}(\ell)$ is the number of entities (including $\ell$ itself)
whose score is the same as $\ell$'s;
and $\metric{faulty}(\ell)$ is the number of entities (including $\ell$ itself)
that tie $\ell$'s score and are faulty (that is $\ell \in \truth(b)$).
Intuitively, the \einspectso rank $\einspect[b][\ell, \locs]$
is thus an average of all possible ranks where tied and faulty entities are shuffled randomly.
When there are no ties, or only one entity among a group of ties is faulty,
\eqref{eq:e-inspect} coincides with the average rank.

Henceforth, we refer to a location's \einspectso rank $\einspect[b][\ell, \locs]$
as simply its \emph{rank}.

\begin{figure}[!tb]
\begin{align}
  \label{eq:einspect-exam}
  \einspect[b][L] &\ = \min_{\ell \in L(b) \cap \truth(b)} \einspect[b][\ell, L(b)]
  &
    \einspect*[b][L] &\ = \min_{\ell \in L^{\infty}(b) \cap \truth(b)} \einspect[b][\ell, L^{\infty}(b)]
  &
  \exam_b(L) &\ = \frac{\einspect[b][L]}{|p_b^-|}
  \\
  \label{eq:metrics-averages}
  L@_Bn &\ = \big| \big\{ b \in B \mid \einspect[b][L] \leq n \big\} \big|
  &
  \einspect*[B][L] &\ = \frac{1}{|B|} \sum_{b \in B} \einspect*[b][L]
  &
  \exam_B(L) &\ = \frac{1}{|B|} \sum_{b \in B} \exam_b(L)
\end{align}
\caption{Definitions of common FL effectiveness metrics.
  The top row shows two variants \einspect, \einspect*
  of the \einspectso metric, and the exam score \exam,
  for a generic bug $b$ and fault localization technique $L$.
  The bottom row shows cumulative metrics for a set $B$ of bugs:
  the ``at $n$'' metric $L@_Bn$, and the average \einspect* and \exam metrics.}
  \label{fig:standard-metrics}
\end{figure}

\paragraph{Better vs.\ worse ranks.}
A clarification about terminology:
a \emph{high} rank is a rank
that is close to the top-1 rank (the first rank), 
whereas a \emph{low} rank is a rank that is further away from the top-1 rank.
Correspondingly,
a high rank corresponds to a small numerical ordinal value;
and a low rank corresponds to a large numerical ordinal value.
Consistently with this standard usage,
the rest of the paper refers to
``better'' ranks to mean ``higher'' ranks (corresponding to smaller ordinals);
and ``worse'' ranks to mean ``lower'' ranks (corresponding to larger ordinals).

\subsection{Fault Localization Effectiveness Metrics}
\label{sec:design:classic-metrics}

\paragraph{\einspectso effectiveness.}
Building on the notion of \emph{rank}%
---defined in \autoref{sec:design:ranking-einspect}---%
we measure the \emph{effectiveness} of a fault localization technique $L$
on a bug $b$ as the rank of the first faulty program entity
in the list $L(b) = \locs$ of entities and suspiciousness scores returned
by $L$ running on $b$---defined as $\einspect[b][L]$ in \eqref{eq:einspect-exam}.
$\einspect[b][L]$ is $L$'s \einspectso rank on bug $b$,
which estimates the number of entities in $L$'s one has to inspect to correctly localize $b$.

\paragraph{Generalized \einspectso effectiveness.}
What happens if a FL technique $L$ cannot localize a bug $b$---that is,
$b$'s faulty entities $\truth(b)$ do not appear at all in $L$'s output?
According to \eqref{eq:e-inspect} and \eqref{eq:einspect-exam},
\einspect[b][L] is \emph{undefined} in these cases.
This is not ideal, as it fails to measure the effort wasted going through
the location list when using $L$ to localize $b$---the original intuition behind
all rank metrics.
Thus, we introduce a generalization $L$'s \einspectso rank on bug $b$ as follows.
Given the list $L(b) = \locs$ of entities and suspiciousness scores returned by $L$ running on $b$,
let $L^\infty(b) = \locs \, \langle \ell_{n+1}, s_0 \rangle \langle \ell_{n+2}, s_0 \rangle \ldots$
be $L(b)$ followed by all \emph{other entities} $\ell_{n+1}, \ell_{n+1}, \ldots$ in program $p_b^-$
that are not returned by $L$, each given a suspiciousness $s_0 < s_n$ lower than
any suspiciousness scores assigned by $L$.

With this definition, $\einspect[b][L] = \einspect*[b][L]$
whenever $L$ can localize $b$---that is some entity from $\truth(b)$ appears in $L$'s output list.
If some technique $L_1$ can localize $b$ whereas another technique $L_2$ cannot,
$\einspect*[b][L_2] > \einspect*[b][L_1]$, thus reflecting that $L_2$ is worse than $L_1$ on $b$.
Finally, if neither $L_1$ nor $L_2$ can localize $b$,
$\einspect*[b][L_2] > \einspect*[b][L_1]$ if $L_2$ returns a longer list
than $L_1$:
all else being equal, a technique that returns a shorter list
is ``better'' than one that returns a longer list since it
requires less of the user's time to inspect the output list.
Accordingly, $\einspect*[b][L]$ denotes $L$'s \emph{generalized}
\einspectso rank on bug $b$---defined as in \eqref{eq:einspect-exam}.

\paragraph{Exam score effectiveness.}
Another commonly used effectiveness metric is the \emph{exam score} $\exam_b(L)$~\cite{Wong:2008},
which is just a FL technique $L$'s \einspectso rank on bug $b$
over the number of program entities $|p_b^-|$ of the analyzed buggy program $p_b^-$%
---as in \eqref{eq:einspect-exam}.
Just like $\einspect[b][L]$, $\exam_b(L)$ is undefined if $L$ cannot localize $b$.

\paragraph{Effectiveness of a technique.}
To assess the overall effectiveness of a FL technique over a set $B$ of bugs,
we aggregate the previously introduced metrics in different ways---as in \eqref{eq:metrics-averages}.
The $L@_Bn$ metric counts the number of bugs in $B$
that $L$ could localize within the top-$n$ positions (according to their \einspectso rank);
$n=1, 3, 5, 10$ are common choices for $n$, reflecting a ``feasible'' number of entities
to inspect.
Then, the $L@_Bn\% = 100 \cdot L@_Bn / |B|$ metric is simply $L@_Bn$ expressed as a percentage of the number $|B|$ of bugs in $B$.
$\einspect*[B][L]$ is $L$'s average generalized \einspectso rank of bugs in $B$.
And $\exam_B(L)$ is $L$'s average exam score of bugs in $B$
(thus ignoring bugs that $L$ cannot localize).

\paragraph{Location list length.}
The $\outputlen{L}[b]$ metric is simply the number of suspicious locations
output by FL technique $L$ when run on bug $b$;
and
$\outputlen{L}[B]$ is the average of $\outputlen{L}[b]$ for all bugs in $B$.
The location list length metric is not, strictly speaking, a measure of effectiveness;
rather, it complements the information provided by other
measures of effectiveness,
as it gives an idea of how much output a technique produces to the user.
All else being equal, a shorter location list length is preferable%
---provided it is not empty.
In practice, we'll compare the location list length to other metrics
of effectiveness, in order to better understand the trade-offs
offered by each FL technique.

Different FL families use different kinds of information to compute
suspiciousness scores; this is also reflected by the entities that may
appear in their output location list.  SBFL techniques include all
locations executed by any tests $T_b$ (passing or failing) even if
their suspiciousness is zero; conversely, they omit all locations that
are \emph{not} executed by the tests.  MBFL techniques include all
locations executed by any \emph{failing} tests $F_b$, since these
locations are the targets of the mutation operators.  PS includes all
locations of \emph{predicates} (branching conditions) that are
executed by any failing tests $F_b$ and that are \emph{critical} (as
defined in \autoref{sec:preliminaries:ps}).  ST includes all locations
of all functions that appear in the stack trace of any crashing test
in $F_b$.

\paragraph{Effectiveness metrics: limitations.}

Despite being commonly used in fault localization research, the
effectiveness metrics presented in this section rely on assumptions
that may not realistically capture the debugging work of developers.
First, they assume that a developer can understand the characteristics
of a bug and devise a suitable fix by examining just one buggy entity;
in contrast, debugging often involves disparate activities, such as
analyzing control and data dependencies and inspecting program states
with different inputs~\cite{Parnin:2011}.  Second, debugging is often
not a \emph{linear} sequence of activities~\cite{KoM08} as simple as
going through the ranked list of entities produced by fault
localization techniques.  Despite these limitations, we still rely on
this section's effectiveness metrics: on the one hand, they are used
in practically all related work on fault localization (in particular,
Zou et al.~\cite{Zhou:2012}); thus, they make our results comparable
to others.  On the other hand, there are no viable, easy-to-measure
alternative metrics that are also fully realistic; devising such
metrics is outside this paper's scope and belongs to future work.

\subsection{Comparison: Statistical Models}
\label{sec:design:statistical-models}

To quantitatively compare the capabilities of
different fault localization techniques,
we consider several standard statistics.

\paragraph{Pairwise comparisons.}
Let $M_b(L)$ be any metric $M$
measuring the capabilities of fault-localization technique $L$
on bug $b$;
$M$ can be
any of \autoref{sec:design:classic-metrics}'s effectiveness metrics,
or $L$'s wall-clock running time $T_b(L)$ on bug $b$ as performance metric.
Similarly, for a fault-localization family $F$,
$M_b(F)$
denotes the average value $\sum_{k \in F} M_b(k) / |F|$
of $M_b$ for all techniques in family $F$.
Given a set $B = \{ b_1, \ldots, b_n \}$ of bugs,
we compare the two vectors
$M_B(F_1) = \langle M_{b_1}(F_1)\,\ldots\,M_{b_n}(F_1) \rangle$
and
$M_B(F_2) = \langle M_{b_1}(F_2)\,\ldots\,M_{b_n}(F_2) \rangle$
using three statistics:
\begin{description}

\item[Correlation $\tau$] between $M_B(F_1)$ and $M_B(F_2)$ computed using Kendall's $\tau$ statistics. The absolute value $|\tau|$ of the correlation
  $\tau$ measures
  how closely changes in the value of metric $M$ for $F_1$
  over different bugs are \emph{associated}
  to changes for $F_2$ over the same bugs:
  if $0 \leq |\tau| \leq 0.3$ the correlation is \emph{negligible};
  if $0.3 < |\tau| \leq 0.5$ the correlation is \emph{weak};
  if $0.5 < |\tau| \leq 0.7$ the correlation is \emph{medium};
  and if $0.7 < |\tau| \leq 1$ the correlation is \emph{strong}.

\item[P-value $p$] of a paired Wilcoxon signed-rank test---a nonparametric
  statistical test comparing $M_B(F_1)$ and $M_B(F_2)$.
  A small value of $p$ is commonly taken as evidence against
  the ``null-hypothesis'' that the distributions underlying
  $M_B(F_1)$ and $M_B(F_2)$ have different medians:\footnote{%
    The practical usefulness of statistical hypothesis tests
    has been seriously questioned in recent years~\cite{ASA-statement,riseup,FFT-TSE19-Bayes2};
    therefore, we mainly report this statistics for conformance with
    standard practices, but we refrain from giving it any serious
    weight as empirical evidence.
  }
  usually, $p \leq 0.05$, $p \leq 0.01$, and $p \leq 0.001$
  are three conventional thresholds of increasing strength.

\item[Cliff's $\delta$] effect size---a nonparametric measure
  of how often the values in $M_B(F_1)$ are larger than those in $M_B(F_2)$.
  The absolute value $|\delta|$ of the effect size $\delta$
  measures how much the values of metric $M$ differ, on the same bugs, between $F_1$ and $F_2$~\cite{romano-cliffdelta}:
  if $0 \leq |\delta| < 0.147$ the differences are \emph{negligible};
  if $0.145 \leq |\delta| < 0.33$ the differences are \emph{small};
  if $0.33 \leq |\delta| < 0.474$ the differences are \emph{medium};
  and if $0.474 \leq |\delta| \leq 1$ the differences are \emph{large}.
\end{description}

\paragraph{Regression models.}
To ferret out the individual impact
of several different factors
(fault localization family, project category, and bug kind)
on the capabilities of fault localization,
we introduce two 
varying effects regression models with
normal likelihood and logarithmic link function.
\begin{small}
\begin{align}
  \left[
  \begin{matrix}
    \var{E}_b
    \\
    \var{T}_b
  \end{matrix}
  \right]
  &\sim
    \dist{MVNormal}
    \left(\!
    \left[
    \begin{matrix}
      \var{e}_b
      \\
      \var{t}_b
    \end{matrix}
    \right]\!,
    S\!
    \right)
  &
  \log(\var{e}_b) &=\ \alpha 
                    + \alpha_{\var{family}[b]}
                    + \alpha_{\var{category}[b]}
  & \log(\var{t}_b) &=
                      \beta
                      + \beta_{\var{family}[b]}
                      + \beta_{\var{category}[b]}
 \label{eq:regmodel}
  \\
    \var{E}_b
  &\sim
    \dist{Normal}\,(\var{e}_b,\ \sigma)
  &
  \log(\var{e}_b) &= \left(
                    \begin{array}{cl}
                      \alpha 
                        & +\ \alpha_{\var{family}[b]}
                          + \alpha_{\var{category}[b]}
                      \\
                   &
                   +\ c_{\var{family}[b]} \,\var{crashing}_b \\
                   &
                   +\ p_{\var{family}[b]} \,\var{predicate}_b \\
                   &
                   +\ m_{\var{family}[b]} \,\log(1 + \var{mutability}_b)
                    \end{array}
                    \right)
\label{eq:regmodel-kinds}
\end{align}
\end{small}

Model \eqref{eq:regmodel} is multivariate,
as it simultaneously captures effectiveness and runtime cost
of fault localization.
For each fault localization experiment on a bug $b$,
\eqref{eq:regmodel}
expresses the vector 
$[ E_b, T_b ]$ 
of standardized\footnote{%
  We standardize the data since this simplifies fitting the model;
  for the same reason,
  we also log-transform the running time in seconds.
}
\einspectso metric $E_b$
and running time $T_b$
as drawn from a multivariate normal distribution
whose means $e_b$ and $t_b$
are log-linear functions of various
\emph{predictors}.
Namely, $\log(e_b)$
is the sum of a base intercept $\alpha$;
a family-specific intercept $\alpha_{\var{family}[b]}$,
for each fault-localization family SBFL, MBFL, PS, and ST;
and a category-specific intercept $\alpha_{\var{category}[b]}$,
for each project category \cli, \dev, \ds, and \web.
The other model component $\log(t_b)$ follows the 
same log-linear relation.

Model \eqref{eq:regmodel-kinds} is univariate,
since it only captures the relation between bug kinds and effectiveness.
For each fault localization experiment on a bug $b$,
\eqref{eq:regmodel-kinds}
expresses the
standardized \einspectso metric $E_b$
as drawn from a normal distribution
whose mean $e_b$
is a log-linear function of
a base intercept $\alpha$;
a family-specific intercept $\alpha_{\var{family}[b]}$;
and a category-specific intercept $\alpha_{\var{category}[b]}$;
a varying intercept $c_{\var{family}[b]} \var{crashing}_b$,
for the interactions between each family and crashing bugs;
a varying intercept $p_{\var{family}[b]} \var{predicate}_b$,
for the interactions between each family and predicate bugs;
and a varying slope $m_{\var{family}[b]} \log(1 + \var{mutability}_b)$,
for the interactions between each family and bugs with
different mutability.\footnote{%
  We log-transform \var{mutability} in this term, since this smooths out the big differences between mutability scores in different experiments (in particular, between zero and non-zero),
  which simplifies modeling the relation statistically. We add one to \var{mutability} before log-transforming it, so that the logarithm is always defined.}
Variables \var{crashing} and \var{predicate} are indicator variables, which are equal to $1$ respectively for crashing or predicate-related bugs, and $0$ otherwise;
variable \var{mutability} is instead the mutability percentage defined in \autoref{sec:design:faults-kinds}.

After completing regression models \eqref{eq:regmodel} and \eqref{eq:regmodel-kinds}
with suitable priors and fitting them on our experimental data\footnote{%
  The replication package includes all details about the regression models,
as well as their validation~\cite{FTF-TOSEM21-Bayes-guidelines}.}
gives a (sampled) distribution of values for the
coefficients $\alpha$'s, $c$, $p$, $m$, and $\beta$'s,
which we can analyze to infer the effects of the various
predictors on the outcome.
For example, if the 95\% probability interval
of $\alpha_{F}$'s distribution lies entirely below zero,
it suggests that FL family $F$ is consistently associated with below-average
values of \einspectso metric \einspect;
in other words, $F$ tends to be more effective than techniques in other families.
As another example, if the 95\% probability interval
of $\beta_{C}$'s distribution
includes zero,
it suggests that bugs in projects of category $C$
are not consistently associated with different-than-average
running times;
in other words, bugs in these projects
do not seem either faster or slower to analyze than those in other projects.

\subsection{Experimental Methodology}
\label{sec:design:experimental-setup}

To answer \autoref{sec:design}'s research questions,
we ran \fp using each of the \n.{number of techniques}
fault localization techniques described in \autoref{sec:preliminaries}
on all \n.{number of subjects} selected bugs (described in \autoref{sec:design:subjects}) from \bip v. b4bfe91,
for a total of
$\n.{number of experiments} = \n.{number of techniques} \times \n.{number of subjects}$
FL experiments.
Henceforth,
the term ``\emph{standalone} techniques'' refers to the
\n.{number of techniques} classic
FL techniques described in \autoref{sec:preliminaries};
whereas ``\emph{combined} techniques''
refers to the \spellout{4} techniques introduced for RQ4.

\paragraph{Test selection.}
The test suites of projects such as \project{keras} (included in \bip)
are very large and can take more than 24~hours to run even once.
Without a suitable test selection strategy,
large-scale FL experiments would be prohibitively time
consuming (especially for MBFL techniques, which rerun the same
test suite hundreds of times).
Therefore, we applied a simple test selection strategy
to only include tests that directly target the parts of a program
that contribute to the failures.\footnote{
  The Defects4J curated collection also includes
  a selection of so-called \emph{relevant} tests~\cite{Defects4J-website}.
}

As we mentioned in \autoref{sec:design:subjects},
each bug $b$ in \bip comes with a selection
of failing tests $F_b$ and passing tests $P_b$.
The failing tests are usually just a few,
and specifically trigger bug $b$.
The passing tests, in contrast, are much more numerous,
as they usually include all non-failing tests available in the project.
In order to cull the number of passing tests
to only include those that expressly target the failing code,
we applied a simple dependency analysis:
for each \bip bug $b$ used in our experiments,
we built the module-level call graph $G(b)$ for the whole of $b$'s project;%
\footnote{To build the call graph we used Python static analysis framework Scalpel~\cite{Li:2022}, which in turn relies on PyCG~\cite{Salis:2021}
  for this task.}
each node in $G(b)$ is a module of the project (including its tests),
and each edge $x_m \to y_m$ means that module $x_m$
directly uses
some entities defined in module $y_m$.
Consider any of $b$'s project \emph{test module} $t_m$;
we run the tests in $t_m$ in our experiments
if and only if:
\begin{enumerate*}
\item $t_m$ includes at least one of the \emph{failing} tests in $F_b$;
\item \emph{or}, $G(b)$ includes an edge $t_m \to f_m$,
where $f_m$ is a module that includes at least one of $b$'s
faulty locations $\truth(b)$ (see \autoref{sec:design:ground-truth}).
\end{enumerate*}
In other words:
we include \emph{all failing} tests for $b$,
as well as the passing tests
that directly
exercise the parts of the project that are faulty.
This simple heuristics substantially reduced the number of
tests that we had to run for the largest projects,
without meaningfully affecting the fault localization's scope.

Our test selection strategy does not include 
test modules that \emph{indirectly} involve failing locations
(unless they include any \emph{failing} tests):
if the tests in a module $t_m$ only call directly an application module
$x_m$, and then some parts of module $x_m$ call another application module $y_m$
(i.e., $t_m \to x_m \to y_m$ in the module-level call graph),
$x_m$ does not include any faulty locations,
and $y_m$ does include some faulty locations,
then we do \emph{not} include the tests in $t_m$ in our test suite;
instead, we will include \emph{other} test modules $u_m$ that directly call
$y_m$ (i.e., $u_m \to y_m$).

To demonstrate that our more aggressive test selection strategy
does not exclude any relevant tests,
and is unlikely to affect the quantitative fault localization results,
we first computed, for each bug $b$ used in our experiments:
\begin{enumerate*}
  \item the set $S^0_b$ of tests selected using the strategy described above;
    and
    \item the set $S^+_b \supseteq S^0_b$ of tests selected
      by including also \emph{indirect} dependencies
      (i.e., by taking the transitive closure of the module-level use relation).
\end{enumerate*}
For \n[0]{zero transitive delta}[100]| of the \n.{number of subjects} bugs
used in our experiments,
$S^+_b = S^0_b$, that is both test selection strategies select
the same tests.
However, there remain a long tail of bugs for which
including indirect dependencies leads to many more tests
being selected;
for example, for \n[0]{num more than threshold transitive delta} bugs in
\n{projects more than threshold transitive delta}! projects,
considering indirect dependencies leads to selecting
more than \n.{num extra tests threshold}! additional tests%
---which would significantly increase the experiments' running time.
Thus, we randomly selected one bug for each project
among those \n[0]{num more than threshold transitive delta} bugs for which indirect dependencies would lead to including
more than \n.{num extra tests threshold}! additional tests.
For each bug $b$ in this sample,
we performed an additional run
of our fault localization experiments with SBFL and MBFL techniques\footnote{
  Since PS and ST only use failing tests, their behavior does not change
  as $S^0_b$ always includes the same failing tests as $S^+_b$.
}
using all tests in $S^+_b$,
for a total of \n.{extra tests new experiments}! new experiments.
We found that none of the key fault localization effectiveness metrics
significantly changed compared to the same experiments using only tests in $S^0_b$.\footnote{Precisely, 
  in \n.{extra experiments no change}
  of these \n.{extra tests new experiments}! experiments
  the \einspectso score did not change at all.
  As for the remaining experiments,
  the \einspectso score changed but only for bugs that were
  not effectively localized:
  the bugs localized in the top-1, top-3, top-5, and top-10
  positions did not change, except for a single bug
  whose \einspect[b][\tech{Metallaxis}] went from
  \n[0]{extra experiments top-10 change einspect nontransitive}
  to \n[0]{extra experiments top-10 change einspect transitive}
  when we added the extra tests.
}
This confirms that our test selection strategy
does not alter the general effectiveness of fault localization,
and hence we adopted it for the rest of the paper's experiments.

\autoref{tab:bugsinpy-test-selection-info}
shows statistics about the fraction of tests that
we selected for our experiments according to the test selection strategy.
Those data indicate that test selection
has a disproportionate impact on projects
that have very large test suites,
such as those in the \ds category.
In these projects, it happens often that the vast majority of
tests are irrelevant for the portion of the project where a failure occurred;
therefore, excluding these tests from our experiments
is instrumental in drastically bringing down execution times without
sacrificing experimental accuracy.

\begin{table}[!tb]

\small
\setlength{\tabcolsep}{4pt}
\centering
\begin{tabular}{llrrrrrrrrrrrrrrrr}
\toprule
\multicolumn{1}{c}{\textsc{category}} & 
\multicolumn{1}{c}{\textsc{project}} & 
\multicolumn{4}{c}{\textsc{min}} & 
\multicolumn{4}{c}{\textsc{median}} &
\multicolumn{4}{c}{\textsc{mean}} &
\multicolumn{4}{c}{\textsc{max}}

\\
\cmidrule(lr){3-6}
\cmidrule(lr){7-10}
\cmidrule(lr){11-14}
\cmidrule(lr){15-18}
& & 
\multicolumn{2}{c}{$C$} & \multicolumn{2}{c}{$P$} &
\multicolumn{2}{c}{$C$} & \multicolumn{2}{c}{$P$} &
\multicolumn{2}{c}{$C$} & \multicolumn{2}{c}{$P$} &
\multicolumn{2}{c}{$C$} & \multicolumn{2}{c}{$P$}
\\

\cmidrule(lr){3-4}\cmidrule(lr){5-6}
\cmidrule(lr){7-8}\cmidrule(lr){9-10}
\cmidrule(lr){11-12}\cmidrule(lr){13-14}
\cmidrule(lr){15-16}\cmidrule(lr){17-18}
  
& & 
\multicolumn{1}{c}{\#} & \multicolumn{1}{r}{\%} &
\multicolumn{1}{c}{\#} & \multicolumn{1}{r}{\%} &
\multicolumn{1}{c}{\#} & \multicolumn{1}{r}{\%} &
\multicolumn{1}{c}{\#} & \multicolumn{1}{r}{\%} &
\multicolumn{1}{c}{\#} & \multicolumn{1}{r}{\%} &
\multicolumn{1}{c}{\#} & \multicolumn{1}{r}{\%} &
\multicolumn{1}{c}{\#} & \multicolumn{1}{r}{\%} &
\multicolumn{1}{c}{\#} & \multicolumn{1}{r}{\%}
\\

\midrule
\multirow{4}{*}{\cli}
& \project{httpie}
& \multirow{4}{*}{\n{test/cli/min/n}} 
& \multirow{4}{*}{\n[1]{test/cli/min/p}} 
& \n{test/httpie/min/n}
& \n[1]{test/httpie/min/p}
& \multirow{4}{*}{\n[1]{test/cli/median/n}} 
& \multirow{4}{*}{\n[1]{test/cli/median/p}} 
& \n[1]{test/httpie/median/n}
& \n[1]{test/httpie/median/p}
& \multirow{4}{*}{\n[1]{test/cli/mean/n}} 
& \multirow{4}{*}{\n[1]{test/cli/mean/p}} 
& \n[1]{test/httpie/mean/n}
& \n[1]{test/httpie/mean/p}
& \multirow{4}{*}{\n{test/cli/max/n}} 
& \multirow{4}{*}{\n[1]{test/cli/max/p}} 
& \n{test/httpie/max/n}
& \n[1]{test/httpie/max/p}
  
\\ 
& \project{thefuck} & & 
& \n{test/thefuck/min/n}
& \n[1]{test/thefuck/min/p}
& & 
& \n[1]{test/thefuck/median/n}
& \n[1]{test/thefuck/median/p} 
& & 
& \n[1]{test/thefuck/mean/n}
& \n[1]{test/thefuck/mean/p} 
& & 
& \n{test/thefuck/max/n}
& \n[1]{test/thefuck/max/p}

\\ 
& \project{tqdm} & & 
& \n{test/tqdm/min/n}
& \n[1]{test/tqdm/min/p}
& & 
& \n[1]{test/tqdm/median/n}
& \n[1]{test/tqdm/median/p} 
& & 
& \n[1]{test/tqdm/mean/n}
& \n[1]{test/tqdm/mean/p} 
& & 
& \n{test/tqdm/max/n}
& \n[1]{test/tqdm/max/p}

\\ 
& \project{youtube-dl} & & 
& \n{test/youtube-dl/min/n}
& \n[1]{test/youtube-dl/min/p}
& & 
& \n[1]{test/youtube-dl/median/n}
& \n[1]{test/youtube-dl/median/p} 
& & 
& \n[1]{test/youtube-dl/mean/n}
& \n[1]{test/youtube-dl/mean/p} 
& & 
& \n{test/youtube-dl/max/n}
& \n[1]{test/youtube-dl/max/p}

\\
\midrule
\multirow{3}{*}{\dev}
& \project{black}
& \multirow{3}{*}{\n{test/dev/min/n}} 
& \multirow{3}{*}{\n[1]{test/dev/min/p}} 
& \n{test/black/min/n}
& \n[1]{test/black/min/p}
& \multirow{3}{*}{\n[1]{test/dev/median/n}} 
& \multirow{3}{*}{\n[1]{test/dev/median/p}} 
& \n[1]{test/black/median/n}
& \n[1]{test/black/median/p}
& \multirow{3}{*}{\n[1]{test/dev/mean/n}} 
& \multirow{3}{*}{\n[1]{test/dev/mean/p}} 
& \n[1]{test/black/mean/n}
& \n[1]{test/black/mean/p}
& \multirow{3}{*}{\n{test/dev/max/n}} 
& \multirow{3}{*}{\n[1]{test/dev/max/p}} 
& \n{test/black/max/n}
& \n[1]{test/black/max/p}

\\ 
& \project{cookiecutter} 
& & 
& \n{test/cookiecutter/min/n}
& \n[1]{test/cookiecutter/min/p}
& & 
& \n[1]{test/cookiecutter/median/n}
& \n[1]{test/cookiecutter/median/p} 
& & 
& \n[1]{test/cookiecutter/mean/n}
& \n[1]{test/cookiecutter/mean/p} 
& & 
& \n{test/cookiecutter/max/n}
& \n[1]{test/cookiecutter/max/p}

\\ 
& \project{luigi} 
& & 
& \n{test/luigi/min/n}
& \n[1]{test/luigi/min/p}
& & 
& \n[1]{test/luigi/median/n}
& \n[1]{test/luigi/median/p} 
& & 
& \n[1]{test/luigi/mean/n}
& \n[1]{test/luigi/mean/p} 
& & 
& \n{test/luigi/max/n}
& \n[1]{test/luigi/max/p}
\\

\midrule
\multirow{3}{*}{\ds}
& \project{keras}
& \multirow{3}{*}{\n{test/ds/min/n}} 
& \multirow{3}{*}{\n[1]{test/ds/min/p}} 
& \n{test/keras/min/n}
& \n[1]{test/keras/min/p}
& \multirow{3}{*}{\n[1]{test/ds/median/n}} 
& \multirow{3}{*}{\n[1]{test/ds/median/p}} 
& \n[1]{test/keras/median/n}
& \n[1]{test/keras/median/p}
& \multirow{3}{*}{\n[1]{test/ds/mean/n}} 
& \multirow{3}{*}{\n[1]{test/ds/mean/p}} 
& \n[1]{test/keras/mean/n}
& \n[1]{test/keras/mean/p}
& \multirow{3}{*}{\n{test/ds/max/n}} 
& \multirow{3}{*}{\n[1]{test/ds/max/p}} 
& \n{test/keras/max/n}
& \n[1]{test/keras/max/p}

\\ 
& \project{pandas} 
& & 
& \n{test/pandas/min/n}
& \n[1]{test/pandas/min/p}
& & 
& \n[1]{test/pandas/median/n}
& \n[1]{test/pandas/median/p} 
& & 
& \n[1]{test/pandas/mean/n}
& \n[1]{test/pandas/mean/p} 
& & 
& \n{test/pandas/max/n}
& \n[1]{test/pandas/max/p}

\\ 
& \project{spaCy} 
& & 
& \n{test/spacy/min/n}
& \n[1]{test/spacy/min/p}
& & 
& \n[1]{test/spacy/median/n}
& \n[1]{test/spacy/median/p} 
& & 
& \n[1]{test/spacy/mean/n}
& \n[1]{test/spacy/mean/p} 
& & 
& \n{test/spacy/max/n}
& \n[1]{test/spacy/max/p}

\\
\midrule
\multirow{3}{*}{\web}
& \project{fastapi}
& \multirow{3}{*}{\n{test/web/min/n}} 
& \multirow{3}{*}{\n[1]{test/web/min/p}} 
& \n{test/fastapi/min/n}
& \n[1]{test/fastapi/min/p}
& \multirow{3}{*}{\n[1]{test/web/median/n}} 
& \multirow{3}{*}{\n[1]{test/web/median/p}} 
& \n[1]{test/fastapi/median/n}
& \n[1]{test/fastapi/median/p}
& \multirow{3}{*}{\n[1]{test/web/mean/n}} 
& \multirow{3}{*}{\n[1]{test/web/mean/p}} 
& \n[1]{test/fastapi/mean/n}
& \n[1]{test/fastapi/mean/p}
& \multirow{3}{*}{\n{test/web/max/n}} 
& \multirow{3}{*}{\n[1]{test/web/max/p}} 
& \n{test/fastapi/max/n}
& \n[1]{test/fastapi/max/p}
  
\\ 
& \project{sanic} 
& & 
& \n{test/sanic/min/n}
& \n[1]{test/sanic/min/p}
& & 
& \n[1]{test/sanic/median/n}
& \n[1]{test/sanic/median/p} 
& & 
& \n[1]{test/sanic/mean/n}
& \n[1]{test/sanic/mean/p} 
& & 
& \n{test/sanic/max/n}
& \n[1]{test/sanic/max/p}

\\ 
& \project{tornado} 
& & 
& \n{test/tornado/min/n}
& \n[1]{test/tornado/min/p}
& & 
& \n[1]{test/tornado/median/n}
& \n[1]{test/tornado/median/p} 
& & 
& \n[1]{test/tornado/mean/n}
& \n[1]{test/tornado/mean/p} 
& & 
& \n{test/tornado/max/n}
& \n[1]{test/tornado/max/p}

\\
\midrule
\multicolumn{2}{r}{\textbf{overall}} 
& \n{test/all/min/n}
& \n[1]{test/all/min/p}
& \n{test/all/min/n}
& \n[1]{test/all/min/p}
& \n[1]{test/all/median/n}
& \n[1]{test/all/median/p}
& \n[1]{test/all/median/n}
& \n[1]{test/all/median/p}
& \n[1]{test/all/mean/n}
& \n[1]{test/all/mean/p}
& \n[1]{test/all/mean/n}
& \n[1]{test/all/mean/p}
& \n{test/all/max/n}
& \n[1]{test/all/max/p}
& \n{test/all/max/n}
& \n[1]{test/all/max/p}
\\
\bottomrule
    \end{tabular}
    \caption{Tests used in the fault localization experiments
      with the bugs of \autoref{tab:bugsinpy-selected-projects}.
      Following the procedure described
      in \autoref{sec:design:experimental-setup},
      we selected
      $s_b$ tests out of the $t_b$ \bip tests for each bug $b$
      among the \n.{number of subjects}! bugs used in our experiments.
      For each \textsc{project}, the table
      reports the \textsc{min}imum, \textsc{median}, \textsc{mean},
      and \textsc{max}imum percentage $100 \cdot s_b/t_b$~\textsc{\%} 
      of selected tests among bugs $b$ in the project (columns $P$);
      similarly, columns \textsc{\#} report the same statistics
      the \textsc{min}imum, \textsc{median}, \textsc{mean},
      and \textsc{max}imum number of selected tests $s_b$
      among all bug $b$ in the project.
      Finally, columns $C$ report the same statistics among all bugs in
      projects of the same \textsc{category};
      and the bottom row reports the \textbf{overall} statistics among all
      \n.{number of subjects}! bugs.
    }
    \label{tab:bugsinpy-test-selection-info}
\end{table}

\paragraph{Experimental setup.}\label{sec:experimental-setup}
Each experiment ran on a node of USI's HPC cluster,\footnote{Managed by USI's Institute of Computational Science (\url{https://intranet.ics.usi.ch/HPC}).}
each equipped with 20-core Intel Xeon E5-2650 processor
and 64~GB of DDR4 RAM, accessing a shared 15~TB RAID 10 SAS3 drive,
and running CentOS 8.2.2004.x86\_64.
We provisioned three CPython Virtualenvs with Python v.~3.6, 3.7, and 3.8;
our scripts chose a version according to the requirements
of each \bip subject.
The experiments took more than two CPU-months to complete%
---not counting the additional time to setup the infrastructure,
fix the execution scripts, and repeat any experiments that failed due to
incorrect configuration.

This paper's detailed replication package includes all scripts
used to ran these experiments,
as well as all raw data that we collected by running them.
The rest of this section
details how we analyzed and summarized the data to answer
the various research questions.\footnote{%
  Research questions RQ1, RQ2, RQ3, RQ4, and RQ6
  only consider \emph{statement-level} granularity;
  in contrast, RQ5 considers all granularities 
  (see~\autoref{sec:preliminaries:granularities}).
}

\subsubsection{RQ1. Effectiveness}
\label{sec:design:experimental-setup:rq1}

To answer RQ1 (fault localization \emph{effectiveness}),
we report the $L@_B1\%$, $L@_B3\%$, $L@_B5\%$, and $L@_B10\%$ counts,
the average generalized \einspectso rank $\einspect*[B][L]$,
the average exam score $\exam_B(L)$,
and the average location list length $\outputlen{L}[B]$
for each technique $L$ among \autoref{sec:preliminaries}'s
\n.{number of techniques}! standalone fault localization \emph{techniques};
as well as the same metrics averaged over each of the \n.{number of families}!
fault localization \emph{families}.
These metrics measure the effectiveness of fault localization
from different angles.
We report these measures for \emph{all} \n.{number of subjects} \bip bugs $B$
selected for our experiments.

To qualitatively
summarize the effectiveness comparison between two FL techniques $A$ and $B$,
we consider their counts
$A@1\% \leq A@3\% \leq A@5\% \leq A@10\%$
and
$B@1\% \leq B@3\% \leq B@5\% \leq B@10\%$
and compare them pairwise:
$A@k\%$ vs.\ $B@k\%$, for the each $k$ among 1, 3, 5, 10.
We say that:
\begin{description}[leftmargin=!,labelwidth=\widthof{$A \gg B\ $}]
\item[$A \gg B$:]
  ``\emph{$A$ is much more effective than $B$}'', 
  if $A@k\% > B@k\%$ for all $k$s, and $A@k\% - B@k\% \geq 10$ for at least three $k$s out of four;
\item[$A > B$:]
  ``\emph{$A$ is more effective than $B$}'',
  if $A@k\% > B@k\%$ for all $k$s, and $A@k\% - B@k\% \geq 5$ for at least one $k$ out of four;
\item[$A \geq B$:]
  ``\emph{$A$ tends to be more effective than $B$}'',
  if $A@k\% \geq B@k\%$ for all $k$s, and
  $A@k\% > B@k\%$ for at least three $k$s out of four;
\item[$A \simeq B$:]
  ``\emph{$A$ is about as effective as $B$}'',
  if none of $A \gg B$, $A > B$, $A \geq B$, $B \gg A$, $B > A$, and $B \geq A$
  holds.
\end{description}

To \emph{visually compare}
the effectiveness of different FL families,
we use \emph{scatterplots}---one for each pair $F_1, F_2$
of families.
The scatterplot comparing $F_1$ to $F_2$
displays one point at coordinates $(x, y)$ for each
bug $b$ analyzed in our experiments.
Coordinate $x = \einspect*[b][F_1]$,
that is the average generalized \einspectso rank that
techniques in family $F_1$ achieved on $b$;
similarly, $y = \einspect*[b][F_2]$, 
that is the average generalized \einspectso rank that
techniques in family $F_2$ achieved on $b$.
Thus, points lying below the diagonal line $x = y$
(such that $x > y$)
correspond
to bugs for which family $F_2$ performed \emph{better}
(remember that a lower \einspectso score means more effective fault localization)
than family $F_1$;
the opposite holds for points lying above the diagonal line.
The location of points in the scatterplot relative
to the diagonal gives a clear idea of which family performed better in
most cases.

To \emph{analytically compare}
the effectiveness of different FL families,
we report the estimates and the 95\% probability intervals
of the coefficients $\alpha_{F}$
in the fitted regression model \eqref{eq:regmodel},
for each FL family $F$.
If the interval of values lies entirely below zero,
it means that family $F$'s effectiveness tends to be \emph{better}
than the other families on average;
if it lies entirely above zero,
it means that family $F$'s effectiveness tends to be \emph{worse}
than the other families;
and if it includes zero,
it means that there is no consistent association
(with above- or below-average effectiveness).

\subsubsection{RQ2. Efficiency}
\label{sec:design:experimental-setup:rq2}

To answer RQ2 (fault localization \emph{efficiency}),
we report 
the average wall-clock running time $T_B(L)$,
for each technique $L$ among \autoref{sec:preliminaries}'s
\n.{number of techniques}! standalone fault localization \emph{techniques}, 
on bugs in $B$;
as well as the same metric averaged over each of the \n.{number of families}!
fault localization \emph{families}.
This basic metric measures how long the various FL techniques take
to perform their analysis.
We report these measures for \emph{all} \n.{number of subjects} \bip bugs $B$
selected for our experiments.

To qualitatively
summarize the efficiency comparison between two FL techniques $A$ and $B$,
we compare pairwise their average running times
$T(A)$ and $T(B)$, and say that:
\begin{description}[leftmargin=!,labelwidth=\widthof{$A \gg B\ $}]
\item[$A \gg B$:]
  ``\emph{$A$ is much more efficient than $B$}'', 
  if $T(A) > 10 \cdot T(B)$;
\item[$A > B$:]
  ``\emph{$A$ is more efficient than $B$}'',
  if $T(A) > 1.1 \cdot T(B)$;
\item[$A \simeq B$:]
  ``\emph{$A$ is about as efficient as $B$}'',
  if none of $A \gg B$, $A > B$, $B \gg A$, and $B > A$ 
  holds.
\end{description}

To \emph{visually compare}
the efficiency of different FL families,
we use \emph{scatterplots}---one for each pair $F_1, F_2$
of families.
The scatterplot comparing $F_1$ to $F_2$
displays one point at coordinates $(x, y)$ for each
bug $b$ analyzed in our experiments.
Coordinate $x = T_b(F_1)$,
that is the average running time of
techniques in family $F_1$ on $b$;
similarly, $y = T_b(F_2)$,
that is the average running time of
techniques in family $F_2$ on $b$.
The interpretation of these scatterplots is as those considered for RQ1.

To \emph{analytically compare}
the efficiency of different FL families,
we report the estimates and the 95\% probability intervals
of the coefficients $\beta_{F}$
in the fitted regression model \eqref{eq:regmodel},
for each FL family $F$.
The interpretation of the regression coefficients' intervals 
is similar to those considered for RQ1:
$\beta_F$'s lies entirely above zero when $F$ tends
to be \emph{slower} (less efficient) than other families;
it lies entirely below zero when $F$ tends to be \emph{faster};
and it includes zero when there is no consistent association with above- or below-average efficiency.

\subsubsection{RQ3. Kinds of Faults and Projects}
\label{sec:design:experimental-setup:rq3}

To answer RQ3
(fault localization behavior for different \emph{kinds of faults}
 and \emph{projects}),
we report
the same effectiveness metrics
considered in RQ1
($F@_X1\%$, $F@_X3\%$, $F@_X5\%$, and $F@_X10\%$ percentages,
average generalized \einspectso ranks $\einspect*[X][F]$,
average exam scores $\exam_X(F)$,
and average location list length $\outputlen{F}[X]$),
as well as the same efficiency metrics considered in RQ2
(average wall-clock running time $T_X(F)$)
for each standalone fault localization family $F$ and separately for
\begin{enumerate*}
\item bugs $X$ of different \emph{kinds}:
crashing bugs, predicate bugs, and mutable bugs
(see \autoref{fig:fault-classification});
\item bugs $X$ from projects of different \emph{category}:
  \cli, \dev, \ds, and \web
  (see \autoref{sec:design:faults-kinds}).
\end{enumerate*}

To \emph{visually compare}
the effectiveness and efficiency of fault localization families
on bugs from projects of different \emph{category},
we color the points in the scatterplots used
to answer RQ1 and RQ2 according to the bug's project category.

To \emph{analytically compare} the effectiveness
of different FL families on bugs of different \emph{kinds},
we report the estimates and the 95\% probability intervals 
of the coefficients $c_F$,
$p_F$, and $m_F$
in the fitted regression model \eqref{eq:regmodel-kinds},
for each FL family $F$.
The interpretation of the regression coefficients' intervals
is similar to those considered for RQ1 and RQ2:
$c_F$, $p_F$, and $m_F$ 
characterize the effectiveness of family $F$ respectively on
crashing, predicate, and mutable bugs, 
\emph{relative} to the average effectiveness of the \emph{same family} $F$
on other kinds of bugs.

Finally, to understand whether bugs from projects of certain categories
are intrinsically harder or easier to localize,
we report the estimates and the 95\% probability intervals
of the coefficients $\alpha_C$ and $\beta_C$
in the fitted regression model \eqref{eq:regmodel},
for each project category $C$.
The interpretation of these regression coefficients' intervals
is like those considered for RQ1 and RQ2;
for example if $\alpha_C$'s interval is entirely below zero,
it means that bugs of projects in category $C$
are easier to localize (higher effectiveness) than the average of bugs in any project.
This sets a baseline useful to interpret the other data that answer RQ3.

\subsubsection{RQ4. Combining Techniques}
\label{sec:design:experimental-setup:rq4}

To answer RQ4 (the effectiveness of \emph{combining} FL techniques),
we consider two additional fault localization techniques:
\combine and \average---both combining the information
collected by some of \autoref{sec:preliminaries}'s standalone techniques
from different families.

\combine was introduced by Zou et al.~\cite{Zou:2021};
it uses a learning-to-rank model
to learn how to combine lists of ranked locations
given by different FL techniques.
After fitting the model on labeled training data,\footnote{
  Since the training time is negligible,
  we ignore it in all measures of running time---consistently
  with Zou et al.~\cite{Zou:2021}.
}
one can use it like any other fault localization technique
as follows:
\begin{enumerate*}
\item Run any combination of techniques $L_1, \ldots, L_n$
  on a bug $b$;
\item Feed the ranked location lists output by each technique
  into the fitted learning-to-rank model;
\item The model's output is a list $\ell_1, \ell_2, \ldots$ of locations,
  which is taken as the FL output of technique \combine.
\end{enumerate*}
We used Zou et al.~\cite{Zou:2021}'s replication package
to run \combine
on the Python bugs that we analyzed using \fp.

To see whether a simpler combination algorithm
can still be effective,
we introduced the combined FL technique
\average, which works as follows:
\begin{enumerate*}
\item Each basic technique $L_k$
  returns a list $\langle \ell_1^k, s_1^k \rangle\,\ldots\,\langle \ell_{n_k}^k, s_{n_k}^k \rangle$ of locations with \emph{normalized}\footnote{
  We used min-max normalization, also known as feature scaling~\cite{Islam:2022}.
}
  suspiciousness scores $0 \leq s_j^k \leq 1$;
\item \average assigns to location $\ell_x$ the weighted average
  $\sum_k w_k s_x^k$, where $k$ ranges over all of FL techniques supported by \fp but Tarantula, and $w_k$ is an integer weight that depends on the FL family of $k$:
  3 for SBFL, 2 for MBFL, and 1 for PS and ST;\footnote{
    These weights roughly reflect the relative effectiveness and applicability of FL techniques
    suggested by our experimental results.
  }
\item The list of locations ranked by their weighted average suspiciousness is taken as the FL output of technique \average.
\end{enumerate*}

Finally, we answer RQ4
by reporting
the same effectiveness metrics
considered in RQ1
(the $L@_B1$\%, $L@_B3$\%, $L@_B5$\%, and $L@_B10$\% counts,
the average generalized \einspectso rank $\einspect*[B][L]$,
the average exam score $\exam_B(L)$,
and the average location list length $\outputlen{L}[B]$)
for techniques \combine and \average.
Precisely, we consider two variants $A$ and $S$ of \combine and of \average,
giving a total of \spellout{4} \emph{combined} fault localization techniques:
variants $A$ ($\combine_A$ and $\average_A$) use the output of 
all FL techniques supported by \fp but Tarantula%
---which was not considered in~\cite{Zou:2021};
variants $S$ ($\combine_S$ and $\average_S$) only use
the Ochiai, DStar, and ST FL techniques
(excluding the more time-consuming MBFL and PS families).

\subsubsection{RQ5. Granularity}
\label{sec:design:experimental-setup:rq5}

To answer RQ5 (how fault localization effectiveness changes with granularity),
we report
the same effectiveness metrics
considered in RQ1
(the $L@_B1$, $L@_B3$, $L@_B5$, and $L@_B10$ counts,
the average generalized \einspectso rank $\einspect*[B][L]$,
the average exam score $\exam_B(L)$,
and the average location list length $\outputlen{L}[B]$)
for all \n.{number of techniques}! standalone techniques,
and for all \spellout{4} combined techniques, 
but targeting
\emph{functions}
and \emph{modules} as suspicious entities.
Similar to Zou et al.~\cite{Zou:2021}, for function-level and module-level granularities, we define the suspiciousness score of an entity as the maximum suspiciousness score computed for the statements in them.

\subsubsection{RQ6. Comparison to Java}
\label{sec:design:experimental-setup:rq6}

To answer RQ6 (comparison between Python and Java),
we quantitatively and qualitatively compare
the main findings of Zou et al.~\cite{Zou:2021}%
---whose empirical study of fault localization in Java
was the basis for our Python replication---%
against our findings for Python.

For the \emph{quantitative} comparison of \emph{effectiveness},
we consider the metrics that are available in both studies:
the percentage of all bugs each technique localized
within the top-$1$, top-$3$, top-$5$, and top-$10$ positions of its output
($L@1$\%, $L@3$\%, $L@5$\%, and $L@10$\%);
and the average exam score.
For Python, we consider all \n{number of subjects} \bip bugs
we selected for our experiments;
the data for Java is about Zou et al.'s
experiments on 357 bugs in Defects4J~\cite{Just:2014}.
We consider all standalone techniques that feature in both studies:
Ochiai and DStar (SBFL),
Metallaxis and \muse (MBFL),
predicate switching (PS),
and stack-trace fault localization (ST).

We also consider the combined techniques $\combine_A$ and $\combine_S$.
The original idea of the \combine technique was introduced by Zou et al.;
however, the variants used in their experiments combine all \spellout{11}
FL techniques they consider, some of which we did not include in our replication
(see \autoref{sec:related-work} for details).
Therefore, we modified \cite{Zou:2021}'s
replication package to
extract from their Java experimental data
the rankings obtained by $\combine_A$ and $\combine_S$
combining the same techniques as in Python
(see \autoref{sec:design:experimental-setup:rq4}).
This way, the quantitative comparison between Python and Java
involves exactly the same techniques and combinations thereof.

Since we did not re-run Zou et al.'s experiments
on the same machines used for our experiments,
we cannot compare efficiency quantitatively.
Anyway, a comparison of this kind between Java and Python
would be outside the scope of our studies,
since any difference would likely merely reflect the different
performance of Java and Python%
---largely independent of fault localization efficiency.

For the \emph{qualitative} comparison between Java and Python,
we consider the union of all findings
presented in this paper or in Zou et al.~\cite{Zou:2021};
we discard all findings from one paper
that are outside the scope of the other paper
(for example,
Java findings about history-based fault localization,
a standalone technique that we did not implement for Python;
or Python findings about \average,
a combined technique that Zou et al.\ did not implement for Java);
for each within-scope finding,
we determine whether it is
confirmed~\confOK (there is evidence corroborating it) %
or refuted~\confNO (there is evidence against it)
for Python and for Java.

\section{Experimental Results}
\label{sec:experimental-results}

This section summarizes
the experimental results
that answer the research questions detailed in \autoref{sec:design:experimental-setup}.
All results except for \autoref{sec:experimental-results:rq5}'s
refer to experiments with statement-level granularity;
results in Sections~\autoref{sec:experimental-results:rq1}--\ref{sec:experimental-results:rq3}
only consider standalone techniques.
To keep the discussion focused,
we mostly comment on the $@n\%$ metrics of effectiveness,
whereas we only touch upon the exam score, \einspectso, and location list length
when they complement other results.

\begin{table}[!tb]
\small
\centering
\setlength{\tabcolsep}{5pt}
\begin{tabular}{lr rr rr rr rr rr rr rr} 
\toprule
\multicolumn{1}{c}{\textsc{family}} & 
\multicolumn{1}{c}{\textsc{technique} $L$} & 
\multicolumn{2}{c}{\einspect*[B][L]} & 
\multicolumn{2}{c}{$L@_B1$\%} & 
\multicolumn{2}{c}{$L@_B3$\%} & 
\multicolumn{2}{c}{$L@_B5$\%} & 
\multicolumn{2}{c}{$L@_B10$\%} & 
\multicolumn{2}{c}{$\exam_B(L)$} & 
\multicolumn{2}{c}{\outputlen{L}[B]} \\

\cmidrule(lr){3-4} 
\cmidrule(lr){5-6} 
\cmidrule(lr){7-8} 
\cmidrule(lr){9-10} 
\cmidrule(lr){11-12} 
\cmidrule(lr){13-14} 
\cmidrule(lr){15-16}  

 &  
 & 
 \multicolumn{1}{c}{\textsc{f}} & 
 \multicolumn{1}{c}{\textsc{t}} & 
 \multicolumn{1}{c}{\textsc{f}} & 
 \multicolumn{1}{c}{\textsc{t}} & 
 \multicolumn{1}{c}{\textsc{f}} & 
 \multicolumn{1}{c}{\textsc{t}} & 
 \multicolumn{1}{c}{\textsc{f}} & 
 \multicolumn{1}{c}{\textsc{t}} & 
 \multicolumn{1}{c}{\textsc{f}} & 
 \multicolumn{1}{c}{\textsc{t}} & 
 \multicolumn{1}{c}{\textsc{f}} & 
 \multicolumn{1}{c}{\textsc{t}} & 
 \multicolumn{1}{c}{\textsc{f}} & 
 \multicolumn{1}{c}{\textsc{t}} \\
 
\midrule

\multirow{2}{*}{MBFL} & 
Metallaxis & 
\multirow{2}{*}{\n{stmt/mbfl/favg/all/einspect}} & 
\n{stmt/mbfl/metallaxis/all/einspect} & 
\multirow{2}{*}{\n.{stmt/mbfl/favg/all/@1}} & 
\n.{stmt/mbfl/metallaxis/all/@1} & 
\multirow{2}{*}{\n.{stmt/mbfl/favg/all/@3}} & 
\n.{stmt/mbfl/metallaxis/all/@3} & 
\multirow{2}{*}{\n.{stmt/mbfl/favg/all/@5}} & 
\n.{stmt/mbfl/metallaxis/all/@5} & 
\multirow{2}{*}{\n.{stmt/mbfl/favg/all/@10}} & 
\n.{stmt/mbfl/metallaxis/all/@10} & 
\multirow{2}{*}{\n[\examdec]{stmt/mbfl/favg/all/javaexam}} & 
\n[\examdec]{stmt/mbfl/metallaxis/all/javaexam} & 
\multirow{2}{*}{\n[\cddec]{stmt/mbfl/favg/all/outputlength}} & 
\n[\cddec]{stmt/mbfl/metallaxis/all/outputlength} \\

 & 
 \muse &  
 & 
 \n{stmt/mbfl/muse/all/einspect} &  
 & 
 \n.{stmt/mbfl/muse/all/@1} &  
 & 
 \n.{stmt/mbfl/muse/all/@3} &  
 & 
 \n.{stmt/mbfl/muse/all/@5} &  
 & 
 \n.{stmt/mbfl/muse/all/@10} &  
 & 
 \n[\examdec]{stmt/mbfl/muse/all/javaexam} &  
 & 
 \n[\cddec]{stmt/mbfl/muse/all/outputlength} \\
 
\cmidrule{2-16}

PS 
&  
&  
\n{stmt/ps/favg/all/einspect} & 
\n{stmt/ps/favg/all/einspect} &  
\n.{stmt/ps/favg/all/@1} & 
\n.{stmt/ps/favg/all/@1} &  
\n.{stmt/ps/favg/all/@3} & 
\n.{stmt/ps/favg/all/@3} &  
\n.{stmt/ps/favg/all/@5} & 
\n.{stmt/ps/favg/all/@5} &  
\n.{stmt/ps/favg/all/@10} & 
\n.{stmt/ps/favg/all/@10} &  
\best{\n[\examdec]{stmt/ps/favg/all/javaexam}} & 
\best{\n[\examdec]{stmt/ps/favg/all/javaexam}} &  
\best{\n[\cddec]{stmt/ps/favg/all/outputlength}} & 
\best{\n[\cddec]{stmt/ps/favg/all/outputlength}} \\

\cmidrule{2-16}

  \multirow{3}{*}{SBFL} & 
 DStar & 
 \multirow{3}{*}{\best{\n{stmt/sbfl/favg/all/einspect}}} & 
 \best{\n{stmt/sbfl/dstar/all/einspect}} &  
 \multirow{3}{*}{\best{\n.{stmt/sbfl/favg/all/@1}}} & 
 \n.{stmt/sbfl/dstar/all/@1} &  
 \multirow{3}{*}{\best{\n.{stmt/sbfl/favg/all/@3}}} & 
 \best{\n.{stmt/sbfl/dstar/all/@3}} &  
 \multirow{3}{*}{\best{\n.{stmt/sbfl/favg/all/@5}}} & 
 \n.{stmt/sbfl/dstar/all/@5} &  
 \multirow{3}{*}{\best{\n.{stmt/sbfl/favg/all/@10}}} & 
 \best{\n.{stmt/sbfl/dstar/all/@10}} &  
 \multirow{3}{*}{\n[\examdec]{stmt/sbfl/favg/all/javaexam}} & 
 \n[\examdec]{stmt/sbfl/dstar/all/javaexam} &  
 \multirow{3}{*}{\n[\cddec]{stmt/sbfl/favg/all/outputlength}} & 
 \n[\cddec]{stmt/sbfl/dstar/all/outputlength} \\

 & 
 Ochiai & 
 & 
 \best{\n{stmt/sbfl/ochiai/all/einspect}} &
 & 
 \best{\n.{stmt/sbfl/ochiai/all/@1}} &  
 & 
 \best{\n.{stmt/sbfl/ochiai/all/@3}} &  
 & 
 \best{\n.{stmt/sbfl/ochiai/all/@5}} &  
 & 
 \best{\n.{stmt/sbfl/ochiai/all/@10}} &  
 & 
 \n[\examdec]{stmt/sbfl/ochiai/all/javaexam} &  
 & 
 \n[\cddec]{stmt/sbfl/ochiai/all/outputlength} \\

 &
Tarantula & 
& 
\n{stmt/sbfl/tarantula/all/einspect} & 
 & 
\best{\n.{stmt/sbfl/tarantula/all/@1}} & 
 & 
\best{\n.{stmt/sbfl/tarantula/all/@3}} & 
 & 
\best{\n.{stmt/sbfl/tarantula/all/@5}} & 
 & 
\best{\n.{stmt/sbfl/tarantula/all/@10}} & 
 & 
\n[\examdec]{stmt/sbfl/tarantula/all/javaexam} & 
 & 
\n[\cddec]{stmt/sbfl/tarantula/all/outputlength} \\

\cmidrule{2-16}

ST 
&  
&  
\n{stmt/st/favg/all/einspect} & 
\n{stmt/st/favg/all/einspect} &  
\n.{stmt/st/favg/all/@1} & 
\n.{stmt/st/favg/all/@1} &  
\n.{stmt/st/favg/all/@3} & 
\n.{stmt/st/favg/all/@3} &  
\n.{stmt/st/favg/all/@5} & 
\n.{stmt/st/favg/all/@5} &  
\n.{stmt/st/favg/all/@10} & 
\n.{stmt/st/favg/all/@10} &  
\n[\examdec]{stmt/st/favg/all/javaexam} & 
\n[\examdec]{stmt/st/favg/all/javaexam} &  
\n[\cddec]{stmt/st/favg/all/outputlength} & 
\n[\cddec]{stmt/st/favg/all/outputlength} \\

\bottomrule

\end{tabular}
\caption{Effectiveness of standalone fault localization techniques
  at the statement-level granularity on all \n{number of subjects}
  selected bugs $B$.
  Each row reports a \textsc{technique} $L$'s
  average generalized \einspectso rank \einspect*[B][L];
  the percentage of all bugs it localized
  within the top-$1$, top-$3$, top-$5$, and top-$10$ positions of its output
  ($L@_B1$\%, $L@_B3$\%, $L@_B5$\%, and $L@_B10$\%);
  its average exam score $\exam_B(L)$;
  and its average suspicious locations length \outputlen{L}[B].
  Columns \textsc{f} report the same metrics averaged for all techniques
  that belong to the same \textsc{family}.
  \best{Highlighted} numbers denote the best technique according
  to each metric.
}
\label{table:effectiveness-family-technique-statement}
\end{table}

\subsection{RQ1. Effectiveness}
\label{sec:experimental-results:rq1}

\paragraph{Family effectiveness.}
Among standalone techniques,
the SBFL fault localization family
achieves the best effectiveness according to several metrics.
\autoref{table:effectiveness-family-technique-statement}
shows that all SBFL techniques have better %
average \einspectso rank \einspect*; and
higher percentages of faulty locations in the top-1,
top-3, top-5, and top-10.
The advantage over MBFL%
---the second most-effective family---%
is consistent and conspicuous.
According to the same metrics,
the MBFL fault localization family achieves clearly better effectiveness
than PS and ST.
Then, PS tends to do better than ST,
but only according to some metrics:
PS has better @1\%, @3\%, and @5\%, and location list length,
whereas ST has better \einspectso and @10\%.

\finding{SBFL is the most effective standalone fault localization family.}
\label{f:eff:sbfl-best}

\finding{Standalone fault localization families ordered by effectiveness: SBFL $>$ MBFL $\gg$ PS $\simeq$ ST, \\where $>$ means better, $\gg$ much better, and $\simeq$ about as good.}
\label{f:eff:standalone-order}

Contrary to these general trends,
PS achieves the best (lowest) exam score and location list length of all families;
and ST is second-best according to these metrics.
As \autoref{sec:experimental-results:rq3} will discuss in more detail,
PS and ST are techniques with a narrower scope than SBFL and MBFL:
they can perform very well on a subset of bugs,
but they fail spectacularly on several others.
They also tend to return shorter lists of suspicious locations,
which is also conducive to achieving a better exam score:
since the exam score is undefined when a technique fails to localize a bug at all
(as explained in \autoref{sec:design:classic-metrics}),
the average exam score of ST and, especially, PS is computed
over the small set of bugs on which they work fairly well.

\finding{PS and ST are specialized fault localization techniques, which may work well only on a small subset of bugs, and thus often return short lists of suspicious locations.}
\label{f:gen:ps-st-specialized}

\begin{figure}[!tb]
  \centering
  \includegraphics[width=\textwidth]{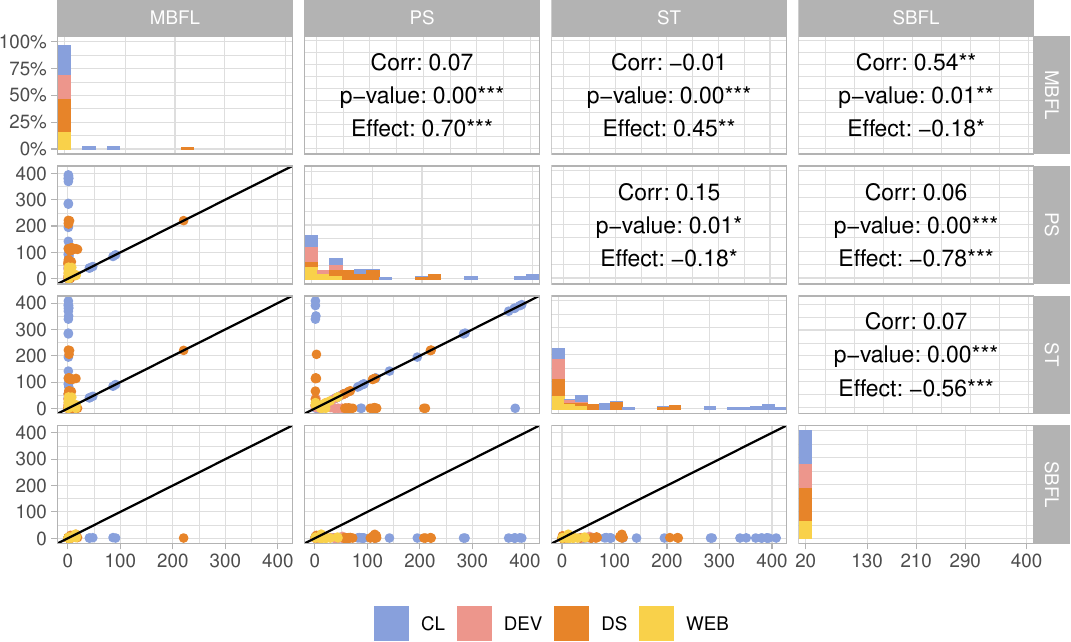}
  \caption{Pairwise visual comparison of \n.{number of families}! FL families for effectiveness. %
    Each point in the scatterplot at row labeled~$R$ and column labeled~$C$ has coordinates $(x, y)$, where $x$ is the generalized \einspectso rank $\einspect*[b][C]$ of FL techniques in family $C$ and $y$ is the rank $\einspect*[b][R]$ of FL techniques in family $R$ on the same bug $b$. Thus, points below (resp.~above) the diagonal line denote bugs on which $R$ had better (resp.~worse) \einspectso ranks. Points are colored according to the bug's project category.
    The opposite box at row labeled~$C$ and column labeled~$R$ displays three statistics
    (correlation, $p$-value, and effect size, see \autoref{sec:design:statistical-models}) quantitatively comparing the same average generalized \einspectso ranks of $C$ and $R$; negative values of effect size mean that $R$ tends to be better, and positive values mean that $C$ tends to be better.
    Each bar plot on the diagonal at row $F$, column $F$
    is a histogram of the distribution of $\einspect*[b][F]$
    for all bugs. Horizontal axes of all diagonal plots
    have the same \einspectso scale as the bottom-right plot's (SBFL);
    their vertical axes have the same 0--100\% scale as the top-left plot (MBFL).
  }
  \label{fig:pairwise-einspect}
\end{figure}

\autoref{fig:pairwise-einspect}'s scatterplots
confirm SBFL's general advantage:
in each scatterplot involving SBFL,
all points are on a straight line corresponding to low ranks for SBFL but increasingly high ranks for the other family.
The plots also indicate that MBFL is often better than PS and ST, although
there are a few hard bugs for which the latter are just as effective
(points on the diagonal line).
The PS-vs-ST scatterplot suggests that these two techniques
are largely complementary:
on several bugs, PS and ST are as effective (points on the diagonal);
on several others, PS is more effective (points above the diagonal);
and on others still, ST is more effective (points below the diagonal).

\begin{figure}[!tb]
  \centering
  \begin{subfigure}[t]{0.49\linewidth}
    \includegraphics[width=\textwidth]{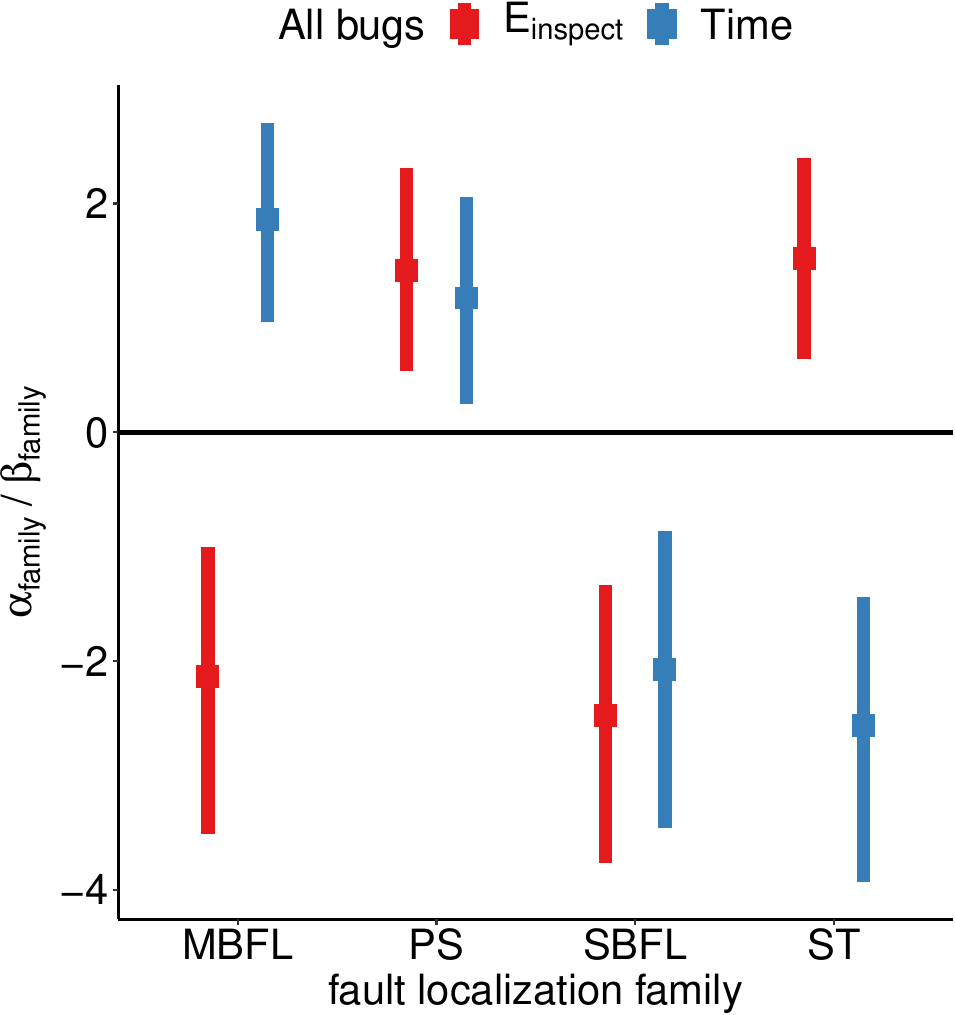}
    \caption{Estimates and 95\% probability intervals for the coefficients {\color{cole}$\alpha_{\var{family}}$} and {\color{colt}$\beta_{\var{family}}$}
      in model \eqref{eq:regmodel} fitted on all bugs,
      for each FL family MBFL, PS, SBFL, and ST.}
    \label{fig:stats-family}
  \end{subfigure}
  \begin{subfigure}[t]{0.49\linewidth}
    \includegraphics[width=\textwidth]{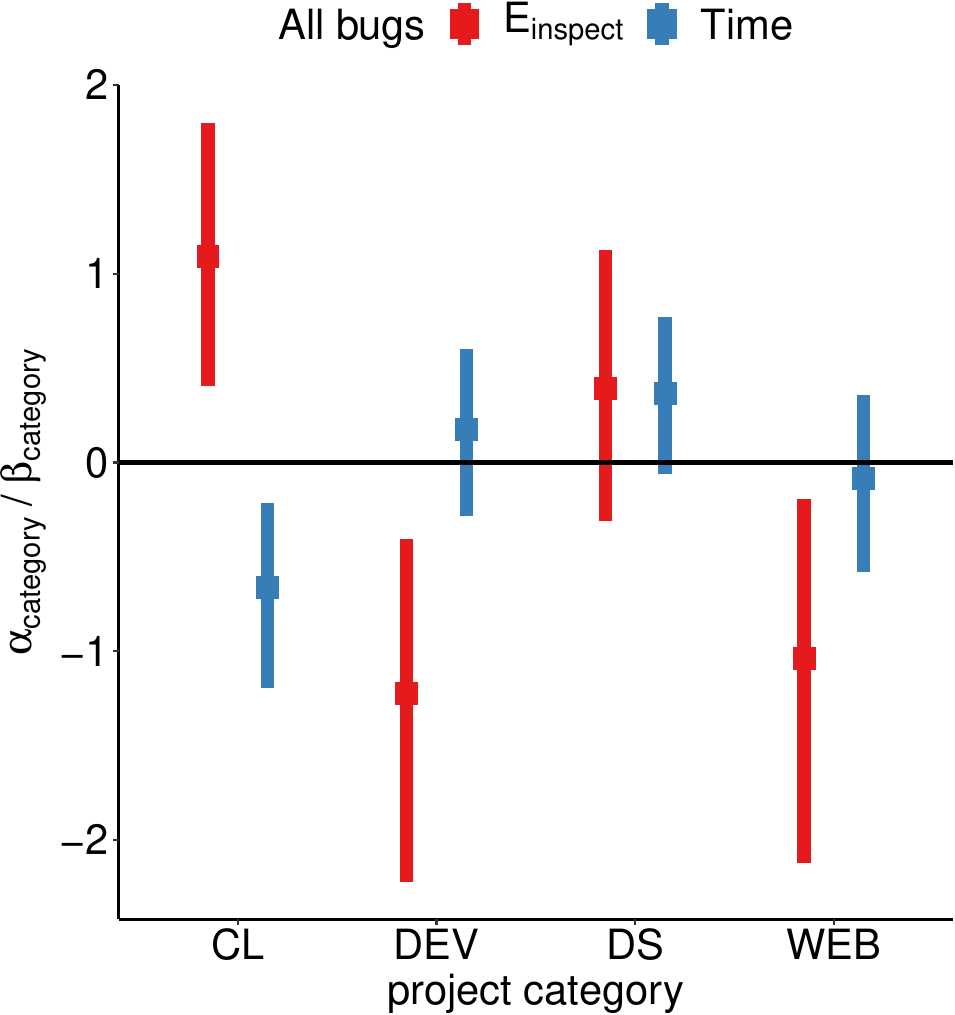}
    \caption{Estimates and 95\% probability intervals for the coefficients {\color{cole}$\alpha_{\var{category}}$} and {\color{colt}$\beta_{\var{category}}$}
      in model \eqref{eq:regmodel} fitted on all bugs,
      for each project category \cli, \dev, \ds, and \web.}
    \label{fig:stats-category}
  \end{subfigure}
  \caption{Point estimates (boxes) and 95\% probability intervals (lines)
  for the regression coefficients of model~\eqref{eq:regmodel}. The scale of the vertical axes is over standard deviation log-units.}
  \label{fig:stats-family-category}
\end{figure}

\autoref{fig:stats-family} confirms these
results based on the statistical model \eqref{eq:regmodel}:
the intervals of coefficients $\alpha_{\text{SBFL}}$ and $\alpha_{\text{MBFL}}$
are clearly below zero,
indicating that SBFL and MBFL have better-than-average effectiveness;
conversely, those of coefficients $\alpha_{\text{PS}}$ and $\alpha_{\text{ST}}$
are clearly above zero,
indicating that PS and ST have worse-than-average effectiveness.

\autoref{fig:stats-family}'s
estimate of $\alpha_{\text{SBFL}}$ is below that of $\alpha_{\text{MBFL}}$,
confirming that SBFL is the most effective family overall.
The bottom-left plot in \autoref{fig:pairwise-einspect}
confirms that SBFL's advantage can be conspicuous
but is observed only on a minority of bugs%
---whereas SBFL and MBFL achieve similar effectiveness on the majority of bugs.
In fact, the effect size comparing SBFL and MBFL is $-\n[2]{statement/einspect/MBFL:SBFL/CliffD}$---weakly in favor of SBFL.

\finding{SBFL and MBFL often achieve similar effectiveness; however, SBFL is strictly better than MBFL on a minority of bugs.}
\label{f:eff:sbfl-mbfl}

\begin{figure}[!bt]
  \centering
  \includegraphics[width=0.8\textwidth]{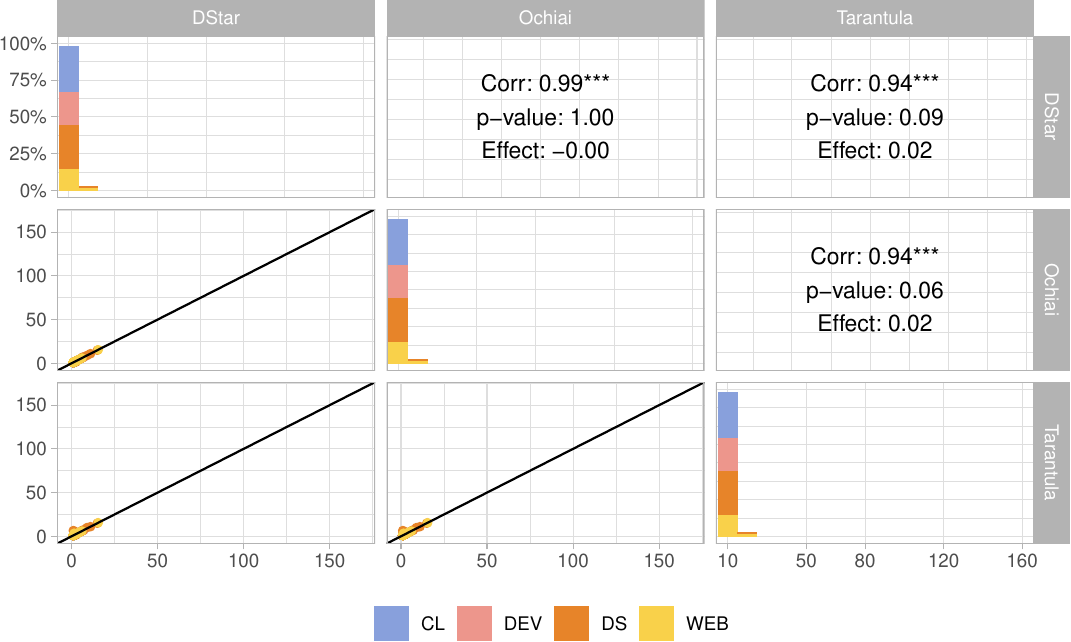}
  \caption{Pairwise visual comparison of 3 SBFL techniques for effectiveness. The interpretation of the plots is the same as in~\autoref{fig:pairwise-einspect}.}
  \label{fig:parwise-sbfl}
\end{figure}

\paragraph{Technique effectiveness.}
FL techniques of the same family
achieve very similar effectiveness.
\autoref{table:effectiveness-family-technique-statement}
shows nearly identical results for the 3 SBFL techniques
Tarantula, Ochiai, and DStar.
The plots and statistics in \autoref{fig:parwise-sbfl}
confirm this: points lie along the diagonal lines in the scatterplots,
and \einspectso ranks for the same bugs
are strongly correlated and differ by a vanishing effect size.

\finding{All techniques in the SBFL family achieve very similar effectiveness.}
\label{f:eff:sbfl-tech-similar}

\begin{figure}[!tb]
  \centering
  \includegraphics[width=0.8\textwidth]{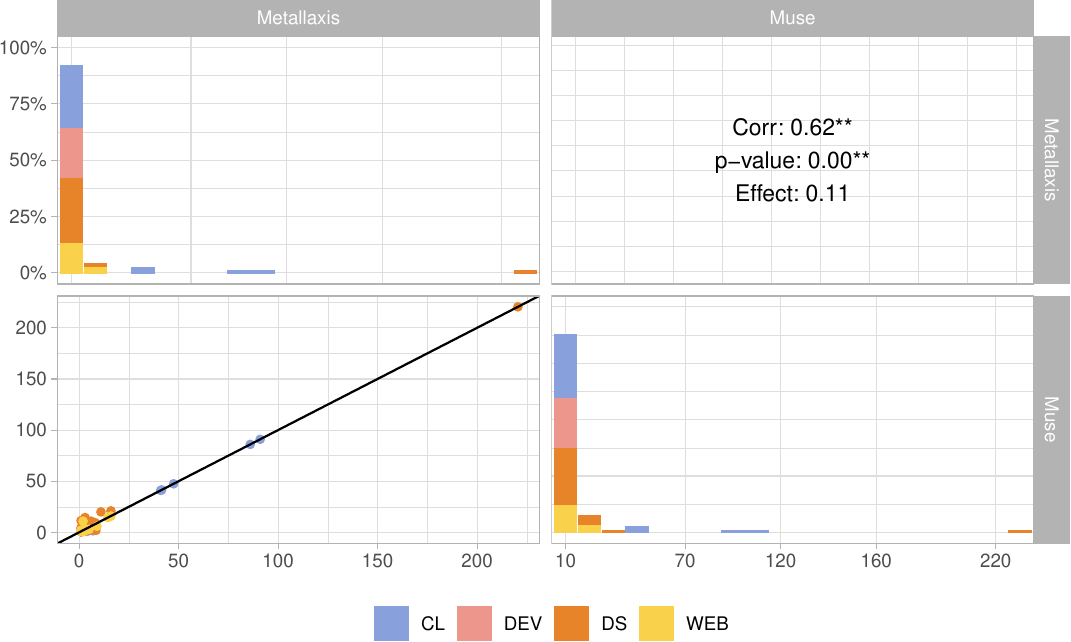}
  \caption{Pairwise visual comparison of 2 MBFL techniques for effectiveness. The interpretation of the plots is the same as in~\autoref{fig:pairwise-einspect}.}
  \label{fig:parwise-mbfl}
\end{figure}

The 2 MBFL techniques also behave similarly, but not quite as closely as the SBFL ones.
Metallaxis has a not huge but consistent advantage over \muse according to \autoref{table:effectiveness-family-technique-statement}.
\autoref{fig:parwise-mbfl} corroborates this observation:
the cloud of points in the scatterplot is centered slightly above the diagonal line;
the correlation between \muse's and Metallaxis's data
is medium (not strong);
and the effect size suggests that Metallaxis is more effective on
around \n[0]{statement-intra-MBFL/einspect/Metallaxis:Muse/CliffD}[-100]|
of subjects.

\muse's lower effectiveness can be traced back
to its stricter definition of ``mutant killing'',
which requires that a failing test becomes passing when run on a mutant
(see \autoref{sec:preliminaries:mbfl}).
As observed elsewhere~\cite{Pearson:2017},
this requirement may be too demanding for fault localization
of real-world bugs,
where it is essentially tantamount to generating a mutant that is similar
to a patch.

\finding{The techniques in the MBFL family achieve generally similar effectiveness, but Metallaxis tends to be better than \muse.}
\label{f:eff:mbfl-tech-similar}

\begin{table}[t]
  \small
  \centering
  \setlength{\tabcolsep}{7.5pt}

\begin{tabular}{lr | r | rrr | rrrr}

\toprule

\multicolumn{1}{c}{\textsc{family}} & 
\multicolumn{1}{c}{\textsc{technique} $L$} & 
\multicolumn{1}{c}{\textsc{all}} & 
\multicolumn{1}{c}{\textsc{crashing}} & 
\multicolumn{1}{c}{\textsc{predicate}} &
\multicolumn{1}{c}{\textsc{mutable}} & 
\multicolumn{1}{c}{\cli} &
\multicolumn{1}{c}{\dev} & 
\multicolumn{1}{c}{\ds} & 
\multicolumn{1}{c}{\web} \\
 
\midrule

\multirow{2}{*}{MBFL} & 
Metallaxis & 
\multirow{2}{*}{\n{stmt/mbfl/metallaxis/all/time}} & 
\multirow{2}{*}{\n{stmt/mbfl/metallaxis/crash/time}} & 
\multirow{2}{*}{\n{stmt/mbfl/metallaxis/pred/time}} & 
\multirow{2}{*}{\n{stmt/mbfl/metallaxis/mut/time}} & 
\multirow{2}{*}{\n{stmt/mbfl/metallaxis/cli/time}} &
\multirow{2}{*}{\n{stmt/mbfl/metallaxis/dev/time}} & 
\multirow{2}{*}{\n{stmt/mbfl/metallaxis/ds/time}} & 
\multirow{2}{*}{\n{stmt/mbfl/metallaxis/web/time}} \\

 & 
 \muse &  
  & 
  & 
  & 
  & 
  & 
  & 
  & 
  \\
 
\cmidrule{2-10}

PS &  
&
\n{stmt/ps/favg/all/time} & 
\n{stmt/ps/favg/crash/time} & 
\n{stmt/ps/favg/pred/time} & 
\n{stmt/ps/favg/mut/time} & 
\n.{stmt/ps/favg/cli/time} &
\n{stmt/ps/favg/dev/time} & 
\n{stmt/ps/favg/ds/time} & 
\n.{stmt/ps/favg/web/time} \\

\cmidrule{2-10}

\multirow{3}{*}{SBFL} & 
 DStar &  
\multirow{3}{*}{\n.{stmt/sbfl/dstar/all/time}} & 
\multirow{3}{*}{\n.{stmt/sbfl/dstar/crash/time}} & 
\multirow{3}{*}{\n{stmt/sbfl/dstar/pred/time}} & 
\multirow{3}{*}{\n.{stmt/sbfl/dstar/mut/time}} & 
\multirow{3}{*}{\n.{stmt/sbfl/dstar/cli/time}} &
\multirow{3}{*}{\n.{stmt/sbfl/dstar/dev/time}} & 
\multirow{3}{*}{\n{stmt/sbfl/dstar/ds/time}} & 
\multirow{3}{*}{\n.{stmt/sbfl/dstar/web/time}} \\

 & 
 Ochiai &  
  & 
  & 
  & 
  & 
  & 
  & 
  & 
  \\

 & 
Tarantula & 
  & 
  & 
  & 
  & 
  & 
  & 
  & 
  \\
 
\cmidrule{2-10}

ST &  
&
\best{\n.{stmt/st/favg/all/time}} & 
\best{\n.{stmt/st/favg/crash/time}} & 
\best{\n.{stmt/st/favg/pred/time}} & 
\best{\n.{stmt/st/favg/mut/time}} & 
\best{\n.{stmt/st/favg/cli/time}} &
\best{\n.{stmt/st/favg/dev/time}} & 
\best{\n.{stmt/st/favg/ds/time}} & 
\best{\n.{stmt/st/favg/web/time}} \\

\bottomrule

\end{tabular}
\caption{Efficiency of fault localization techniques
  at the statement-level granularity.
  Each row reports a \textsc{technique} $L$'s
  per-bug average wall-clock running time $T_X(L)$ in seconds on:
  \textsc{all} \n{number of subjects} bugs
  selected for the experiments ($X = B$);
  \textsc{crashing},
  \textsc{predicate}-related,
  and \textsc{mutable} bugs;
  bugs in projects of category \cli, \dev, \ds, and \web
  (see \autoref{sec:design:faults-kinds}).
  The running time is the same for all techniques of the same \textsc{family}.
  \best{Highlighted} numbers denote the fastest technique for bugs
  in each group.
}
\label{table:efficiency-family-bug-type}
\end{table}

\subsection{RQ2. Efficiency}
\label{sec:experimental-results:rq2}

As demonstrated in \autoref{table:efficiency-family-bug-type}, the four FL families differ greatly in
their efficiency---measured as their wall-clock running time.
ST is by far the fastest,
taking a mere \n[0]{stmt/st/favg/all/time} seconds per bug on average;
SBFL is second-fastest,
taking around \n[0]{stmt/sbfl/favg/all/time}[0.01666] \emph{minutes} on average;
PS is one order of magnitude slower,
taking approximately \n[1]{stmt/ps/favg/all/time}[0.000277] \emph{hours} on average;
and MBFL is slower still,
taking over \n[0]{stmt/mbfl/favg/all/time}[0.000277] hours per bug on average.

\finding{Standalone fault localization families ordered by efficiency: ST $\gg$ SBFL $\gg$ PS $>$ MBFL, \\where $>$ means faster, and $\gg$ much faster.\footnote{\scriptsize As we discuss at the end of \autoref{sec:experimental-results:rq2}, these results are largely expected given how the different fault localization techniques work algorithmically.}}
\label{f:time:standalone-order}

\begin{figure}[!tb]
  \centering
  \includegraphics[width=\textwidth]{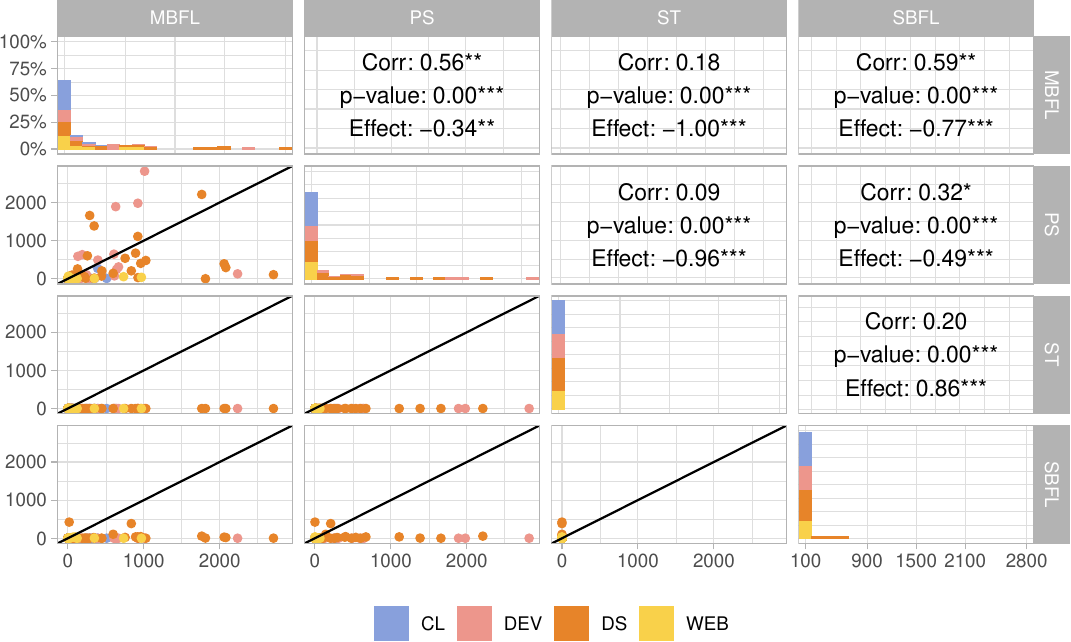}
  \caption{Pairwise visual comparison of \n.{number of families}! FL families for efficiency. Each point in the scatterplot at row labeled~$R$ and column labeled~$C$ has coordinates $(x, y)$, where $x$ is the average per-bug wall-clock running time of FL techniques in family $C$ and $y$ average per-bug wall-clock running time of FL techniques in family $R$. Points are colored according to the bug's project category.
    The opposite box at row labeled~$C$ and column labeled~$R$ displays three statistics
    (correlation, $p$-value, and effect size, see \autoref{sec:design:statistical-models}) quantitatively comparing the same per-bug average running times of $C$ and $R$; negative values of effect size mean that $R$ tends to be better, and positive values that $C$ tends to be better.}
  \label{fig:pairwise-time}
\end{figure}

\autoref{fig:pairwise-time}'s scatterplots
confirm that ST outperforms all other techniques,
and that SBFL is generally second-fastest.
It also shows
that MBFL and PS have similar overall performance
but can be slower or faster on different bugs:
a narrow majority of points lies below the diagonal line in the scatterplot
(meaning PS is faster than MBFL),
but there are also several points that are on the opposite side of the diagonal%
---and their effect size (\n[2]{statement/time/MBFL:PS/CliffD})
is medium, lower than all other pairwise effect sizes
in the comparison of efficiency.

\finding{PS is more efficient than MBFL on average;
  however, the two families tend to be faster or slower on different bugs.}
\label{f:time:ps-mbfl}

Based on the statistical model \eqref{eq:regmodel},
\autoref{fig:stats-family} clearly confirms
the differences of efficiency:
the intervals of coefficients $\beta_{\text{ST}}$ and $\beta_{\text{SBFL}}$
are well below zero,
indicating that ST and SBFL are faster than average
(with ST the fastest, as its estimated $\beta_{\text{ST}}$ is lower);
conversely, the intervals of coefficients $\beta_{\text{MBFL}}$ and $\beta_{\text{PS}}$
are entirely above zero,
indicating that MBFL and PS stand out
as slower than average compared to the other families.

These major differences in efficiency are unsurprising
if one remembers that the various FL families differ in what kind of information they collect
for localization.
ST only needs the stack-trace information,
which only requires to run once the failing tests;
SBFL compares the traces of passing and failing runs,
which involves running \emph{all} tests once.
PS
dynamically tries out a large number of
different branch changes in a program,
each of which runs the failing tests;
in our experiments,
PS tried \numprint{4588} different ``switches''
on average for each bug---up to a whopping \numprint{101454}
switches for project \project{black}'s bug~\#6.
MBFL 
generates hundreds of different mutations
of the program under analysis,
each of which has to be run against \emph{all} tests;
in our experiments,
MBFL generated 461 mutants
on average for each bug---up to \numprint{2718} mutants
for project \project{black}'s bug~\#6.
After collecting this information,
the additional running time to compute suspiciousness scores
(using the formulas presented in \autoref{sec:preliminaries})
is negligible for all techniques%
---which explains why the running times of techniques of the same family are
practically indistinguishable.

\begin{table}[!tb]
\small
\centering
\setlength{\tabcolsep}{9pt}
\begin{tabular}{l r r rrrr r r}
\toprule

\multicolumn{1}{c}{\textsc{bugs} $X$} &
\multicolumn{1}{c}{\textsc{family} $F$} &
\multicolumn{1}{c}{\einspect*[X][F]} &
\multicolumn{1}{c}{$F@_X1$\%} &
\multicolumn{1}{c}{$F@_X3$\%} &
\multicolumn{1}{c}{$F@_X5$\%} &
\multicolumn{1}{c}{$F@_X10$\%} &
\multicolumn{1}{c}{$\exam_X(F)$} &
\multicolumn{1}{c}{\outputlen{F}[X]} \\
\midrule

\multirow{4}{*}{\textsc{all}}  & 
 MBFL &
 \n{stmt/mbfl/favg/all/einspect} &
 \n.{stmt/mbfl/favg/all/@1} &
 \n.{stmt/mbfl/favg/all/@3} &
 \n.{stmt/mbfl/favg/all/@5} &
 \n.{stmt/mbfl/favg/all/@10} &
 \n[\examdec]{stmt/mbfl/favg/all/javaexam} &
 \n[\cddec]{stmt/mbfl/favg/all/outputlength} \\
 
 &
 PS &
 \n{stmt/ps/favg/all/einspect} &
 \n.{stmt/ps/favg/all/@1} & 
 \n.{stmt/ps/favg/all/@3} & 
 \n.{stmt/ps/favg/all/@5} & 
 \n.{stmt/ps/favg/all/@10} & 
 \best{\n[\examdec]{stmt/ps/favg/all/javaexam}} & 
 \best{\n[\cddec]{stmt/ps/favg/all/outputlength}} \\

  &
SBFL &
\best{\n{stmt/sbfl/favg/all/einspect}} &
\best{\n.{stmt/sbfl/favg/all/@1}} &
\best{\n.{stmt/sbfl/favg/all/@3}} &
\best{\n.{stmt/sbfl/favg/all/@5}} &
\best{\n.{stmt/sbfl/favg/all/@10}} &
\n[\examdec]{stmt/sbfl/favg/all/javaexam} &
\n[\cddec]{stmt/sbfl/favg/all/outputlength} \\
 
 &
 ST &
 \n{stmt/st/favg/all/einspect} &
 \n.{stmt/st/favg/all/@1} &
 \n.{stmt/st/favg/all/@3} & 
 \n.{stmt/st/favg/all/@5} & 
 \n.{stmt/st/favg/all/@10} & 
 \n[\examdec]{stmt/st/favg/all/javaexam} & 
 \n[\cddec]{stmt/st/favg/all/outputlength} \\
 
\midrule

\multirow{4}{*}{\textsc{crashing}} & 
 MBFL & 
 \n{stmt/mbfl/favg/crash/einspect} & 
 \n.{stmt/mbfl/favg/crash/@1} & 
 \n.{stmt/mbfl/favg/crash/@3} & 
 \n.{stmt/mbfl/favg/crash/@5} & 
 \n.{stmt/mbfl/favg/crash/@10} & 
 \best{\n[\examdec]{stmt/mbfl/favg/crash/javaexam}} & 
 \n[\cddec]{stmt/mbfl/favg/crash/outputlength} \\
 
 & 
 PS & 
 \n{stmt/ps/favg/crash/einspect} & 
 \n.{stmt/ps/favg/crash/@1} & 
 \n.{stmt/ps/favg/crash/@3} & 
 \n.{stmt/ps/favg/crash/@5} & 
 \n.{stmt/ps/favg/crash/@10} & 
-- & 
\best{\n[\cddec]{stmt/ps/favg/crash/outputlength}} \\

& 
SBFL & 
\best{\n.{stmt/sbfl/favg/crash/einspect}} & 
\best{\n.{stmt/sbfl/favg/crash/@1}} & 
\best{\n.{stmt/sbfl/favg/crash/@3}} & 
\best{\n.{stmt/sbfl/favg/crash/@5}} & 
\best{\n.{stmt/sbfl/favg/crash/@10}} & 
\n[\examdec]{stmt/sbfl/favg/crash/javaexam} & 
\n[\cddec]{stmt/sbfl/favg/crash/outputlength} \\
 
 & 
 ST & 
 \n{stmt/st/favg/crash/einspect} & 
 \n.{stmt/st/favg/crash/@1} & 
 \n.{stmt/st/favg/crash/@3} & 
 \n.{stmt/st/favg/crash/@5} & 
 \n.{stmt/st/favg/crash/@10} & 
 \n[\examdec]{stmt/st/favg/crash/javaexam} & 
 \n[\cddec]{stmt/st/favg/crash/outputlength} \\
 
\cmidrule{2-9}

\multirow{4}{*}{\textsc{predicate}} &
 MBFL & 
 \n{stmt/mbfl/favg/pred/einspect} & 
 \n.{stmt/mbfl/favg/pred/@1} & 
\best{\n.{stmt/mbfl/favg/pred/@3}} & 
\best{\n.{stmt/mbfl/favg/pred/@5}} & 
\best{\n.{stmt/mbfl/favg/pred/@10}} & 
\n[\examdec]{stmt/mbfl/favg/pred/javaexam} & 
\n[\cddec]{stmt/mbfl/favg/pred/outputlength} \\
 
 & 
 PS & 
 \n{stmt/ps/favg/pred/einspect} & 
 \n.{stmt/ps/favg/pred/@1} & 
 \n.{stmt/ps/favg/pred/@3} & 
 \n.{stmt/ps/favg/pred/@5} & 
 \n.{stmt/ps/favg/pred/@10} & 
 \best{\n[\examdec]{stmt/ps/favg/pred/javaexam}} & 
 \best{\n[\cddec]{stmt/ps/favg/pred/outputlength}} \\

 &
SBFL &
\best{\n.{stmt/sbfl/favg/pred/einspect}} &
\best{\n.{stmt/sbfl/favg/pred/@1}} &
\n.{stmt/sbfl/favg/pred/@3} &
\n.{stmt/sbfl/favg/pred/@5} &
\n.{stmt/sbfl/favg/pred/@10} &
\n[\examdec]{stmt/sbfl/favg/pred/javaexam} &
\n[\cddec]{stmt/sbfl/favg/pred/outputlength} \\
 
 & 
 ST &
 \n{stmt/st/favg/pred/einspect} & 
 \n.{stmt/st/favg/pred/@1} & 
 \n.{stmt/st/favg/pred/@3} & 
 \n.{stmt/st/favg/pred/@5} & 
 \n.{stmt/st/favg/pred/@10} & 
 \n[\examdec]{stmt/st/favg/pred/javaexam} & 
 \n[\cddec]{stmt/st/favg/pred/outputlength} \\
 
\cmidrule{2-9}
 
\multirow{4}{*}{\textsc{mutable}} & 
 MBFL & 
\best{\n.{stmt/mbfl/favg/mut/einspect}} & 
\best{\n.{stmt/mbfl/favg/mut/@1}} & 
\best{\n.{stmt/mbfl/favg/mut/@3}} & 
\best{\n.{stmt/mbfl/favg/mut/@5}} & 
\best{\n.{stmt/mbfl/favg/mut/@10}} & 
\n[\examdec]{stmt/mbfl/favg/mut/javaexam} & 
 \n[\cddec]{stmt/mbfl/favg/mut/outputlength} \\
 
 & 
 PS & 
 \n{stmt/ps/favg/mut/einspect} & 
 \n.{stmt/ps/favg/mut/@1} & 
 \n.{stmt/ps/favg/mut/@3} & 
 \n.{stmt/ps/favg/mut/@5} & 
 \n.{stmt/ps/favg/mut/@10} & 
 \best{\n[\examdec]{stmt/ps/favg/mut/javaexam}} & 
 \best{\n[\cddec]{stmt/ps/favg/mut/outputlength}} \\

  & 
SBFL & 
\n.{stmt/sbfl/favg/mut/einspect} & 
\n.{stmt/sbfl/favg/mut/@1} & 
\n.{stmt/sbfl/favg/mut/@3} & 
\best{\n.{stmt/sbfl/favg/mut/@5}} & 
\n.{stmt/sbfl/favg/mut/@10} & 
\n[\examdec]{stmt/sbfl/favg/mut/javaexam} & 
\n[\cddec]{stmt/sbfl/favg/mut/outputlength} \\
 
 & 
 ST & 
 \n{stmt/st/favg/mut/einspect} & 
 \n.{stmt/st/favg/mut/@1} & 
 \n.{stmt/st/favg/mut/@3} & 
 \n.{stmt/st/favg/mut/@5} & 
 \n.{stmt/st/favg/mut/@10} & 
 \n[\examdec]{stmt/st/favg/mut/javaexam} & 
 \n[\cddec]{stmt/st/favg/mut/outputlength} \\

\midrule

\multirow{4}{*}{\cli} & 
 MBFL & 
 \n{stmt/mbfl/favg/cli/einspect} & 
 \best{\n.{stmt/mbfl/favg/cli/@1}} & 
 \n.{stmt/mbfl/favg/cli/@3} & 
 \n.{stmt/mbfl/favg/cli/@5} & 
 \n.{stmt/mbfl/favg/cli/@10} & 
 \n[\examdec]{stmt/mbfl/favg/cli/javaexam} & 
 \n[\cddec]{stmt/mbfl/favg/cli/outputlength} \\
 
 & 
 PS & 
 \n{stmt/ps/favg/cli/einspect} & 
 \n.{stmt/ps/favg/cli/@1} & 
 \n.{stmt/ps/favg/cli/@3} & 
 \n.{stmt/ps/favg/cli/@5} & 
 \n.{stmt/ps/favg/cli/@10} & 
 \best{\n[\examdec]{stmt/ps/favg/cli/javaexam}} & 
 \best{\n[\cddec]{stmt/ps/favg/cli/outputlength}} \\

 & 
SBFL & 
\best{\n{stmt/sbfl/favg/cli/einspect}} & 
\best{\n.{stmt/sbfl/favg/cli/@1}} & 
\best{\n.{stmt/sbfl/favg/cli/@3}} & 
\best{\n.{stmt/sbfl/favg/cli/@5}} & 
\best{\n.{stmt/sbfl/favg/cli/@10}} & 
\n[\examdec]{stmt/sbfl/favg/cli/javaexam} & 
\n[\cddec]{stmt/sbfl/favg/cli/outputlength} \\
 
 & 
 ST & 
 \n{stmt/st/favg/cli/einspect} & 
 \n.{stmt/st/favg/cli/@1} & 
 \n.{stmt/st/favg/cli/@3} & 
 \n.{stmt/st/favg/cli/@5} & 
 \n.{stmt/st/favg/cli/@10} & 
 \n[\examdec]{stmt/st/favg/cli/javaexam} & 
 \n[\cddec]{stmt/st/favg/cli/outputlength} \\

\cmidrule{2-9}

\multirow{4}{*}{\dev} & 
 MBFL & 
 \n{stmt/mbfl/favg/dev/einspect} & 
 \n.{stmt/mbfl/favg/dev/@1} & 
 \n.{stmt/mbfl/favg/dev/@3} & 
 \n.{stmt/mbfl/favg/dev/@5} & 
 \n.{stmt/mbfl/favg/dev/@10} & 
 \n[\examdec]{stmt/mbfl/favg/dev/javaexam} & 
 \n[\cddec]{stmt/mbfl/favg/dev/outputlength} \\
 
 & 
 PS & 
 \n{stmt/ps/favg/dev/einspect} & 
 \n.{stmt/ps/favg/dev/@1} & 
 \n.{stmt/ps/favg/dev/@3} & 
 \n.{stmt/ps/favg/dev/@5} & 
 \n.{stmt/ps/favg/dev/@10} & 
 \best{\n[\examdec]{stmt/ps/favg/dev/javaexam}} & 
 \best{\n[\cddec]{stmt/ps/favg/dev/outputlength}} \\

 & 
SBFL & 
\best{\n{stmt/sbfl/favg/dev/einspect}} & 
\best{\n.{stmt/sbfl/favg/dev/@1}} & 
\best{\n.{stmt/sbfl/favg/dev/@3}} & 
\best{\n.{stmt/sbfl/favg/dev/@5}} & 
\best{\n.{stmt/sbfl/favg/dev/@10}} & 
\n[\examdec]{stmt/sbfl/favg/dev/javaexam} & 
\n[\cddec]{stmt/sbfl/favg/dev/outputlength} \\
 
 & 
 ST & 
 \n{stmt/st/favg/dev/einspect} & 
 \n.{stmt/st/favg/dev/@1} & 
 \n.{stmt/st/favg/dev/@3} & 
 \n.{stmt/st/favg/dev/@5} & 
 \n.{stmt/st/favg/dev/@10} & 
 \n[\examdec]{stmt/st/favg/dev/javaexam} & 
 \n[\cddec]{stmt/st/favg/dev/outputlength} \\

\cmidrule{2-9}

 \multirow{4}{*}{\ds} & 
 MBFL & 
 \n{stmt/mbfl/favg/ds/einspect} & 
 \n.{stmt/mbfl/favg/ds/@1} & 
 \n.{stmt/mbfl/favg/ds/@3} & 
 \n.{stmt/mbfl/favg/ds/@5} & 
 \n.{stmt/mbfl/favg/ds/@10} & 
 \n[\examdec]{stmt/mbfl/favg/ds/javaexam} & 
 \n[\cddec]{stmt/mbfl/favg/ds/outputlength} \\
 
 & 
 PS & 
 \n{stmt/ps/favg/ds/einspect} & 
 \n.{stmt/ps/favg/ds/@1} & 
 \n.{stmt/ps/favg/ds/@3} & 
 \n.{stmt/ps/favg/ds/@5} & 
 \n.{stmt/ps/favg/ds/@10} & 
 \best{\n[\examdec]{stmt/ps/favg/ds/javaexam}} & 
 \best{\n[\cddec]{stmt/ps/favg/ds/outputlength}} \\

 & 
SBFL & 
\best{\n.{stmt/sbfl/favg/ds/einspect}} & 
\best{\n.{stmt/sbfl/favg/ds/@1}} & 
\best{\n.{stmt/sbfl/favg/ds/@3}} & 
\best{\n.{stmt/sbfl/favg/ds/@5}} & 
\best{\n.{stmt/sbfl/favg/ds/@10}} & 
\n[\examdec]{stmt/sbfl/favg/ds/javaexam} & 
\n[\cddec]{stmt/sbfl/favg/ds/outputlength} \\
 
 & 
 ST & 
 \n{stmt/st/favg/ds/einspect} & 
 \n.{stmt/st/favg/ds/@1} & 
 \n.{stmt/st/favg/ds/@3} & 
 \n.{stmt/st/favg/ds/@5} & 
 \n.{stmt/st/favg/ds/@10} & 
 \n[\examdec]{stmt/st/favg/ds/javaexam} & 
 \n[\cddec]{stmt/st/favg/ds/outputlength} \\

\cmidrule{2-9}

\multirow{4}{*}{\web} & 
 MBFL & 
 \n{stmt/mbfl/favg/web/einspect} & 
 \n.{stmt/mbfl/favg/web/@1} & 
 \best{\n.{stmt/mbfl/favg/web/@3}} & 
 \n.{stmt/mbfl/favg/web/@5} & 
 \n.{stmt/mbfl/favg/web/@10} & 
 \n[\examdec]{stmt/mbfl/favg/web/javaexam} & 
 \n[\cddec]{stmt/mbfl/favg/web/outputlength} \\
 
 & 
 PS & 
 \n{stmt/ps/favg/web/einspect} & 
 \n.{stmt/ps/favg/web/@1} & 
 \n.{stmt/ps/favg/web/@3} & 
 \n.{stmt/ps/favg/web/@5} & 
 \n.{stmt/ps/favg/web/@10} & 
 \best{\n[\examdec]{stmt/ps/favg/web/javaexam}} & 
 \best{\n[\cddec]{stmt/ps/favg/web/outputlength}} \\

 & 
SBFL & 
\best{\n.{stmt/sbfl/favg/web/einspect}} & 
\best{\n.{stmt/sbfl/favg/web/@1}} & 
\n.{stmt/sbfl/favg/web/@3} & 
\best{\n.{stmt/sbfl/favg/web/@5}} & 
\best{\n.{stmt/sbfl/favg/web/@10}} & 
\n[\examdec]{stmt/sbfl/favg/web/javaexam} & 
\n[\cddec]{stmt/sbfl/favg/web/outputlength} \\
 
 & 
 ST & 
 \n{stmt/st/favg/web/einspect} & 
 \n.{stmt/st/favg/web/@1} & 
 \n.{stmt/st/favg/web/@3} & 
 \n.{stmt/st/favg/web/@5} & 
 \n.{stmt/st/favg/web/@10} & 
 \n[\examdec]{stmt/st/favg/web/javaexam} & 
 \n[\cddec]{stmt/st/favg/web/outputlength} \\

\bottomrule

\end{tabular}
\caption{%
  Effectiveness of fault localization families
  at the statement-level granularity on
  different \emph{kinds} of bugs and \emph{categories} of projects.
  Each row reports a \textsc{family} $F$'s
  average generalized \einspectso rank \einspect*[X][F];
  the percentage of all bugs it localized
  within the top-$1$, top-$3$, top-$5$, and top-$10$ positions of its output
  ($F@_X1$\%, $F@_X3$\%, $F@_X5$\%, and $F@_X10$\%);
  its average exam score $\exam_X(F)$
  and the length \outputlen{F}[X] of the output list of locations
  on different groups $X$ of bugs:
  \textsc{all} bugs selected for the experiments
  (same results as in \autoref{table:effectiveness-family-technique-statement});
  bugs of different \emph{kinds}
  (\textsc{crashing}, \textsc{predicate}-related, and \textsc{mutable} bugs);
  and bugs from projects of different \emph{categories}
  (\cli, \dev, \ds, and \web).
  \best{Highlighted} numbers denote the best family
  on each group of bugs according to each metric.
}
\label{table:effectiveness-family-statement-bug-type}
\end{table}

\begin{figure}[!tb]
  \centering
  \begin{subfigure}[t]{0.33\linewidth}
    \includegraphics[width=\textwidth]{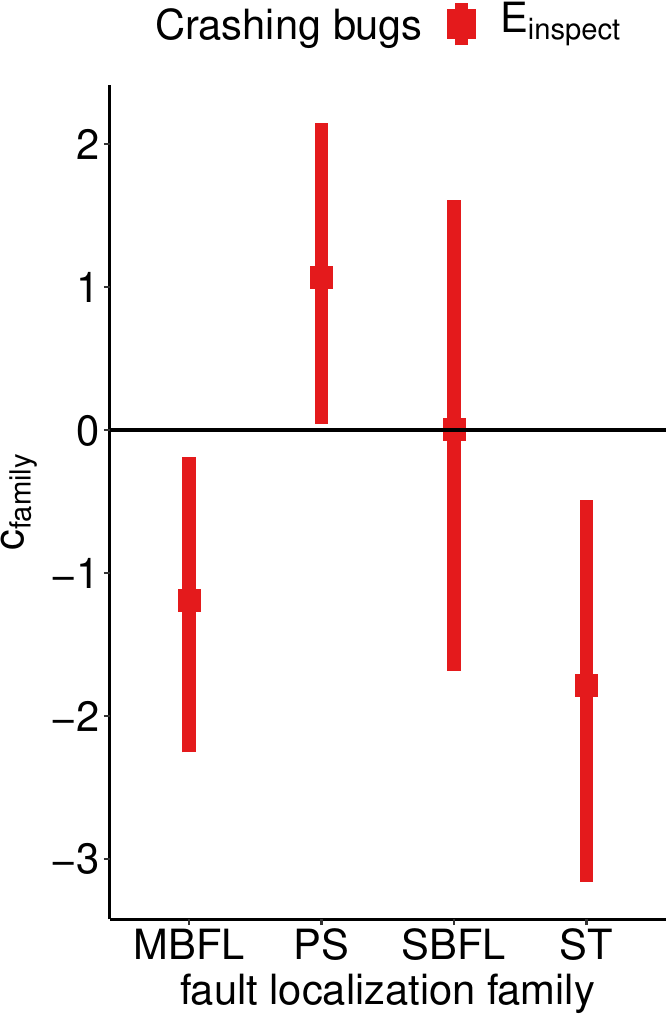}
    \caption{Estimates and 95\% probability intervals for the coefficients
      $c_{\var{family}}$ in model \eqref{eq:regmodel-kinds},
      for each FL family MBFL, PS, SBFL, and ST.}
    \label{fig:stats-crashing}
  \end{subfigure}
  \begin{subfigure}[t]{0.33\linewidth}
    \includegraphics[width=\textwidth]{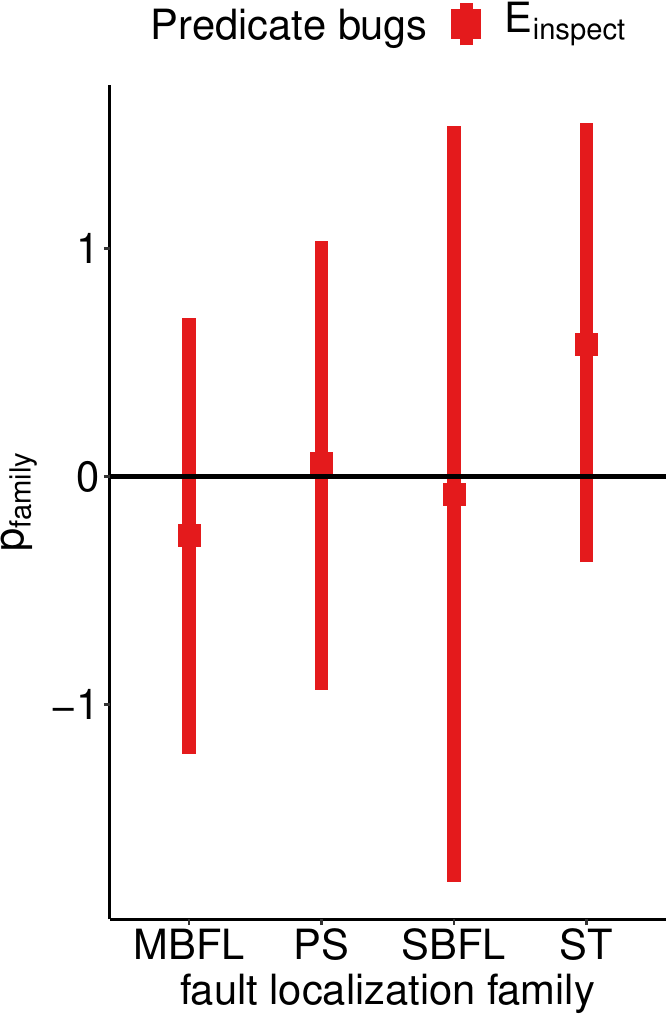}
    \caption{Estimates and 95\% probability intervals for the coefficients
      $p_{\var{family}}$ in model \eqref{eq:regmodel-kinds},
      for each FL family MBFL, PS, SBFL, and ST.}
    \label{fig:stats-predicate}
  \end{subfigure}
  \begin{subfigure}[t]{0.33\linewidth}
    \includegraphics[width=\textwidth]{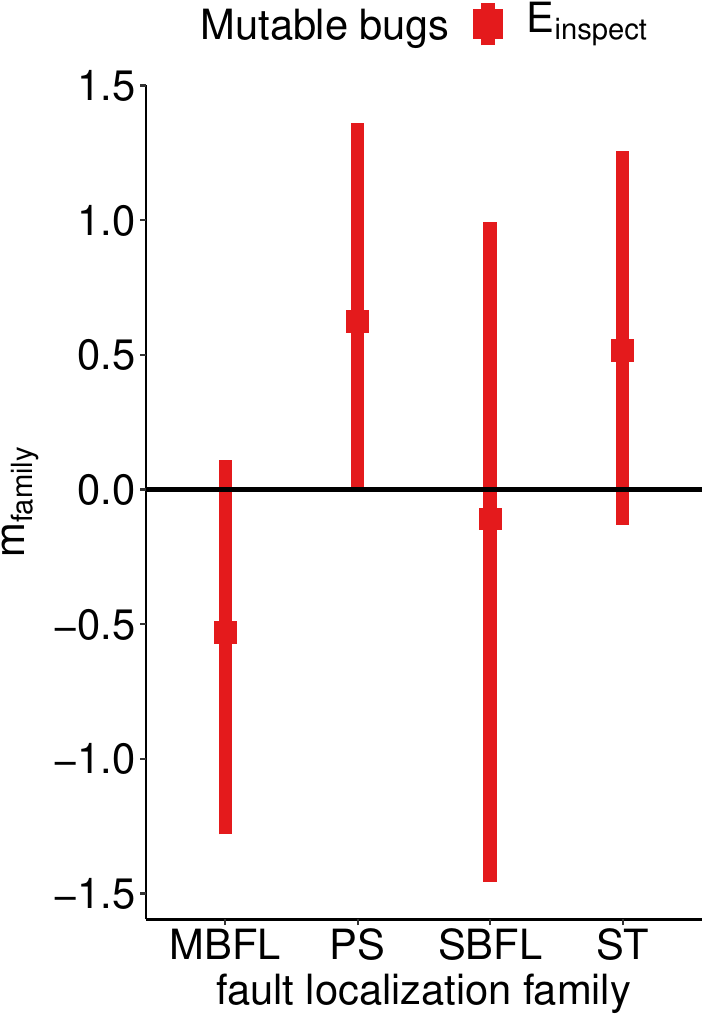}
    \caption{Estimates and 95\% probability intervals for the coefficients
      $m_{\var{family}}$ in model \eqref{eq:regmodel-kinds},
      for each FL family MBFL, PS, SBFL, and ST.}
    \label{fig:stats-mutable}
  \end{subfigure}
  \caption{Point estimates (boxes) and 95\% probability intervals (lines)
  for the regression coefficients of model~\eqref{eq:regmodel-kinds}. The scale of the vertical axes is over standard deviation log-units.}
  \label{fig:stats-bug-kinds}
\end{figure}

\subsection{RQ3. Kinds of Faults and Projects}
\label{sec:experimental-results:rq3}

\paragraph{Project category: effectiveness.}
\autoref{fig:stats-family-category}'s
intervals of coefficients $\alpha_{\var{category}}$
in model~\eqref{eq:regmodel}
indicate that fault localization tends to be more accurate on
projects in categories \dev and \web,
and less accurate on projects in categories \cli and \ds.

This finding is consistent with
the observations that data science programs, their bugs, and their fixes
are often different compared to traditional programs~\cite{Islam:2019, Islam:2020}.
For instance, bug~\#38 in project \project{keras}
is an example of what Islam et al. call ``structural data flow'' bugs~\cite{Islam:2019}:
its root cause is passing an incorrect input shape setting to
a neural network layer.
These characteristics also determine long spectra
(i.e., execution traces) that span several functions---which are required
to construct the various layer objects;
as a result, SBFL techniques struggle to effectively localize this bug.
Bugs~\#68 and \#137 in project \project{pandas}
are instead examples of API bugs,
whose root causes are incorrect import statements.
While such bugs may occur in any kind of project,
they are common in data science programs~\cite{Islam:2019}
due to their complex dependencies.
In Python, import statements are usually top-level declarations;
therefore, FL techniques that can only target locations inside functions
end up being ineffective at localizing these API bugs.
As yet another example,
the overall mutability of
bugs in \ds projects is 0.7\%,
whereas it is 1.3\% for bugs in other categories of projects.
This indicates that the standard mutation operators,
used by MBFL, are a poor fit for the kinds of bugs
that are most commonly found in data science projects.

\finding{Bugs in data science projects challenge fault localization's effectiveness (that is, they are harder to localize correctly) more than bugs in other categories of projects.}
\label{f:eff:ds-challenging}

The data in \autoref{table:effectiveness-family-statement-bug-type}'s
bottom section
confirm that SBFL remains the most effective FL family,
largely independent of the category of projects it analyzes.
MBFL ranks second for effectiveness in every project category;
it is not that far from SBFL for projects in categories \dev and \cli
(for example, MBFL and SBFL both localize
\n{stmt/sbfl/favg/cli/@1}\% of \cli bugs in the first position;
and both localize over
\n{stmt/mbfl/favg/dev/@10}\% of \dev bugs in the top-10 positions).
In contrast, SBFL's advantage over MBFL is more conspicuous
for projects in categories \ds and \web.
Given that bugs in categories \cli are generally harder to localize,
this suggests that the characteristics of bugs in these projects seem to
be a good fit for MBFL.
As we have seen in \autoref{sec:experimental-results:rq2},
MBFL is the slowest FL family by far;
since it reruns the available tests hundreds, or even thousands,
of times, projects with a large number of tests
are near impossible to analyze efficiently with MBFL.
As we'll discuss below,
MBFL is considerably faster on projects in category \cli
than on projects in other categories;
this is probably the main reason
why MBFL is also more effective on these projects:
it simply generates a more manageable number of mutants,
which sharpen the dynamic analysis.

\finding{SBFL remains the most effective standalone fault localization family on all categories of projects.}
\label{f:eff:sbfl-best-project}

\autoref{fig:pairwise-einspect}'s plots 
confirm some of these trends.
In most plots, we see that the points positioned
far apart from the diagonal line correspond to projects in the \cli and \ds
categories,
confirming that these ``harder'' bugs exacerbate the different effectiveness
of the various FL families.

\paragraph{Project category: efficiency.}
\autoref{fig:stats-family-category}'s
intervals of coefficients $\beta_{\var{category}}$
in model~\eqref{eq:regmodel}
indicate that fault localization tends to be more efficient
(i.e., faster)
on projects in category \cli,
and less efficient
(i.e., slower)
on projects in category \ds
($\beta_{\ds}$ barely touches zero).
In contrast, projects in categories \dev and \web
do not have a consistent association with faster or slower fault localization.
\autoref{tab:bugsinpy-selected-projects}
shows that projects in category \ds
have the largest number of tests by far
(mostly because of outlier project \project{pandas});
furthermore,
some of their tests involve training and testing different machine learning models,
or other kinds of time-consuming tasks.
Since FL invariably requires to run tests,
this explains why bugs in \ds projects
tend to take longer to localize.

\finding{Bugs in data science projects challenge fault localization's efficiency (that is, they take longer to localize)  more than bugs in other categories of projects.}
\label{f:time:ds-challenging}

The data in \autoref{table:efficiency-family-bug-type}'s
right-hand side
generally confirm the same rankings of efficiency among FL families,
largely regardless of what category of projects we consider:
ST is by far the most efficient, followed by SBFL,
and then---at a distance---PS and MBFL.
The difference of performance between SBFL and ST is
largest for projects in category \ds (\spellout{3} orders of magnitude),
large for projects in category \web (\spellout{2} orders of magniture),
and more moderate for projects in categories \cli and \dev (\spellout{1} order of magnitude).
PS is slower than MBFL only for projects in category \dev,
although their absolute difference of running times is not very big
(around $7.5$\%);
in contrast,
it is one order of magnitude faster for projects in categories \cli and \web.

\finding{The difference in efficiency between MBFL and SBFL is largest for data science projects.}
\label{f:time:mbfl-sbfl-ds}

In most of \autoref{fig:pairwise-time}'s plots,
we see that the points most frequently positioned
far apart from the diagonal line
correspond to projects in category \ds,
confirming that these bugs take longer to analyze
and aggravate performance differences among techniques.
In the scatterplot comparing MBFL to PS,
points corresponding to projects in categories \web and \cli
are mostly below the diagonal line,
which corroborates the advantage of PS over MBFL for bugs of projects
in these two categories.

\paragraph{Crashing bugs: effectiveness.}
According to \autoref{fig:stats-crashing},
both FL families ST and MBFL
are more effective on \emph{crashing} bugs than
on other kinds of bugs.
Still, their \emph{absolute}
effectiveness on crashing bugs remains limited
compared to SBFL's, as shown by the results in \autoref{table:effectiveness-family-statement-bug-type}'s
middle part;
for example, $@_{\textsc{crashing}}10\%$ is
\n{stmt/st/favg/crash/@10}| for ST,
\n{stmt/mbfl/favg/crash/@10}| for MBFL,
and \n{stmt/sbfl/favg/crash/@10}| for SBFL,
whereas ST localizes \n.{stmt/st/favg/crash/@1}!
(crashing) bugs in the top rank.
Remember that ST assigns that same
suspiciousness to all statements within the same function
(see~\autoref{sec:preliminaries:st});
thus, it cannot be as accurate as SBFL even on the minority of crashing bugs.

\finding{ST and MBFL are more effective on crashing bugs than on other kinds of bugs \\(but they remain overall less effective than SBFL even on crashing bugs).} \label{f:eff:st-mbfl-crashing}

On the other hand,
PS is \emph{less} effective on crashing bugs
than on other kinds of bugs;
in fact, it localizes \n.{stmt/ps/favg/crash/@10}! bugs among the top-10 ranks.
PS has a chance to work only
if it can find a so-called \emph{critical predicate}
(see \autoref{sec:preliminaries:ps});
only three
of the crashing bugs included critical predicates,
and hence PS was a bust.

\finding{PS is the least effective on crashing bugs.}
\label{f:eff:ps-crashing-worst}

\paragraph{Predicate-related bugs: effectiveness.}
\autoref{fig:stats-predicate}
says that no FL family achieves consistently better or worse effectiveness
on predicate-related bugs.
\autoref{table:effectiveness-family-statement-bug-type}
complements this observation;
the ranking of families by effectiveness is
different for predicate-related bugs
than it is for all bugs:
MBFL is about as effective as SBFL,
whereas PS is clearly more effective than ST.

\finding{On predicate-related bugs, MBFL is about as effective as SBFL, and PS is more effective than ST.}
\label{f:eff:each-tech-predicate}

This outcome is somewhat unexpected for PS:
predicate-related bugs are bugs
whose ground truth includes at least
a branching predicate (see \autoref{sec:design:faults-kinds}),
and yet PS is still clearly less effective than SBFL or MBFL.
Indeed, the presence of a faulty predicate is not sufficient
for PS to work:
the predicate must also be \emph{critical},
which means that flipping its value turns a failing test into a passing one.
When a program has no critical predicates, PS simply returns an empty list of locations.
In contrast, when a program has a critical predicate,
PS is highly effective:
$\textrm{PS}@_{\chi}1\% = \n[0]{critpred/PS/@1p}\%$,
$\textrm{PS}@_{\chi}3\% = \n[0]{critpred/PS/@3p}\%$,
and $\textrm{PS}@_{\chi}5\% = \n[0]{critpred/PS/@5p}\%$
for PS on the \n{critpred/number of subjects}
bugs $\chi$ with a critical predicate%
---even better than SBFL's
results for the same bugs
($\textrm{SBFL}@_{\chi}1\% = \n[0]{critpred/SBFL/@1p}\%$,
$\textrm{SBFL}@_{\chi}3\% = \n[0]{critpred/SBFL/@3p}\%$,
and $\textrm{SBFL}@_{\chi}5\% = \n[0]{critpred/SBFL/@5p}\%$).
In all, PS is a highly specialized FL technique,
which works quite well for a narrow category of bugs,
but is inapplicable in many other cases.

\finding{On the few bugs that it can analyze successfully, PS is the most effective standalone fault localization technique.}
\label{f:eff:ps-critical-best}

\paragraph{Mutable bugs: effectiveness.}
According to \autoref{fig:stats-mutable},
FL family MBFL
tends to be more effective
on \emph{mutable} bugs than
on other kinds of bugs:
$m_{\textrm{MBFL}}$ 95\%
probability interval is mostly below zero
(and the \n[0]{m6/mutability/mbfl/significant}[100]\%
probability interval would be entirely below zero).
Furthermore,
\autoref{table:effectiveness-family-statement-bug-type}
shows that MBFL is the most effective technique 
on mutable bugs, where it tends to outperform even SBFL.
Intuitively, a bug is mutable if
the syntactic mutation operators used for MBFL
``match'' the fault in a way that it affects program behavior.
Thus, the capabilities of MBFL ultimately depend
on the nature of faults it analyzes and on the selection of mutation operators
it employs.

\finding{MBFL is more effective on mutable bugs than on other kinds of bugs;
  in fact,
  it is the most effective standalone fault localization family on these bugs.}
\label{f:eff:mbfl-mutable-best}

\autoref{fig:stats-mutable} also suggests
that PS and ST
are less effective on mutable bugs than
on other kinds of bugs.
Possibly, this is because mutable bugs
tend to be more complex, ``semantic'' bugs,
whereas ST works well only for ``simple'' crashing bugs,
and PS is highly specialized to work on a narrow group of bugs.

\finding{PS and ST are less effective on mutable bugs than on other kinds of bugs.}
\label{f:eff:ps-st-mutable}

\paragraph{Bug kind: efficiency.}
\autoref{table:efficiency-family-bug-type}
does not suggest any consistent changes in the efficiency of
FL families when they work on crashing, predicate-related, or mutable bugs%
---as opposed to all bugs.
In other words, for every kind of bugs:
ST is orders of magnitude faster than SBFL,
which is one order of magnitude faster than PS,
which is 14--37\% faster than MBFL.
As discussed above, the kind of information that a FL technique
collects is the main determinant of its overall efficiency;
in contrast, different kinds of bugs do not seem to
have any significant impact.

\finding{The relative efficiency of each fault localization family does not depend on the kinds of bugs that are analyzed.}
\label{f:time:family-kind-dependency}

\begin{table}[!bt]
\small
\centering
\begin{tabular}{l l r rrrr r r r}
\toprule

\multicolumn{2}{c}{\textsc{technique} $L$} &
\multicolumn{1}{c}{\einspect*[B][L]} &
\multicolumn{1}{c}{$L@_B1$\%} &
\multicolumn{1}{c}{$L@_B3$\%} &
\multicolumn{1}{c}{$L@_B5$\%} &
\multicolumn{1}{c}{$L@_B10$\%} &
\multicolumn{1}{c}{$\exam_B(L)$} &
\multicolumn{1}{c}{\outputlen{L}[B]} &
\multicolumn{1}{c}{$\timecost_B(L)$} \\

\midrule

\multirow{2}{*}{\average} &
$\average_A$ &
\best{\n{stmt/avfl/alfa/all/einspect}} &
\n.{stmt/avfl/alfa/all/@1} &
\n.{stmt/avfl/alfa/all/@3} &
\n.{stmt/avfl/alfa/all/@5} &
\n.{stmt/avfl/alfa/all/@10} &
\best{\n[\examdec]{stmt/avfl/alfa/all/javaexam}} &
\best{\n[\cddec]{stmt/comb/alfa/all/outputlength}} &
\n{stmt/comb/alfa/all/time} \\

 & 
$\average_S$ &
 \n{stmt/avfl/sbst/all/einspect} &
 \n.{stmt/avfl/sbst/all/@1} &
 \n.{stmt/avfl/sbst/all/@3} &
 \n.{stmt/avfl/sbst/all/@5} &
 \n.{stmt/avfl/sbst/all/@10} &
 \n[\examdec]{stmt/avfl/sbst/all/javaexam} &
 \best{\n[\cddec]{stmt/comb/sbst/all/outputlength}} &
 \best{\n.{stmt/comb/sbst/all/time}} \\
  
\cmidrule{2-10}

\multirow{2}{*}{\combine} &
$\combine_A$ &
\n{stmt/comb/alfa/all/einspect} &
\best{\n.{stmt/comb/alfa/all/@1}} &
\best{\n.{stmt/comb/alfa/all/@3}} &
\best{\n.{stmt/comb/alfa/all/@5}} &
\best{\n.{stmt/comb/alfa/all/@10}} &
\best{\n[\examdec]{stmt/comb/alfa/all/javaexam}} &
\best{\n[\cddec]{stmt/comb/alfa/all/outputlength}} & 
\n{stmt/comb/alfa/all/time} \\

 & 
$\combine_S$ &
 \n{stmt/comb/sbst/all/einspect} &
 \n.{stmt/comb/sbst/all/@1} &
 \n.{stmt/comb/sbst/all/@3} &
 \n.{stmt/comb/sbst/all/@5} &
 \n.{stmt/comb/sbst/all/@10} &
 \n[\examdec]{stmt/comb/sbst/all/javaexam} &
 \best{\n[\cddec]{stmt/comb/sbst/all/outputlength}} &
 \best{\n.{stmt/comb/sbst/all/time}} \\

\bottomrule

\end{tabular}
\caption{%
  Effectiveness and efficiency of fault localization techniques
  \average and \combine 
  at the statement-level granularity
  on all \n{number of subjects} selected bugs $B$.
  Each row reports a \textsc{technique} $L$'s
  average generalized \einspectso rank \einspect*[B][L];
  the percentage of all bugs it localized
  within the top-$1$, top-$3$, top-$5$, and top-$10$ positions of its output
  ($L@_B1$\%, $L@_B3$\%, $L@_B5$\%, and $L@_B10$\%);
  its average exam score $\exam_B(L)$;
  its average suspicious locations length $\outputlen{L}[B]$;
  and its average per-bug wall-clock running time $T_B(L)$ in seconds.
  The four rows correspond to two variants $\average_A$ and $\combine_A$
  that combine the information of all FL techniques but Tarantula,
  and two variants $\average_S$ and $\combine_S$
  that combine the information of SBFL and ST techniques but Tarantula.
  \best{Highlighted} numbers denote the best technique according to each metric.
}
\label{table:effectiveness-combined-statement}
\end{table}

\subsection{RQ4. Combining Techniques}
\label{sec:experimental-results:rq4}

\paragraph{Effectiveness.}
\autoref{table:effectiveness-combined-statement}
clearly indicates that the combined FL techniques \average and \combine
achieve high ef\-fec\-tive\-ness---especially according to the fundamental $@n\%$ metrics.
$\combine_A$ and $\average_A$,
combining the information from all other FL techniques,
beat every other technique.
For example, $\average_A$ localizes in the top position
18\% of all bugs, $\combine_A$ localizes 20\% of all bugs,
whereas the next-best technique is SBFL,
which localizes 12\% of all bugs (\autoref{table:effectiveness-family-technique-statement}).
$\combine_S$ and $\average_S$,
combining the information from only SBFL and ST techniques,
do at least as well as every other standalone technique.

\finding{Combined fault localization techniques $\average_A$ and $\combine_A$,
which combine all baseline techniques, achieve better effectiveness than any other techniques.}
\label{f:eff:avg-comb-all}

While $\combine_A$ is strictly more effective than $\average_A$,
their difference is usually modest (at most \spellout{3} percentage points).
Similarly, the difference between $\combine_S$, $\average_S$, and SBFL
is generally limited;
however, SBFL tends to be less effective than $\average_S$,
whereas $\combine_S$ is never strictly more effective than $\average_S$. 
In all, \average is a simpler approach to combining techniques than \combine,
but both are quite successful at boosting FL effectiveness.

\finding{Fault localization families ordered by effectiveness: \\
  $\combine_A$ $\geq$ $\average_A$ $>$ $\combine_S$ $\simeq$ $\average_S$ $>$ SBFL $>$ MBFL $\gg$ PS $\simeq$ ST, \\
  where $>$ means better,
  $\geq$ better or as good, $\gg$ much better, and $\simeq$ about as good.}
\label{f:eff:family-comb-order}

The suspicious location length is the very same for \average and \combine,
and higher than for every other technique.
This is simply because all variants of \average and \combine
consider a location as suspicious if and only if
any of the techniques they combine considers it so.
Therefore, they end up with long location lists---at least as long
as any combined technique's.

\paragraph{Efficiency.}
The running time of \average and \combine
is essentially just the sum of
running times of the FL families they combine,
because merging the output list of locations 
and training \combine's machine learning model
take negligible time.
This makes
$\average_A$ and $\combine_A$
the least efficient FL techniques in our experiments;
and $\average_S$ and $\combine_S$
barely slower than SBFL.

\finding{Combined fault localization techniques $\average_A$ and $\combine_A$,
which combine all baseline techniques, achieve worse efficiency than any other techniques.}
\label{f:time:avg-comb-worst}

Combining these results with those about effectiveness,
we conclude that $\average_A$ and $\combine_A$
exclusively favor effectiveness;
whereas $\average_S$ and $\combine_S$ 
promise a modest improvement in effectiveness
in exchange for a modest performance loss.

\finding{Fault localization families ordered by efficiency: \\
  ST $\gg$ SBFL $\geq$ $\average_S$ $\simeq$ $\combine_S$ $\gg$ PS $>$ MBFL $>$ $\average_A$ $\simeq$ $\combine_A$, \\
  where $>$ means faster, $\geq$ faster or as fast, $\gg$ much faster,
  and $\simeq$ about as fast.}
\label{f:time:family-comb-order}

\subsection{RQ5. Granularity}
\label{sec:experimental-results:rq5}

\begin{table}[tb]

\small
\centering
\setlength{\tabcolsep}{5pt}
\begin{tabular}{lr rr rr rr rr rr rr rr}

\toprule
\multicolumn{1}{c}{\textsc{family}} & 
\multicolumn{1}{c}{\textsc{technique} $L$} & 
\multicolumn{2}{c}{\einspect*[B][L]} & 
\multicolumn{2}{c}{$L@_B1$\%} & 
\multicolumn{2}{c}{$L@_B3$\%} & 
\multicolumn{2}{c}{$L@_B5$\%} & 
\multicolumn{2}{c}{$L@_B10$\%} & 
\multicolumn{2}{c}{$\exam_B(L)$} &
\multicolumn{2}{c}{\outputlen{L}[B]} \\

\cmidrule(lr){3-4} 
\cmidrule(lr){5-6} 
\cmidrule(lr){7-8} 
\cmidrule(lr){9-10} 
\cmidrule(lr){11-12} 
\cmidrule(lr){13-14} 
\cmidrule(lr){15-16} 

 &  
 & 
 \multicolumn{1}{c}{\textsc{f}} & 
 \multicolumn{1}{c}{\textsc{t}} & 
 \multicolumn{1}{c}{\textsc{f}} & 
 \multicolumn{1}{c}{\textsc{t}} & 
 \multicolumn{1}{c}{\textsc{f}} & 
 \multicolumn{1}{c}{\textsc{t}} & 
 \multicolumn{1}{c}{\textsc{f}} & 
 \multicolumn{1}{c}{\textsc{t}} & 
 \multicolumn{1}{c}{\textsc{f}} & 
 \multicolumn{1}{c}{\textsc{t}} & 
 \multicolumn{1}{c}{\textsc{f}} & 
 \multicolumn{1}{c}{\textsc{t}} &
 \multicolumn{1}{c}{\textsc{f}} & 
 \multicolumn{1}{c}{\textsc{t}} \\

\midrule

\multirow{2}{*}{} & 
$\average_A$ & 
\multirow{2}{*}{\best{\n[0]{func/alfafl/favg/all/einspect}}} & 
\best{\n.{func/avfl/alfa/all/einspect}} & 
\multirow{2}{*}{\best{\n[0]{func/alfafl/favg/all/@1}}} & 
\best{\n.{func/avfl/alfa/all/@1}} & 
\multirow{2}{*}{\best{\n[0]{func/alfafl/favg/all/@3}}} & 
\best{\n.{func/avfl/alfa/all/@3}} & 
\multirow{2}{*}{\best{\n[0]{func/alfafl/favg/all/@5}}} & 
\n.{func/avfl/alfa/all/@5} & 
\multirow{2}{*}{\best{\n[0]{func/alfafl/favg/all/@10}}} & 
\best{\n.{func/avfl/alfa/all/@10}} & 
\multirow{2}{*}{\n[\examdec]{func/alfafl/favg/all/javaexam}} & 
\n[\examdec]{func/avfl/alfa/all/javaexam} &
\multirow{2}{*}{\n[\cddec]{func/alfafl/favg/all/outputlength}} & 
\n[\cddec]{func/avfl/alfa/all/outputlength} \\

 & 
 $\combine_A$ &  
 & 
 \best{\n.{func/comb/alfa/all/einspect}} &  
 & 
 \best{\n.{func/comb/alfa/all/@1}} &  
 & 
 \n.{func/comb/alfa/all/@3} &  
 & 
 \best{\n.{func/comb/alfa/all/@5}} &  
 & 
 \best{\n.{func/comb/alfa/all/@10}} &  
 & 
 \n[\examdec]{func/comb/alfa/all/javaexam} &
 &
 \n[\cddec]{func/comb/alfa/all/outputlength} \\

\cmidrule{2-16}

\multirow{2}{*}{} & 
$\average_S$ & 
\multirow{2}{*}{\n[0]{func/sbstfl/favg/all/einspect}} & 
\best{\n.{func/avfl/sbst/all/einspect}} & 
\multirow{2}{*}{\n[0]{func/sbstfl/favg/all/@1}} & 
\n.{func/avfl/sbst/all/@1} & 
\multirow{2}{*}{\n[0]{func/sbstfl/favg/all/@3}} & 
\n.{func/avfl/sbst/all/@3} & 
\multirow{2}{*}{\n[0]{func/sbstfl/favg/all/@5}} & 
\n.{func/avfl/sbst/all/@5} & 
\multirow{2}{*}{\n[0]{func/sbstfl/favg/all/@10}} & 
\n.{func/avfl/sbst/all/@10} & 
\multirow{2}{*}{\n[\examdec]{func/sbstfl/favg/all/javaexam}} & 
\n[\examdec]{func/avfl/sbst/all/javaexam} &
\multirow{2}{*}{\n[\cddec]{func/sbstfl/favg/all/outputlength}} & 
\n[\cddec]{func/avfl/sbst/all/outputlength} \\

 & 
 $\combine_S$ &  
 & 
 \n.{func/comb/sbst/all/einspect} &  
 & 
 \n.{func/comb/sbst/all/@1} &  
 & 
 \n.{func/comb/sbst/all/@3} &  
 & 
 \n.{func/comb/sbst/all/@5} &  
 & 
 \n.{func/comb/sbst/all/@10} &  
 & 
 \n[\examdec]{func/comb/sbst/all/javaexam} &
 &
 \n[\cddec]{func/comb/sbst/all/outputlength} \\

 \cmidrule{2-16}

\multirow{2}{*}{MBFL} & 
Metallaxis & 
\multirow{2}{*}{\n.{func/mbfl/favg/all/einspect}} & 
\n.{func/mbfl/metallaxis/all/einspect} & 
\multirow{2}{*}{\n.{func/mbfl/favg/all/@1}} & 
\n.{func/mbfl/metallaxis/all/@1} & 
\multirow{2}{*}{\n.{func/mbfl/favg/all/@3}} & 
\n.{func/mbfl/metallaxis/all/@3} & 
\multirow{2}{*}{\n.{func/mbfl/favg/all/@5}} & 
\n.{func/mbfl/metallaxis/all/@5} & 
\multirow{2}{*}{\n.{func/mbfl/favg/all/@10}} & 
\n.{func/mbfl/metallaxis/all/@10} & 
\multirow{2}{*}{\n[\examdec]{func/mbfl/favg/all/javaexam}} & 
\n[\examdec]{func/mbfl/metallaxis/all/javaexam} &
\multirow{2}{*}{\n[\cddec]{func/mbfl/favg/all/outputlength}} & 
\n[\cddec]{func/mbfl/metallaxis/all/outputlength} \\

 & 
 \muse &  
 & 
 \n.{func/mbfl/muse/all/einspect} &  
 & 
 \n.{func/mbfl/muse/all/@1} &  
 & 
 \n.{func/mbfl/muse/all/@3} &  
 & 
 \n.{func/mbfl/muse/all/@5} &  
 & 
 \n.{func/mbfl/muse/all/@10} &  
 & 
 \n[\examdec]{func/mbfl/muse/all/javaexam} &
 &
 \n[\cddec]{func/mbfl/muse/all/outputlength} \\
 
\cmidrule{2-16}

PS 
&  
&  
\n.{func/ps/favg/all/einspect} &
\n.{func/ps/favg/all/einspect} &  
\n.{func/ps/favg/all/@1} &  
\n.{func/ps/favg/all/@1} &  
\n.{func/ps/favg/all/@3} &  
\n.{func/ps/favg/all/@3} &  
\n.{func/ps/favg/all/@5} &  
\n.{func/ps/favg/all/@5} &  
\n.{func/ps/favg/all/@10} & 
\n.{func/ps/favg/all/@10} &  
\best{\n[\examdec]{func/ps/favg/all/javaexam}} &
\best{\n[\examdec]{func/ps/favg/all/javaexam}} &
\best{\n[\cddec]{func/ps/favg/all/outputlength}} &
\best{\n[\cddec]{func/ps/favg/all/outputlength}} \\

\cmidrule{2-16}

\multirow{3}{*}{SBFL} 

 & 
 DStar & 
 \multirow{3}{*}{\n.{func/sbfl/favg/all/einspect}} & 
 \n.{func/sbfl/dstar/all/einspect} &  
 \multirow{3}{*}{\n.{func/sbfl/favg/all/@1}} & 
 \n.{func/sbfl/dstar/all/@1} &  
 \multirow{3}{*}{\n.{func/sbfl/favg/all/@3}} & 
 \n.{func/sbfl/dstar/all/@3} &  
 \multirow{3}{*}{\n.{func/sbfl/favg/all/@5}} & 
 \n.{func/sbfl/dstar/all/@5} &  
 \multirow{3}{*}{\n.{func/sbfl/favg/all/@10}} & 
 \n.{func/sbfl/dstar/all/@10} &  
 \multirow{3}{*}{\n[\examdec]{func/sbfl/favg/all/javaexam}} & 
 \n[\examdec]{func/sbfl/dstar/all/javaexam} &
 \multirow{3}{*}{\n[\cddec]{func/sbfl/favg/all/outputlength}} &
 \n[\cddec]{func/sbfl/dstar/all/outputlength} \\

 & 
 Ochiai & 
 & 
 \n.{func/sbfl/ochiai/all/einspect} &
 & 
 \n.{func/sbfl/ochiai/all/@1} &  
 & 
 \n.{func/sbfl/ochiai/all/@3} &  
 & 
 \n.{func/sbfl/ochiai/all/@5} &  
 & 
 \n.{func/sbfl/ochiai/all/@10} &  
 & 
 \n[\examdec]{func/sbfl/ochiai/all/javaexam} &
 &
 \n[\cddec]{func/sbfl/ochiai/all/outputlength} \\

&
Tarantula & 
 & 
\n.{func/sbfl/tarantula/all/einspect} & 
 & 
\n.{func/sbfl/tarantula/all/@1} & 
 & 
\n.{func/sbfl/tarantula/all/@3} & 
 & 
\n.{func/sbfl/tarantula/all/@5} & 
 & 
\n.{func/sbfl/tarantula/all/@10} & 
 & 
\n[\examdec]{func/sbfl/tarantula/all/javaexam} &
 & 
\n[\cddec]{func/sbfl/tarantula/all/outputlength} \\
 
\cmidrule{2-16}

ST 
&  
&  
\n.{func/st/favg/all/einspect} &  
\n.{func/st/favg/all/einspect} &  
\n.{func/st/favg/all/@1} &  
\n.{func/st/favg/all/@1} &  
\n.{func/st/favg/all/@3} &  
\n.{func/st/favg/all/@3} &  
\n.{func/st/favg/all/@5} &  
\n.{func/st/favg/all/@5} &  
\n.{func/st/favg/all/@10} & 
\n.{func/st/favg/all/@10} &  
\n[\examdec]{func/st/favg/all/javaexam} &
\n[\examdec]{func/st/favg/all/javaexam} &
\n[\cddec]{func/st/favg/all/outputlength} &
\n[\cddec]{func/st/favg/all/outputlength} \\

\bottomrule

\end{tabular}
\caption{Effectiveness of fault localization techniques
  at the \emph{function}-level granularity on all \n{number of subjects}
  selected bugs $B$.
  The table reports the same metrics as \autoref{table:effectiveness-family-technique-statement}
  and \autoref{table:effectiveness-combined-statement}
  but targeting functions as suspicious entities.
  \best{Highlighted} numbers denote the best technique according
  to each metric.
}
\label{table:effectiveness-family-technique-function}
\end{table}

\begin{table}[tb]

\small
\centering
\setlength{\tabcolsep}{5pt}
\begin{tabular}{lr rr rr rr rr rr rr rr}

\toprule
\multicolumn{1}{c}{\textsc{family}} & 
\multicolumn{1}{c}{\textsc{technique} $L$} & 
\multicolumn{2}{c}{\einspect*[B][L]} & 
\multicolumn{2}{c}{$L@_B1$\%} & 
\multicolumn{2}{c}{$L@_B3$\%} & 
\multicolumn{2}{c}{$L@_B5$\%} & 
\multicolumn{2}{c}{$L@_B10$\%} & 
\multicolumn{2}{c}{$\exam_B(L)$} & 
\multicolumn{2}{c}{\outputlen{L}[B]} \\

\cmidrule(lr){3-4} 
\cmidrule(lr){5-6} 
\cmidrule(lr){7-8} 
\cmidrule(lr){9-10} 
\cmidrule(lr){11-12} 
\cmidrule(lr){13-14} 
\cmidrule(lr){15-16} 

 &  
 & 
 \multicolumn{1}{c}{\textsc{f}} & 
 \multicolumn{1}{c}{\textsc{t}} & 
 \multicolumn{1}{c}{\textsc{f}} & 
 \multicolumn{1}{c}{\textsc{t}} & 
 \multicolumn{1}{c}{\textsc{f}} & 
 \multicolumn{1}{c}{\textsc{t}} & 
 \multicolumn{1}{c}{\textsc{f}} & 
 \multicolumn{1}{c}{\textsc{t}} & 
 \multicolumn{1}{c}{\textsc{f}} & 
 \multicolumn{1}{c}{\textsc{t}} & 
 \multicolumn{1}{c}{\textsc{f}} & 
 \multicolumn{1}{c}{\textsc{t}} &
 \multicolumn{1}{c}{\textsc{f}} & 
 \multicolumn{1}{c}{\textsc{t}} \\
 
\midrule

\multirow{2}{*}{} & 
$\average_A$ & 
\multirow{2}{*}{\best{\n[0]{mod/alfafl/favg/all/einspect}}} & 
\best{\n.{mod/avfl/alfa/all/einspect}} & 
\multirow{2}{*}{\best{\n[0]{mod/alfafl/favg/all/@1}}} & 
\best{\n.{mod/avfl/alfa/all/@1}} & 
\multirow{2}{*}{\best{\n[0]{mod/alfafl/favg/all/@3}}} & 
\best{\n.{mod/avfl/alfa/all/@3}} & 
\multirow{2}{*}{\best{\n[0]{mod/alfafl/favg/all/@5}}} & 
\best{\n.{mod/avfl/alfa/all/@5}} & 
\multirow{2}{*}{\best{\n[0]{mod/alfafl/favg/all/@10}}} & 
\best{\n.{mod/avfl/alfa/all/@10}} & 
\multirow{2}{*}{\n[\examdec]{mod/alfafl/favg/all/javaexam}} & 
\n[\examdec]{mod/avfl/alfa/all/javaexam} & 
\multirow{2}{*}{\n[\cddec]{mod/alfafl/favg/all/outputlength}} & 
\n[\cddec]{mod/avfl/alfa/all/outputlength} \\

 & 
 $\combine_A$ &  
 & 
 \best{\n.{mod/comb/alfa/all/einspect}} &  
 & 
 \best{\n.{mod/comb/alfa/all/@1}} &  
 & 
 \best{\n.{mod/comb/alfa/all/@3}} &  
 & 
 \best{\n.{mod/comb/alfa/all/@5}} &  
 & 
 \best{\n.{mod/comb/alfa/all/@10}} &  
 & 
 \n[\examdec]{mod/comb/alfa/all/javaexam} &
 &
 \n[\cddec]{mod/comb/alfa/all/outputlength} \\

 \cmidrule{2-16}

 \multirow{2}{*}{} & 
$\average_S$ & 
\multirow{2}{*}{\best{\n[0]{mod/sbstfl/favg/all/einspect}}} & 
\best{\n.{mod/avfl/sbst/all/einspect}} & 
\multirow{2}{*}{\n[0]{mod/sbstfl/favg/all/@1}} & 
\n.{mod/avfl/sbst/all/@1} & 
\multirow{2}{*}{\n[0]{mod/sbstfl/favg/all/@3}} & 
\n.{mod/avfl/sbst/all/@3} & 
\multirow{2}{*}{\best{\n[0]{mod/sbstfl/favg/all/@5}}} & 
\best{\n.{mod/avfl/sbst/all/@5}} & 
\multirow{2}{*}{\n[0]{mod/sbstfl/favg/all/@10}} & 
\n.{mod/avfl/sbst/all/@10} & 
\multirow{2}{*}{\n[\examdec]{mod/sbstfl/favg/all/javaexam}} & 
\n[\examdec]{mod/avfl/sbst/all/javaexam} &
\multirow{2}{*}{\n[\cddec]{mod/sbstfl/favg/all/outputlength}} & 
\n[\cddec]{mod/avfl/sbst/all/outputlength} \\

 & 
 $\combine_S$ &  
 & 
 \best{\n.{mod/comb/sbst/all/einspect}} &  
 & 
 \n.{mod/comb/sbst/all/@1} &  
 & 
 \n.{mod/comb/sbst/all/@3} &  
 & 
 \best{\n.{mod/comb/sbst/all/@5}} &  
 & 
 \n.{mod/comb/sbst/all/@10} &  
 & 
 \n[\examdec]{mod/comb/sbst/all/javaexam} &
 &
 \n[\cddec]{mod/comb/sbst/all/outputlength} \\

\cmidrule{2-16} 

\multirow{2}{*}{MBFL} & 
Metallaxis & 
\multirow{2}{*}{\n.{mod/mbfl/favg/all/einspect}} & 
\n.{mod/mbfl/metallaxis/all/einspect} & 
\multirow{2}{*}{\n.{mod/mbfl/favg/all/@1}} & 
\n.{mod/mbfl/metallaxis/all/@1} & 
\multirow{2}{*}{\n.{mod/mbfl/favg/all/@3}} & 
\n.{mod/mbfl/metallaxis/all/@3} & 
\multirow{2}{*}{\n.{mod/mbfl/favg/all/@5}} & 
\n.{mod/mbfl/metallaxis/all/@5} & 
\multirow{2}{*}{\n.{mod/mbfl/favg/all/@10}} & 
\n.{mod/mbfl/metallaxis/all/@10} & 
\multirow{2}{*}{\n[\examdec]{mod/mbfl/favg/all/javaexam}} & 
\n[\examdec]{mod/mbfl/metallaxis/all/javaexam} &
\multirow{2}{*}{\n[\cddec]{mod/mbfl/favg/all/outputlength}} & 
\n[\cddec]{mod/mbfl/metallaxis/all/outputlength} \\

 & 
 \muse &  
 & 
 \n.{mod/mbfl/muse/all/einspect} &  
 & 
 \n.{mod/mbfl/muse/all/@1} &  
 & 
 \n.{mod/mbfl/muse/all/@3} &  
 & 
 \n.{mod/mbfl/muse/all/@5} &  
 & 
 \n.{mod/mbfl/muse/all/@10} &  
 & 
 \n[\examdec]{mod/mbfl/muse/all/javaexam} &
 &
 \n[\cddec]{mod/mbfl/muse/all/outputlength} \\
 
\cmidrule{2-16}

PS 
&  
&  
\n.{mod/ps/favg/all/einspect} & 
\n.{mod/ps/favg/all/einspect} &  
\n.{mod/ps/favg/all/@1} & 
\n.{mod/ps/favg/all/@1} &  
\n.{mod/ps/favg/all/@3} & 
\n.{mod/ps/favg/all/@3} &  
\n.{mod/ps/favg/all/@5} & 
\n.{mod/ps/favg/all/@5} &  
\n.{mod/ps/favg/all/@10} & 
\n.{mod/ps/favg/all/@10} &  
\best{\n[\examdec]{mod/ps/favg/all/javaexam}} & 
\best{\n[\examdec]{mod/ps/favg/all/javaexam}} &
\best{\n[\cddec]{mod/ps/favg/all/outputlength}} &
\best{\n[\cddec]{mod/ps/favg/all/outputlength}} \\

\cmidrule{2-16}

\multirow{3}{*}{SBFL} &
 DStar & 
\multirow{3}{*}{\best{\n.{mod/sbfl/favg/all/einspect}}} &
 \best{\n.{mod/sbfl/dstar/all/einspect}} &  
\multirow{3}{*}{\n.{mod/sbfl/favg/all/@1}} & 
 \n.{mod/sbfl/dstar/all/@1} &  
\multirow{3}{*}{\n.{mod/sbfl/favg/all/@3}} & 
 \n.{mod/sbfl/dstar/all/@3} &  
\multirow{3}{*}{\n.{mod/sbfl/favg/all/@5}} & 
 \best{\n.{mod/sbfl/dstar/all/@5}} &  
\multirow{3}{*}{\n.{mod/sbfl/favg/all/@10}} &
 \n.{mod/sbfl/dstar/all/@10} &  
\multirow{3}{*}{\n[\examdec]{mod/sbfl/favg/all/javaexam}} & 
 \n[\examdec]{mod/sbfl/dstar/all/javaexam} &
\multirow{3}{*}{\n[\cddec]{mod/sbfl/favg/all/outputlength}} &
 \n[\cddec]{mod/sbfl/dstar/all/outputlength} \\

 & 
 Ochiai & 
 & 
 \best{\n.{mod/sbfl/ochiai/all/einspect}} &
 & 
 \n.{mod/sbfl/ochiai/all/@1} &  
 & 
 \n.{mod/sbfl/ochiai/all/@3} &  
 & 
 \best{\n.{mod/sbfl/ochiai/all/@5}} &  
 & 
 \n.{mod/sbfl/ochiai/all/@10} &  
 & 
 \n[\examdec]{mod/sbfl/ochiai/all/javaexam} &
 &
 \n[\cddec]{mod/sbfl/ochiai/all/outputlength} \\

&
Tarantula & 
 & 
\best{\n.{mod/sbfl/tarantula/all/einspect}} & 
 & 
\n.{mod/sbfl/tarantula/all/@1} & 
 & 
\n.{mod/sbfl/tarantula/all/@3} & 
 & 
\n.{mod/sbfl/tarantula/all/@5} & 
 & 
\n.{mod/sbfl/tarantula/all/@10} & 
 & 
\n[\examdec]{mod/sbfl/tarantula/all/javaexam} &
 & 
\n[\cddec]{mod/sbfl/tarantula/all/outputlength} \\

\cmidrule{2-16}

ST 
&  
&  
\n.{mod/st/favg/all/einspect} &  
\n.{mod/st/favg/all/einspect} &  
\n.{mod/st/favg/all/@1} &  
\n.{mod/st/favg/all/@1} &  
\n.{mod/st/favg/all/@3} &  
\n.{mod/st/favg/all/@3} &  
\n.{mod/st/favg/all/@5} &  
\n.{mod/st/favg/all/@5} &  
\n.{mod/st/favg/all/@10} & 
\n.{mod/st/favg/all/@10} &  
\n[\examdec]{mod/st/favg/all/javaexam} &
\n[\examdec]{mod/st/favg/all/javaexam} &
\n[\cddec]{mod/st/favg/all/outputlength} &
\n[\cddec]{mod/st/favg/all/outputlength} \\

\bottomrule

\end{tabular}
\caption{Effectiveness of fault localization techniques
  at the \emph{module}-level granularity on all \n{number of subjects}
  selected bugs $B$.
  The table reports the same metrics as \autoref{table:effectiveness-family-technique-statement}
  and \autoref{table:effectiveness-combined-statement}
  but targeting modules (files in Python) as suspicious entities.
  \best{Highlighted} numbers denote the best technique according
  to each metric.
}
\label{table:effectiveness-family-technique-module}
\end{table}

\paragraph{Function-level granularity.}
\autoref{table:effectiveness-family-technique-function}'s
data about function-level effectiveness of the various FL techniques and families
lead to very similar high-level conclusions
as for statement-level effectiveness:
combination techniques 
$\combine_A$ and $\average_A$
achieves the best effectiveness,
followed by
$\combine_S$ and $\average_S$,
then SBFL, and finally MBFL;
differences among techniques in the same family are modest (often negligible).

ST is the only technique whose relative effectiveness
changes considerably from statement-level to function-level:
ST is the least effective at the level of statements,
but becomes considerably better than PS at the level of functions.
This change is no surprise,
as ST is precisely geared towards localizing \emph{functions}
responsible for crashes%
---and cannot distinguish among statements belonging to the same function.
ST's overall effectiveness remains limited, since the technique
is simple and can only work on crashing bugs.

\paragraph{Module-level granularity.}
\autoref{table:effectiveness-family-technique-module}
leads to the same conclusions for module-level granularity:
the relative effectiveness of the various techniques
is very similar as for statement-level granularity,
except that ST gains effectiveness simply because it is designed
for coarser granularities.

\finding{ST is more effective than PS both at the function-level and module-level granularity;
however, it remains considerably less effective than other fault localization techniques even at these coarser granularities.}
\label{f:eff:st-ps-granularity}

\paragraph{Comparisons between granularities.}
It is apparent that fault localization's absolute effectiveness
strictly \emph{increases} as we target coarser granularities%
---from statements, to functions, to modules.
This happens simply because the number of entities at a coarser granularity
is considerably less than the number of entities at a finer granularity:
each function consists of several statements,
and each module consists of several functions.
Therefore, it does not make sense to
directly compare the same effectiveness metric measured
at two different granularity levels,
since each granularity level refers to different entities%
---and inspecting different entities involves
incomparable effort.

We do not discuss efficiency
(i.e., running time)
in relation to granularity:
the running time of our fault localization techniques
does not depend on the chosen level of granularity,
which only affects how the collected information is combined
(see \autoref{sec:preliminaries}).

\subsection{RQ6. Comparison to Java}
\label{sec:experimental-results:rq6}

\autoref{table:java-vs-python-effectiveness-statement}
collects the main quantitative results for Python fault localization effectiveness
that we presented in detail in previous parts of the paper,
and displays them next to the corresponding
results for Java.
The results are selected so that they can be directly compared:
they exclude any technique (e.g., Tarantula)
or family (e.g., history-based fault localization)
that was not experimented within both our paper and Zou et al.~\cite{Zou:2021};
and the rows about \combine were computed using \cite{Zou:2021}'s replication
package so that they combine exactly the same techniques
(DStar, Ochiai, Metallaxis, \muse, PS, and ST for $\combine_A$;
and DStar, Ochiai, and ST for $\combine_S$).

\begin{table}[!bt]

\small
\centering
\setlength{\tabcolsep}{5pt}
\begin{tabular}{lr rr rr rr rr rr} 

\toprule

\multicolumn{1}{c}{\textsc{family}} & 
\multicolumn{1}{c}{\textsc{technique} $L$} & 
\multicolumn{2}{c}{$L@1$\%} & 
\multicolumn{2}{c}{$L@3$\%} & 
\multicolumn{2}{c}{$L@5$\%} & 
\multicolumn{2}{c}{$L@10$\%} & 
\multicolumn{2}{c}{$\exam(L)$} \\

\cmidrule(lr){3-4} 
\cmidrule(lr){5-6} 
\cmidrule(lr){7-8} 
\cmidrule(lr){9-10} 
\cmidrule(lr){11-12}

 &  
 & 
 \multicolumn{1}{c}{\textsl{Python}} & 
 \multicolumn{1}{c}{\textsl{Java}} & 
 \multicolumn{1}{c}{\textsl{Python}} & 
 \multicolumn{1}{c}{\textsl{Java}} &
 \multicolumn{1}{c}{\textsl{Python}} & 
 \multicolumn{1}{c}{\textsl{Java}} &
 \multicolumn{1}{c}{\textsl{Python}} & 
 \multicolumn{1}{c}{\textsl{Java}} &
 \multicolumn{1}{c}{\textsl{Python}} & 
 \multicolumn{1}{c}{\textsl{Java}} \\

\cmidrule{2-12}

\multirow{2}{*}{\combine} & 
$\combine_A$ & 
\best{\n.{stmt/comb/alfa/all/@1}} & 
\best{\n.{stmt/comb/alfa/all/@1/java}} & 
\best{\n.{stmt/comb/alfa/all/@3}} & 
\best{\n.{stmt/comb/alfa/all/@3/java}} & 
\best{\n.{stmt/comb/alfa/all/@5}} & 
\best{\n.{stmt/comb/alfa/all/@5/java}} & 
\best{\n.{stmt/comb/alfa/all/@10}} & 
\best{\n.{stmt/comb/alfa/all/@10/java}} & 
\n[\examdecjava]{stmt/comb/alfa/all/javaexam} & 
\best{\n[\examdecjava]{stmt/comb/alfa/all/javaexam/java}} \\

 & 
$\combine_S$ & 
\n.{stmt/comb/sbst/all/@1} & 
\n.{stmt/comb/sbst/all/@1/java} & 
\n.{stmt/comb/sbst/all/@3} & 
\n.{stmt/comb/sbst/all/@3/java} & 
\n.{stmt/comb/sbst/all/@5} & 
\n.{stmt/comb/sbst/all/@5/java} & 
\n.{stmt/comb/sbst/all/@10} & 
\n.{stmt/comb/sbst/all/@10/java} & 
\n[\examdecjava]{stmt/comb/sbst/all/javaexam} & 
\n[\examdecjava]{stmt/comb/sbst/all/javaexam/java} \\

\cmidrule{2-12}

\multirow{2}{*}{MBFL} & 
Metallaxis & 
\n.{stmt/mbfl/metallaxis/all/@1} & 
\n.{stmt/mbfl/metallaxis/all/@1/java} & 
\n.{stmt/mbfl/metallaxis/all/@3} & 
\n.{stmt/mbfl/metallaxis/all/@3/java} & 
\n.{stmt/mbfl/metallaxis/all/@5} & 
\n.{stmt/mbfl/metallaxis/all/@5/java} & 
\n.{stmt/mbfl/metallaxis/all/@10} & 
\n.{stmt/mbfl/metallaxis/all/@10/java} & 
\n[\examdecjava]{stmt/mbfl/metallaxis/all/javaexam} & 
\n[\examdecjava]{stmt/mbfl/metallaxis/all/javaexam/java} \\

 & 
\muse & 
\n.{stmt/mbfl/muse/all/@1} & 
\n.{stmt/mbfl/muse/all/@1/java} & 
\n.{stmt/mbfl/muse/all/@3} & 
\n.{stmt/mbfl/muse/all/@3/java} & 
\n.{stmt/mbfl/muse/all/@5} & 
\n.{stmt/mbfl/muse/all/@5/java} & 
\n.{stmt/mbfl/muse/all/@10} & 
\n.{stmt/mbfl/muse/all/@10/java} & 
\n[\examdecjava]{stmt/mbfl/muse/all/javaexam} & 
\n[\examdecjava]{stmt/mbfl/muse/all/javaexam/java} \\
 
\cmidrule{2-12}

PS & 
 & 
\n.{stmt/ps/favg/all/@1} & 
\n.{stmt/ps/favg/all/@1/java} & 
\n.{stmt/ps/favg/all/@3} & 
\n.{stmt/ps/favg/all/@3/java} & 
\n.{stmt/ps/favg/all/@5} & 
\n.{stmt/ps/favg/all/@5/java} & 
\n.{stmt/ps/favg/all/@10} & 
\n.{stmt/ps/favg/all/@10/java} & 
\best{\n[\examdecjava]{stmt/ps/favg/all/javaexam}} & 
\n[\examdecjava]{stmt/ps/favg/all/javaexam/java} \\

\cmidrule{2-12}

\multirow{2}{*}{SBFL} & 
DStar & 
\n.{stmt/sbfl/dstar/all/@1} & 
\n.{stmt/sbfl/dstar/all/@1/java} & 
\n.{stmt/sbfl/dstar/all/@3} & 
\n.{stmt/sbfl/dstar/all/@3/java} & 
\n.{stmt/sbfl/dstar/all/@5} & 
\n.{stmt/sbfl/dstar/all/@5/java} & 
\n.{stmt/sbfl/dstar/all/@10} & 
\n.{stmt/sbfl/dstar/all/@10/java} & 
\n[\examdecjava]{stmt/sbfl/dstar/all/javaexam} & 
\n[\examdecjava]{stmt/sbfl/dstar/all/javaexam/java} \\

 &
Ochiai & 
\n.{stmt/sbfl/ochiai/all/@1} & 
\n.{stmt/sbfl/ochiai/all/@1/java} & 
\n.{stmt/sbfl/ochiai/all/@3} & 
\n.{stmt/sbfl/ochiai/all/@3/java} & 
\n.{stmt/sbfl/ochiai/all/@5} & 
\n.{stmt/sbfl/ochiai/all/@5/java} & 
\n.{stmt/sbfl/ochiai/all/@10} & 
\n.{stmt/sbfl/ochiai/all/@10/java} & 
\n[\examdecjava]{stmt/sbfl/ochiai/all/javaexam} & 
\n[\examdecjava]{stmt/sbfl/ochiai/all/javaexam/java} \\
 
\cmidrule{2-12}

ST & 
 & 
\n.{stmt/st/favg/all/@1} & 
\n.{stmt/st/favg/all/@1/java} & 
\n.{stmt/st/favg/all/@3} & 
\n.{stmt/st/favg/all/@3/java} & 
\n.{stmt/st/favg/all/@5} & 
\n.{stmt/st/favg/all/@5/java} & 
\n.{stmt/st/favg/all/@10} & 
\n.{stmt/st/favg/all/@10/java} & 
\n[\examdecjava]{stmt/st/favg/all/javaexam} & 
\n[\examdecjava]{stmt/st/favg/all/javaexam/java} \\

\bottomrule

\end{tabular}
\caption{Effectiveness of fault localization techniques in Python and Java.
  Each row reports a \textsc{technique} $L$'s
  percentage of all bugs it localized
  within the top-$1$, top-$3$, top-$5$, and top-$10$ positions of its output
  ($L@1$\%, $L@3$\%, $L@5$\%, and $L@10$\%);
  and its average exam score $\exam(L)$.
  Python's data corresponds to the experiments discussed in the rest of the paper
  on the \n{number of subjects} bugs from \bip;
  Java's data is taken from Zou et al.'s empirical study~\cite{Zou:2021}
  or computed from its replication package.
  \best{Highlighted} numbers denote each language's
  best technique according to each metric.
}
\label{table:java-vs-python-effectiveness-statement}
\end{table}

\begin{table}
  \centering
  \footnotesize
  \renewcommand{\arraystretch}{1.4}
  \begin{tabularx}{1.0\linewidth}{r X cl cl}
    \toprule
    &
    \multicolumn{1}{c}{\textsc{finding}}
    & \multicolumn{2}{c}{\textsc{python}}
    & \multicolumn{2}{c}{\textsc{java}} \\
    \midrule
    \newrow &
    SBFL is the most effective standalone fault localization family.
    & \confOK & \tabref{f:eff:sbfl-best}
    & \confOK & \tabref*{1.1}
    \\
    \newrow &
    Standalone fault localization families ordered by effectiveness:\newline
    SBFL $>$ MBFL $\gg$ PS, ST
    & \confOK & \tabref{f:eff:standalone-order}
    & \confOK & \tabref*[T]{3}
    \\
    \newrow \label{d:1} &
    Regarding effectiveness, PS $\simeq$ ST.
    & \confOK & \tabref{f:eff:standalone-order}
    & \confNO & \tabref[T]{table:java-vs-python-effectiveness-statement}
    \\
    \newrow &
    All techniques in the SBFL family achieve very similar effectiveness.
    & \confOK & \tabref{f:eff:sbfl-tech-similar}
    & \confOK & \tabref*[T]{3}
    \\
    \newrow &
    The techniques in the MBFL family achieve generally similar effectiveness.
    & \confOK & \tabref{f:eff:mbfl-tech-similar}
    & \confOK & \tabref*[T]{3}
    \\
    \newrow &
    Metallaxis tends to be better than \muse.
    & \confOK & \tabref{f:eff:mbfl-tech-similar}
    & \confOK & \tabref*[T]{3}
    \\
    \newrow &
    Standalone fault localization families ordered by efficiency:\newline
    ST $\gg$ SBFL $>$ PS $>$ MBFL
    & \confOK & \tabref{f:time:standalone-order}
    & \confOK & \tabref*{4.2}
    \\
    \newrow &
    PS is more efficient than MBFL on average.
    & \confOK & \tabref{f:time:ps-mbfl}
    & \confOK & \tabref*[T]{9}
    \\
    \newrow &
    ST is more effective on crashing bugs than on other kinds of bugs.
    & \confOK & \tabref{f:eff:st-mbfl-crashing}
    & \confOK & \tabref*{1.3}
    \\
    \newrow &
    MBFL is more effective on crashing bugs than on other kinds of bugs.
    & \confOK & \tabref{f:eff:st-mbfl-crashing}
    & \confOK & \tabref*[T]{3}, \tabref*[T]{4}
    \\
    \newrow &
    PS is the least effective on crashing bugs.
    & \confOK & \tabref{f:eff:ps-crashing-worst}
    & \confOK & \tabref*[T]{4}
    \\
    \newrow &
    On predicate-related bugs, MBFL is about as effective as SBFL.
    & \confOK & \tabref[T]{tab:pred-python-java}, \tabref{f:eff:each-tech-predicate}
    & \confOK & \tabref[T]{tab:pred-python-java}, \tabref*[T]{5}
    \\
    \newrow &
    On predicate-related bugs, PS tends to be more effective than ST.
    & \confOK & \tabref{f:eff:each-tech-predicate}
    & \confOK & \tabref*[T]{5}
    \\
    \newrow &
    Combined fault localization technique $\combine_A$, which combines all baseline techniques, achieves better effectiveness than any other techniques.
    & \confOK & \tabref{f:eff:avg-comb-all}
    & \confOK & \tabref[T]{table:java-vs-python-effectiveness-statement}
    \\
    \newrow &
    Fault localization families ordered by effectiveness:\newline
    $\combine_A$ $>$ $\combine_S$ $>$ SBFL $>$ MBFL $\gg$ PS, ST
    & \confOK & \tabref{f:eff:family-comb-order}
    & \confOK & \tabref[T]{table:java-vs-python-effectiveness-statement}
    \\
    \newrow &
    Combined fault localization technique $\combine_A$, which combines all baseline techniques, achieves worse efficiency than any other technique.
    & \confOK & \tabref{f:time:avg-comb-worst}
    & \confOK & \tabref*[T]{10}
    \\
    \newrow &
    Fault localization families ordered by efficiency:\newline
    ST $\gg$ SBFL $\geq$ $\combine_S$ $>$ PS $>$ MBFL $>$ $\combine_A$
    & \confOK & \tabref{f:time:family-comb-order}
    & \confOK & \tabref*[T]{10}
    \\
    \newrow &
    ST is more effective than PS at the function-level granularity; however, it remains considerably less effective than other fault localization techniques even at this coarser granularity.
    & \confOK & \tabref{f:eff:st-ps-granularity}
    & \confOK & \tabref*[T]{11}
    \\
    \newrow \label{d:2} &
    ST is the most effective technique for crashing bugs.
    & \confNO & \tabref[T]{table:effectiveness-family-statement-bug-type}
    & \confOK & \tabref*{1.3}
    \\
    \newrow &
    PS is not the most effective technique for predicate-related faults.
    & \confOK & \tabref[T]{table:effectiveness-family-statement-bug-type}
    & \confOK & \tabref*{1.4}
    \\
    \newrow &
              Different correlation patterns exist between
              the effectiveness of different pairs of techniques.
    & \confOK & \tabref[F]{fig:pairwise-einspect}, \tabref[F]{fig:parwise-sbfl}
    & \confOK & \tabref*{2.1}
    \\
    \newrow &
              The effectiveness of most techniques from different families
              is weakly correlated.
    & \confOK & \tabref[F]{fig:pairwise-einspect}
    & \confOK & \tabref*{2.2}
    \\
    \newrow \label{d:3} &
              The SBFL family's effectiveness
              has medium correlation with the MBFL family's.
    & \confOK & \tabref[F]{fig:pairwise-einspect}
    & \confNO & \tabref*[T]{6}
    \\
    \newrow &
    The effectiveness of SBFL techniques is strongly correlated.
    & \confOK & \tabref[F]{fig:parwise-sbfl}
    & \confOK & \tabref*[T]{6}
    \\
    \newrow \label{d:4} &
    The effectiveness of MBFL techniques is weakly correlated.
    & \confNO & \tabref[F]{fig:parwise-mbfl}
    & \confOK & \tabref*[T]{6}
    \\
    \newrow &
    Techniques with strongly correlated effectiveness only exist in the same family.
    & \confOK & \tabref[F]{fig:pairwise-einspect}, \tabref[F]{fig:parwise-sbfl}, \tabref[F]{fig:parwise-mbfl}
    & \confOK & \tabref*{2.3}
    \\
    \newrow &
    Not all techniques in the same family have strongly correlated effectiveness.
    & \confOK & \tabref[F]{fig:parwise-sbfl}, \tabref[F]{fig:parwise-mbfl}
    & \confOK & \tabref*{2.3}
    \\
    \newrow\label{cmp:last} &
    The main findings about the relative effectiveness of fault localization families at statement-level granularity still hold at function-level granularity.
    & \confOK & \tabref[T]{table:effectiveness-family-technique-function}
    & \confOK & \tabref*{5.1}
    \\
    \bottomrule
  \end{tabularx}
  \caption{A comparison of findings about fault localization in Python vs.\ Java.
    Each row lists a \textsc{finding} discussed in the present paper or in Zou et al.~\cite{Zou:2021},
    whether the finding was confirmed \confOK or refuted \confNO for \textsc{python} and for \textsc{java}, and the reported evidence that confirms or refutes it (a reference to a numbered \underline{f}inding, \underline{F}igure, or \underline{T}able in our paper or in \cite{Zou:2021}).}
  \label{tab:java-py-findings}
\end{table}

Then, \autoref{tab:java-py-findings}
lists all claims about fault localization
made in our paper or in \cite{Zou:2021} that are within the scope of both papers,
and shows which were confirmed or refuted for Python and for Java.
Most of the findings (25/\ref{cmp:last}) were confirmed consistently
for both Python and Java.
Thus, the big picture about the effectiveness and efficiency of fault localization
is the same for Python programs and bugs as it is for Java programs and bugs.

There are, however, a few interesting discrepancies;
let's discuss possible explanations for them.
The most marked difference is about the effectiveness of
ST, which was mediocre on Python programs but competitive on Java programs
(row~\ref{d:1} in \autoref{tab:java-py-findings}).
We think the main reason for these differences is that
there were more Java experimental subjects that were an ideal target
for ST:
20 out of the 357 Defects4J bugs used in \cite{Zou:2021}'s experiments
consisted of short failing methods
whose programmer-written fixes
entirely replaced or removed the method body.\footnote{%
For example, project \project{Chart}'s bug \#17 in Defects4J v1.0.1.}
In these cases, the ground truth consists of all locations
within the method;
thus, ST easily ranks the fault location at the top
by simply reporting all lines of the crashing method with the same suspiciousness.
As a result, \autoref{table:java-vs-python-effectiveness-statement}
shows that 
ST was consistently more effective than PS in the Java experiments%
---whereas there was no consistent difference between ST and PS in our Python experiments.
For the same reason, the difference between Java and Python
is even more evident on crashing bugs:
ST outperformed all other techniques on such bugs in Java but not in Python
(row~\ref{d:2} in \autoref{tab:java-py-findings}).
We still confirmed that ST works better on crashing bugs than on
other kinds of bugs in Python as well,
but the nature of our experimental subjects
did not allow ST to reach an overall competitive effectiveness on crashing bugs.

Other findings about MBFL were different in Python compared to Java,
but the differences were more nuanced in this case.
In particular, Zou et al.\ found that the correlation
between the effectiveness of SBFL and MBFL techniques
is negligible, whereas we found a medium correlation
($\tau = \n[2]{statement/einspect/MBFL:SBFL/corrTau}$).
It is plausible that the discrepancy
(reflected in \autoref{tab:java-py-findings}'s row~\ref{d:3})
is simply a result
of several details of how this correlation was measured:
we use Kendall's $\tau$, they use the coefficient of determination $r^2$;
we use a generalized \einspectso measure \einspect* that applies to all bugs,
they exclude experiments where a technique completely fails to localize the bug (\einspect);
we compare the average effectiveness of SBFL vs.\ MBFL techniques,
they pairwise compare individual SBFL and MBFL techniques.
Even if the correlation patterns were actually different between Python and Java,
this would still have limited practical consequences:
MBFL and SBFL techniques still have clearly different characteristics,
and hence they remain largely complementary.
The same analysis applies to the other correlation discrepancy
(reflected in \autoref{tab:java-py-findings}'s row~\ref{d:4}):
in Python, we found a medium correlation between the effectiveness
of the Metallaxis and \muse MBFL techniques ($\tau = \n[2]{statement-intra-MBFL/einspect/Metallaxis:Muse/corrTau}$);
in Java, Zou et al.\ found negligible correlation.

\begin{table}[!bt]
  \centering
  \setlength{\tabcolsep}{8pt}
  \begin{tabular}{l rrrr | rrrr}
    \toprule
    & \multicolumn{4}{c}{\textsc{python}} & \multicolumn{4}{c}{\textsc{java}}
    \\
    \cmidrule(lr){2-5} \cmidrule(lr){6-9}
    \multicolumn{1}{c}{\textsc{family} $F$}
    & \multicolumn{1}{c}{$F@1\%$}
                                          & \multicolumn{1}{c}{$F@3\%$}
    & \multicolumn{1}{c}{$F@5\%$}
                                          & \multicolumn{1}{c}{$F@10\%$}
    & \multicolumn{1}{c}{$F@1\%$}
                                          & \multicolumn{1}{c}{$F@3\%$}
    & \multicolumn{1}{c}{$F@5\%$}
                                          & \multicolumn{1}{c}{$F@10\%$}
    \\
    \midrule
    MBFL &
           \n.{stmt/mbfl/favg/pred/@1} & 
                                         \best{\n.{stmt/mbfl/favg/pred/@3}} & 
                                                                              \best{\n.{stmt/mbfl/favg/pred/@5}} & 
                                                                                                                   \best{\n.{stmt/mbfl/favg/pred/@10}} &
                                                                                                                                                         \best{9} & \best{21} & \best{29} & 34
    \\
    SBFL &
           \best{\n.{stmt/sbfl/favg/pred/@1}} &
                                                \n.{stmt/sbfl/favg/pred/@3} &
                                                                              \n.{stmt/sbfl/favg/pred/@5} &
                                                                                                            \n.{stmt/sbfl/favg/pred/@10} &
                                                                                                                                           4 & 18 & 26 & \best{37}
    \\
    \bottomrule
  \end{tabular}
  \caption{A comparison of MBFL's and SBFL's effectiveness on Python and Java \emph{predicate-related} bugs.
    The left part of the table reports a portion of the same data as \autoref{table:effectiveness-family-statement-bug-type}: each column $@k\%$ reports the average percentage of the \n.{predicate bugs} predicate bugs in \bip Python projects used in our experiments that techniques in the MBFL or SBFL family ranked within the top-$k$.
    The right part of the table averages some of the data in~\cite[Table~5]{Zou:2021} by family: each column $@k\%$ reports the average percentage of the 115 predicate bugs in Defects4J Java projects used in Zou et al.'s experiments that techniques in the MBFL or SBFL family ranked within the top-$k$. \best{Highlighted} numbers denote each language's
  best family according to each metric.}
  \label{tab:pred-python-java}
\end{table}

Finally, a clarification about
the finding that 
``\emph{On predicate-related bugs, MBFL is about as effective as SBFL}'',
which \autoref{tab:java-py-findings} reports as confirmed for both Python and Java.
This claim hinges on the definition of ``about as effective'',
which we rigorously introduced in \autoref{sec:design:experimental-setup:rq1}.
To clarify the comparison, \autoref{tab:pred-python-java}
displays the Python and Java data about the effectiveness
of MBFL and SBFL on predicate bugs.
On Python predicate-related bugs (left part of \autoref{tab:pred-python-java}), 
MBFL achieves better $@3\%$, $@5\%$, and $@10\%$ than SBFL
but a worse $@1\%$ (by only one percentage point);
similarly, on Java predicate-related bugs
(right part of \autoref{tab:pred-python-java}),
MBFL achieves better $@1\%$, $@3\%$, and $@5\%$ than SBFL
but a worse $@10\%$ (by three percentage points).
In both cases, MBFL is not strictly better than SBFL,
but one could argue that a clear tendency exists.
Regardless of the definition of ``more effective''
(which can be arbitrary),
the conclusion we can draw remain very similar in Python as in Java.

\finding{
  Our experiments 
  confirmed for Python programs 
  most of Zou et al.~\cite{Zou:2021}'s findings about fault localization techniques on Java programs.
}
\label{f:final-finding}

\subsection{Threats to Validity}
\label{sec:threats}

\paragraph{Construct validity} refers to whether the experimental metrics adequately operationalize the quantities of interest.
Since we generally used widely adopted and well-understood metrics of
effectiveness and efficiency, threats of this kind are limited.

The metrics of effectiveness are all based on the
assumption that users of a fault localization technique
process its output list of program entities
in the order in which the technique ranked them.
This model has been criticized as unrealistic~\cite{parnin-orso-11};
nevertheless, the metrics of effectiveness remain the standard
for fault localization studies, and hence are at least adequate to
compare the capabilities of different techniques and on different programs.

Using \bip's curated collection of Python bugs
helps reduce the risks involved with our selection of subjects;
as we detail in \autoref{sec:design:subjects},
we did not blindly reuse \bip's bugs but we first
verified which bugs we could reliably reproduce on our machines.

\paragraph{Internal validity} can be threatened
by factors such as implementation bugs or inadequate statistics,
which may jeopardize the reliability of our findings.
We implemented the tool \fp to enable large-scale
experimenting with Python fault localization;
we applied the usual best practices of software development
(testing, incremental development,
refactoring to improve performance and design, and so on)
to reduce the chance that it contains fundamental bugs
that affect our overall experimental results.
To make it a robust and scalable tool,
\fp's implementation uses external libraries for
tasks, such as coverage collection and mutant generation,
for which high-quality open-source implementations are available.

The scripts that we used to process and summarize the
experimental results may also include mistakes;
we checked the scripts several times,
and validated the consistency between different data representations.

We did our best to validate the test-selection process
(described in \autoref{sec:design:experimental-setup}),
which was necessary to make feasible the experiments with the largest projects;
in particular, we ran fault localization experiments on about 30 bugs
without test selection,
and checked that the results did not change after we applied test selection.

Our statistical analysis (\autoref{sec:design:statistical-models})
follows best practices~\cite{FTF-TOSEM21-Bayes-guidelines},
including validations and comparisons of the chosen statistical models
(detailed in the replication package).
To further help
future replications and
internal validity,
we make available all our experimental artifacts and data in a detailed
replication package.

\paragraph{External validity} is about generalizability of our findings.
Using bugs from real-world open-source projects
substantially mitigates the threat
that our findings do not apply to realistic scenarios.
Precisely, we analyzed \n.{number of subjects} bugs
in \n.{number of project subjects} projects
from the curated \bip collection,
which ensures a variety of bugs and project types.

As usual, we cannot make strong claims that our findings
generalize to different application scenarios,
or to different programming languages.
Nevertheless, our study successfully confirmed
a number of findings about fault localization in Java~\cite{Zou:2021}
(see \autoref{sec:experimental-results:rq6}),
which further mitigates any major threats to external validity.

Zou et al.'s study used the Defects4J~\cite{Just:2014}
curated collection of real-world Java faults
as their experimental subjects;
we used the \bip~\cite{Widyasari:2020}
curated collection of real-world Python faults.
This invariably limits %
the generalizability of our findings to \emph{all}
Python programs, and the generalizability of
our comparison %
to all Python vs.\ Java programs:
the two curated collections of bugs
may not represent all programs and faults
in Python or Java.
While there is always a risk that
any selection of experimental subjects
is not fully representative of the whole population,
choosing standard well-known benchmarks such as Defects4J
and \bip helps mitigate this threat.
First, \bip was explicitly inspired by Defects4J,
and was built following a very similar approach but applied to
real-world open-source Python programs.
Second, \bip projects were ``selected as they represent
the diverse domains [\ldots] that Python is used for''~\cite[Sec.~1]{Widyasari:2020},
which bodes well for generalizability.
Third, \bip and Defects4J are extensible frameworks,
which have been and will be extended with new projects and bugs;
thus, using them as the basis of FL studies helps
to make future research in this area comparable to previous results.
While \bip and Defects4J are only imperfect proxies
for a fully general comparison of FL in Java and Python,
they are a sensible basis given the current state of the art.

\section{Conclusions}
\label{sec:conclusions}

This paper described an extensive empirical study
of fault localization in Python,
based on a differentiated conceptual replication
of Zou et al.'s recent Java empirical study~\cite{Zou:2021}.
Besides replicating for Python several of their results for Java,
we shed light on some nuances,
and released detailed experimental data
that can support further replications and analyses.

As a concluding discussion,
let's highlight a few points relevant for possible follow-up work.
\autoref{sec:future-work-widyasari} discusses a different
angle for a comparison with other studies,
suggested by Widyasari et al.'s recent work~\cite{Widyasari:2022}.
\autoref{sec:future-work} describes broader ideas
to improve the capabilities of fault localization in Python.

\subsection{Other Fault Localization Studies}
\label{sec:future-work-widyasari}

As we discussed in \autoref{sec:related-work},
Widyasari et al.'s recent work~\cite{Widyasari:2022}
is the only other large-scale study targeting fault localization
in real-world Python projects.
We also explained how our study's goals and methodology
is quite different from theirs;
as a result, we cannot directly compare most
of their findings to ours.
Now that we have presented our results in detail,
we are in a better position to discuss how
Widyasari et al.'s methodology
suggests future work that complements our own.

Widyasari et al.\ directly compare FL effectiveness metrics
(such as exam score)
between their experiments on Python subjects from \bip
and Pearson et al.'s experiments on Java subjects from Defects4J~\cite{Pearson:2017}.
\autoref{tab:extra:python-java}
displays the key results of their comparison,
alongside a roughly similar comparison 
between our experiments on Python subjects from \bip
and Zou et al.'s experiments on Java subjects from Defects4J~\cite{Zou:2021}.
The picture that emerges from these comparisons is somewhat inconclusive:
in our comparison,
there is a significant difference, with large effect size, between Python and Java
with respect to exam scores, but not with respect to the \einspectso metric;
conversely, in their comparison,
there is a significant difference, with large/medium effect size,
between Python and Java
with respect to
the top-$k$ ranks in the best-case debugging scenarios
(roughly analogous to the \einspectso ranking metric),
whereas the differences with respect to exam scores
are significant but with small effect sizes.
Furthermore, the \emph{sign} of the effect sizes is opposite:
in our comparison, fault localization is more effective on Python programs
(negative effect sizes);
in their comparison, it is more effective on Java programs
(positive effect sizes).
It is plausible to surmise
that these inconsistencies reflect
differences between the effectiveness metrics,
how they are measured in each study,
and---most important---differences between the experimental subjects;
the exam score metric, in particular, also depends on 
the size of the programs under analysis.
As we discussed in \autoref{sec:threats},
even though both benchmarks \bip and Defects4J
are carefully curated and of significant size,
there is the risk that they do not necessarily
represent \emph{all} Python and Java real-world projects
and their faults.
This suggests that follow-up studies targeting different projects
in Python and Java (or different selections of projects from
\bip and Defects4J) could help validate the generalizability of
any results.
Conversely, applying stricter project and bug selection
criteria could also be useful not to generalize findings,
but to strengthen their validity in more specific settings
(for example, with projects of certain characteristics).
Without provisioning stricter experimental controls,
directly comparing,
fault localization effectiveness
metrics on sundry programs
in two different programming languages,
as we did in \autoref{tab:extra:python-java} for the sake of illustration,
is unlikely to lead to clear-cut, robust findings.

Even though Widyasari et al.'s study found
some statistically significant differences of effectiveness
between SBFL techniques,
those differences tend to be
modest or insignificant.
As shown in \autoref{tab:extra:pairwise},
this is largely consistent with our findings:
even though we found some weakly statistically significant differences
between SBFL techniques
(between DStar and Tarantula for $p < 0.1$,
and between Ochiai and Tarantula for $p < 0.06$)
these have little practical consequence 
as the effect sizes of the differences are vanishing small.

Our study did not consider 
two dimensions of analysis that play an important role in
Widyasari et al.'s study:
different debugging scenarios,
and a classification of faults according to their syntactic characteristics.
Debugging scenarios determine how
we classify a fault as localized
when it affects multiple lines.
In our paper,
we only considered the ``best-case'' scenario:
as long as \emph{any} of the ground-truth locations
is localized, we consider the fault localized.
Widyasari et al.\ also consider other scenarios such as the 
worst-case scenario
(\emph{all} ground-truth locations must be localized).
While they did not find any significant differences
in the various findings under different debugging scenarios,
investigating the robustness of our empirical findings
in different scenarios remains a viable direction for future work.

\newcommand{\effmagnitude}[1]{{\footnotesize\textsc{\n.{#1}}}}

\begin{table}[!bt]
  \centering
  \begin{subtable}{\textwidth}
    \centering
    \begin{tabular}{c l r r r rl rrll}
      \toprule
      &
      & \multicolumn{3}{c}{\textsc{this paper}}
      & \multicolumn{4}{c}{\cite{Widyasari:2022}}
      \\
      
      \cmidrule(lr){3-5}
      \cmidrule(lr){6-9}
      
      \multicolumn{1}{c}{\textsc{metric}} 
      & \multicolumn{1}{c}{\textsc{technique} $L$} 
      & \multicolumn{1}{c}{$p$} 
      & \multicolumn{2}{c}{\textsc{effect}} 
      & \multicolumn{1}{c}{$p$} 
      & \multicolumn{2}{c}{\textsc{effect}} 
      & \multicolumn{1}{c}{\textsc{reference}}
      
      \\
      
      \midrule
      
      \multirow{2}{*}{$\exam(L)$} 
      & DStar 
      & \n[4]{discussion/python/java/exam/dstar/pval} 
      & \n[2]{discussion/python/java/exam/dstar/cohend} 
      & \effmagnitude{discussion/python/java/exam/dstar/esmagnitude} 
      &
0.000000 & 0.32 & \textsc{s} & \multirow{2}{*}{\cite[Tab.~5]{Widyasari:2022}}
      \\
& Ochiai 
& \n[4]{discussion/python/java/exam/ochiai/pval} 
& \n[2]{discussion/python/java/exam/ochiai/cohend} 
& \effmagnitude{discussion/python/java/exam/ochiai/esmagnitude} 
&
0.000093 & 0.15 & \textsc{s}
      \\
      \midrule
      \multirow{2}{*}{\einspect[][L]} & DStar &
\n[4]{discussion/python/java/e_inspect/dstar/pval} & \n[2]{discussion/python/java/e_inspect/dstar/cohend} & \effmagnitude{discussion/python/java/e_inspect/dstar/esmagnitude} &
0.000000 
& 0.54 
& \textsc{l} 
& \multirow{2}{*}{\cite[Tab.~3]{Widyasari:2022}}
      \\
& Ochiai 
& \n[4]{discussion/python/java/e_inspect/ochiai/pval} 
& \n[2]{discussion/python/java/e_inspect/ochiai/cohend} 
& \effmagnitude{discussion/python/java/e_inspect/ochiai/esmagnitude} &
0.000000 
& 0.41 
& \textsc{m}
      \\
      \bottomrule
    \end{tabular}
    \caption{Comparison of SBFL techniques on Python vs.\ Java programs.
      Each row compares the same SBFL \textsc{technique} $L$
      applied to \textsc{python} and to \textsc{java} programs,
      reporting the $p$-value of a Wilcoxon rank-sum test,
      and Cliff's delta \textsc{effect} size; a letter gives a qualitative assessment of the effect size: \textsc{n} for negligible, \textsc{s} for small, \textsc{m} for medium, and \textsc{l} for large.
      The data for \textsc{this paper}
      is each technique $L$'s exam score $\exam(L)$
      and \einspectso rank $\einspect[][L]$
      for each bug among
      all \n{number of subjects} Python bugs
      used in the rest of the paper's experiments,
      and for each Java bug in
      Zou et al.'s replication package data~\cite{Zou:2021};
      to reflect the behavior on all bugs in these statistics, 
      bugs that were not localized are assigned an \einspect rank
      and an exam score of $-1$
      (unlike the rest of the paper where this value is undefined).
      The statistics of \cite{Widyasari:2022}
      (in the four rightmost columns)
      are taken from its Table~5 (exam score,
      which they compute based on their top-$k$ ranks)
      and Table~3 (best-case debugging scenario top-$k$ ranks).}
    \label{tab:extra:python-java}
  \end{subtable}
  \\[3mm]
  \begin{subtable}{0.55\textwidth}
    \centering
    \setlength{\tabcolsep}{3pt}
    \begin{tabular}{ll rrl c rc}
      \toprule
      && \multicolumn{3}{c}{\textsc{this paper}}
       && \multicolumn{2}{c}{\cite[Tab.~14]{Widyasari:2022}}
      \\
      \cmidrule(lr){3-5} \cmidrule(lr){7-8}
      \multicolumn{1}{c}{\textsc{technique} $L_1$} & \multicolumn{1}{c}{\textsc{technique} $L_2$} & \multicolumn{1}{c}{$p$} & \multicolumn{2}{c}{\textsc{effect}} & \quad & \multicolumn{2}{c}{\textsc{effect}} \\
      \midrule
      DStar & Ochiai
            & \n[3]{discussion/dstar/ochiai/exam/pval}
            & \n[2]{discussion/dstar/ochiai/exam/cohend}
            & \effmagnitude{discussion/dstar/ochiai/exam/esmagnitude}
      && 0.14 & {\footnotesize \textsc{n}}
      \\
      DStar & Tarantula
      & \n[3]{discussion/dstar/tarantula/exam/pval}
      & \n[2]{discussion/dstar/tarantula/exam/cohend}
      & \effmagnitude{discussion/dstar/tarantula/exam/esmagnitude}
            && 0.19 & {\footnotesize \textsc{s}}
      \\
      Ochiai & Tarantula
      & \n[3]{discussion/ochiai/tarantula/exam/pval}
      & \n[2]{discussion/ochiai/tarantula/exam/cohend}
      & \effmagnitude{discussion/ochiai/tarantula/exam/esmagnitude}
            && 0.04 & {\footnotesize \textsc{n}}
      \\
      \bottomrule
    \end{tabular}
    \caption{Pairwise comparison of SBFL techniques according to exam score.
      Each row compares the exam scores of two \textsc{technique}s $L_1$ and $L_2$
      for significant differences,
      reporting the $p$-value of a Wilcoxon signed-rank test,
      and Cliff's delta \textsc{effect} size; a letter gives a qualitative assessment of the effect size: \textsc{n} for negligible, \textsc{s} for small, \textsc{m} for medium, and \textsc{l} for large.
      The data for \textsc{this paper}
      is each technique $L$'s exam score $\exam(L)$
      for each bug among
      all \n{number of subjects} Python bugs
      used in the rest of the paper's experiments;
      to reflect the behavior on all bugs in these statistics, 
      bugs that were not localized are assigned 
      an exam score of $-1$
      (unlike the rest of the paper where this value is undefined).
      The statistics of \cite{Widyasari:2022}
      (in the two rightmost columns)
      are taken from its Table~14.
    }
    \label{tab:extra:pairwise}
  \end{subtable}

  \caption{A summary of some data presented in Widyasari et al.'s fault localization study~\cite{Widyasari:2022} vis-{\`a}-vis analogous data presented
  in this paper.}
  \label{tab:widyasary-like-comparison}
\end{table}

\subsection{Future Work}
\label{sec:future-work}

One of the dimensions of analysis that we included in our empirical study
was the classification of projects (and their bugs) in categories,
which led to the finding that faults in data science projects
tend to be harder and take longer to localize.
This is not a surprising finding if we consider
the sheer size of some of these projects
(and of their test suites).
However, it also highlights an important category of projects
that are much more popular in Python as opposed to
more ``traditional'' languages like Java.
In fact, a lot of the exploding popularity of Python
in the last decade has been connected to its many usages for
statistics, data analysis, and machine learning.
Furthermore, there is growing
evidence that these applications have distinctive characteristics%
---especially when it comes to faults~\cite{Islam:2019,HumbatovaJBR0T20,RF-JSS23-aNNoTest}.
Thus, investigating how fault localization
can be made more effective for certain categories of projects
is an interesting direction for related work
(which we briefly discussed in \autoref{sec:related-work}).

It is remarkable that SBFL techniques,
proposed nearly two decades ago~\cite{Jones:2005},
still remain formidable in terms of both effectiveness and efficiency.
As we discussed in \autoref{sec:related-work},
MBFL was introduced expressly
to overcome some limitations of SBFL.
In our experiments (similarly to Java projects~\cite{Zou:2021})
MBFL performed generally well but not always on par with SBFL;
furthermore, MBFL is much more expensive to run than SBFL,
which may put its practical applicability into question.
Our empirical analysis of ``mutable'' bugs (\autoref{sec:experimental-results:rq3})
indicated that MBFL loses to SBFL
usually when its mutation operators
are not applicable to the faulty statements
(which happened for nearly half of the bugs we used in our experiments);
in these cases,
the mutation analysis will not bring relevant information
about the faulty parts of the program.
These observations raise the question
of whether it is possible to predict
the effectiveness of MBFL based on preliminary information about
a failure;
and whether one can develop new mutation operators
that extend the practical capabilities of MBFL to new kinds of bugs.
More generally, one could try to relate the
various kinds of source-code edits (add, remove, modify)~\cite{dissection-defects4j}
introduced to fix a fault
to the effectiveness of different fault localization algorithms.
We leave answering these questions to future research in this area.

\section*{Acknowledgements}

Work partially supported by SNF grant 200021-182060 (Hi-Fi).

\section*{Data Availability}

A replication package with data, analysis scripts, and other artifacts related to the research described in this paper are available: \url{https://doi.org/10.6084/m9.figshare.23254688}.

\section*{Declaration of Competing Interest}

The authors declare that they have no competing interests that are related to the work described in this paper.

\printendnotes[itemize]

\end{document}